\documentclass{article}

 \usepackage[main, final]{neurips_2025}

\usepackage[utf8]{inputenc} 
\usepackage[T1]{fontenc}    
\usepackage{hyperref}       
\usepackage{url}            
\usepackage{booktabs}       
\usepackage{amsfonts}       
\usepackage{nicefrac}       
\usepackage{microtype}      
\usepackage{xcolor}         

\usepackage{graphicx}
\usepackage{subcaption}
\usepackage{tikz}
\usetikzlibrary{positioning, shapes.geometric, arrows.meta, decorations.pathreplacing, calc, shapes.arrows}
\usepackage{comment}
\usepackage{amsmath}
\tikzset{
mybrace/.style={decorate,decoration={brace,aspect=#1}}
}

\title{Emulating Radiative Transfer in Astrophysical Environments}
\workshoptitle{Machine Learning and the Physical Sciences}

\author{
  \setlength{\tabcolsep}{2pt}
  
  \vspace{-1mm}
  Rune Rost$^{1}$ \quad
  Lorenzo Branca$^{1}$ \quad
  Tobias Buck$^{1}$\\[2pt]
  $^{1}$\footnotesize{Interdisciplinary Center for Scientific Computing, Heidelberg University, Germany}\\
  \footnotesize\texttt{rune.rost@stud.uni-heidelberg.de}\\
  \footnotesize\texttt{lorenzo.branca@iwr.uni-heidelberg.de}\\
  \footnotesize\texttt{tobias.buck@iwr.uni-heidelberg.de} 
}

\begin{document}

\maketitle

\begin{abstract} 
Radiative transfer is a fundamental process in astrophysics, essential for both interpreting observations and modeling thermal and dynamical feedback in simulations via ionizing radiation and photon pressure. However, numerically solving the underlying radiative transfer equation is computationally intensive due to the complex interaction of light with matter and the disparity between the speed of light and the typical gas velocities in astrophysical environments, making it particularly expensive to include the effects of on-the-fly radiation in hydrodynamic simulations. This motivates the development of surrogate models that can significantly accelerate radiative transfer calculations while preserving high accuracy. We present a surrogate model based on a Fourier Neural Operator architecture combined with U-Nets. Our model approximates three-dimensional, monochromatic radiative transfer in time-dependent regimes, in absorption-emission approximation, achieving speedups of more than 2 orders of magnitude while maintaining an average relative error below 3\%, demonstrating our approach's potential to be integrated into state-of-the-art hydrodynamic simulations. 
\end{abstract}

\section{Introduction}

Radiative transfer is a cornerstone of computational astrophysics, providing the essential link between physical models and observational diagnostics \citep[e.g.][]{Rosdahl2013,Buck2017}. Simulating the propagation of radiation through astrophysical media, such as stellar atmospheres, interstellar clouds, or galaxy clusters, requires solving the radiative transfer equation (RTE), shown in Equation~\eqref{eq:radiative_transfer}. 
\begin{equation}
\label{eq:radiative_transfer}
\frac{1}{c} \frac{\partial I_\nu}{\partial t} + \omega \cdot \nabla I_\nu + (k_{\nu, s} + k_{\nu, a}) \rho I_\nu = j_\nu \rho + \frac{k_{\nu, s} \rho}{4\pi}  \int_{\mathcal{S}} I_\nu \, dw^{\prime}
\end{equation}
The RTE is a  Partial Differential Equation (PDE) that is often coupled to hydrodynamics, thermochemistry, and dust physics. The quantity of interest is the spectral radiative intensity $I_\nu$, whose evolution is determined by the scattering and absorption coefficients $k_{\nu, s}$ and $k_{\nu, a}$, the density $\rho$, the emission $j$, and the integral term representing scattering. 
Due to its high dimensionality (dependence on time $t$, spatial position $x$, direction $\omega$, and frequency $\nu$),
the RTE is highly complex and 
computationally expensive to solve numerically.
This has motivated the development of a wide range of numerical techniques, including Monte Carlo methods~\cite{Noebauer2019}, ray tracing~\cite{2006ApJ...645..920S}, moment-based approximations~\cite{Rosdahl2013}, and RT approximations using the gravity tree \citep{Obreja2019,Kannan2014}.
Unfortunately, numerical methods often suffer from high computational costs, dimensionality issues, or instability.
%
In particular, computational cost becomes a major concern when incorporating on-the-fly radiative transport into hydrodynamic simulations to account for the effects of radiation on gas thermodynamics (via photoheating) and dynamics (via radiation pressure). In moment-based approaches, the large value of the speed of light necessitates much smaller time steps ($c>>v_{gas}$); in ray-tracing approaches, memory requirements are the primary bottleneck, scaling as $\mathcal{O}(N_{sources}N_{cells})$.

While  traditional deep learning approaches and Physics-Informed Neural Networks have been explored for developing surrogate models for the simulation of radiative transfer~\cite{PINNRT,LU2024105282,COemu}, 
they  often struggle with generalization across discretizations and parameter settings, 
and are further known to face challenges with stability and accuracy in such high-dimensional PDE problems~\cite{2025arXiv250521573W,2022arXiv220313181D}. Furthermore, to the best of our knowledge, no prior work has presented an emulator for time-dependent radiative transfer.
%
We therefore propose a data-driven approach based on
Neural Operators, 
a recently introduced class of deep learning architectures inspired by the Universal Approximation Theorem for Operators~\cite{chen2:1995}, to develop a surrogate model for simulating radiative transfer.
Operating on infinite-dimensional function spaces, Neural Operators  extend traditional neural networks, enabling the efficient approximation of PDE solution operators with improved generalization across grid discretizations and parameter variations.
We present a Neural Operator-based surrogate model prototype 
designed to replace
conventional on-the-fly radiative transfer solvers
within hydrodynamic simulations, achieving significant speedups while maintaining high accuracy. 

\section{Surrogate Model Architecture}

\begin{figure}[h]
    \centering
    \begin{minipage}{\textwidth}
    \centering
    \resizebox{\textwidth}{!}{
    \begin{tikzpicture}[
        node distance=0.4cm and 0.4cm,,
        layer/.style={rectangle, draw=black, rounded corners, minimum width=1.0cm, minimum height=1.0cm, text centered},
        img/.style={inner sep=0pt},
        timearrow/.style={single arrow, draw, fill=blue!20, minimum height=16cm, minimum width=0.5cm, single arrow head extend=0.25cm,},
        ]

        \node[img, label=left:\textbf{j(x)}, thick] (img1) {\includegraphics[width=1.5cm]{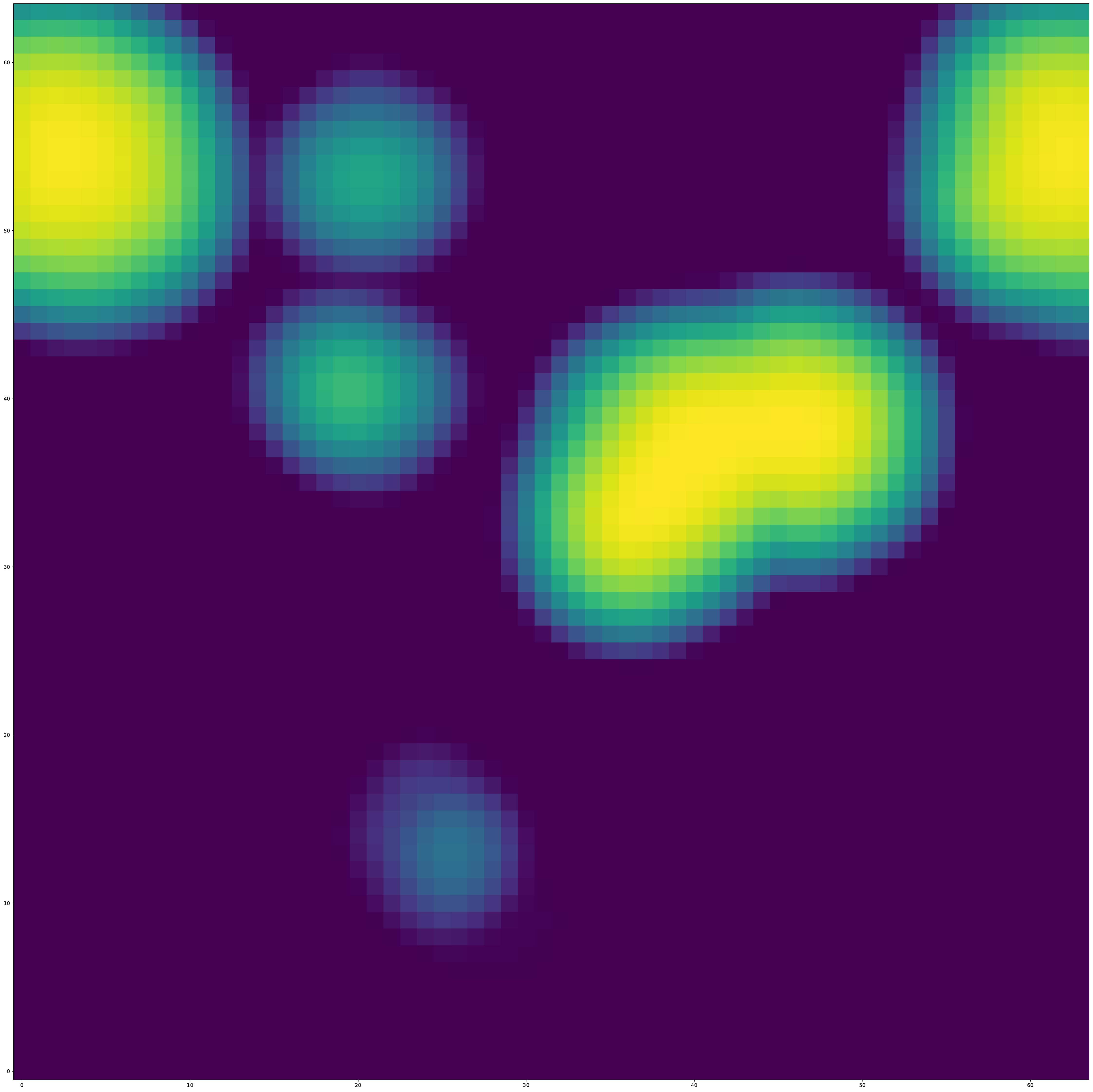}};
        \node[img, above=of img1, label=left:\textbf{a(x)}, thick] (img2) {\includegraphics[width=1.5cm]{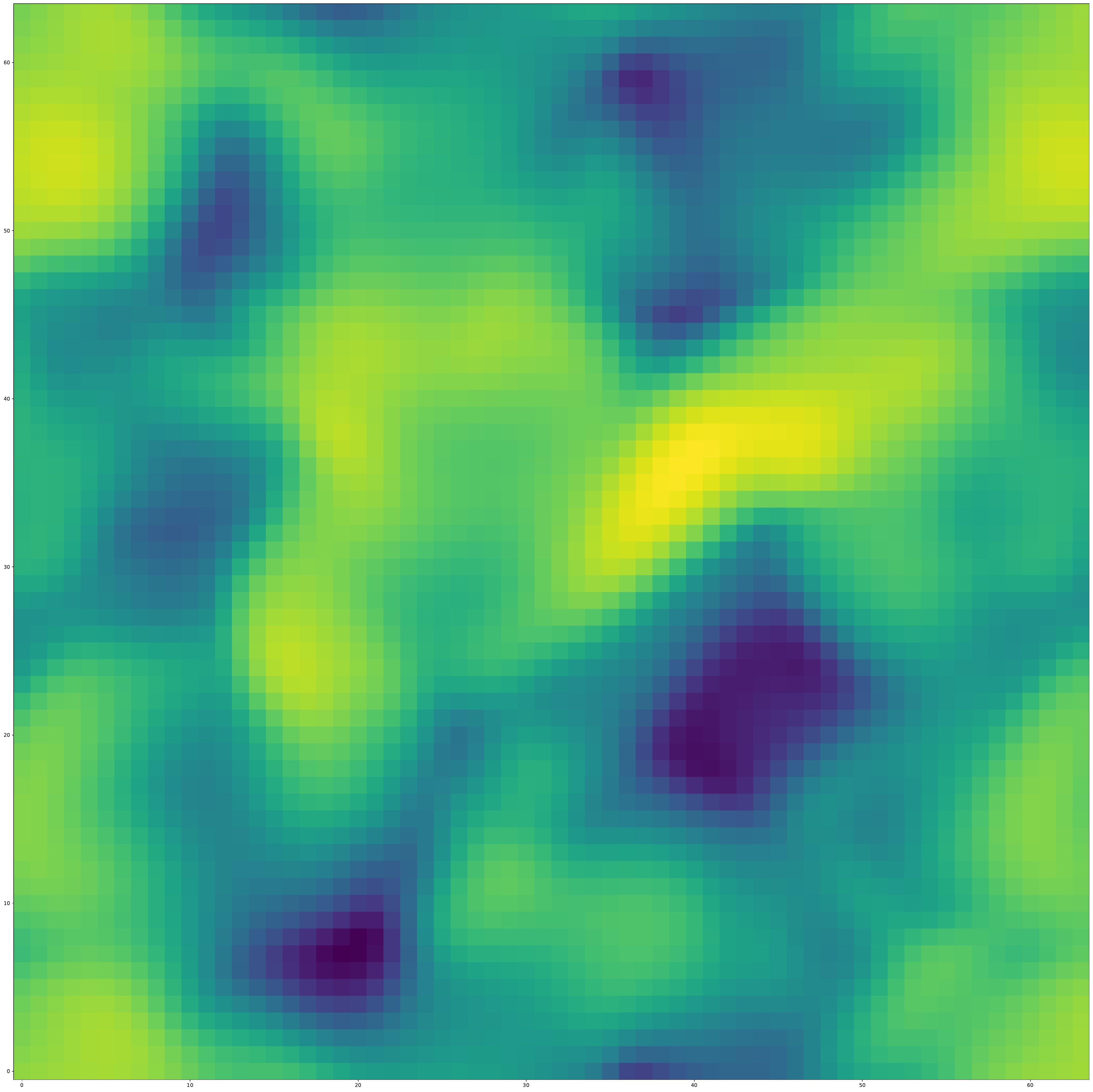}};
        \node[img, below=of img1, label=left:{$\mathbf{I}_{\nu,t}(\mathbf{x})$}, thick] (img3) {\includegraphics[width=1.5cm]{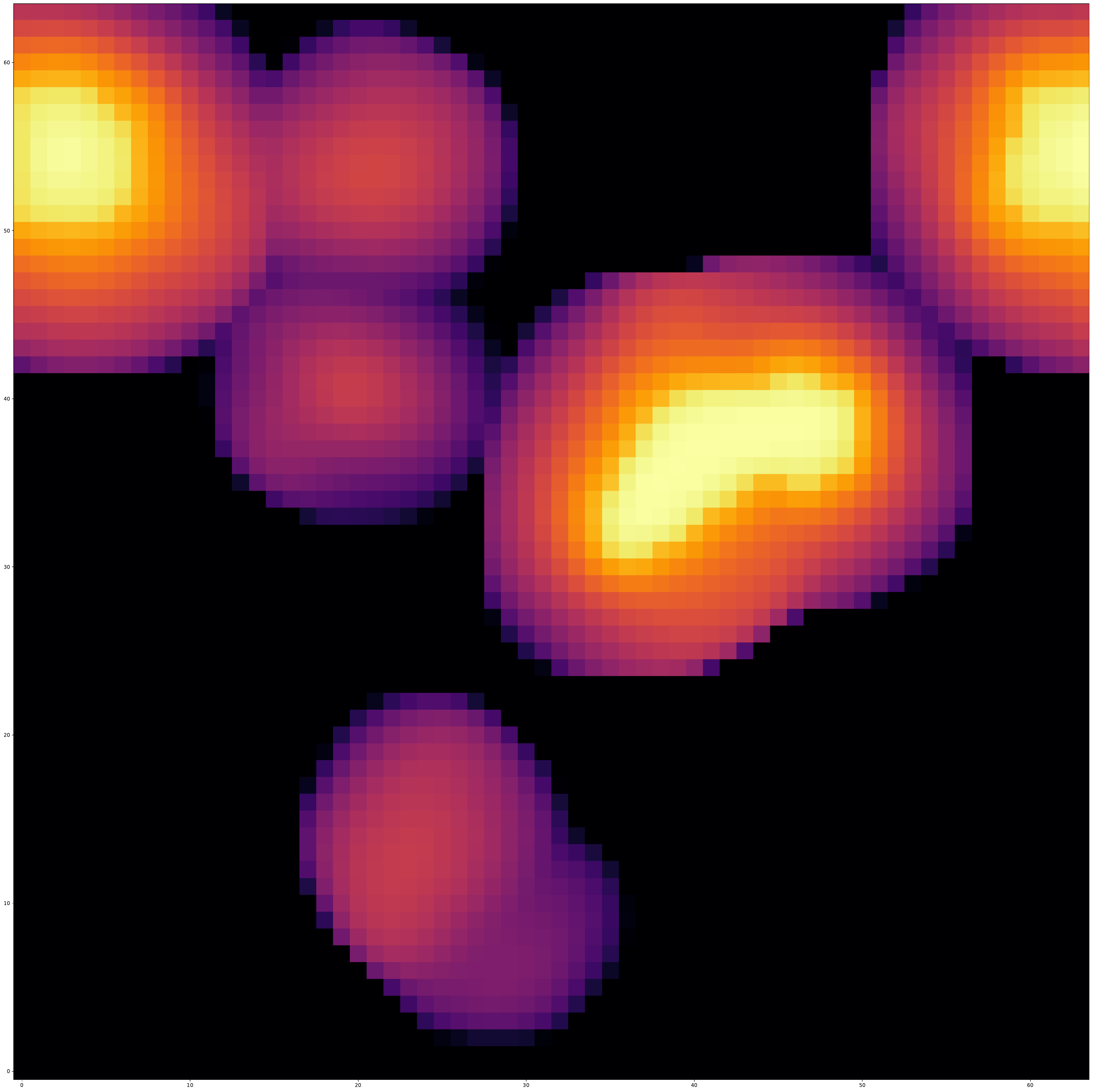}};

        \def\offset{0.075cm}
        \def\imgwidth{1.5cm} 
        \def\imgheight{1.5cm} 

        \draw[mybrace=0.75, decoration={brace, mirror, amplitude=6pt}, thick]
        (0.8, -2.5) -- (0.8, 2.5) 
        node[layer, pos=0.75, right=20pt, yshift=0pt] (lifting){$P$}
        ;

        \draw[->, thick] ($(lifting.west) - (0.4, 0.0)$) -- (lifting.west);
        \def \n{8}       
        \def\m{3}        

        \def\offset{0.075cm}
        \def\imgwidth{1.5cm} 
        \def\imgheight{1.5cm} 
    
        \foreach \i in {1,...,\n} {
            \draw[fill=gray!20, draw=black]
            ($(lifting.east) + (0.4, 0.0) + (\i*\offset, -\i*\offset)$)
             rectangle ++(\imgwidth, \imgwidth);
        };

        \draw[decorate, decoration={brace, mirror, amplitude=6pt}, thick]
        (5.10, 2.10) -- (4.55, 2.65)node[midway, right=5pt, yshift=7pt] (lift dim){lifting dimension};

        \draw[->, thick] (lifting.east) -- ++(0.4cm, 0);

        \node[layer, align=center, thick] (fourier1) at ($(lifting.east) + (3.8, 0.0)$) {U-Fourier \\ Layer 1};
        \draw[->, thick] ($(fourier1.west) - (0.4, 0.0)$) -- (fourier1.west);
        \draw[->, thick] (fourier1.east) -- ++(0.4cm, 0);
        \node at ($(fourier1.east) + (0.8, 0)$) {\Large $\cdots$};
        \node[layer, align=center, thick] (fourierN) at ($(fourier1.east) + (2.4, 0.0)$) {U-Fourier \\ Layer N};
        \draw[->, thick] (fourierN.east) -- ++(0.4cm, 0);
        \draw[->, thick] ($(fourierN.west) - (0.4, 0.0)$) -- (fourierN.west);

        \foreach \i in {1,...,\n} {
            \draw[fill=gray!20, draw=black]
            ($(fourierN.east) + (0.4, 0.0) + (\i*\offset, -\i*\offset)$)
            rectangle ++(\imgwidth, \imgwidth);

        }

        \node[layer, align=center, thick] (projection) at ($(fourierN.east) + (3.6, 0.0)$) {Q};

        \draw[->, thick] ($(projection.west) - (0.4, 0.0)$) -- (projection.west);

        \node[img, align=center, label=right:{$\mathbf{I}_{\nu,t+1}(\mathbf{x})$}, thick](output) at ($(fourierN.east) + (5.4, 0.0)$){\includegraphics[width=1.5cm]{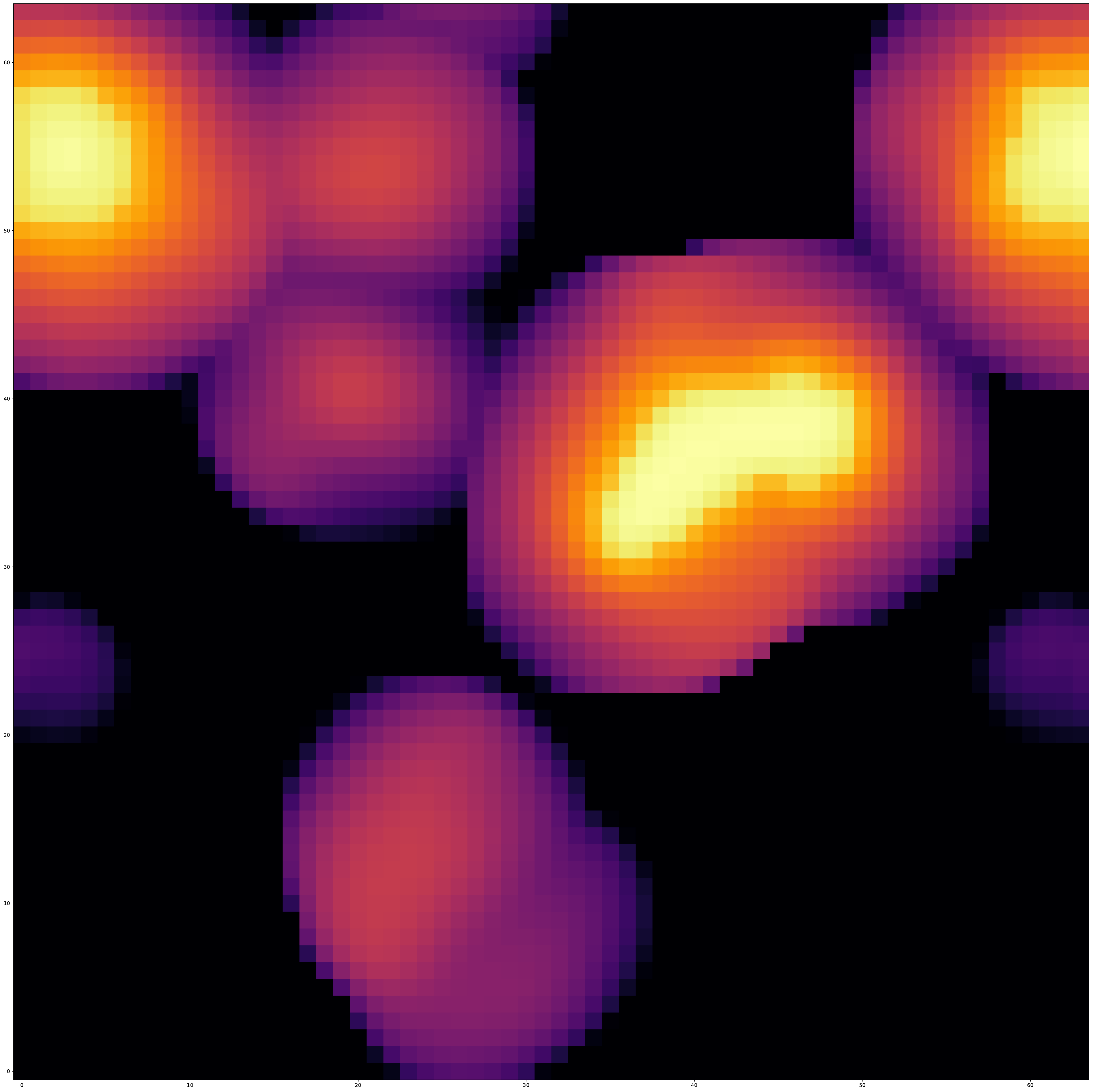}};

        \draw[->, thick] (projection.east) -- (output.west);
    \node[draw, thick, fill=gray!5, rounded corners, inner sep=10pt] (box) at (10.5, -2.25)
            {
              \begin{tikzpicture}
                \foreach \i in {1,...,\n} {

                    \draw[fill=gray!20, draw=black, rounded corners=0pt]
                    ($(0.0, -0.1) + (\i*\offset, -\i*\offset)$)rectangle ++(\imgwidth, \imgwidth);
                
                \node[layer, align=center, thick] (FFT) at ($(3.5, 1.5)$) {$\mathcal{F}$};

                \node[layer, align=center, thick] (uNet) at ($(5.5, 0.0)$) {U-Net};

                \node[layer, align=center, thick] (W) at ($(5.5, -1.5)$) {W, b};

                \node[layer, align=center, thick] (R) at ($(5.5, 1.5)$) {$\mathcal{R}$};

                \node[layer, align=center, thick] (FFT1) at ($(7.5, 1.5)$) {$\mathcal{F}^{-1}$};

                \node[layer, align=center, thick] (add) at ($(9.5, 0.0)$) {$+$};

                \node[layer, align=center, thick] (sigma) at ($(11.5, 0.0)$) {$\mathcal{\sigma}$};

                \draw[->, thick] (FFT.east) -- (R.west);
                \draw[->, thick] (R.east) -- (FFT1.west);
                \draw[->, thick] (FFT1.east) -- (add.north);
                \draw[->, thick] (uNet.east) -- (add.west);
                \draw[->, thick] (W.east) -- (add.south);
                \draw[->, thick] (add.east) -- (sigma.west);
                \draw[->, thick] ($(2.2, 0.3)$) -- (FFT.west);
                \draw[->, thick] ($(2.2, 0.0)$) -- (uNet.west);
                \draw[->, thick] ($(2.2, -0.3)$) -- (W.west);                
        };
              \end{tikzpicture}
            };
          \node[left=4pt of box, font=\bfseries] {U-Fourier Layer};
        
    \draw[->, thick] (fourier1.south) -- (box.north);
    \draw[->, thick] (fourierN.south) -- (box.north);

    \end{tikzpicture}
    }

    \caption{
    Schematic overview of U-FNO architecture. Input fields are lifted to a higher-dimensional latent space via a lifting layer $\mathcal{P}$, processed by U-Fourier Layers that combine a convolution integral (via Fast Fourier Transform (FFT) $\mathcal{F}$, multiplication with weights $\mathcal{R}$, and inverse FFT $\mathcal{F}^{-1}$), a U-Net, and an affine transformation $(W, b)$, and finally projected to the output field via $\mathcal{Q}$. 
    }
    \label{fig:UFNO}
    \end{minipage}
\end{figure}
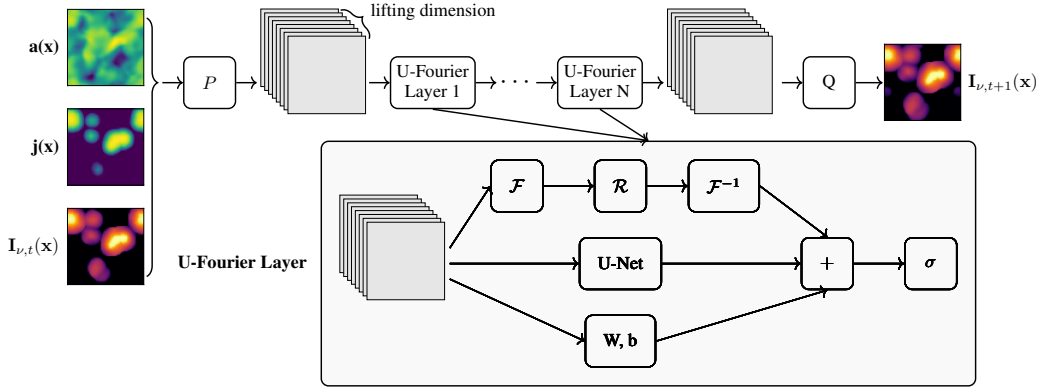

Neural Operators are deep learning architectures that, unlike traditional neural networks, learn mappings between infinite-dimensional function spaces, capturing intrinsic properties rather than discretization-specific details, and thus enable more robust and broadly applicable models.
%
%
This work employs a specific class of Neural Operators known as the Fourier Neural Operator (FNO)~\citep{2020arXiv201008895L} and combines it with a U-Net architecture, following the approach chosen in~\cite{2022AdWR..16304180W}. The high-level architecture is shown in Figure~\ref{fig:UFNO}. 
For simplicity, all visualizations in this figure use 2D slices, even though the model in this work is trained and evaluated using full 3D data. 

In brief, a U-FNO comprises a lifting layer $\mathcal{P}$, which maps the input function to a higher-dimensional latent space, thereby enhancing the model's expressivity. This is followed up by a series of U-Fourier Layers, which combine a convolutional integral operator with an affine transformation and a U-Net. 
To compute the convolutional integral operator efficiently, inputs are transformed into Fourier space via FFT, 
multiplied with learnable weights $\mathcal{R}$, and afterward transformed back 
via inverse FFT, allowing the network to capture global dependencies.
A U-Net is an encoder–decoder convolutional network with symmetric downsampling and upsampling paths linked by skip connections, enabling precise localization of 
fine-scale details~\cite{2015arXiv150504597R},
particularly if kernels are small.
This is especially relevant given that the data used in this study is highly complex, including sharp gradients and discontinuities, and U-Nets have been shown to excel in such settings \citep[e.g.][]{Buck2021}.
Finally, a projection layer $\mathcal{Q}$ maps the latent representation to the desired output dimension.

\section{Results}
In this study, we  present a U-FNO-based surrogate model designed as a drop-in replacement 
for on-the-fly radiative transfer solvers 
in hydrodynamic simulations. The model predicts the temporal evolution of radiative intensity 
through recurrent application over time steps, providing an efficient alternative to computationally expensive numerical solvers.
To align with state-of-the-art hydrodynamic simulations, we consider radiative transfer in a scattering-free regime, as scattering is typically neglected in on-the-fly computations. Moreover, the angular dependence of the radiative intensity is currently omitted but planned for future inclusion.
Accordingly, the model is trained to predict $I_{\nu,t+1}(\mathbf{x})$ from $I_{\nu,t}(\mathbf{x})$, $a(\mathbf{x})=k_{\nu,a}\rho(\mathbf{x})$, and $j(\mathbf{x})=j_{\nu}\rho(\mathbf{x})$,
as illustrated in Figure~\ref{fig:UFNO}. 
While the presented model focuses on a single frequency $\nu$, modern hydrodynamic simulations typically partition the radiation spectrum into a small number of frequency (energy) bins ($\sim$10), centered on the photoionization and photodissociation energies of key
gas-phase species \citep{arepo-RT, Rosdahl2013, davide2020}. 
Since these bins are independent in the absence of scattering,
our approach naturally scales to such multi-frequency setups by training one model per frequency bin, ensuring consistent generalization regardless of the number of bins used in the simulation. 
Alternatively, a unified model can be trained to jointly predict the evolution across all frequencies (see Appendix~\ref{outlook} for details).
The model is implemented in  JAX~\cite{jax2018github}, leveraging its efficient support for just-in-time compilation and GPU acceleration.

\begin{figure}[h]
    \centering

    \makebox[\textwidth][c]{
    \begin{subfigure}[b]{0.22\textwidth}
        \captionsetup{labelformat=empty}
        \includegraphics[width=\textwidth]{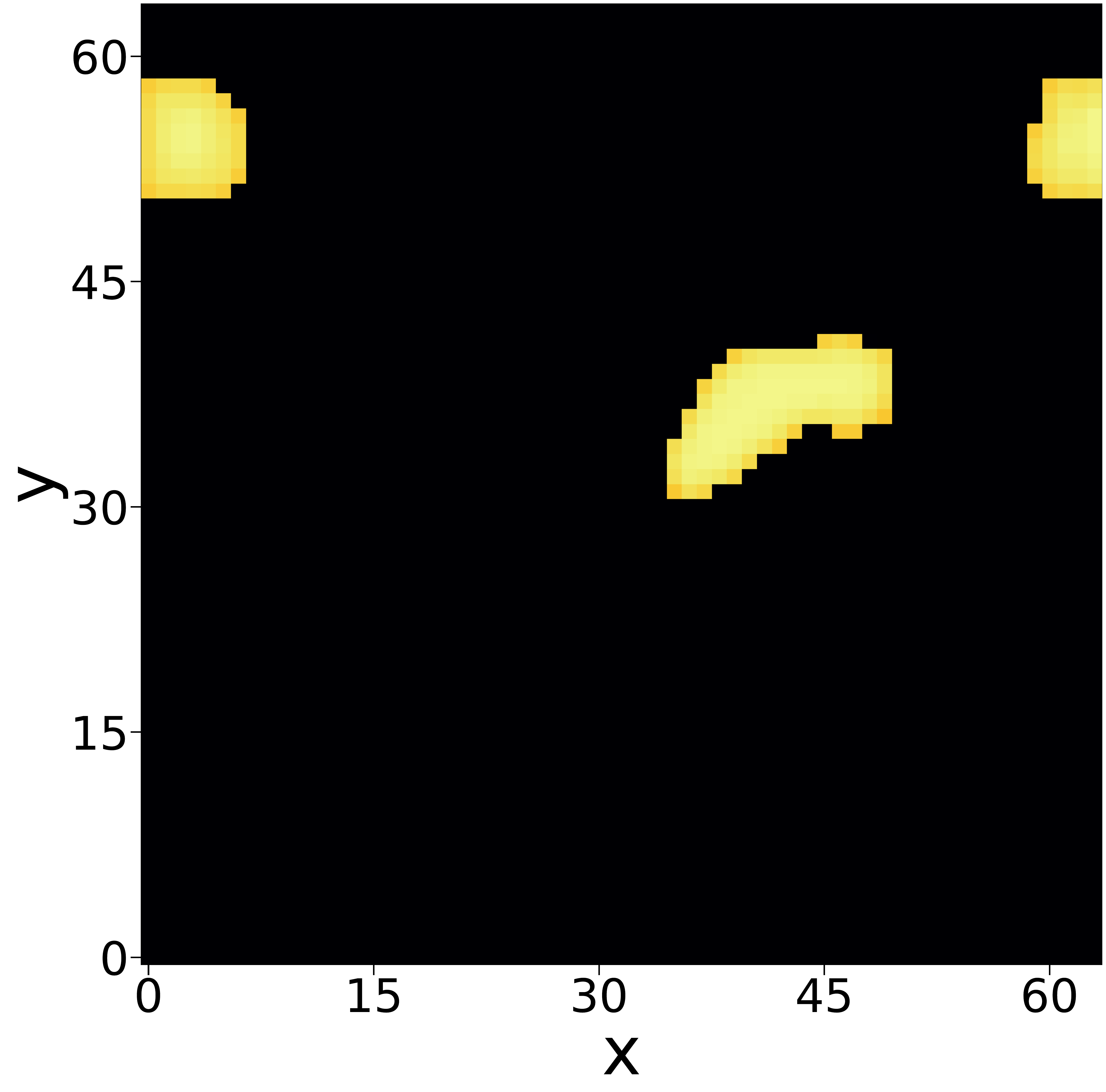}
    \end{subfigure}
    \begin{subfigure}[b]{0.22\textwidth}
        \captionsetup{labelformat=empty}
        \includegraphics[width=\textwidth]{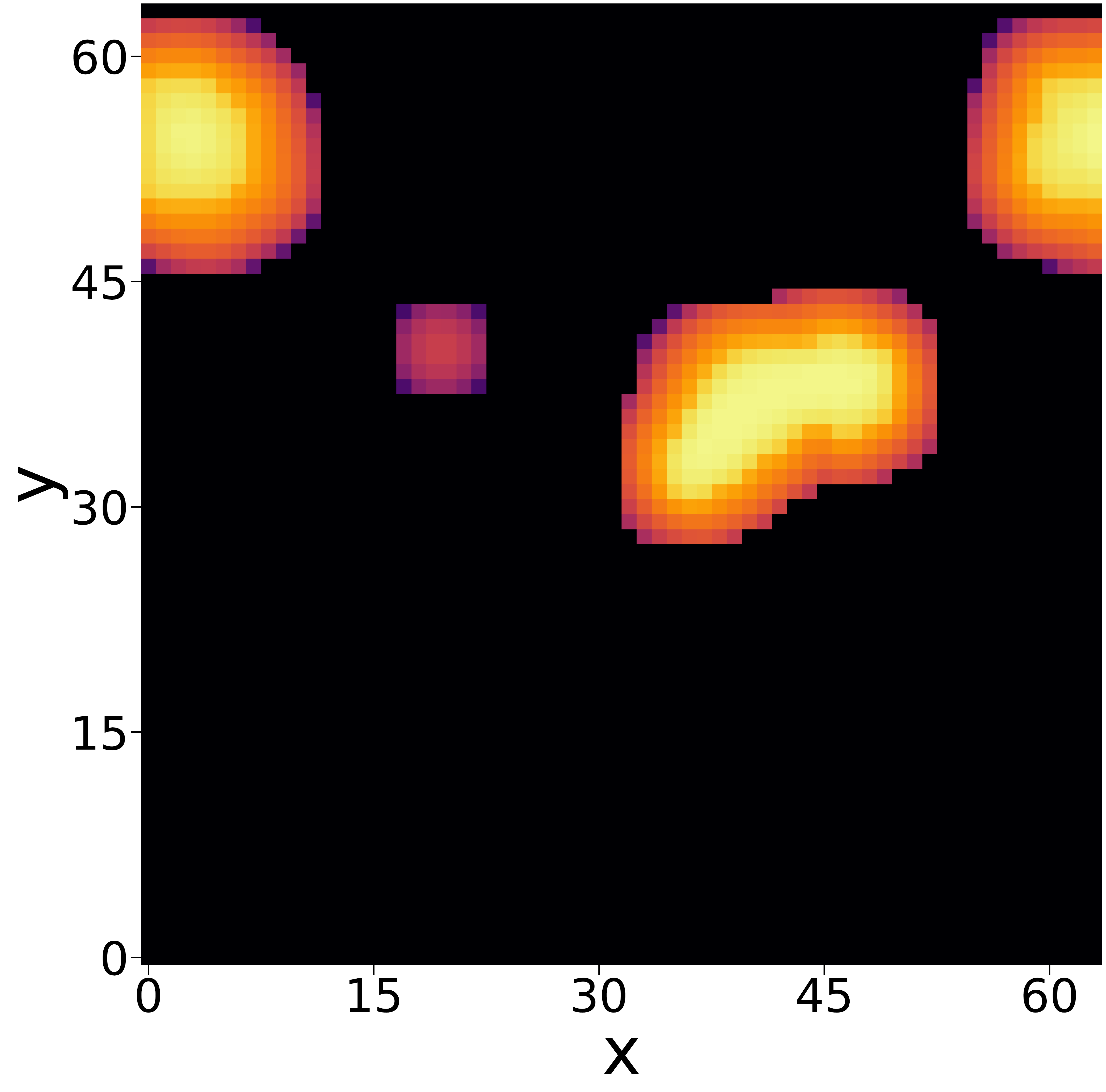}
    \end{subfigure}
    \begin{subfigure}[b]{0.22\textwidth}
        \captionsetup{labelformat=empty}
        \includegraphics[width=\textwidth]{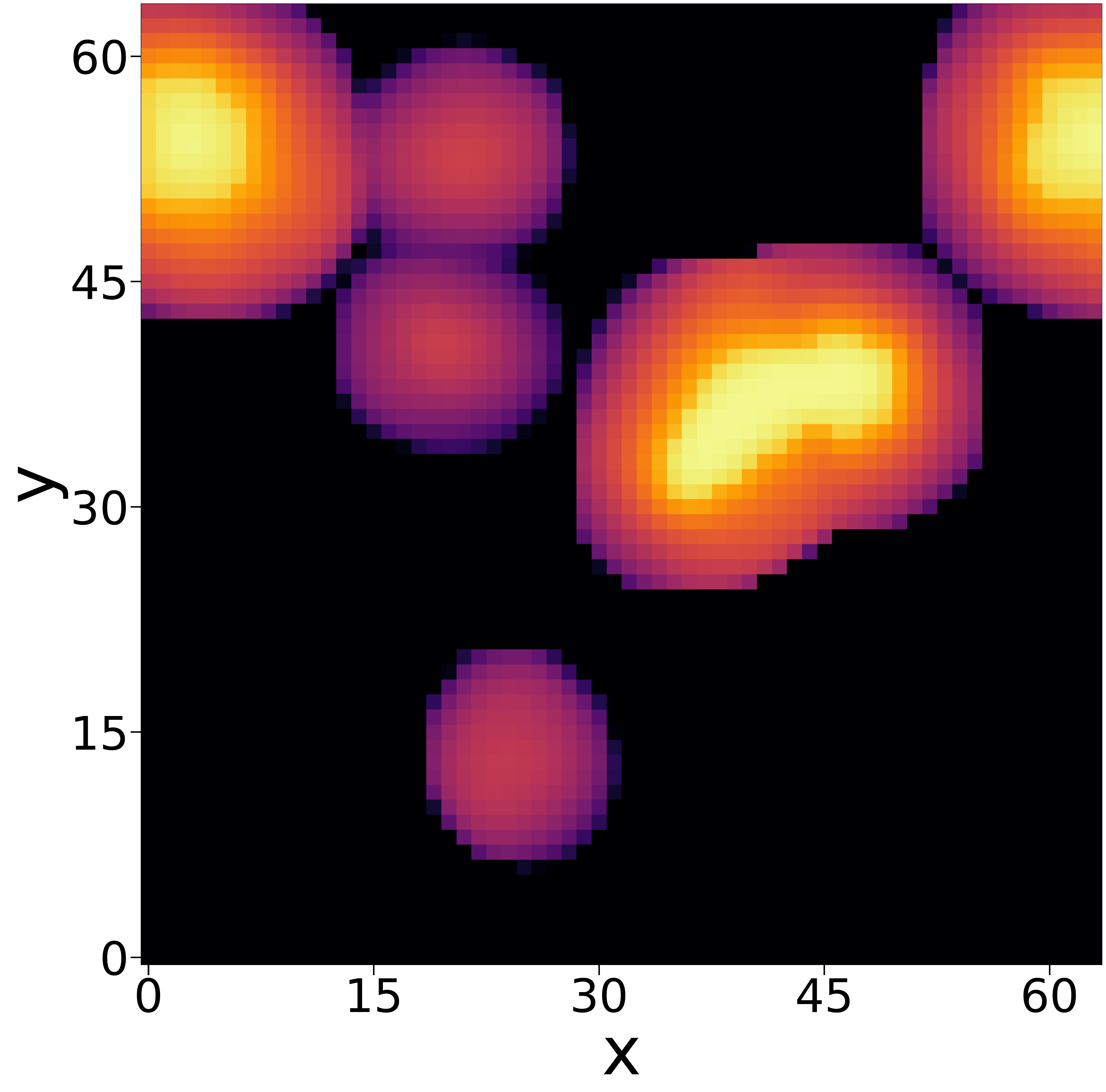}
    \end{subfigure}
    \begin{subfigure}[b]{0.22\textwidth}
        \captionsetup{labelformat=empty}
        \includegraphics[width=\textwidth]{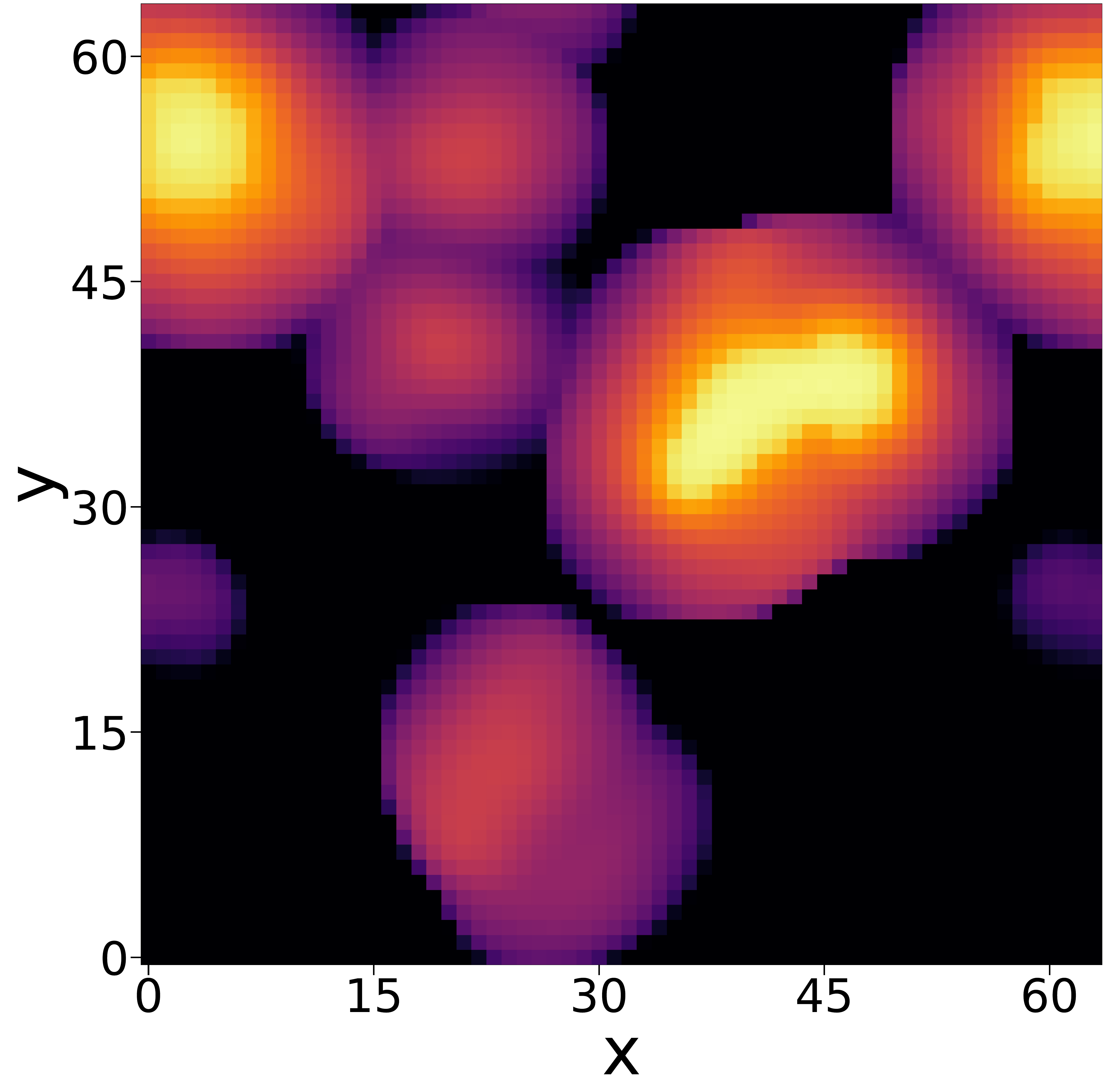}
    \end{subfigure}

    \begin{subfigure}[b]{0.0585\textwidth}
        \captionsetup{labelformat=empty}
        \includegraphics[width=\textwidth]{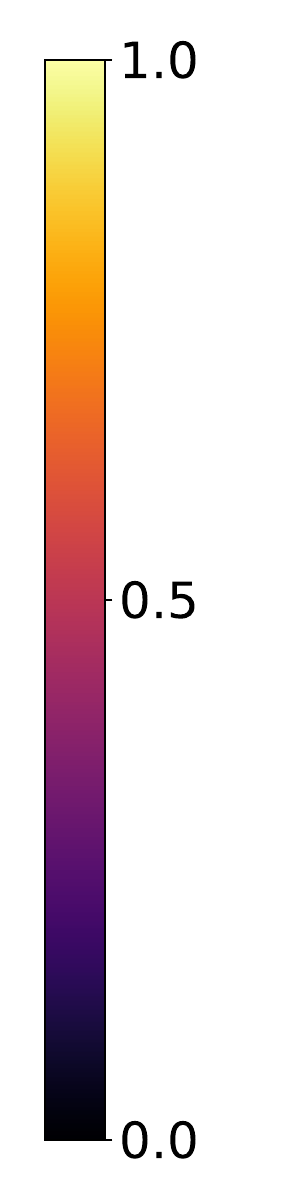}
    \end{subfigure}
    }


    \makebox[\textwidth][c]{
    \begin{subfigure}[b]{0.22\textwidth}
        \captionsetup{labelformat=empty}
        \includegraphics[width=\textwidth]{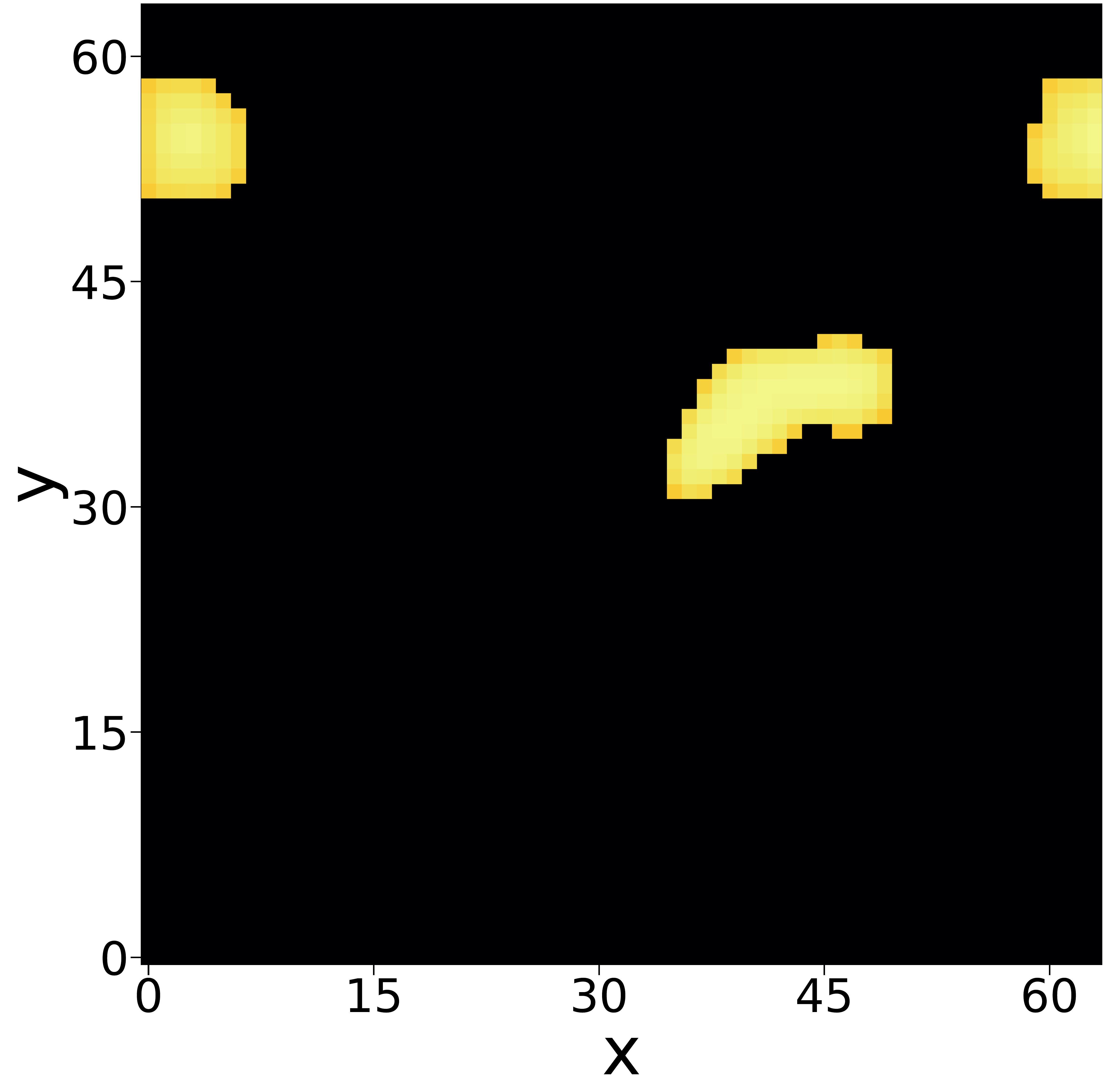}
    \end{subfigure}
    \begin{subfigure}[b]{0.22\textwidth}
        \captionsetup{labelformat=empty}
        \includegraphics[width=\textwidth]{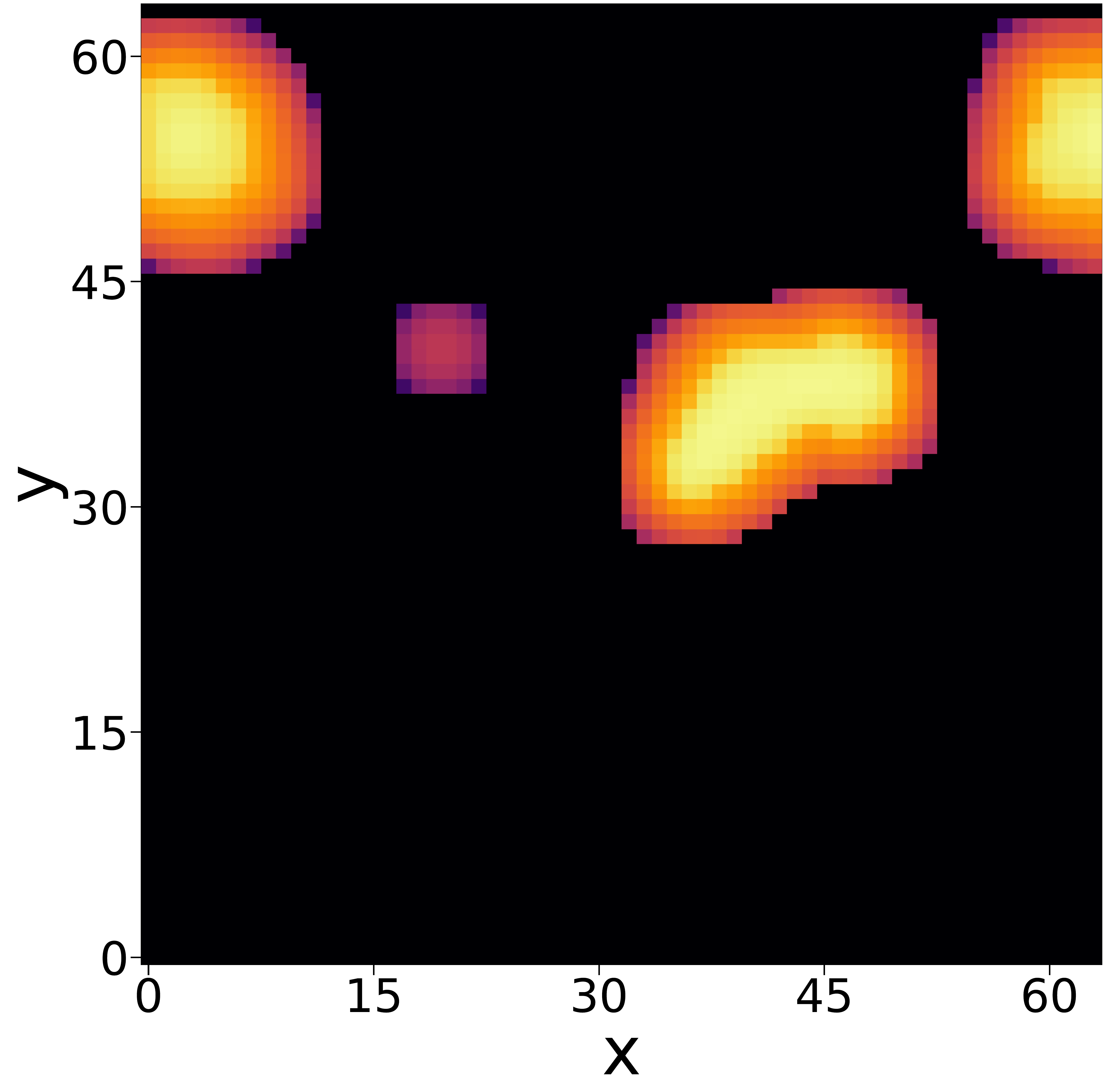}
    \end{subfigure}
    \begin{subfigure}[b]{0.22\textwidth}
        \captionsetup{labelformat=empty}
        \includegraphics[width=\textwidth]{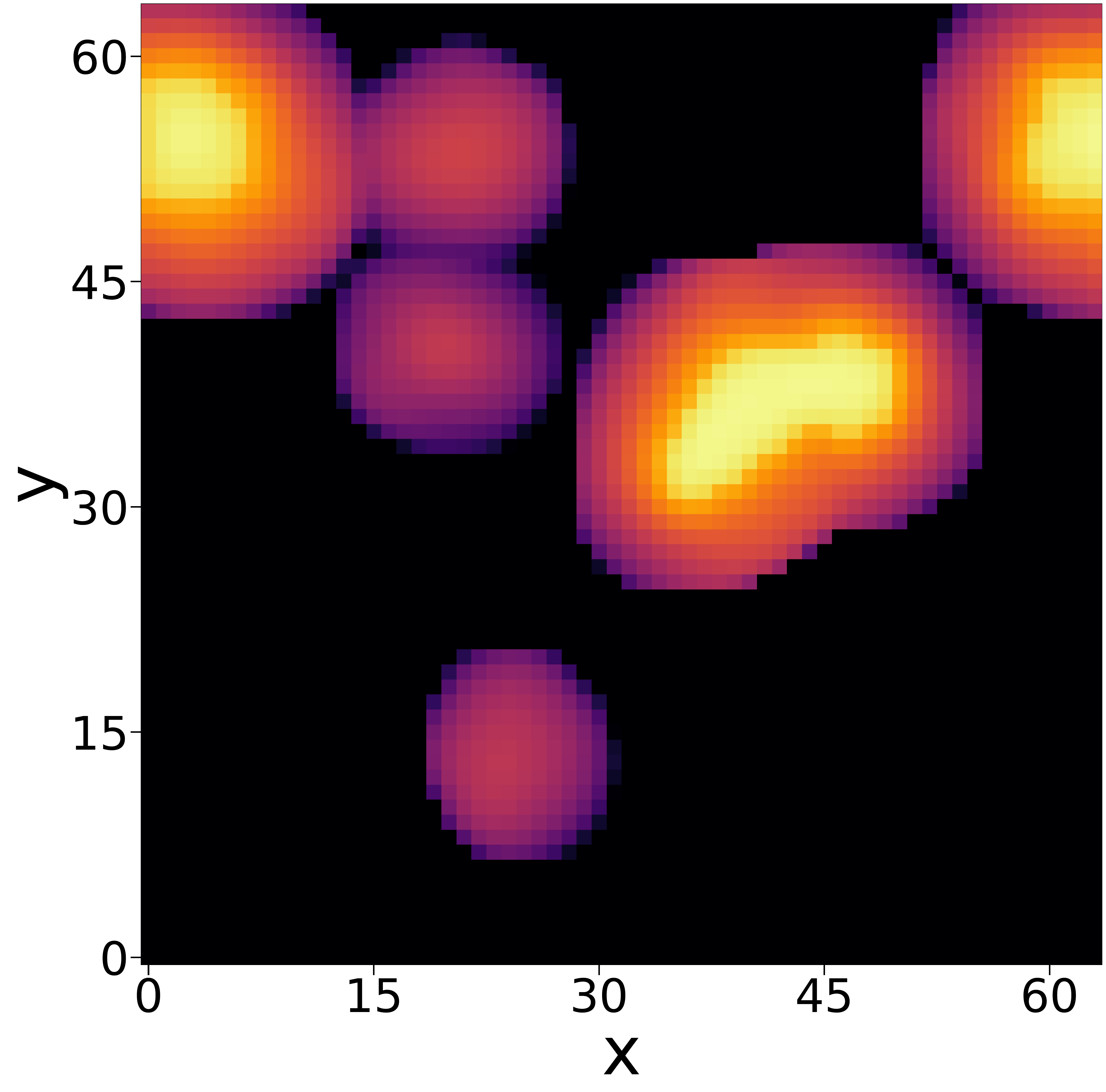}
    \end{subfigure}
    \begin{subfigure}[b]{0.22\textwidth}
        \captionsetup{labelformat=empty}
        \includegraphics[width=\textwidth]{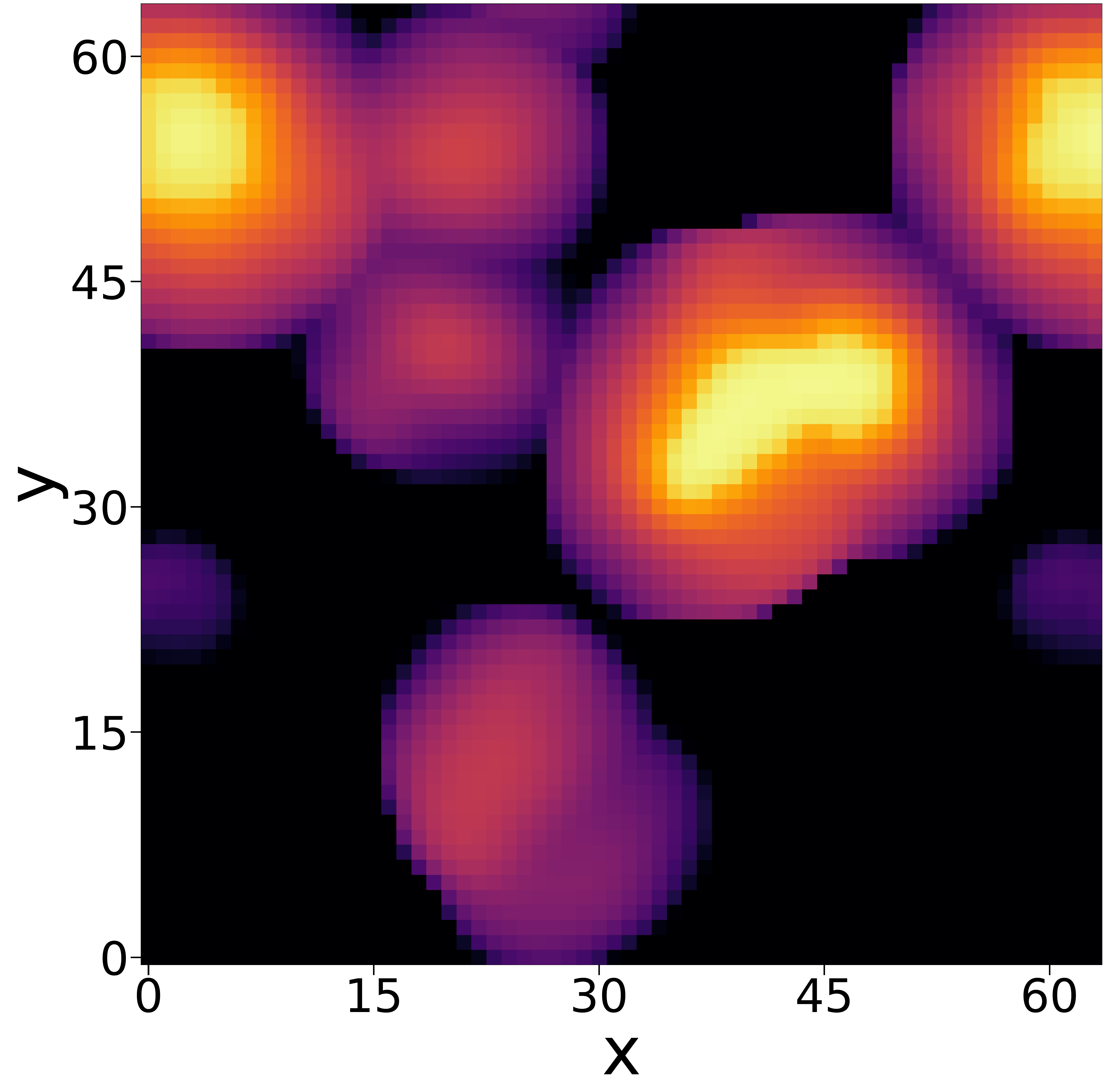}
    \end{subfigure}
    \begin{subfigure}[b]{0.0585\textwidth}
        \captionsetup{labelformat=empty}
        \includegraphics[width=\textwidth]{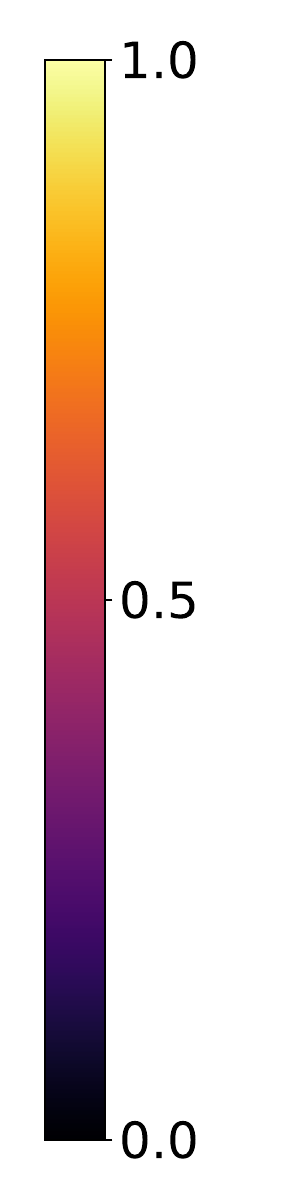}
    \end{subfigure}
    }


    \makebox[\textwidth][c]{
    \begin{subfigure}[b]{0.22\textwidth}
        \captionsetup{labelformat=empty}
        \includegraphics[width=\textwidth]{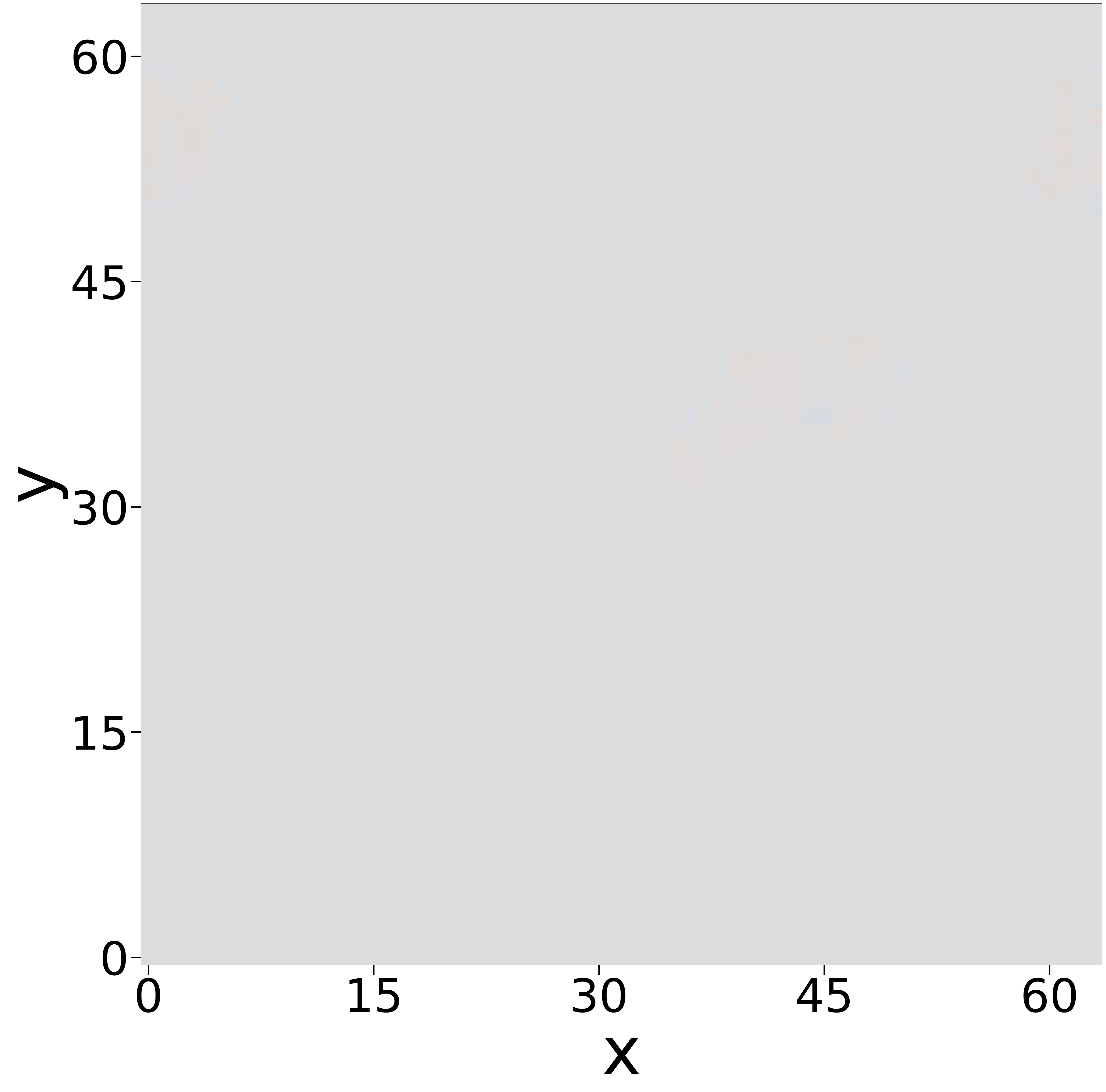}
    \end{subfigure}
    \begin{subfigure}[b]{0.22\textwidth}
        \captionsetup{labelformat=empty}
        \includegraphics[width=\textwidth]{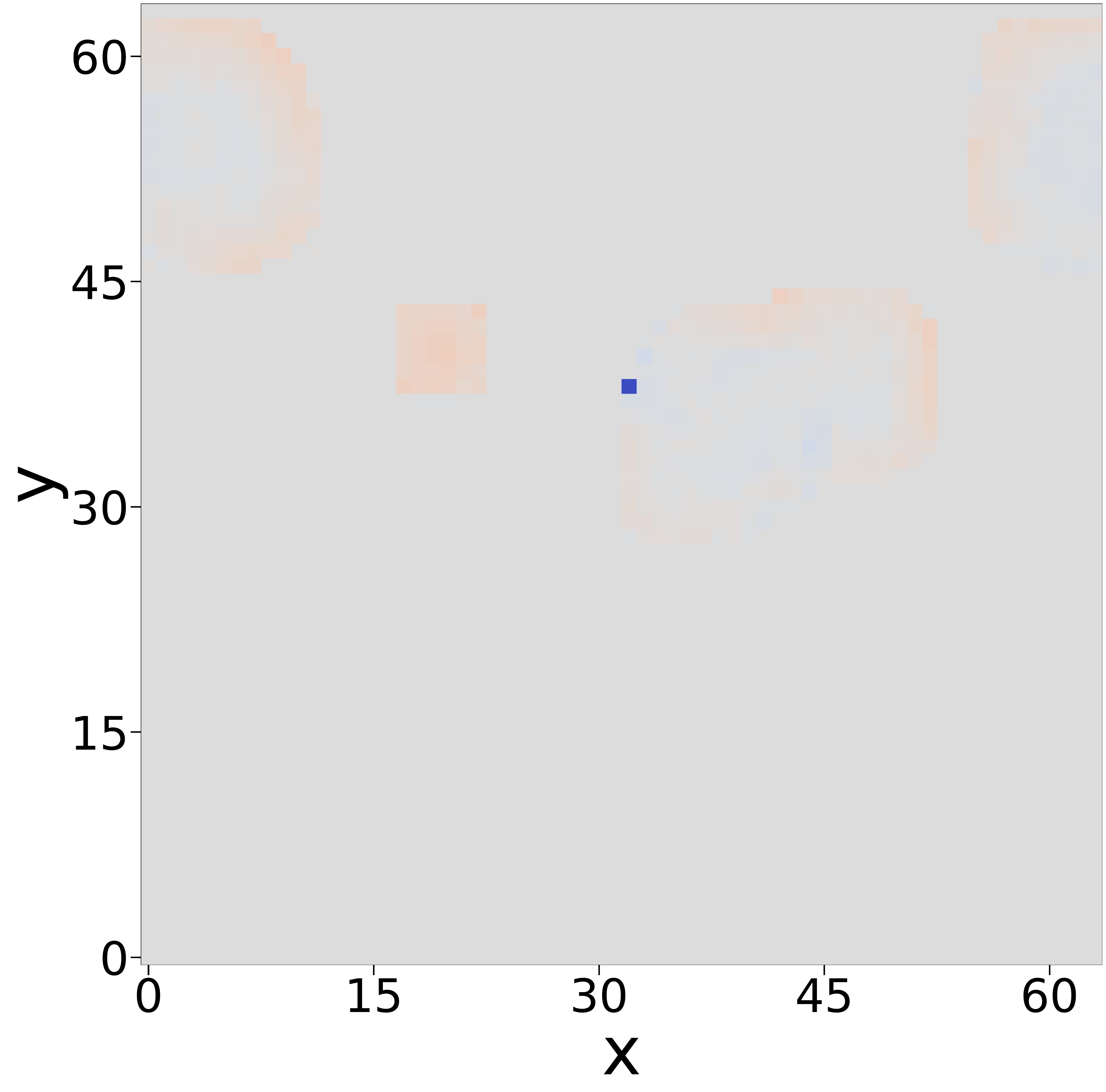}
    \end{subfigure}
    \begin{subfigure}[b]{0.22\textwidth}
        \captionsetup{labelformat=empty}
        \includegraphics[width=\textwidth]{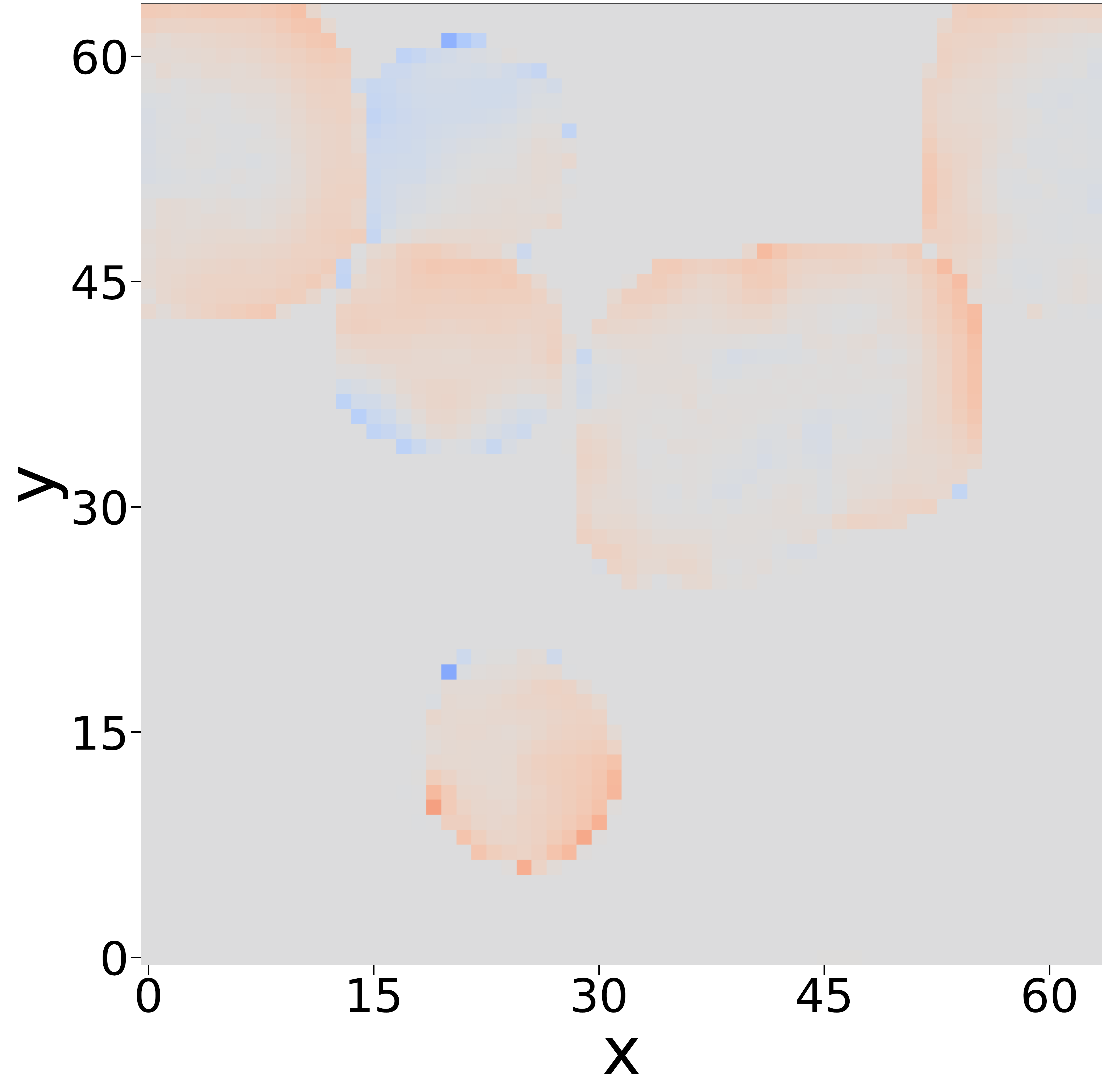}
    \end{subfigure}
    \begin{subfigure}[b]{0.22\textwidth}
        \includegraphics[width=\textwidth]{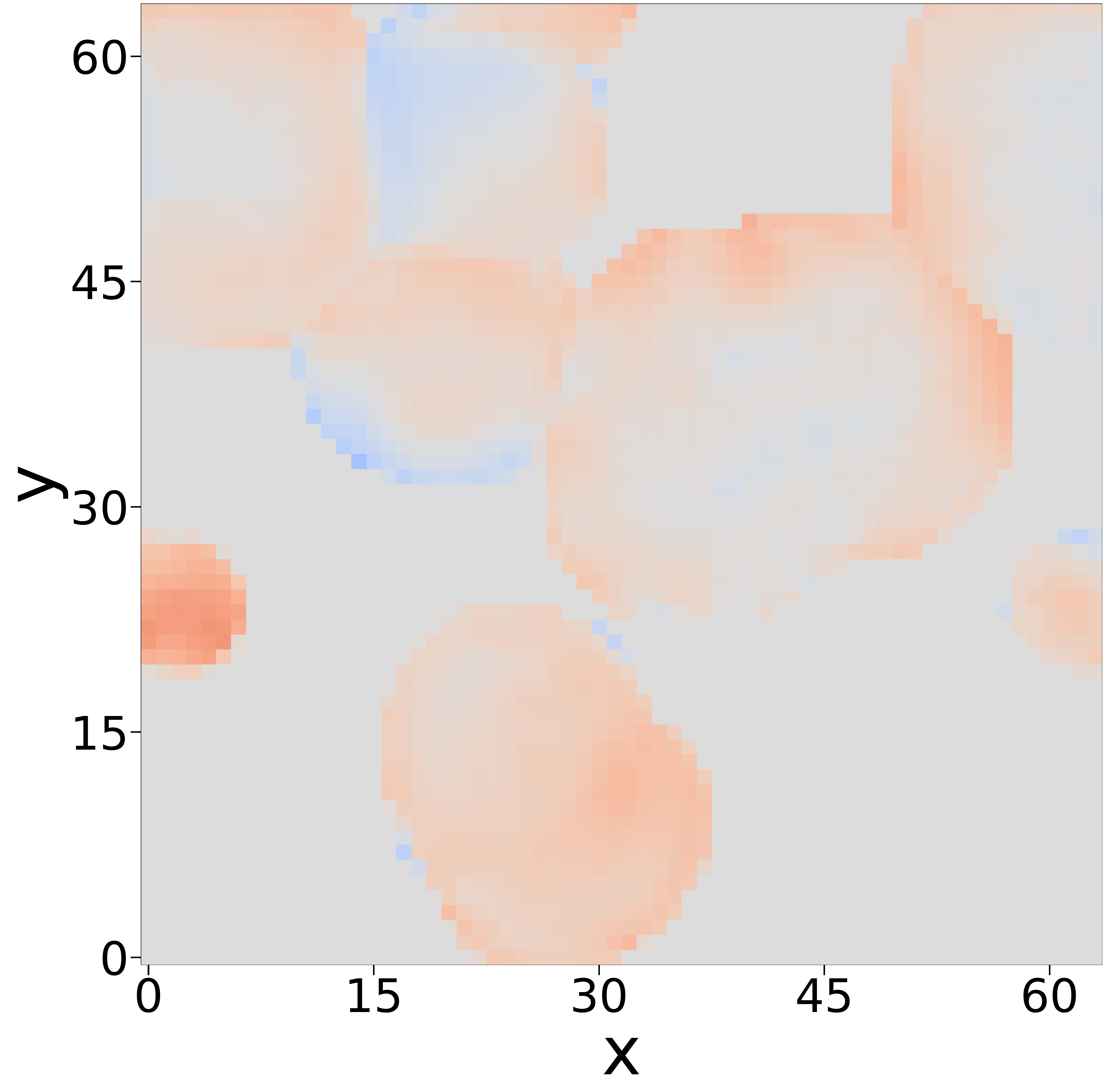}
    \end{subfigure}
    \begin{subfigure}[b]{0.0585\textwidth}
        \includegraphics[width=\textwidth]{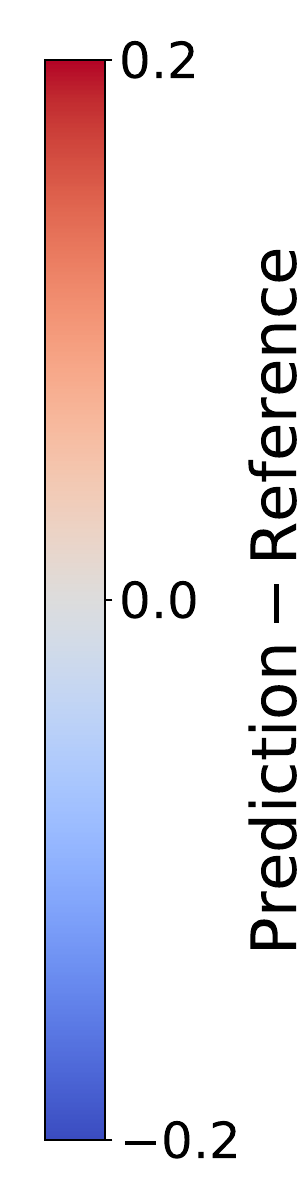}
    \end{subfigure}
    }

    \begin{tikzpicture}
        \def\arrowLength{8}
        \def\arrowHeight{0.15}
        \def\arrowTip{0.3} 
    
        \draw[fill=black!30, draw=black] 
            (0, -\arrowHeight/2) -- 
            ({\arrowLength - \arrowTip}, -\arrowHeight/2) -- 
            ({\arrowLength - \arrowTip}, -\arrowHeight) -- 
            (\arrowLength, 0) -- 
            ({\arrowLength - \arrowTip}, \arrowHeight) -- 
            ({\arrowLength - \arrowTip}, \arrowHeight/2) -- 
            (0, \arrowHeight/2) -- 
            cycle;
        \node at (\arrowLength/2, 0.45) {\large \textbf{Temporal evolution}};
    \end{tikzpicture}

    \caption{Selected timesteps of the temporal evolution of radiative intensity at  z=32: preprocessed numerical reference (top row), model prediction (middle row), and residual (bottom row). 
    }
    \label{fig:3d_time}
\end{figure}
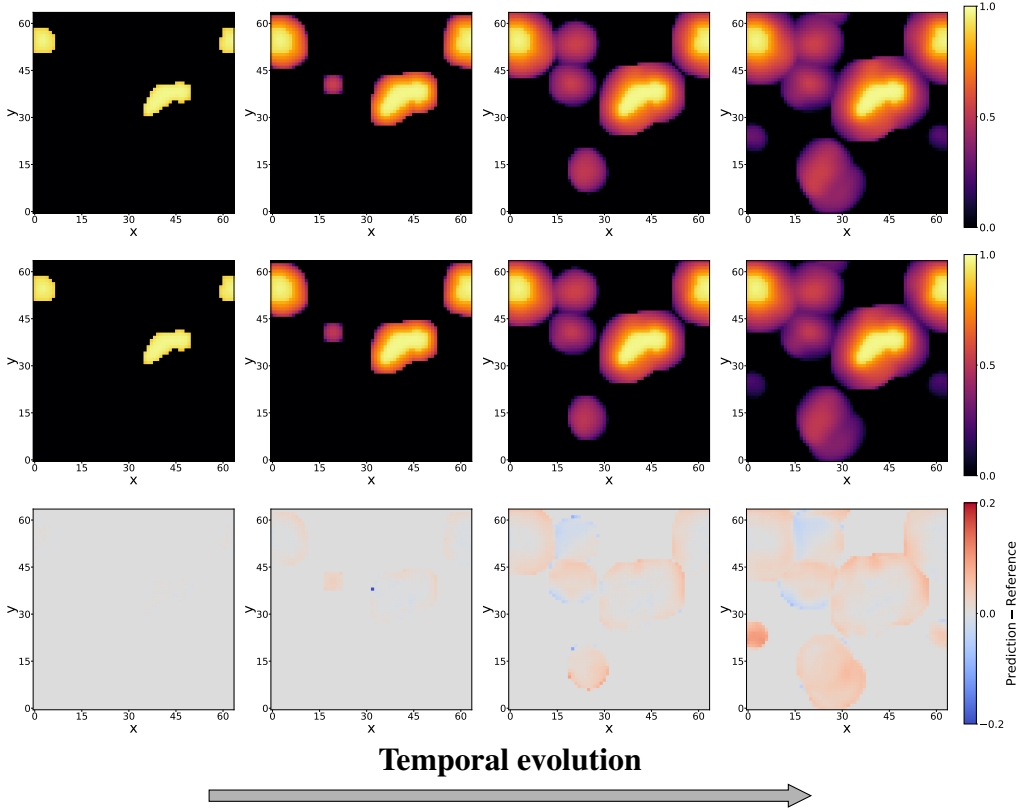

To train and evaluate the model, we generate a diverse dataset of absorption and emission fields  
based on turbulent periodic boxes produced with the hydrodynamic code \texttt{jf1uids}~\cite{2024arXiv241023093S} in a $64^3$ 
domain. To ensure sufficiently heterogeneous density-- and thus opacity--fields, we vary the turbulence random seed, the amplitude of velocity fluctuations, and the slope of the turbulent kinetic-energy power spectrum across simulations,
running each simulation until the turbulence spectrum reaches a stable equilibrium. 
This setup approximates conditions closely resembling those in giant molecular clouds (star-forming 
regions) that are among the most extensively studied numerically for their radiative effects \citep{davide2020}.
The absorption field $a(\mathbf{x})$ is correlated with the density field, 
and radiation sources $j(\mathbf{x})$ are placed 
in the top 1.5\% 
of the density field. 
For each pair ($a(\mathbf{x})$, $j(\mathbf{x})$), the corresponding radiative intensity evolution is computed over ten time steps, starting from $I_{\nu,0}(\textbf{x})$=0, using ray tracing
to complete the dataset (see Appendix~\ref{datasets}). 
The dataset is split into a training (70\%), validation (10\%), and test (20\%) set.
Training leverages pairs of consecutive intensities ($I_{\nu,t}(\textbf{x}), I_{\nu,t+1}(\textbf{x})$) alongside the corresponding absorption and emission fields to train the model for next-state intensity predictions, and key hyperparameters are optimized using Optuna~\cite{2019arXiv190710902A}. 
To stabilize the training process, all fields are log-transformed and min–max normalized to the range $[0,1]$. Additionally, we employ a relative loss combining pixelwise and spatial gradient differences  to encourage sharp feature reconstruction (see Appendix~\ref{parameters} for training details).
Training and evaluation are performed on an NVIDIA H100 GPU.
During inference, full temporal evolution is obtained by recursively feeding predictions back as input.
Figure~\ref{fig:3d_time} compares the 
preprocessed numerically computed (top row) and the predicted (middle row) temporal evolution of the radiative intensity at the midplane (z=32) for a random test set sample. Additional cross-sections as well as the absorption and emission fields are shown in Appendixes~\ref{datasets} and~\ref{extended_crosssections}. 


The 
model 
accurately captures the temporal evolution of the radiative intensity. Predictions closely match the reference solutions, 
and residuals remain low, with deviations mainly near edges
around evolving structures. 
The shown sample reflects the model's overall test performance, with an average relative error of 2.9\% 
per pixel for next-state predictions,
highlighting its ability to generalize across varying scenarios. This accuracy aligns with state-of-the-art works~\cite{COemu, PINNRT}. 
Notably, the integration of the U-Net significantly improves performance, as a pure FNO model only reaches a relative error of 60\%. 
Due to the recurrent prediction scheme,  errors accumulate over time but remain small, as shown by the residuals (see Appendix~\ref{quantanalysis} for a detailed assessment). 
Crucially, the surrogate model achieves a speedup of \textasciitilde$600\times$, producing full spatiotemporal predictions in 0.1s compared to 59.2s for the numerical solver. 
Additionally, memory costs do not scale with the number of sources, addressing a key computational bottleneck of ray-tracing algorithms and further highlighting the surrogate's potential as a drop-in replacement for on-the-fly solvers in hydrodynamic simulations.


To contextualize the performance of our approach relative to existing emulators, we also apply it to monochromatic three-dimensional static  radiative transfer, neglecting scattering and angular dependence.
Our model achieves a speedup of more than three orders of magnitude while maintaining a relative error of 2.6\%, matching or surpassing the results reported in other 3D static RT emulators~\cite{COemu}. Further details are provided in Appendix~\ref{steady-state}.

%


\section{Discussion}

\textbf{Summary.} 
Simulating radiative transfer numerically is computationally demanding, especially when included on-the-fly in hydrodynamic simulations.
We present a U-FNO-based surrogate model prototype, 
designed as a drop-in replacement for  conventional on-the-fly 
solvers. The model predicts the temporal evolution of radiative intensity via recurrent application, closely matching numerical references with an average relative error of 2.9\% 
for next-state predictions and achieving a speedup of \textasciitilde$600\times$ over state-of-the-art 
ray tracing, with memory costs independent of the number of sources.
Demonstrated for a single frequency, this approach generalizes naturally to multi-frequency setups by training one model per frequency bin.
All code required to reproduce our experiments is publicly available\footnote{\url{https://github.com/RuneRost/Astro-RT.git} 
}. 

\textbf{Limitations.}
Due to the integrated U-Nets, the current U-FNO model is not invariant to discretization or resolution, lacking the flexibility of simpler FNO-based architectures. 
In fixed-grid scenarios, however,  
this limitation is not critical, and additional techniques also offer resolution-independent formulations~\cite{2022AdWR..16304180W}. 
Moreover, as the U-FNO predicts the entire field, localized predictions cannot be made in isolation, leading to unnecessary overhead when only specific regions are of interest. Our current model demonstrates predictions for only a single frequency, however, separate models can be trained for each frequency bin 
for integration into multi-frequency simulations. 

\textbf{Outlook.}
Future research will extend our models to more general approximations of the RTE solution operator by also considering the angular dependency of the radiative intensity.
Moreover, it will focus on training a unified model capable of predicting the temporal evolution of radiative intensity across multiple frequency bins simultaneously. Preliminary results are shown in Appendix~\ref{outlook}. 
Furthermore, we plan to integrate the 
developed surrogate model into full simulation pipelines to accelerate computations and reduce memory costs. 


\begin{ack}
This work is funded by the Carl-Zeiss-Stiftung through the NEXUS program. This work was supported by the Deutsche Forschungsgemeinschaft (DFG, German Research Foundation) under Germany’s Excellence Strategy EXC 2181/1 - 390900948 (the Heidelberg STRUCTURES Excellence Cluster). We acknowledge the usage of the AI-clusters Tom and Jerry funded by the Field of Focus 2 of Heidelberg University.
\end{ack}

\bibliographystyle{plain}
\bibliography{main}

@ARTICLE{Rosdahl2013,
       author = {{Rosdahl}, J. and {Blaizot}, J. and {Aubert}, D. and {Stranex}, T. and {Teyssier}, R.},
        title = "{RAMSES-RT: radiation hydrodynamics in the cosmological context}",
      journal = {Monthly Notices of the Royal Astronomical Society},
         year = 2013,
        month = dec,
       volume = {436},
       number = {3},
        pages = {2188-2231},
          doi = {10.1093/mnras/stt1722},
archivePrefix = {arXiv},
       eprint = {1304.7126},
 primaryClass = {astro-ph.CO},
       adsurl = {https://ui.adsabs.harvard.edu/abs/2013MNRAS.436.2188R}
}

@ARTICLE{Buck2022,
       author = {{Buck}, Tobias and {Pfrommer}, Christoph and {Girichidis}, Philipp and {Corobean}, Bogdan},
        title = "{Escaping the maze: a statistical subgrid model for cloud-scale density structures in the interstellar medium}",
      journal = {Monthly Notices of the Royal Astronomical Society},
         year = 2022,
        month = jun,
       volume = {513},
       number = {1},
        pages = {1414-1428},
          doi = {10.1093/mnras/stac952},
archivePrefix = {arXiv},
       eprint = {2204.02053},
 primaryClass = {astro-ph.GA},
       adsurl = {https://ui.adsabs.harvard.edu/abs/2022MNRAS.513.1414B}
}

@ARTICLE{Buck2021,
       author = {{Buck}, Tobias and {Wolf}, Steffen},
        title = "{Predicting resolved galaxy properties from photometric images using convolutional neural networks}",
      journal = {arXiv e-prints},
         year = 2021,
        month = nov,
          eid = {arXiv:2111.01154},
        pages = {arXiv:2111.01154},
          doi = {10.48550/arXiv.2111.01154},
archivePrefix = {arXiv},
       eprint = {2111.01154},
 primaryClass = {astro-ph.GA},
       adsurl = {https://ui.adsabs.harvard.edu/abs/2021arXiv211101154B}
}

@ARTICLE{Buck2017,
       author = {{Buck}, Tobias and {Macci{\`o}}, Andrea V. and {Obreja}, Aura and {Dutton}, Aaron A. and {Dom{\'\i}nguez-Tenreiro}, Rosa and {Granato}, Gian Luigi},
        title = "{NIHAO XIII: Clumpy discs or clumpy light in high-redshift galaxies?}",
      journal = {Monthly Notices of the Royal Astronomical Society},
         year = 2017,
        month = jul,
       volume = {468},
       number = {3},
        pages = {3628-3649},
          doi = {10.1093/mnras/stx685},
archivePrefix = {arXiv},
       eprint = {1612.05277},
 primaryClass = {astro-ph.GA},
       adsurl = {https://ui.adsabs.harvard.edu/abs/2017MNRAS.468.3628B}
}

@ARTICLE{Kannan2014,
       author = {{Kannan}, R. and {Stinson}, G.~S. and {Macci{\`o}}, A.~V. and {Hennawi}, J.~F. and {Woods}, R. and {Wadsley}, J. and {Shen}, S. and {Robitaille}, T. and {Cantalupo}, S. and {Quinn}, T.~R. and {Christensen}, C.},
        title = "{Galaxy formation with local photoionization feedback - I. Methods}",
      journal = {Monthly Notices of the Royal Astronomical Society},
         year = 2014,
        month = jan,
       volume = {437},
       number = {3},
        pages = {2882-2893},
          doi = {10.1093/mnras/stt2098},
archivePrefix = {arXiv},
       eprint = {1310.6748},
 primaryClass = {astro-ph.GA},
       adsurl = {https://ui.adsabs.harvard.edu/abs/2014MNRAS.437.2882K}
}

@ARTICLE{Obreja2019,
       author = {{Obreja}, Aura and {Macci{\`o}}, Andrea V. and {Moster}, Benjamin and {Udrescu}, Silviu M. and {Buck}, Tobias and {Kannan}, Rahul and {Dutton}, Aaron A. and {Blank}, Marvin},
        title = "{Local photoionization feedback effects on galaxies}",
      journal = {Monthly Notices of the Royal Astronomical Society},
         year = 2019,
        month = dec,
       volume = {490},
       number = {2},
        pages = {1518-1538},
          doi = {10.1093/mnras/stz2639},
archivePrefix = {arXiv},
       eprint = {1909.00832},
 primaryClass = {astro-ph.GA},
       adsurl = {https://ui.adsabs.harvard.edu/abs/2019MNRAS.490.1518O}
}

@ARTICLE{COemu,
       author = {{Su}, Shiqi and {De Ceuster}, Frederik and {Cha}, Jaehoon and {Wilkinson}, Mark I. and {Thiyagalingam}, Jeyan and {Yates}, Jeremy and {Zhu}, Yi-Hang and {Bolte}, Jan},
        title = "{Emulating CO Line Radiative Transfer with Deep Learning}",
      journal = {arXiv e-prints},
         year = 2025,
        month = jul,
          eid = {arXiv:2507.11398},
        pages = {arXiv:2507.11398},
          doi = {10.48550/arXiv.2507.11398},
archivePrefix = {arXiv},
       eprint = {2507.11398},
 primaryClass = {astro-ph.IM},
       adsurl = {https://ui.adsabs.harvard.edu/abs/2025arXiv250711398S}
}

@ARTICLE{PINNRT,
       author = {{Mishra}, Siddhartha and {Molinaro}, Roberto},
        title = "{Physics informed neural networks for simulating radiative transfer}",
      journal = {Journal of Quantitative Spectroscopy and Radiative Transfer},
         year = 2021,
        month = aug,
       volume = {270},
          eid = {107705},
        pages = {107705},
          doi = {10.1016/j.jqsrt.2021.107705},
archivePrefix = {arXiv},
       eprint = {2009.13291},
 primaryClass = {cs.LG},
       adsurl = {https://ui.adsabs.harvard.edu/abs/2021JQSRT.27007705M}
}

@article{LU2024105282,
title = {Surrogate modeling for radiative heat transfer using physics-informed deep neural operator networks},
journal = {Proceedings of the Combustion Institute},
volume = {40},
number = {1},
pages = {105282},
year = {2024},
issn = {1540-7489},
doi = {https://doi.org/10.1016/j.proci.2024.105282},
url = {https://www.sciencedirect.com/science/article/pii/S1540748924000920},
author = {Xiaoyi Lu and Yi Wang}
}

@article{Noebauer2019,
  author       = {Noebauer, Ulrich M. and Sim, Stuart A.},
  title        = {Monte Carlo radiative transfer},
  journal      = {Living Reviews in Computational Astrophysics},
  volume       = {5},
  number       = {1},
  pages        = {1},
  year         = {2019},
  doi          = {10.1007/s41115-019-0004-9},
  url          = {https://doi.org/10.1007/s41115-019-0004-9},
  issn         = {2365-0524}
}

@article{chen2:1995,
author = {Chen, Tianping and Chen, Hong},
year = {1995},
month = {08},
pages = {911 - 917},
vol   = {6},
title = {Universal approximation to nonlinear operators by neural networks with arbitrary activation functions and its applications to dynamic systems},
journal = {Neural Networks, IEEE Transactions on},
doi = {10.1109/72.392253}
}

@ARTICLE{2022AdWR..16304180W,
       author = {{Wen}, Gege and {Li}, Zongyi and {Azizzadenesheli}, Kamyar and {Anandkumar}, Anima and {Benson}, Sally M.},
        title = "{U-FNO-An enhanced Fourier neural operator-based deep-learning model for multiphase flow}",
      journal = {Advances in Water Resources},
         year = 2022,
        month = may,
       volume = {163},
          eid = {104180},
        pages = {104180},
          doi = {10.1016/j.advwatres.2022.104180},
archivePrefix = {arXiv},
       eprint = {2109.03697},
 primaryClass = {physics.geo-ph},
       adsurl = {https://ui.adsabs.harvard.edu/abs/2022AdWR..16304180W}
}

@ARTICLE{2015arXiv150504597R,
       author = {{Ronneberger}, Olaf and {Fischer}, Philipp and {Brox}, Thomas},
        title = "{U-Net: Convolutional Networks for Biomedical Image Segmentation}",
      journal = {arXiv e-prints},
         year = 2015,
        month = may,
          eid = {arXiv:1505.04597},
        pages = {arXiv:1505.04597},
          doi = {10.48550/arXiv.1505.04597},
archivePrefix = {arXiv},
       eprint = {1505.04597},
 primaryClass = {cs.CV},
       adsurl = {https://ui.adsabs.harvard.edu/abs/2015arXiv150504597R}
}

@ARTICLE{2025arXiv250521573W,
       author = {{Wan}, Han and {Zhang}, Rui and {Sun}, Hao},
        title = "{Spectral-inspired Neural Operator for Data-efficient PDE Simulation in Physics-agnostic Regimes}",
      journal = {arXiv e-prints},
         year = 2025,
        month = may,
          eid = {arXiv:2505.21573},
        pages = {arXiv:2505.21573},
          doi = {10.48550/arXiv.2505.21573},
archivePrefix = {arXiv},
       eprint = {2505.21573},
 primaryClass = {stat.ML},
       adsurl = {https://ui.adsabs.harvard.edu/abs/2025arXiv250521573W}
}

@ARTICLE{2022arXiv220313181D,
       author = {{de Hoop}, Maarten V. and {Zhengyu Huang}, Daniel and {Qian}, Elizabeth and {Stuart}, Andrew M.},
        title = "{The Cost-Accuracy Trade-Off In Operator Learning With Neural Networks}",
      journal = {arXiv e-prints},
         year = 2022,
        month = mar,
          eid = {arXiv:2203.13181},
        pages = {arXiv:2203.13181},
          doi = {10.48550/arXiv.2203.13181},
archivePrefix = {arXiv},
       eprint = {2203.13181},
 primaryClass = {math.NA},
       adsurl = {https://ui.adsabs.harvard.edu/abs/2022arXiv220313181D}
}

@ARTICLE{2020arXiv201008895L,
       author = {{Li}, Zongyi and {Kovachki}, Nikola and {Azizzadenesheli}, Kamyar and {Liu}, Burigede and {Bhattacharya}, Kaushik and {Stuart}, Andrew and {Anandkumar}, Anima},
        title = "{Fourier Neural Operator for Parametric Partial Differential Equations}",
      journal = {arXiv e-prints},
         year = 2020,
        month = oct,
          eid = {arXiv:2010.08895},
        pages = {arXiv:2010.08895},
          doi = {10.48550/arXiv.2010.08895},
archivePrefix = {arXiv},
       eprint = {2010.08895},
 primaryClass = {cs.LG},
       adsurl = {https://ui.adsabs.harvard.edu/abs/2020arXiv201008895L}
}

@software{jax2018github,
  author = {James Bradbury and Roy Frostig and Peter Hawkins and Matthew James Johnson and Chris Leary and Dougal Maclaurin and George Necula and Adam Paszke and Jake Vander{P}las and Skye Wanderman-{M}ilne and Qiao Zhang},
  title = {{JAX}: composable transformations of {P}ython+{N}um{P}y programs},
  url = {http://github.com/jax-ml/jax},
  version = {0.3.13},
  year = {2018},
}

@ARTICLE{2019arXiv190710902A,
       author = {{Akiba}, Takuya and {Sano}, Shotaro and {Yanase}, Toshihiko and {Ohta}, Takeru and {Koyama}, Masanori},
        title = "{Optuna: A Next-generation Hyperparameter Optimization Framework}",
      journal = {arXiv e-prints},
         year = 2019,
        month = jul,
          eid = {arXiv:1907.10902},
        pages = {arXiv:1907.10902},
          doi = {10.48550/arXiv.1907.10902},
archivePrefix = {arXiv},
       eprint = {1907.10902},
 primaryClass = {cs.LG},
       adsurl = {https://ui.adsabs.harvard.edu/abs/2019arXiv190710902A}
}

@ARTICLE{2019JCoPh.384....1F,
       author = {{Fan}, Yuwei and {Orozco Bohorquez}, Cindy and {Ying}, Lexing},
        title = "{BCR-Net: A neural network based on the nonstandard wavelet form}",
      journal = {Journal of Computational Physics},
         year = 2019,
        month = may,
       volume = {384},
        pages = {1-15},
          doi = {10.1016/j.jcp.2019.02.002},
archivePrefix = {arXiv},
       eprint = {1810.08754},
 primaryClass = {math.NA},
       adsurl = {https://ui.adsabs.harvard.edu/abs/2019JCoPh.384....1F}
}

@inproceedings{
jiang2019enforcing,
title={Enforcing Physical Constraints in {CNN}s through Differentiable {PDE} Layer},
author={Chiyu ''Max'' Jiang and Karthik Kashinath and Prabhat and Philip Marcus},
booktitle={ICLR 2020 Workshop on Integration of Deep Neural Models and Differential Equations},
year={2019},
url={https://openreview.net/forum?id=q2noHUqMkK}
}

@ARTICLE{2024arXiv240215715K,
       author = {{Kovachki}, Nikola B. and {Lanthaler}, Samuel and {Stuart}, Andrew M.},
        title = "{Operator Learning: Algorithms and Analysis}",
      journal = {arXiv e-prints},
         year = 2024,
        month = feb,
          eid = {arXiv:2402.15715},
        pages = {arXiv:2402.15715},
          doi = {10.48550/arXiv.2402.15715},
archivePrefix = {arXiv},
       eprint = {2402.15715},
 primaryClass = {cs.LG},
       adsurl = {https://ui.adsabs.harvard.edu/abs/2024arXiv240215715K}
}

@ARTICLE{2011ApJ...730...40P,
       author = {{Padoan}, Paolo and {Nordlund}, {\r{A}}ke},
        title = "{The Star Formation Rate of Supersonic Magnetohydrodynamic Turbulence}",
      journal = {The Astrophysical Journal},
         year = 2011,
        month = mar,
       volume = {730},
       number = {1},
          eid = {40},
        pages = {40},
          doi = {10.1088/0004-637X/730/1/40},
archivePrefix = {arXiv},
       eprint = {0907.0248},
 primaryClass = {astro-ph.GA},
       adsurl = {https://ui.adsabs.harvard.edu/abs/2011ApJ...730...40P}
}

@ARTICLE{2024arXiv241023093S,
       author = {{Storcks}, Leonard and {Buck}, Tobias},
        title = "{Differentiable Conservative Radially Symmetric Fluid Simulations and Stellar Winds -- jf1uids}",
      journal = {arXiv e-prints},
         year = 2024,
        month = oct,
          eid = {arXiv:2410.23093},
        pages = {arXiv:2410.23093},
          doi = {10.48550/arXiv.2410.23093},
archivePrefix = {arXiv},
       eprint = {2410.23093},
 primaryClass = {physics.flu-dyn},
       adsurl = {https://ui.adsabs.harvard.edu/abs/2024arXiv241023093S}
}

@ARTICLE{2006ApJ...645..920S,
       author = {{Steinacker}, J. and {Bacmann}, A. and {Henning}, T.},
        title = "{Ray Tracing for Complex Astrophysical High-opacity Structures}",
      journal = {The Astrophysical Journal},
         year = 2006,
        month = jul,
       volume = {645},
       number = {2},
        pages = {920-927},
          doi = {10.1086/504367},
archivePrefix = {arXiv},
       eprint = {astro-ph/0603463},
 primaryClass = {astro-ph},
       adsurl = {https://ui.adsabs.harvard.edu/abs/2006ApJ...645..920S}
}

@ARTICLE{2012ApJ...761..156F,
       author = {{Federrath}, Christoph and {Klessen}, Ralf S.},
        title = "{The Star Formation Rate of Turbulent Magnetized Clouds: Comparing Theory, Simulations, and Observations}",
      journal = {The Astrophysical Journal},
         year = 2012,
        month = dec,
       volume = {761},
       number = {2},
          eid = {156},
        pages = {156},
          doi = {10.1088/0004-637X/761/2/156},
archivePrefix = {arXiv},
       eprint = {1209.2856},
 primaryClass = {astro-ph.SR},
       adsurl = {https://ui.adsabs.harvard.edu/abs/2012ApJ...761..156F}
}

@ARTICLE{2014ApJ...781...91G,
       author = {{Girichidis}, Philipp and {Konstandin}, Lukas and {Whitworth}, Anthony P. and {Klessen}, Ralf S.},
        title = "{On the Evolution of the Density Probability Density Function in Strongly Self-gravitating Systems}",
      journal = {The Astrophysical Journal},
         year = 2014,
        month = feb,
       volume = {781},
       number = {2},
          eid = {91},
        pages = {91},
          doi = {10.1088/0004-637X/781/2/91},
archivePrefix = {arXiv},
       eprint = {1310.4346},
 primaryClass = {astro-ph.GA},
       adsurl = {https://ui.adsabs.harvard.edu/abs/2014ApJ...781...91G}
}

@ARTICLE{2017arXiv171105101L,
       author = {{Loshchilov}, Ilya and {Hutter}, Frank},
        title = "{Decoupled Weight Decay Regularization}",
      journal = {arXiv e-prints},
         year = 2017,
        month = nov,
          eid = {arXiv:1711.05101},
        pages = {arXiv:1711.05101},
          doi = {10.48550/arXiv.1711.05101},
archivePrefix = {arXiv},
       eprint = {1711.05101},
 primaryClass = {cs.LG},
       adsurl = {https://ui.adsabs.harvard.edu/abs/2017arXiv171105101L}
}

@ARTICLE{arepo-RT,
       author = {{Kannan}, Rahul and {Vogelsberger}, Mark and {Marinacci}, Federico and {McKinnon}, Ryan and {Pakmor}, R{\"u}diger and {Springel}, Volker},
        title = "{AREPO-RT: radiation hydrodynamics on a moving mesh}",
      journal = {Monthly Notices of the Royal Astronomical Society},
         year = 2019,
        month = may,
       volume = {485},
       number = {1},
        pages = {117-149},
          doi = {10.1093/mnras/stz287},
archivePrefix = {arXiv},
       eprint = {1804.01987},
 primaryClass = {astro-ph.IM},
       adsurl = {https://ui.adsabs.harvard.edu/abs/2019MNRAS.485..117K}
}

@ARTICLE{davide2020,
       author = {{Decataldo}, D. and {Lupi}, A. and {Ferrara}, A. and {Pallottini}, A. and {Fumagalli}, M.},
        title = "{Shaping the structure of a GMC with radiation and winds}",
      journal = {Monthly Notices of the Royal Astronomical Society},
         year = 2020,
        month = oct,
       volume = {497},
       number = {4},
        pages = {4718-4732},
          doi = {10.1093/mnras/staa2326},
archivePrefix = {arXiv},
       eprint = {2009.07860},
 primaryClass = {astro-ph.GA},
       adsurl = {https://ui.adsabs.harvard.edu/abs/2020MNRAS.497.4718D}
}

\newpage
\appendix
\section*{\Large Appendix}

The following appendix contains additional material complementing the main text, including architectural details, dataset descriptions, model and training configurations, and additional evaluation results. 
Moreover, we provide an outlook on future research by presenting preliminary results from a unified model currently being trained to predict the temporal evolution of radiative intensity across multiple frequency bins simultaneously.

\section{Outlook}
\label{outlook}

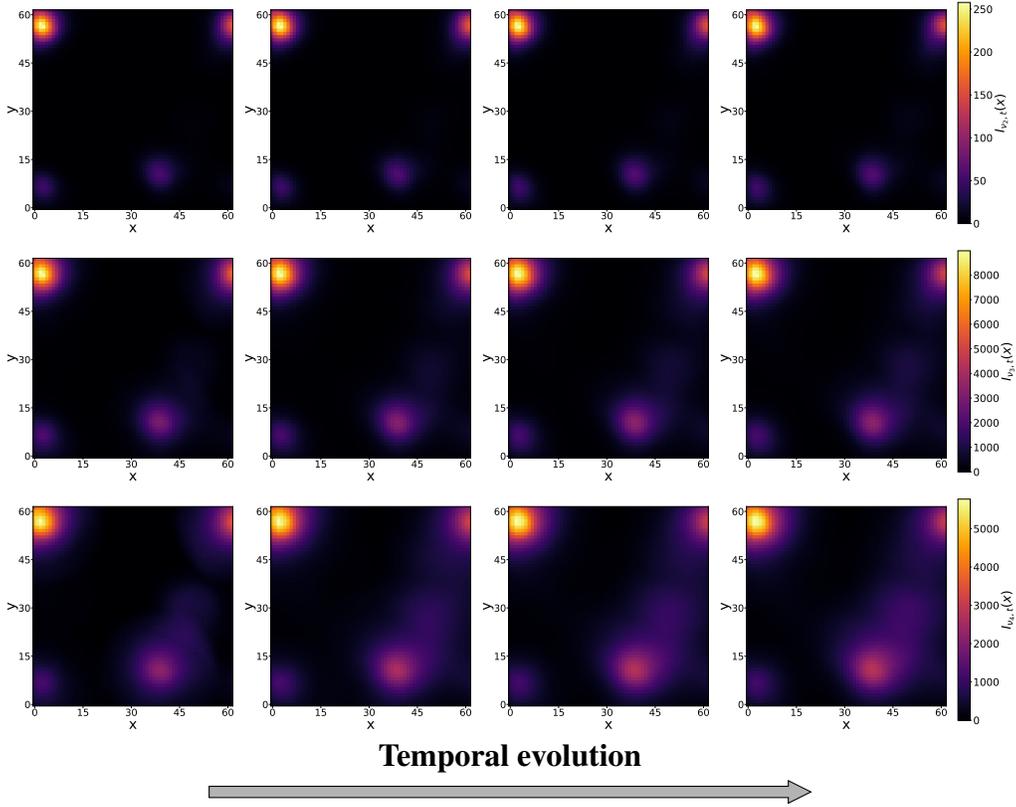
\begin{figure}[h]
    \centering

    \makebox[\textwidth][c]{
    \begin{subfigure}[b]{0.22\textwidth}
        \captionsetup{labelformat=empty}
        \includegraphics[width=\textwidth]{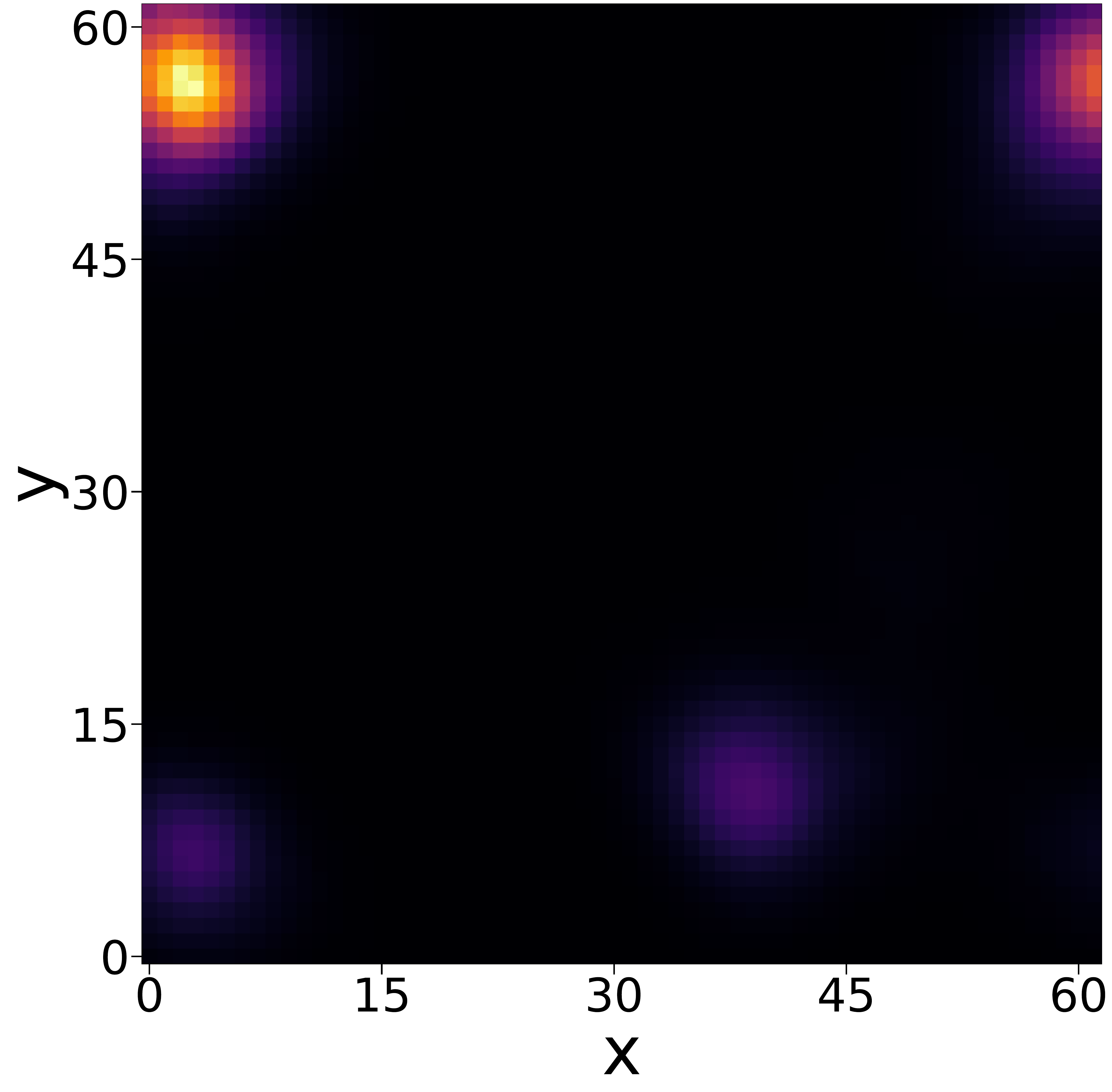}
    \end{subfigure}
    \begin{subfigure}[b]{0.22\textwidth}
        \captionsetup{labelformat=empty}
        \includegraphics[width=\textwidth]{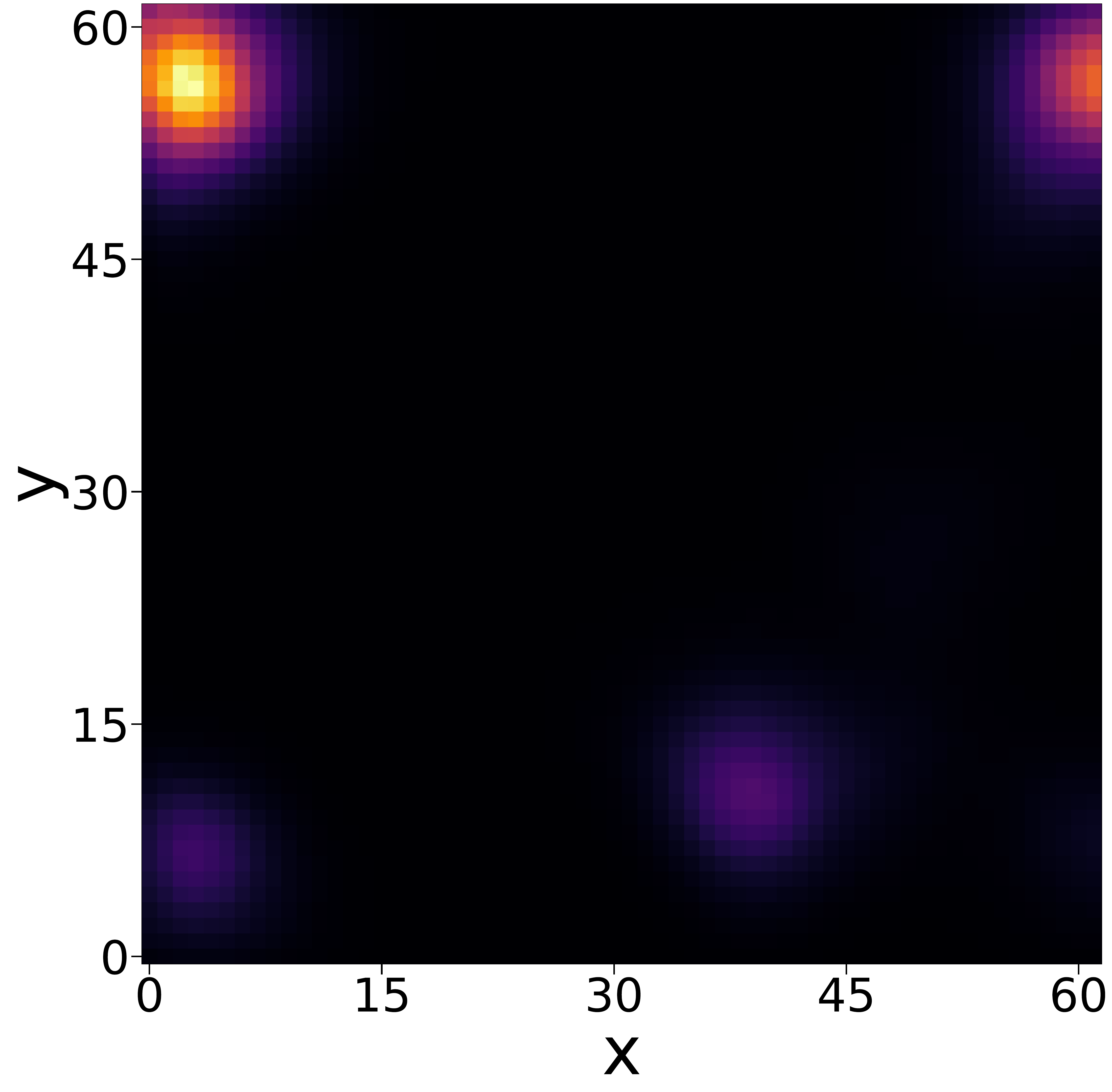}
    \end{subfigure}
    \begin{subfigure}[b]{0.22\textwidth}
        \captionsetup{labelformat=empty}
        \includegraphics[width=\textwidth]{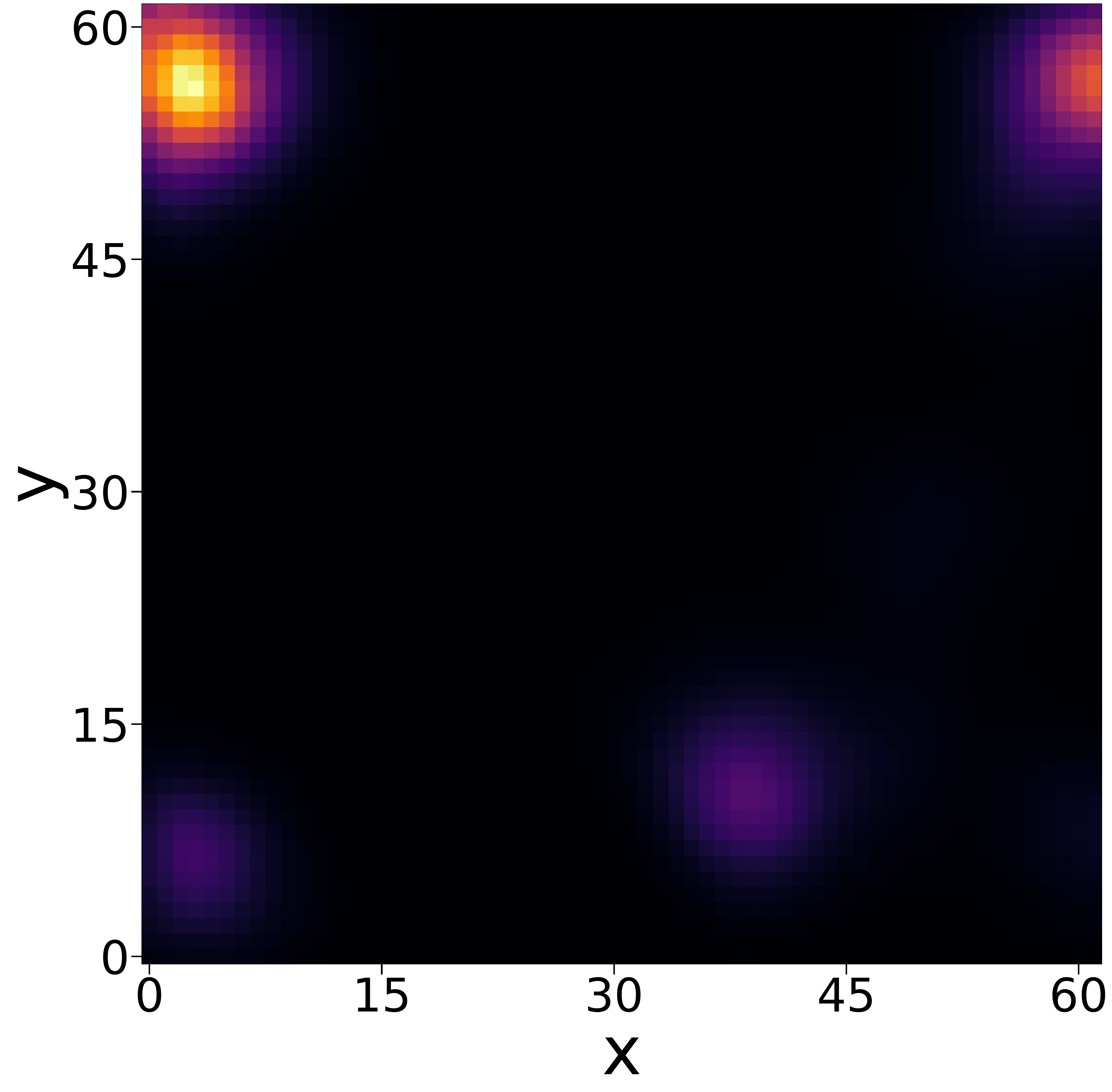}
    \end{subfigure}
    \begin{subfigure}[b]{0.22\textwidth}
        \captionsetup{labelformat=empty}
        \includegraphics[width=\textwidth]{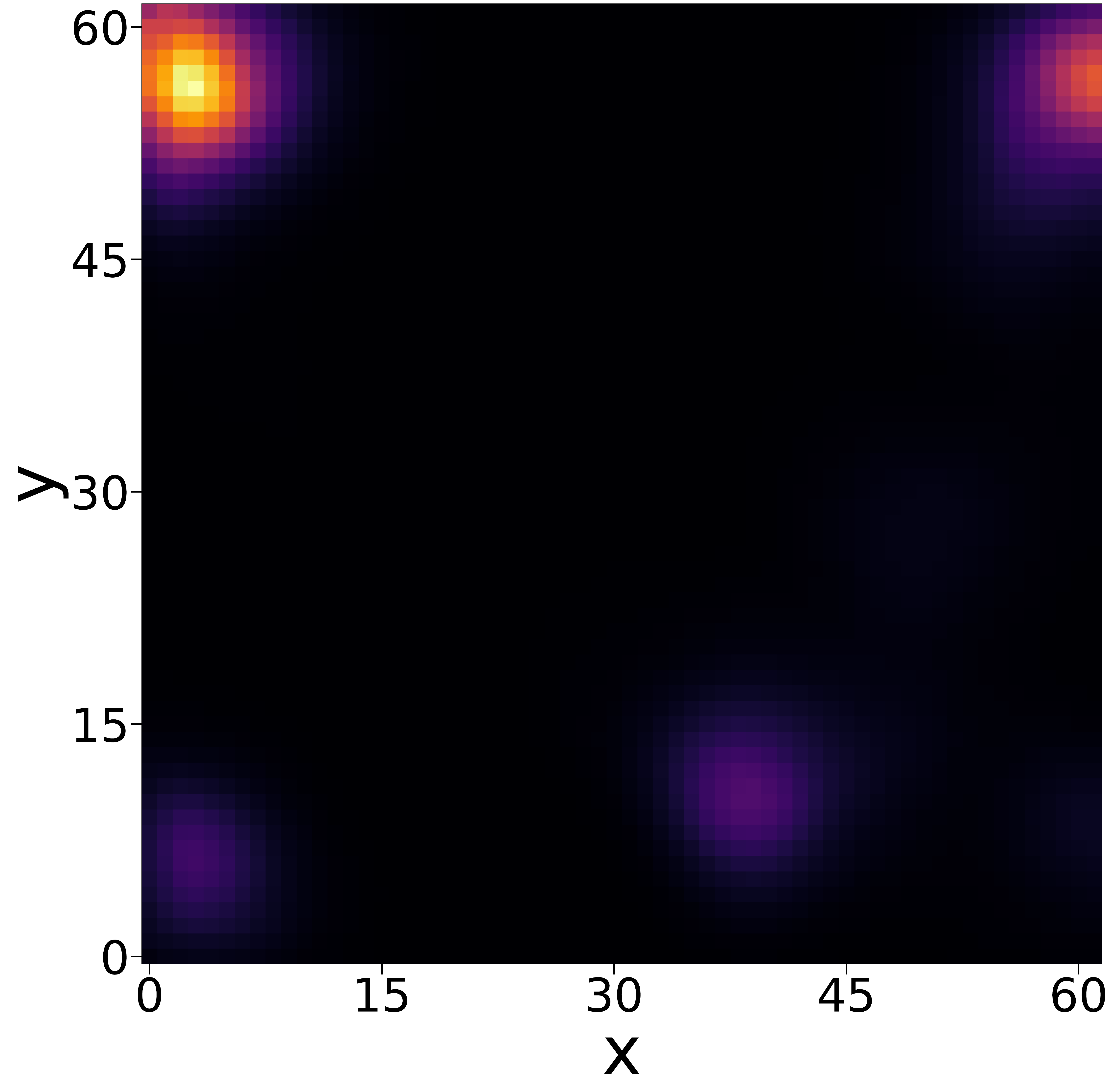}
    \end{subfigure}

    \begin{subfigure}[b]{0.0585\textwidth}
        \captionsetup{labelformat=empty}
        \includegraphics[width=\textwidth]{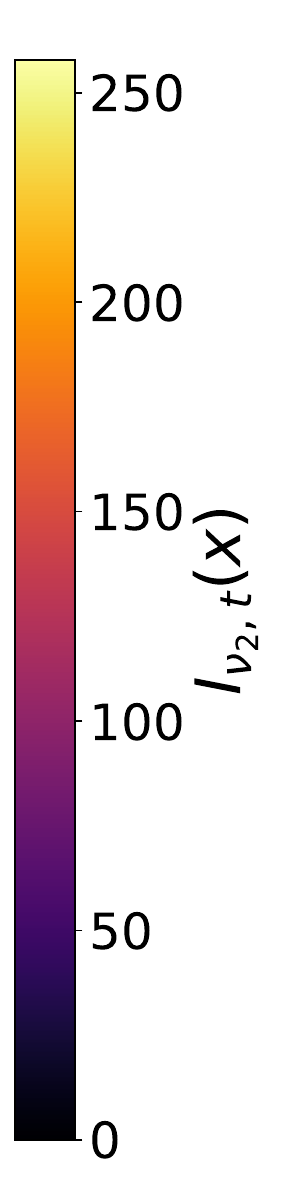}
    \end{subfigure}
    }


    \makebox[\textwidth][c]{
    \begin{subfigure}[b]{0.22\textwidth}
        \captionsetup{labelformat=empty}
        \includegraphics[width=\textwidth]{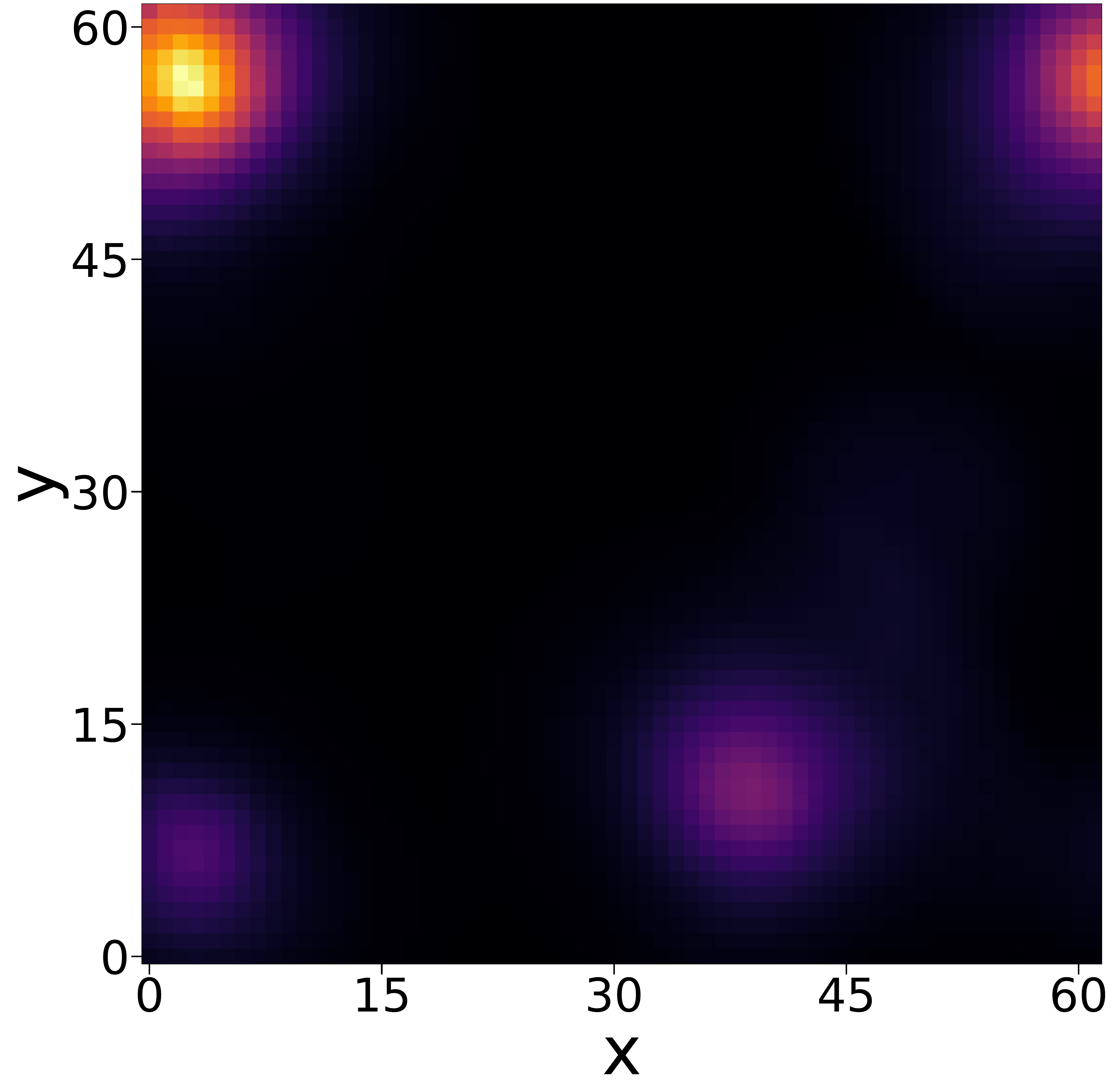}
    \end{subfigure}
    \begin{subfigure}[b]{0.22\textwidth}
        \captionsetup{labelformat=empty}
        \includegraphics[width=\textwidth]{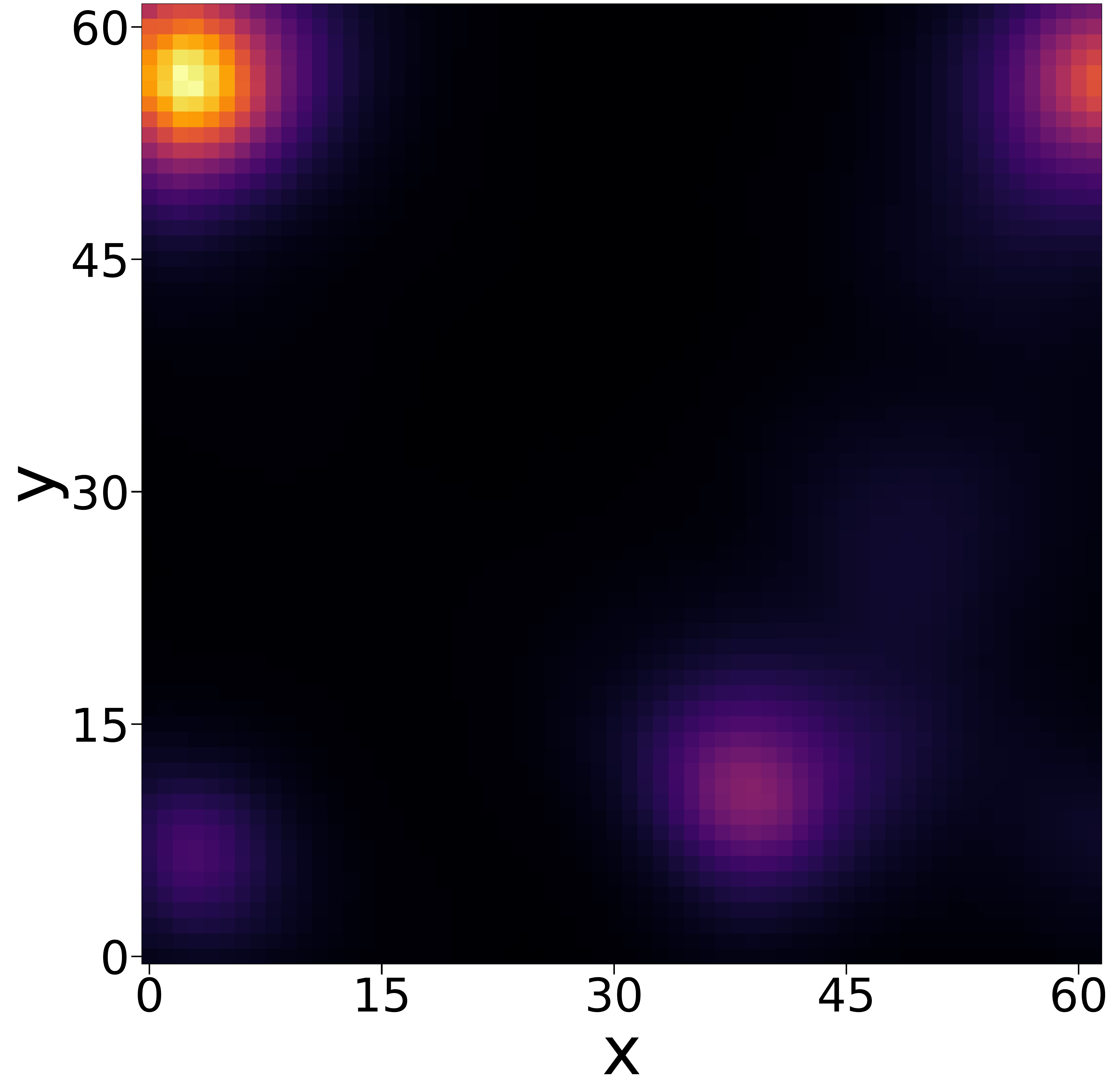}
    \end{subfigure}
    \begin{subfigure}[b]{0.22\textwidth}
        \captionsetup{labelformat=empty}
        \includegraphics[width=\textwidth]{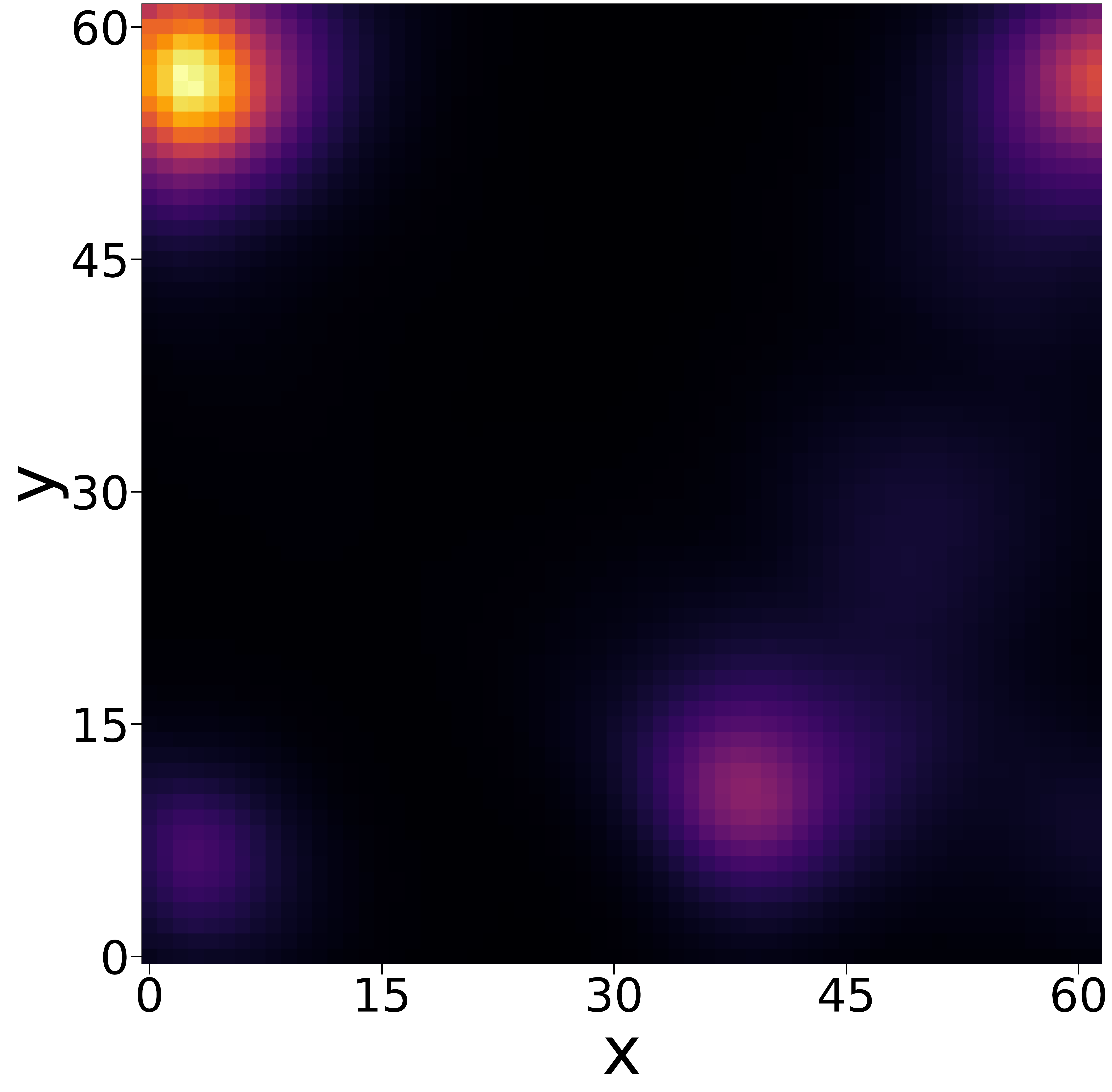}
    \end{subfigure}
    \begin{subfigure}[b]{0.22\textwidth}
        \captionsetup{labelformat=empty}
        \includegraphics[width=\textwidth]{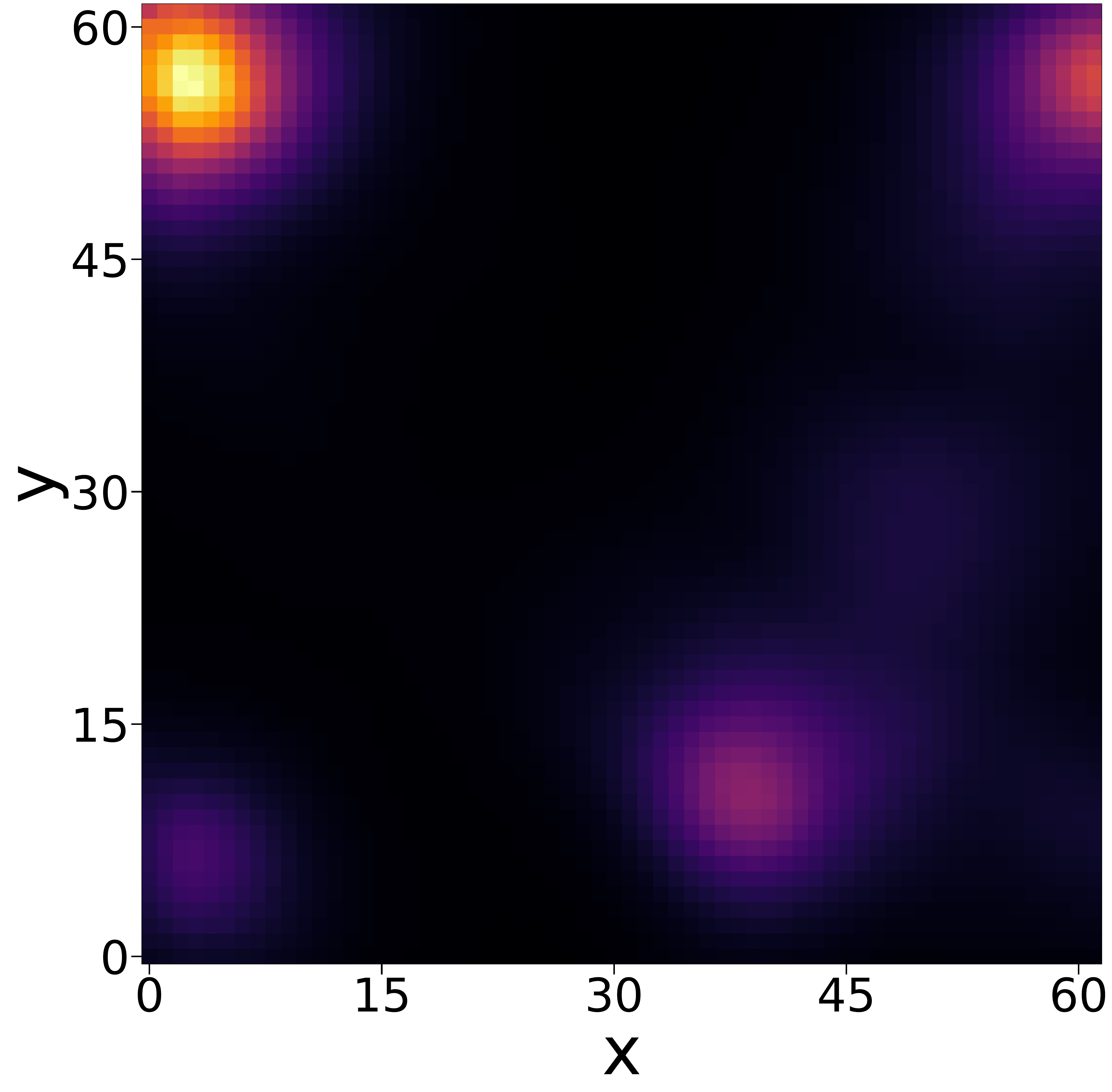}
    \end{subfigure}
    \begin{subfigure}[b]{0.0585\textwidth}
        \captionsetup{labelformat=empty}
        \includegraphics[width=\textwidth]{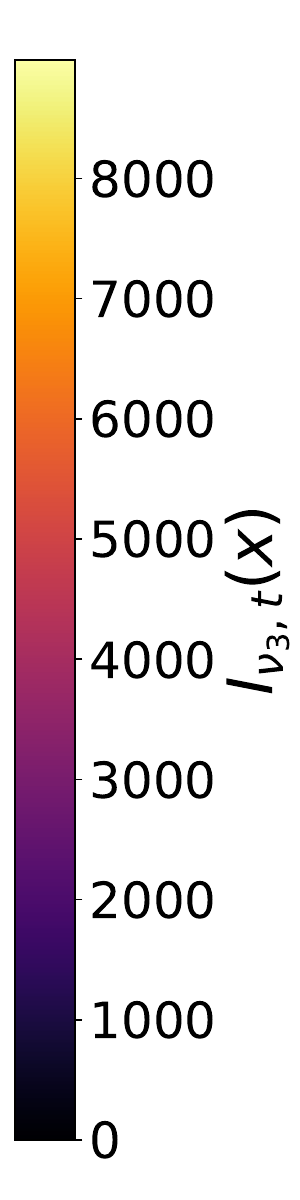}
    \end{subfigure}
    }


    \makebox[\textwidth][c]{
    \begin{subfigure}[b]{0.22\textwidth}
        \captionsetup{labelformat=empty}
        \includegraphics[width=\textwidth]{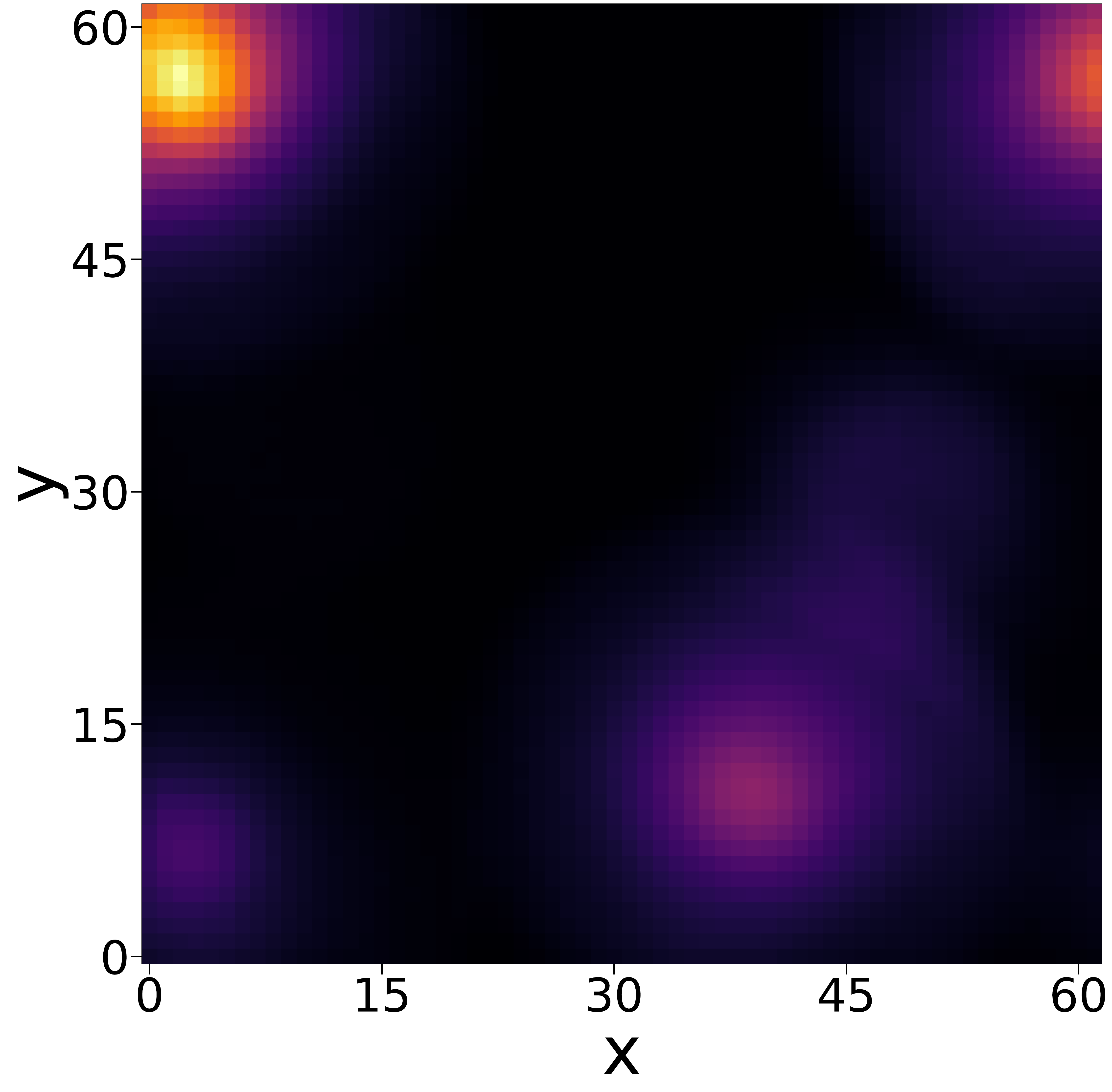}
    \end{subfigure}
    \begin{subfigure}[b]{0.22\textwidth}
        \captionsetup{labelformat=empty}
        \includegraphics[width=\textwidth]{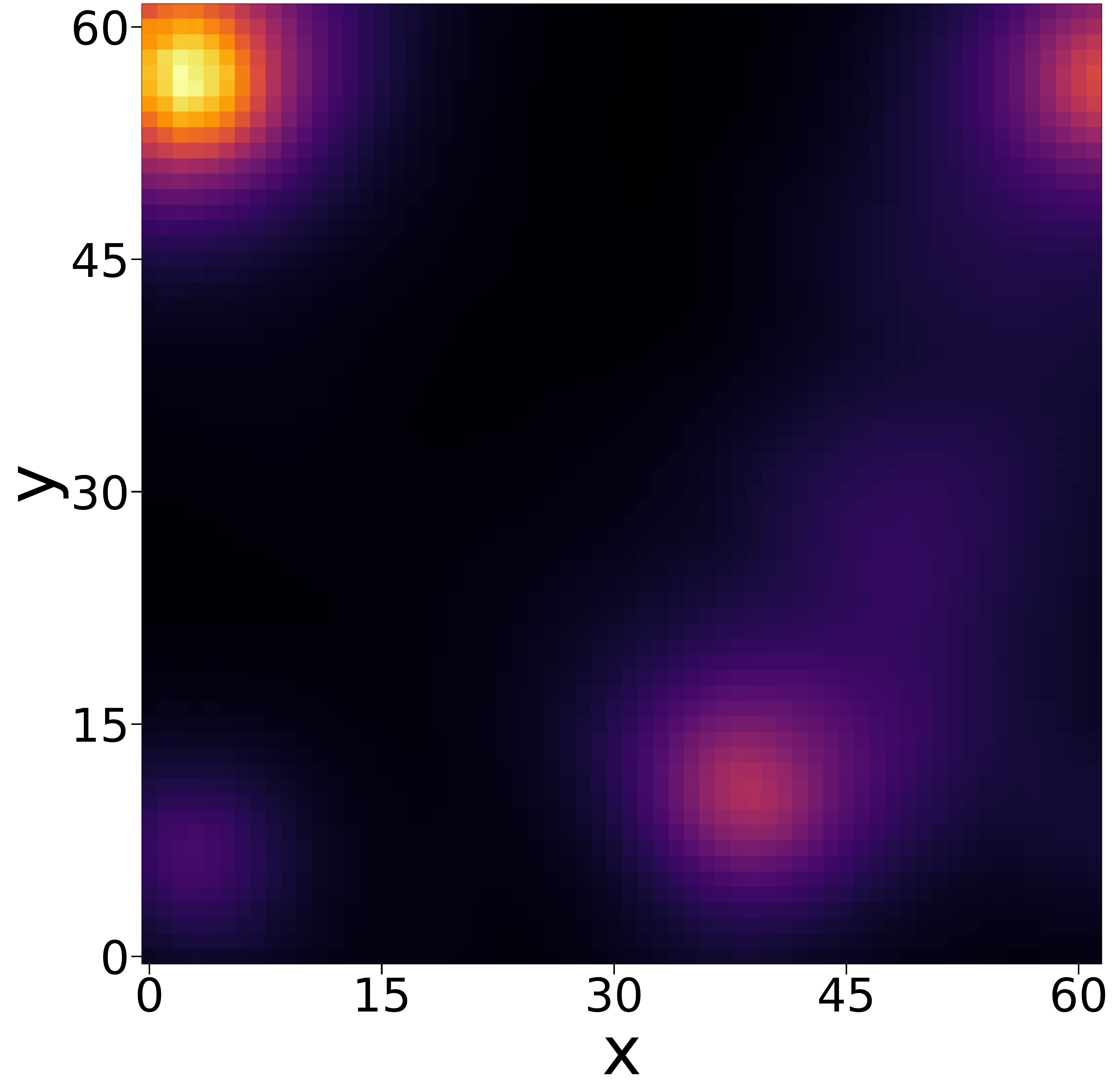}
    \end{subfigure}
    \begin{subfigure}[b]{0.22\textwidth}
        \captionsetup{labelformat=empty}
        \includegraphics[width=\textwidth]{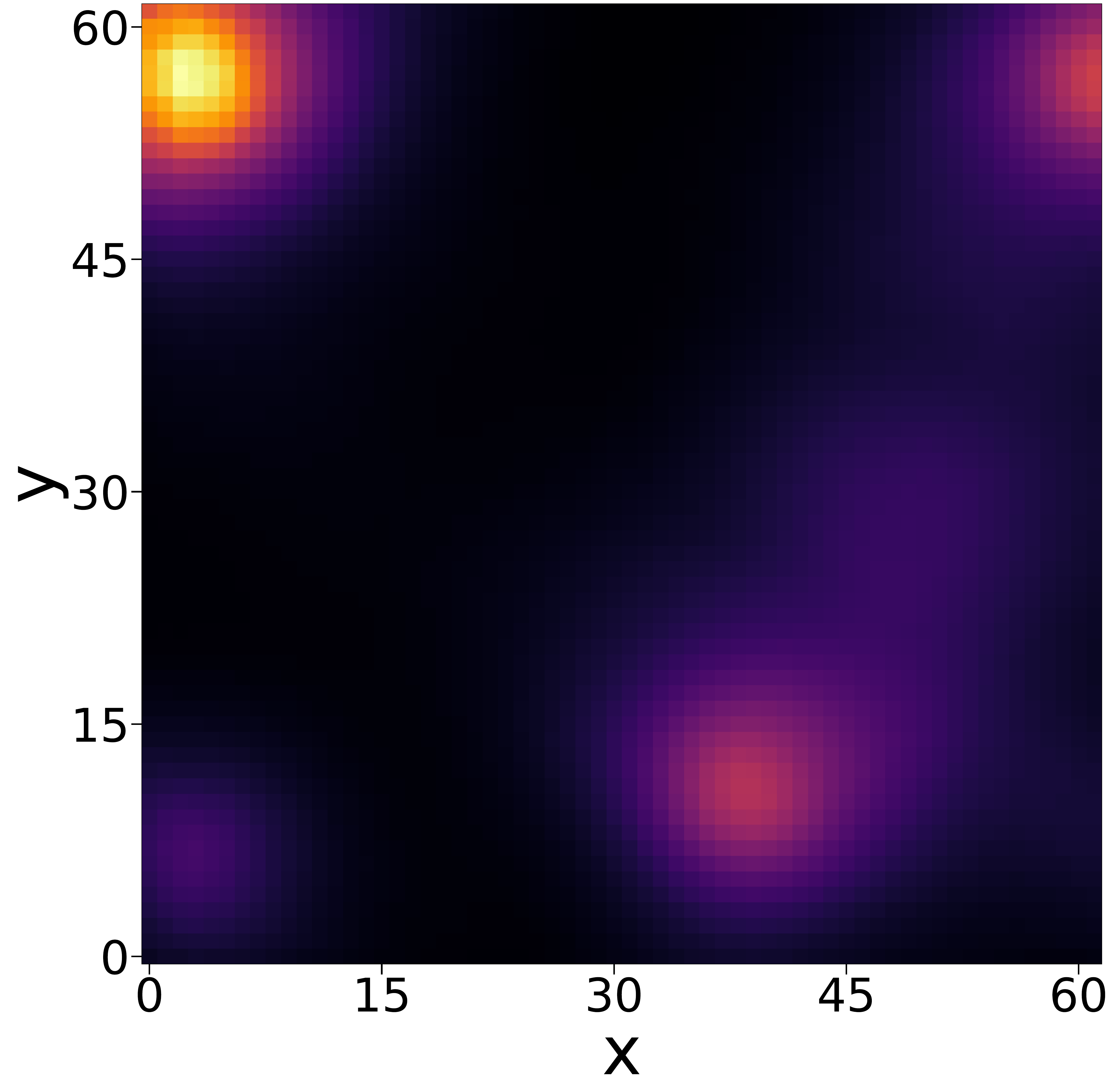}
    \end{subfigure}
    \begin{subfigure}[b]{0.22\textwidth}
        \includegraphics[width=\textwidth]{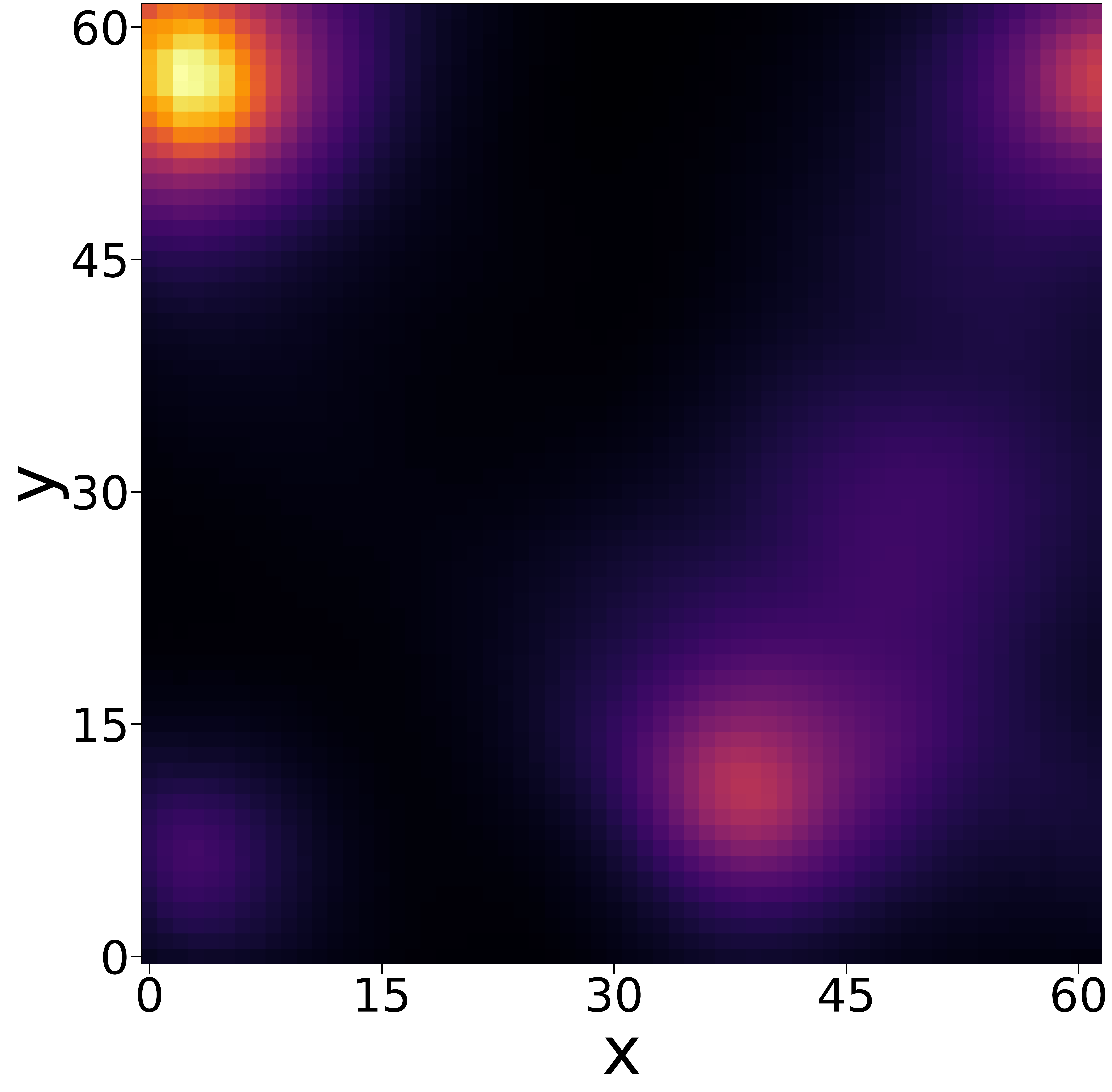}
    \end{subfigure}
    \begin{subfigure}[b]{0.0585\textwidth}
        \includegraphics[width=\textwidth]{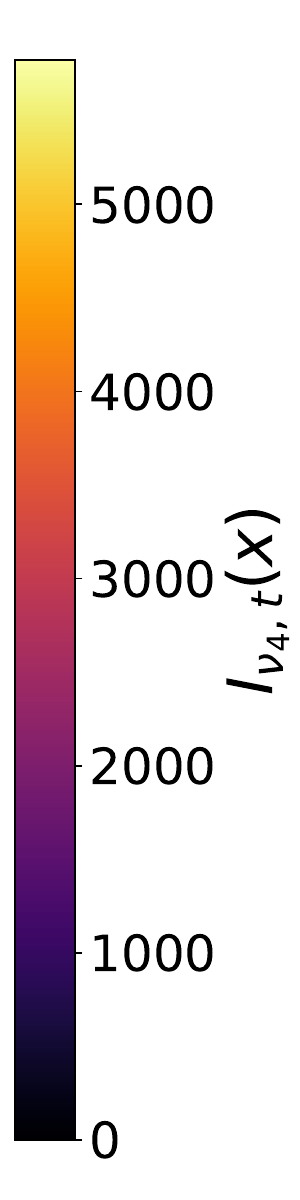}
    \end{subfigure}
    }

    \begin{tikzpicture}
        \def\arrowLength{8}
        \def\arrowHeight{0.15}
        \def\arrowTip{0.3} 
    
        \draw[fill=black!30, draw=black] 
            (0, -\arrowHeight/2) -- 
            ({\arrowLength - \arrowTip}, -\arrowHeight/2) -- 
            ({\arrowLength - \arrowTip}, -\arrowHeight) -- 
            (\arrowLength, 0) -- 
            ({\arrowLength - \arrowTip}, \arrowHeight) -- 
            ({\arrowLength - \arrowTip}, \arrowHeight/2) -- 
            (0, \arrowHeight/2) -- 
            cycle;
        \node at (\arrowLength/2, 0.45) {\large \textbf{Temporal evolution}};
    \end{tikzpicture}

    \caption{Selected timesteps of the temporal evolution of radiative intensity at  z=32 for three representative frequency bins: each row corresponds to a different frequency bin. 
    }
    \label{fig:3d_time_frequency}
\end{figure}

Our current research focuses on training a unified model that jointly predicts the temporal evolution of radiative intensity across multiple frequency bins simultaneously. Similar to the monochromatic setup presented in the main part, we generate absorption and emission fields based on turbulent periodic boxes produced with the hydrodynamic code \texttt{jf1uids}~\cite{2024arXiv241023093S}in a $64^3$ domain. However, instead of a single frequency, we now consider six frequency bins and vary  $j(\mathbf{x}) = j_{\nu} \rho$ and $a(\mathbf{x}) = k_{\nu, a} \rho$ across bins. For each frequency bin, the corresponding radiative intensity evolution is computed via ray tracing. Training is then performed using pairs of consecutive intensity fields
($\{I_{\nu_k,t}(\mathbf{x})\}_{k=1}^6$, $\{I_{\nu_k,t+1}(\mathbf{x})\}_{k=1}^6$) 
Figure~\ref{fig:3d_time_frequency} illustrates selected time steps of the models' predicted temporal evolution of radiative intensity for three representative frequency bins, where each row corresponds to a different frequency bin.

As shown, the frequency dependence of the absorption results in different temporal evolutions of the radiative intensity across the different frequency bins. This is intended purely as an outlook on ongoing work, as the relative error is roughly two magnitudes higher than for the single-frequency model presented in the main part. Hence, further architecture-- and hyperparameter--optimization is necessary to account for the increased data complexity. This is also the reason why we do not include a comparison to the numerical reference here.
Due to the large size of the data ($ 6 \times 64 \times 64 \times 64 \approx 1.5$ million pixels for a single time step), training such a unified model is highly computationally demanding. Consequently, it still may often remain advantageous to generalize the approach from the main part by training separate models for individual frequency bins.

\section{Details of the U-FNO Architecture.}
\label{architecture}
Due to the page limit of the main part, a comprehensive description of the basic model architecture used in this work is provided here. 
For our models, we combined a typical FNO architecture with additional U-Nets in each Fourier layer, following the approach in~\cite{2022AdWR..16304180W} to build models capable of capturing global dependencies as well as modeling fine-scale features.

The original FNO architecture was proposed in 2020 in~\cite{2020arXiv201008895L} and relies on the use of the Fourier Transformation. Motivated by the previous extension of PDE-solving Fourier spectral methods to neural networks~\cite{2019JCoPh.384....1F, jiang2019enforcing}, FNOs have since then proven to be very effective in solving high-dimensional PDEs.
As a class of Neural Operators, they can be seen as a generalization of traditional neural networks that enable mapping between infinite-dimensional function spaces. Consequently, unlike in traditional neural networks, their layers do not define an affine transformation but instead rely on integral operators that capture intrinsic properties of the input function.
A key feature of FNOs is that the integral operator can be represented as a convolution, which allows for computation via the  Fourier Transformation. Specifically, the input is transformed to Fourier space, multiplied with learnable weights, and then transformed back using inverse Fourier Transformation.
To implement this in practice, functions are typically discretized on a uniform, equidistant grid, resulting in a finite-dimensional tensor representation. This discretization enables efficient evaluation of the integral operator via the Fast Fourier Transform (FFT), as implemented in modern deep learning libraries~\cite{2024arXiv240215715K}. To further improve computational efficiency, the resulting Fourier coefficients are truncated beyond a fixed number of modes $K$ in each dimension, with the remaining entries set to zero. This reduces model complexity and mitigates overfitting by concentrating on the most relevant frequency components and thereby encourages smoother, more stable predictions. The convolutional integral operator is combined with an affine transformation. Together, this makes up a so-called Fourier layer.


As data gets more complex, FNOs often struggle to model fine-scale details, resulting in blurry predictions that lack sharpness and thus fail to accurately model steep gradients and discontinuities. Therefore, in our model the previously described Fourier layers  are extended with U-Nets~\cite{2015arXiv150504597R}, resulting in the U-Fourier Layers. 
As mentioned in the main part, a U-Net is an encoder–decoder convolutional network with symmetric downsampling and upsampling paths linked by skip connections. By choosing small kernels with sizes of 2 or 3 in each spatial dimension, the additional U-Nets allow our model to precisely locate fine-scale details in the prediction of the intensity patterns.


In the final architecture, a concatenation of U-Fourier Layers is preceded by a lifting layer that maps the input into a higher-dimensional latent space and followed by a projection layer that maps it back to the desired output dimension. 

\section{Dataset}
\label{datasets}
In the following, we provide a detailed description of the dataset used to train the surrogate model we present in the main part. As described there, this dataset consists of samples that each contain an absorption field and an emission field, along with ten snapshots representing the respective temporal evolution of the radiative intensity starting from $I_{\nu,0}(\textbf{x})=0$ and evolving in fixed-size time steps.
As described in the main part, the absorption and emission fields are generated based on turbulent periodic boxes produced with the hydrodynamics code \texttt{jf1luids}~\cite{2024arXiv241023093S} in a $64^3$ domain. 
To produce sufficiently heterogeneous density-- and thus opacity--fields, we vary the turbulence random seed, the amplitude of velocity fluctuations, and the slope of the turbulent kinetic-energy power spectrum across simulations and run each simulation until the turbulence spectrum reaches a stable equilibrium.
This setup allows us to mimic conditions typical of  giant molecular clouds (star-forming 
regions) that are among the most extensively studied numerically for their radiative effects~\citep{davide2020}.
Subsequently, the absorption field $a(\mathbf{x})$ is correlated with the density field, and radiation sources $j(\mathbf{x})$
are placed in the top 1.5\% of the density field.
As a result, the network inputs span a wide variety of structures with statistical properties directly inherited from the underlying fluid dynamics.

To complete the dataset, we employ a ray tracing algorithm\footnote{Code available at \url{https://github.com/lorenzobranca/Ray-trax.git}.} to compute the corresponding radiative intensity evolution across ten time steps for each pair ($a(\mathbf{x})$, $j(\mathbf{x})$).
The code determines the radiative intensity by following rays emitted from the sources as they propagate through the domain and interact with matter.
To align with state-of-the-art hydrodynamic simulations, we consider radiative transfer in a scattering-free regime, as scattering is typically
neglected in on-the-fly computations. Moreover, the angular dependence of the radiative intensity is currently omitted by summing the total intensity in each pixel but is planned for future inclusion.
The entire radiation code is implemented in JAX~\cite{jax2018github} and exploits parallelization to achieve highly efficient computations.  
Starting from $I_{\nu,0}(\textbf{x})$=0, the ray tracing code computes the radiative intensity $I_{\nu_k,t}(\textbf{x})$ based on ($a(\mathbf{x})$, $j(\mathbf{x})$) for ten consecutive, equally spaced time steps.
In total,  400 
pairs of absorption and emission fields are simulated, and for each pair, a full ray-tracing simulation 
is performed to compute the corresponding temporal evolution of the radiative intensity.
%

\begin{figure}[h]
  \centering
      \includegraphics[width=\linewidth]{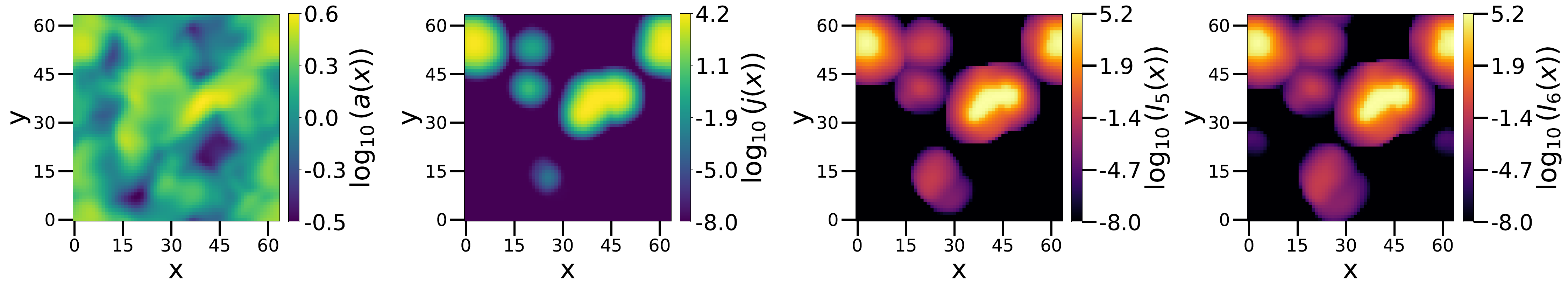}
  \caption{Absorption field (first image), emission field (second image), and the corresponding numerically computed  radiative intensity at z=32 at time step t=8 (third image) and t=9 (fourth image) of the test sample for which the comparison between prediction and numerical reference is shown in Figure~\ref{fig:3d_time}.
  }
  \label{fig:a_j_3d_time}
\end{figure}

%
For training the surrogate model on next-state intensity prediction, 
this dataset needs to be further processed.
Each  sample ($a(\textbf{x})$, $j(\textbf{x})$, $I_{\nu,0}(\textbf{x})$, $\ldots$, $I_{\nu,9}(\textbf{x})$) is split up into 9 new samples of the form ($a(\textbf{x})$, $j(\textbf{x})$, $I_{\nu,t}(\textbf{x})$, $I_{\nu,t+1}(\textbf{x})$), 
with each new sample containing two consecutive radiative intensity fields, as well as the corresponding absorption and emission fields. Thus, the network can be trained to predict the radiative intensity at a given time t (in the sample shown in Figure~\ref{fig:a_j_3d_time}: t=9) based on an absorption and an emission field as well as the radiative intensity at time t-1 (in the sample shown in Figure~\ref{fig:a_j_3d_time}: t-1=8).
The final dataset therefore contains 3600 samples, which proves to be sufficient given 
Neural Operators' ability to leverage the high functional diversity in the training data.

Figure~\ref{fig:a_j_3d_time} shows the absorption field and the emission field, as well as the radiative intensity at the midplane (z=32) at time t=8 and t=9 from the test sample whose temporal intensity evolution is shown in Figure~\ref{fig:3d_time}.
During inference, full temporal evolution is obtained by recursively feeding predictions back as input. 
Unlike in the main part, we show the unpreprocessed fields, so the intensity map appears different than in Figure~\ref{fig:3d_time}.

As this illustrates, both the inputs and outputs represent highly complex functions, posing a particularly challenging learning task that highlights the effectiveness of the proposed architecture.


\section{Steady-state predictions} 
\label{steady-state}

As mentioned in the main part, to contextualize the performance of our approach relative to existing emulators, we also apply it to monochromatic three-dimensional static radiative transfer, neglecting scattering and angular dependence. 

Therefore, we generate a dataset consisting of diverse samples, each comprising an absorption field $a(\textbf{x})=k_{\nu,a}  \rho(\textbf{x})$  and an emission field $j(\textbf{x})=j_{\nu}\rho(\mathbf{x})$ as input, and the resulting steady-state radiative intensity $I_\nu(\textbf{x})$ as target.
To construct these fields, the density in a $64^3$ domain is modeled as a log-normal random field, mimicking the structure of a turbulent star-forming region \citep[e.g.][]{2011ApJ...730...40P, 2012ApJ...761..156F, 2014ApJ...781...91G, Buck2022}. 
High-density regions (top 1\% of pixels) are masked as belonging to a star in order to assign them as centers for Gaussian emissivity profiles to mimic stellar emission for $j(\textbf{x})$. Additionally, the absorption field is defined as $a = k_{\nu,a} \rho(\textbf{x})$, where $k_{\nu,a}$, which determines the medium’s opacity, is set to one for simplicity. 
Varying the random seed of the density field initialization ensures diverse, non-redundant samples. 



To determine the radiative intensity for each pair of absorption and emission fields, we employ a tracing algorithm\footnote{Code available at \url{https://github.com/lorenzobranca/Ray-trax.git}.}. 
As before, we restrict the calculation to monochromatic radiation and neglect scattering while also omitting angular dependence by summing the total intensity in each pixel.
For each pair of absorption and emission fields, the radiative intensity is computed in the limit for $t \to \infty$, i.e. the equilibrium setting in after a long time.
In total, 1000 samples are generated, a number sufficient given the ability of Neural Operators to leverage the high functional diversity in the training data.
Modeling the density as a log-normal random field ensures that input fields span a wide variety of structurally different functions 
rather than mere parametrized variants of one profile. Consequently, learning a mapping from inputs to their corresponding outputs represents a highly complex task, thereby demonstrating the Neural Operators' capability to process and generalize across complex function spaces.

\begin{figure}[h]
  \centering
      \includegraphics[width=\linewidth]{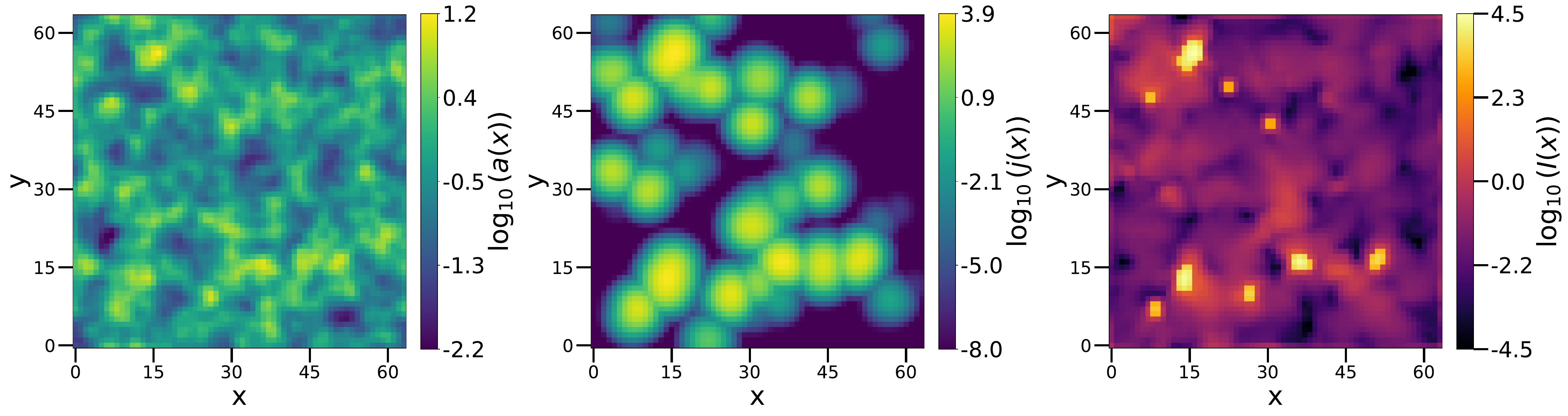}
  \caption{Absorption field (left), emission field (middle), and the corresponding  numerically computed steady-state radiative intensity (right) of the test sample for which the comparison between prediction and numerical reference is shown in Figure~\ref{fig:3d}.
  }
  \label{fig:a_j_3d}
\end{figure}

\begin{figure}[h]
  \centering
      \includegraphics[width=\linewidth]{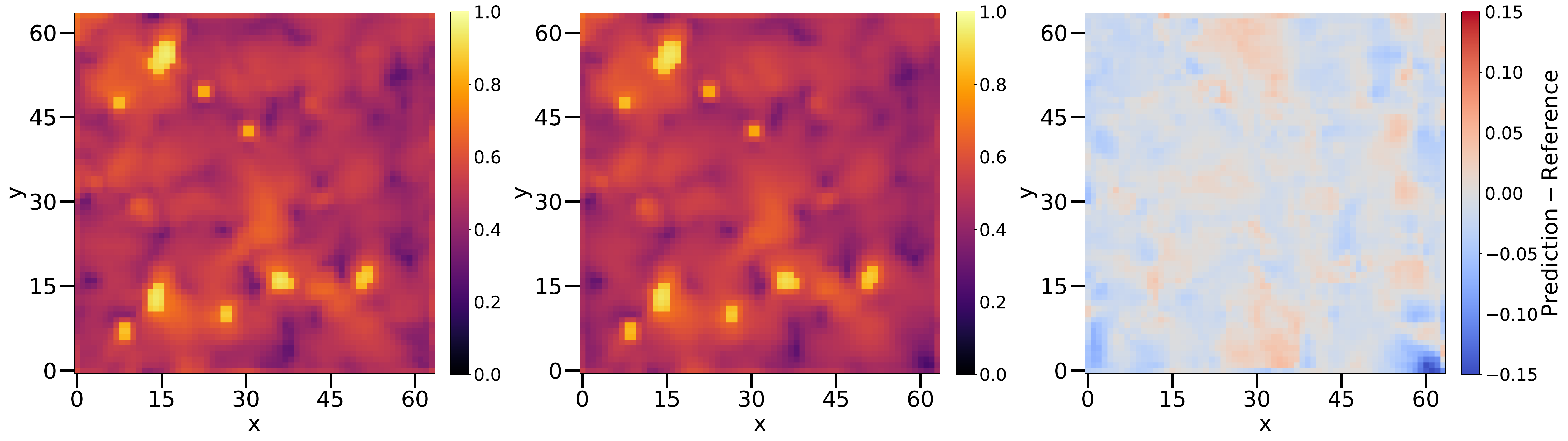}
  \caption{
  Comparison of steady-state radiative intensity at 
  z=32: preprocessed numerical reference (left), model prediction (middle), and corresponding residual (right). 
}
  \label{fig:3d}
\end{figure}

Figure~\ref{fig:a_j_3d} shows an example input, namely an absorption field (left panel) and an emission field (middle panel),  together with the corresponding target, i.e. the numerically computed steady-state radiative intensity field (right panel). This is the same sample for which the comparison between numerical reference  and prediction  is shown in Figure~\ref{fig:3d}. 
Similar to this figure, the cross-section at the mid-plane at z=32 is displayed. Note that Figure~\ref{fig:a_j_3d} displays the unpreprocessed fields, so the intensity map appears different than in Figure~\ref{fig:3d}.
As Figure~\ref{fig:a_j_3d} shows, the inputs represent highly complex functions, producing intricate intensity patterns that the surrogate model must learn to reproduce.


For the training, the dataset is split into a training (70\%), validation (10\%), and test (20\%) set, and key hyperparameters are optimized using Optuna~\cite{2019arXiv190710902A} (see Appendix~\ref{parameters} for details). 
To stabilize the training process, all fields are log-transformed and subsequently min–max normalized to the range $[0,1]$. Additionally, a relative loss combining pixelwise and spatial gradient differences is used to encourage sharp feature reconstruction. 
Training and evaluation are performed on a single NVIDIA A100 GPU.

Figure~\ref{fig:3d} shows a comparison of the preprocessed (log-transformed and min-max normalized) numerically  computed (left panel) and the predicted (middle panel) radiative intensity for a random sample from the test set at the mid-plane (z=32). Additional cross-sections are shown in Appendix~\ref{extended_crosssections}.
%
%
The predicted intensity 
closely matches the numerical reference. 
The sample reflects the model's overall test performance, demonstrating its ability to accurately approximate the underlying PDE, with a mean relative error of 2.6\% per pixel across the test set.
Notably, the surrogate model offers a speedup of \textasciitilde$6750\times$, achieving predictions in 0.003s compared to 20.3s for the numerical solver, and memory costs are independent of the number of sources.
The results match or even surpass the results reported in other 3D static RT emulators~\citep{COemu}.



\section{Hyperparameters of Models and Training Procedure} 
\label{parameters}
The following section provides additional implementation details on the hyperparameters of the surrogate model we present in the main part 
as well as the training setup and procedure. We additionally include this information for the previously mentioned model for steady-state radiative intensity predictions.
Both models are trained using the Adam optimizer with decoupled weight decay (AdamW)~\cite{2017arXiv171105101L}. The learning rate follows an exponential schedule.
Since the pixel values  in the inputs and outputs spanned over multiple orders of magnitude, the data is preprocessed prior to training. First, the logarithm is applied, followed by a min-max normalization to scale the data to the interval of $[0,1]$. This is performed independently for absorption, emission, and intensity fields.
Following the approach in~\cite{2022AdWR..16304180W}, we employ a loss function that additionally considers the spatial gradients of the predictions to encourage sharp feature reconstruction, resulting in a relative loss that accounts for deviations both in pixel values and in their spatial gradients. For a single prediction, its mathematical formulation is given by:

\begin{equation}
\label{eq:loss}
\mathcal{L} =\frac{\sqrt{\sum_{i=1}^N (\hat{f}_i - f_i)^2}}{ \sqrt{\sum_{i=1}^N (f_i)^2}} 
+ \lambda   \frac{ \sqrt{\sum_{i=1}^N (\nabla \hat{f}_i - \nabla f_i)^2}}{ \sqrt{\sum_{i=1}^N (\nabla f_i)^2}}
\end{equation}

Here, $i$ indexes the $N$ pixels, which are flattened into a one-dimensional vector. $f_i$ is the target value for pixel $i$, $\hat{f}_i$ the corresponding prediction, and $\nabla$ represents the spatial derivative.

To achieve optimal performance, most hyperparameters for both the  training procedure and the model architectures are optimized using Optuna. Optuna is an open source library for automatic hyperparameter optimization in machine learning, which enables efficient and systematic identification of the best hyperparameters for a given model~\cite{2019arXiv190710902A}.
The following tables list the values of the final key hyperparameters used for the model architectures and the model training.
Table~\ref{steady-table} shows the hyperparameters for the steady-state model, 
while Table~\ref{time_dep-table} lists those for the recurrent model presented in the main part. 
Among these hyperparameters, only the number of layers, the U-Net kernel size, and the loss-function parameter $\lambda$ are set manually rather than optimized with Optuna. The number of layers and the U-Net kernel size are fixed manually due to architectural and memory constraints, and $\lambda$ is set based on previous studies indicating that the chosen value provides a good compromise between predictive accuracy and sharpness.


For both models, we observe that training is robust to small variations in these hyperparameters. The ranges of Fourier modes and network widths tested were deliberately limited, as increasing them substantially raises computational costs while yielding only marginal improvements. Consequently, the models use a relatively low number of modes and a modest width, which proves sufficient to accurately capture the underlying mappings and produce highly precise predictions.
Additionally, manual tests of U-Net architectures with varying kernel sizes confirmed that smaller kernels better enable the models to resolve fine-scale details.
%
%
%
%
%
%

The steady-state model is trained on 700 samples (70\% of the dataset for the scenario of steady-state radiative intensity) for 40 epochs, taking approximately 80 minutes. The recurrent model from the main part is trained for 20 epochs. Since each of the 280 full-simulation samples (70\% of the dataset for the scenario of the temporal evolution of radiative intensity) is split into 9 training samples, the recurrent model is trained on substantially more training data. Combined with the model's increased width, this results in a considerably longer training time of approximately 320 minutes. 

\begin{table}[h]
  \caption{Hyperparameters of Architecture and Training Procedure for the Steady-State Model.}
  \vspace{1em} 
  \label{steady-table}
  \centering
  
  \begin{tabular}{llll}
    \toprule
    \multicolumn{2}{c}{Model} & \multicolumn{2}{c}{Training} \\
    \midrule
    Number of Layers & 6  & Initial Learning Rate  &  0.0005 \\
    Layer Width     & 16 & Decay Rate   &   0.9000\\
    Number of Modes     & 4       & Weight Decay  & 0.0050  \\
    U-Net Kernel Size     &    3    & Dropout Probability  & 0.08 \\
    U-Net Width     &    16    & $\lambda$ in Loss & 0.5  \\
    \bottomrule
  \end{tabular}
\end{table}

\begin{table}[h]
  \caption{Hyperparameters of Architecture and Training Procedure for the Recurrent Model.}
  \vspace{1em} 
  \label{time_dep-table}
  \centering
  \begin{tabular}{llll}
    \toprule
    \multicolumn{2}{c}{Model} & \multicolumn{2}{c}{Training} \\
    \midrule
    Number of Layers & 6  & Initial Learning Rate  &  0.0006 \\
    Layer Width     & 32 & Decay Rate   &   0.9120\\
    Number of Modes     & 4       & Weight Decay  & 0.0052  \\
    U-Net Kernel Size     &    2    & Dropout Probability  & 0.08 \\
    U-Net Width     &    32    &   $\lambda$ in Loss & 0.5  \\
    \bottomrule
  \end{tabular}
\end{table}


\section{Quantitative Analysis of Prediction Accuracy} 
\label{quantanalysis}
\label{quant_analysis}
\begin{figure}[h!]
  \centering
  \includegraphics[width=0.44\textwidth]{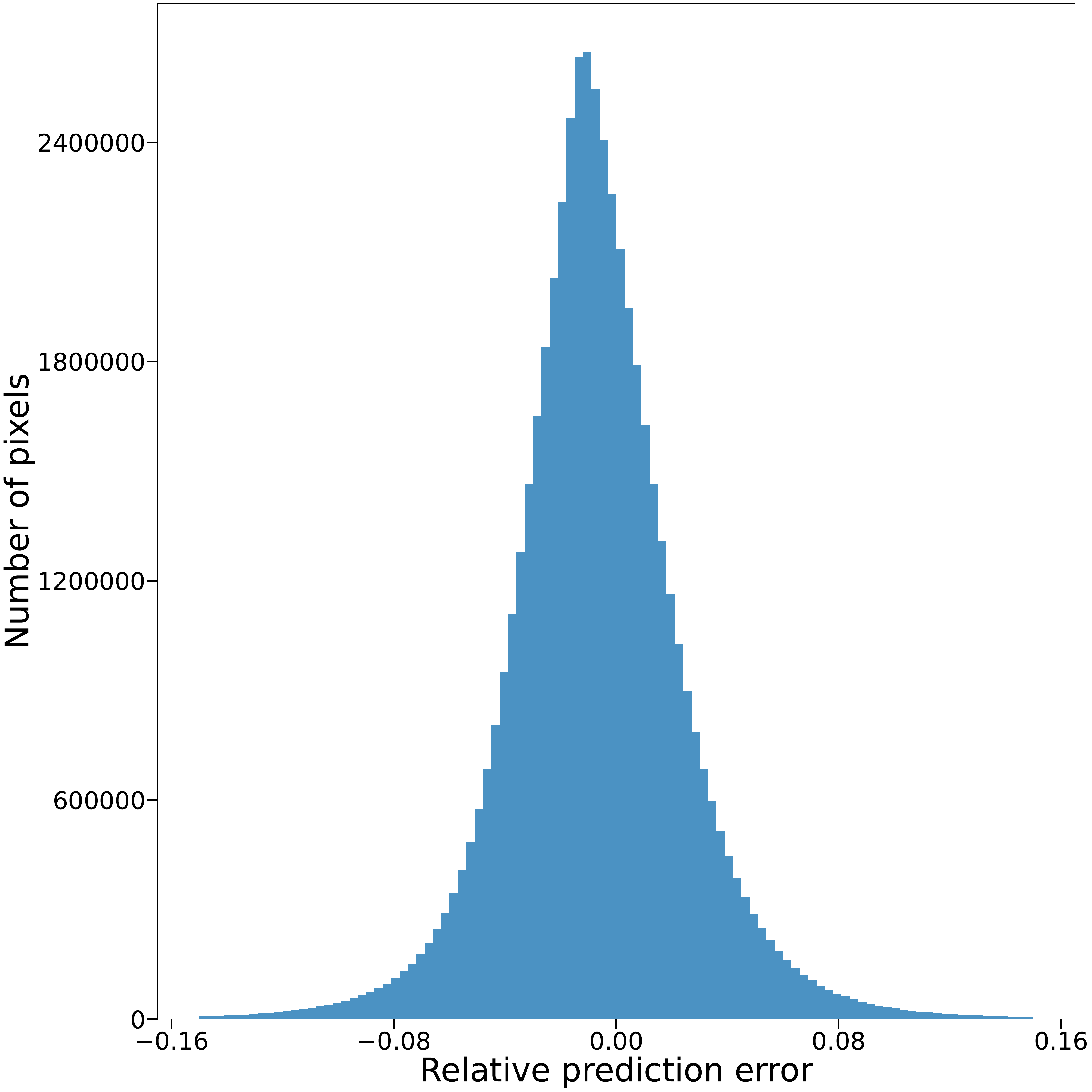}
  \caption{Pixelwise relative deviations between  the preprocessed numerically computed steady-state radiative intensity and the model prediction computed over all test samples.
  }
  \label{fig:hist_3d}
\end{figure}

To evaluate the overall predictive accuracy across the entire test set, we provide histograms of the pixelwise 
relative deviations. 
Similar to the previous section, we do this for both the model presented in the main part and the steady-state model.
Figure~\ref{fig:hist_3d} shows the relative deviations between  the preprocessed numerically computed steady-state radiative intensity and the steady-state model prediction, computed over all test samples.
The histogram indicates that the relative deviations approximately follow a Gaussian distribution. 
Overall, relative errors remain low across all pixels in the test set, with a slight bias towards underestimation. A plausible explanation is that the network produces smoother predictions, particularly when modeling sharp gradients and discontinuities, leading to reduced extrema. Since the intensity is generally high throughout the domain, these less extreme predictions result in predictions being too small on average and cause a negative bias. Nevertheless, the accuracy is very high, with only a small portion of pixels deviating by more than 8\% and an average per-pixel deviation of only 2.6\%.

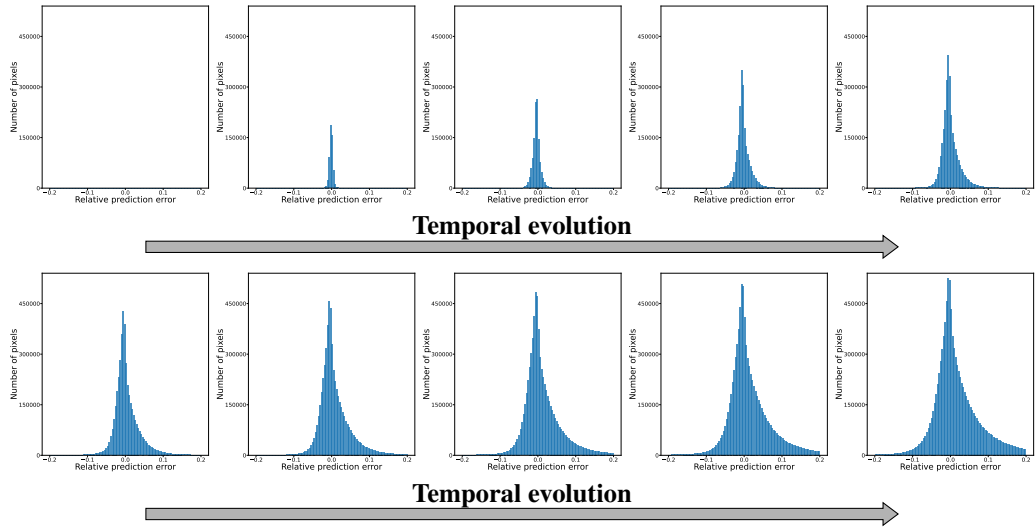
\begin{figure}[h!]
    \centering
    \makebox[\textwidth][c]{
    \begin{subfigure}[b]{0.19\textwidth}
        \captionsetup{labelformat=empty}
        \includegraphics[width=\textwidth]{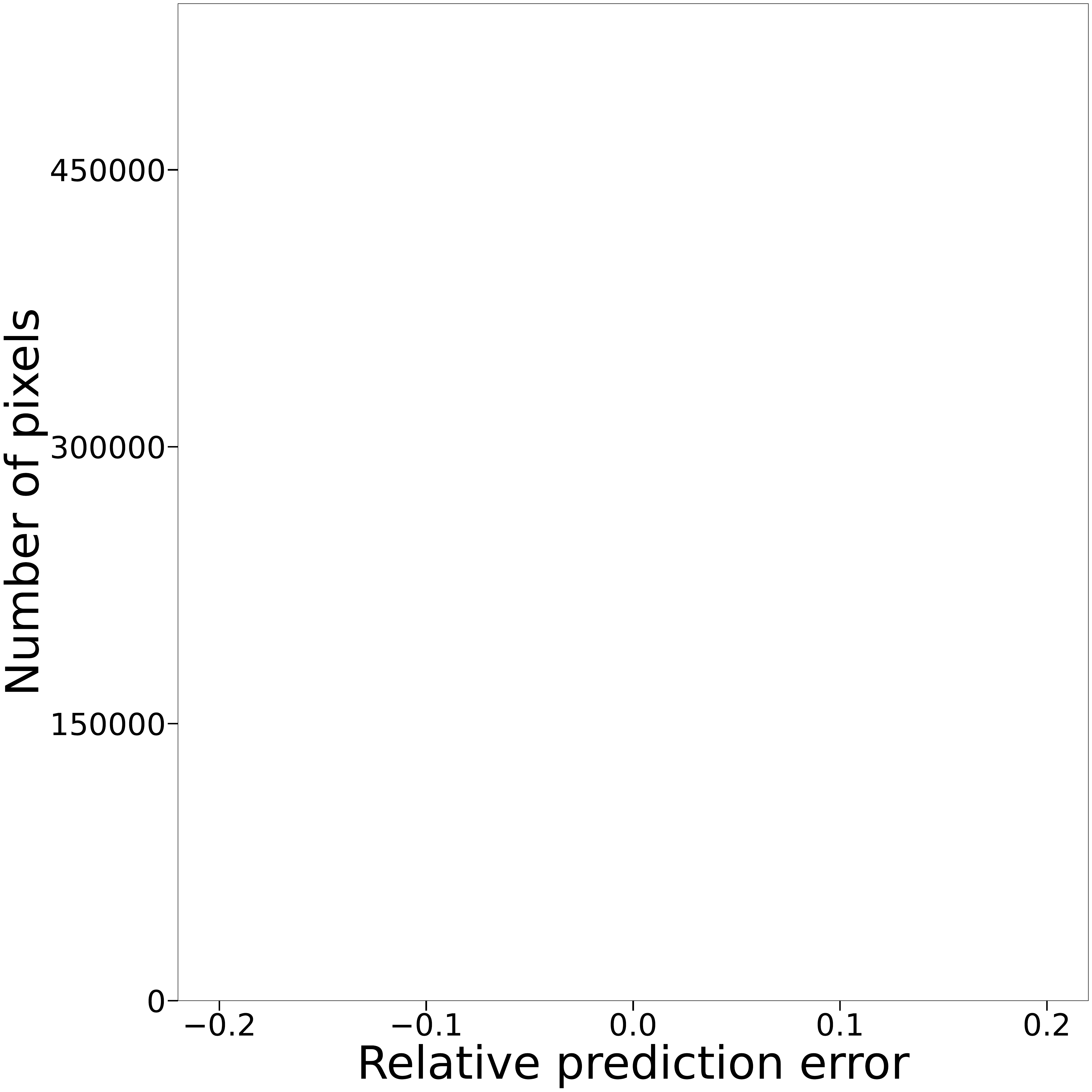}
    \end{subfigure}
    \begin{subfigure}[b]{0.19\textwidth}
        \captionsetup{labelformat=empty}
        \includegraphics[width=\textwidth]{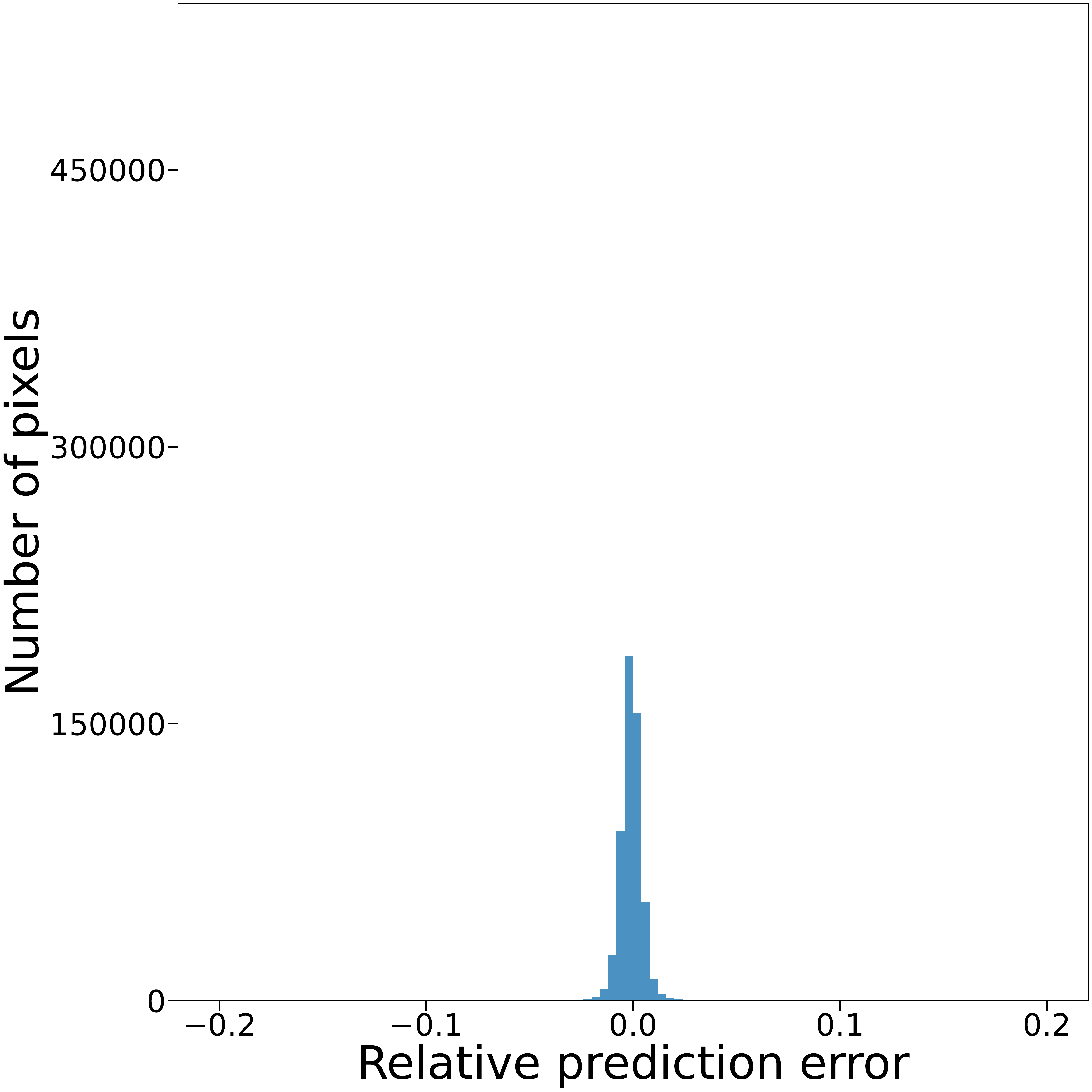}
    \end{subfigure}
    \begin{subfigure}[b]{0.19\textwidth}
        \captionsetup{labelformat=empty}
        \includegraphics[width=\textwidth]{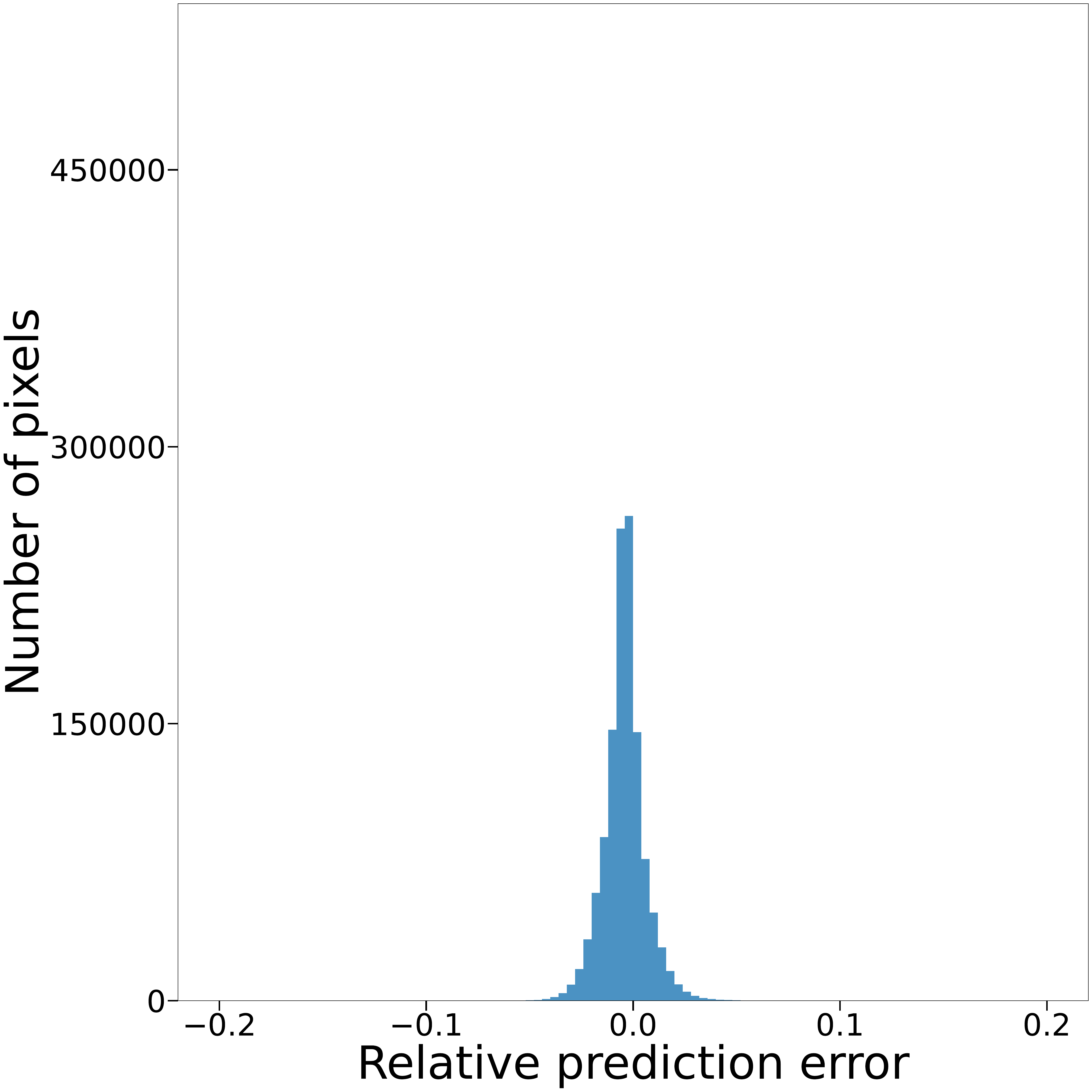}
    \end{subfigure}
    \begin{subfigure}[b]{0.19\textwidth}
        \captionsetup{labelformat=empty}
        \includegraphics[width=\textwidth]{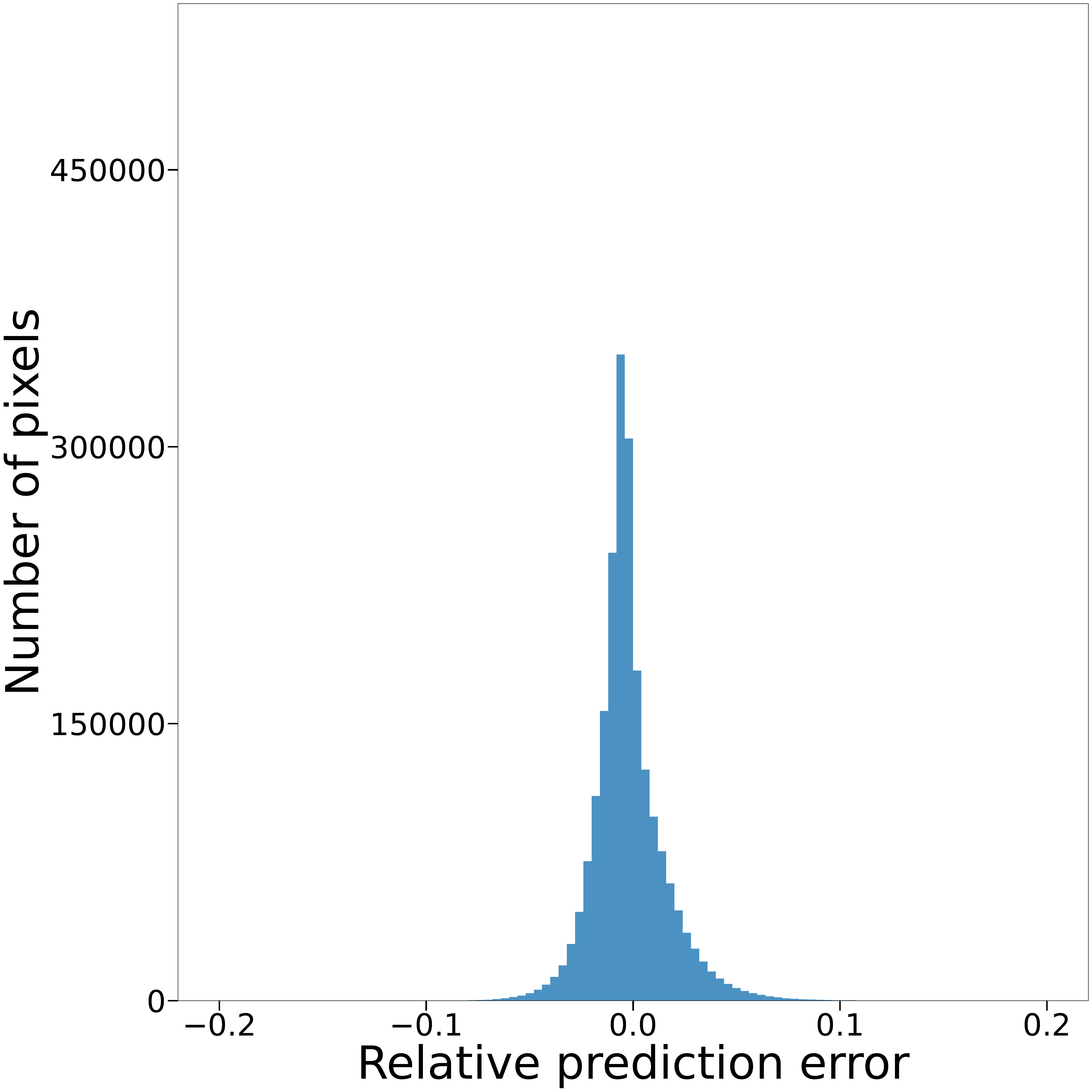}
    \end{subfigure}    
    \begin{subfigure}[b]{0.19\textwidth}
        \captionsetup{labelformat=empty}
        \includegraphics[width=\textwidth]{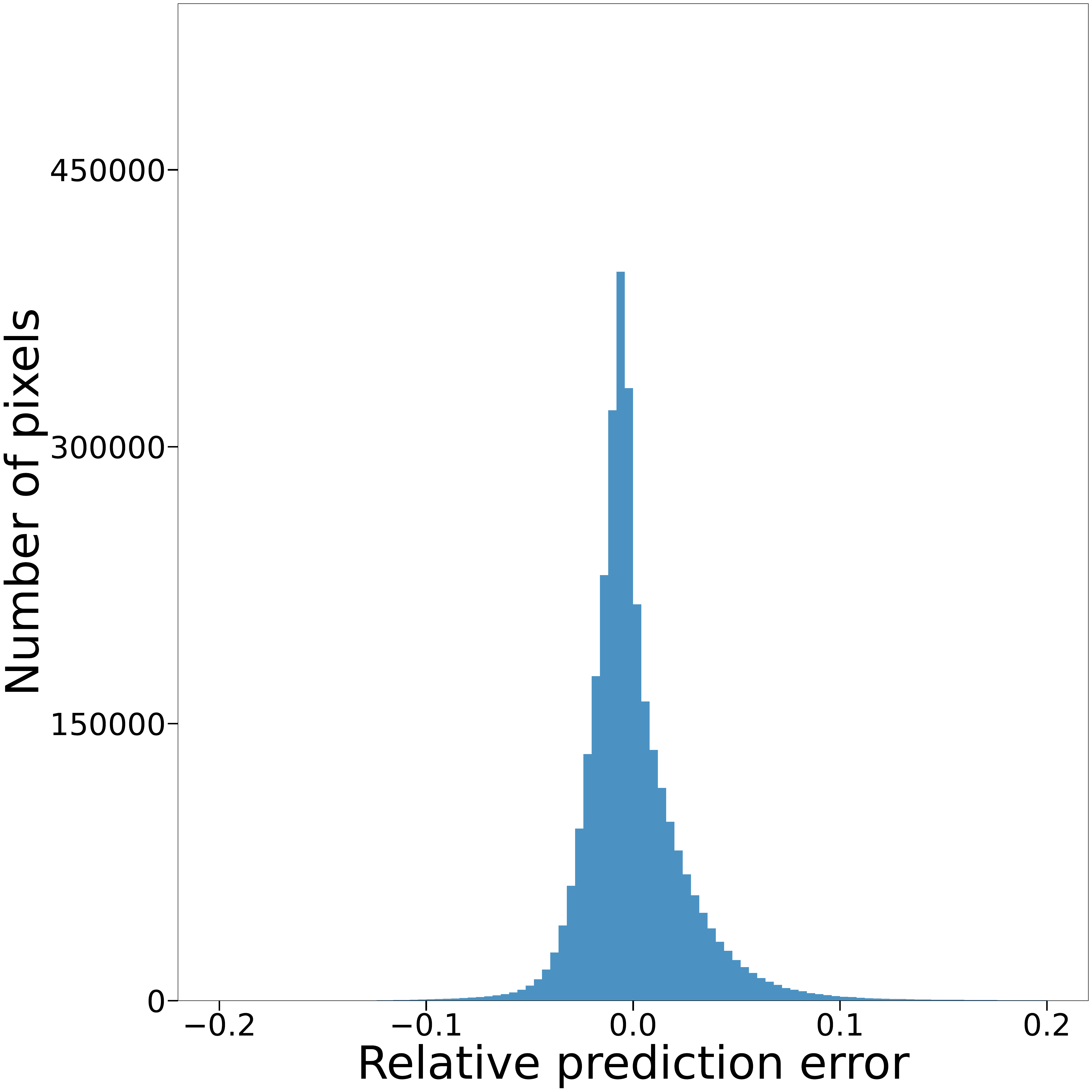}
    \end{subfigure}
    }
    \begin{tikzpicture}
        \def\arrowLength{10}
        \def\arrowHeight{0.15}
        \def\arrowTip{0.2} 
    
        \draw[fill=black!30, draw=black] 
            (0, -\arrowHeight/2) -- 
            ({\arrowLength - \arrowTip}, -\arrowHeight/2) -- 
            ({\arrowLength - \arrowTip}, -\arrowHeight) -- 
            (\arrowLength, 0) -- 
            ({\arrowLength - \arrowTip}, \arrowHeight) -- 
            ({\arrowLength - \arrowTip}, \arrowHeight/2) -- 
            (0, \arrowHeight/2) -- 
            cycle;
        \node at (\arrowLength/2, 0.25) {\normalsize \textbf{Temporal evolution}};
    \end{tikzpicture}

    \vspace{0.15cm}

    \makebox[\textwidth][c]{
    \begin{subfigure}[b]{0.19\textwidth}
        \captionsetup{labelformat=empty}
        \includegraphics[width=\textwidth]{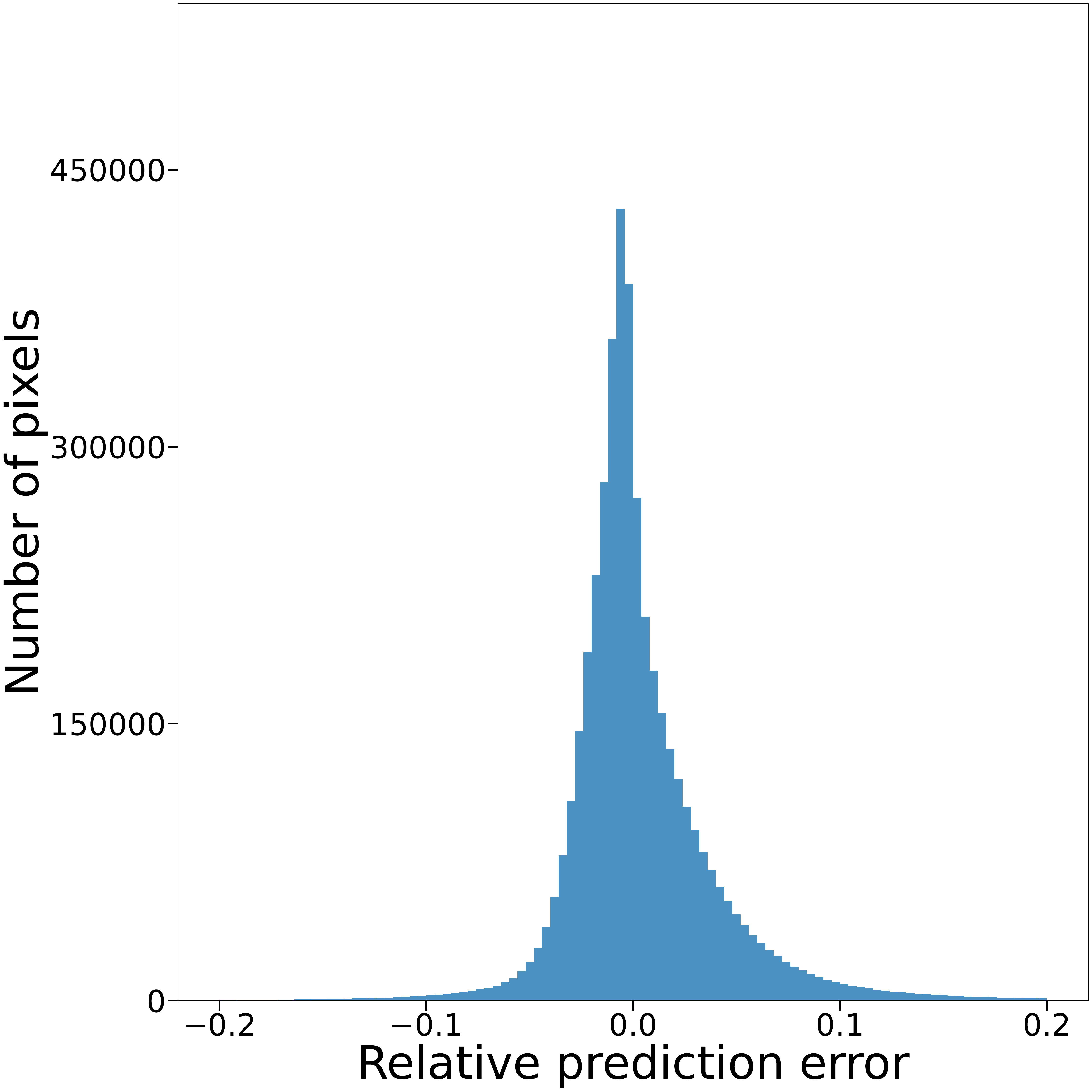}
    \end{subfigure}
    \begin{subfigure}[b]{0.19\textwidth}
        \captionsetup{labelformat=empty}
        \includegraphics[width=\textwidth]{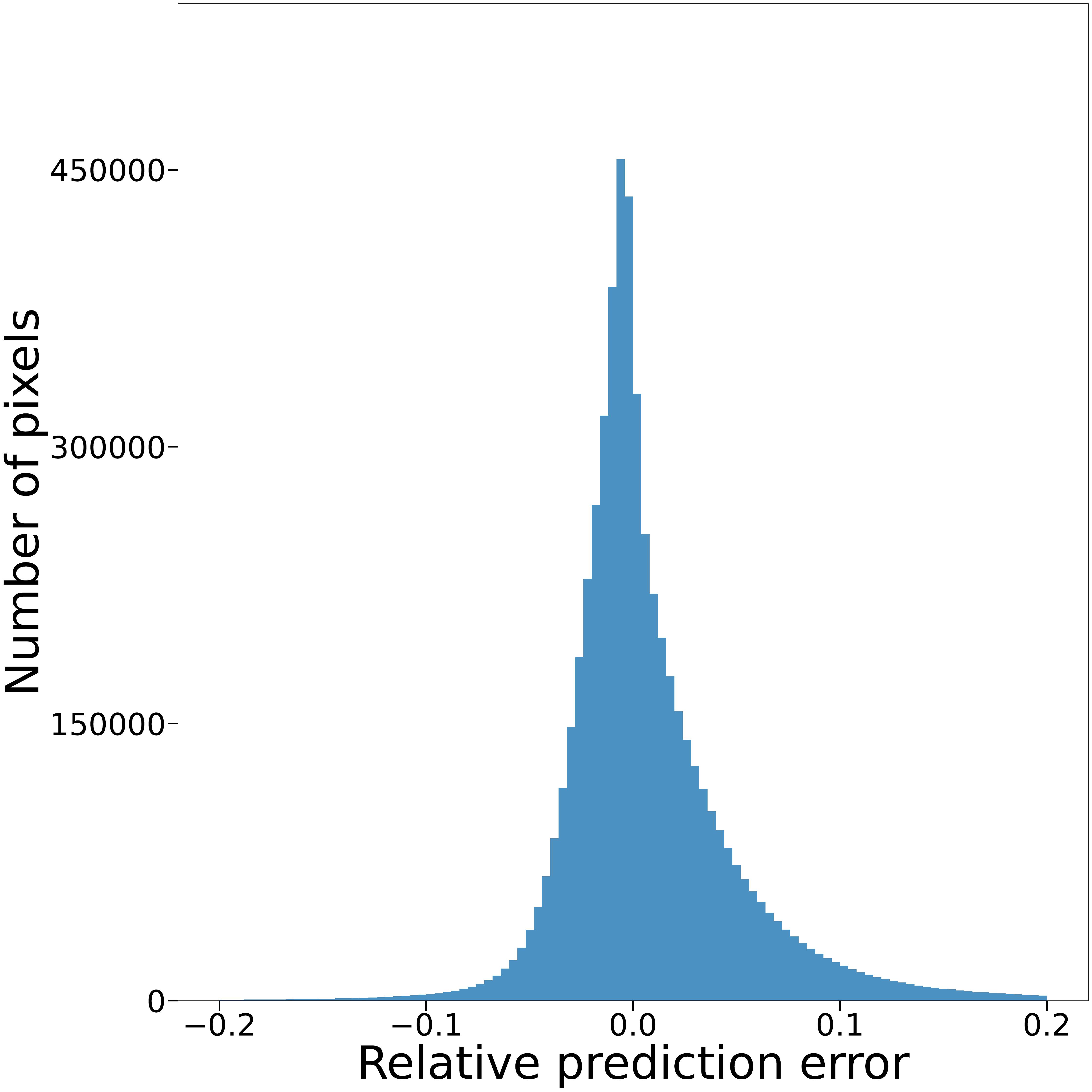}
    \end{subfigure}
    \begin{subfigure}[b]{0.19\textwidth}
        \captionsetup{labelformat=empty}
        \includegraphics[width=\textwidth]{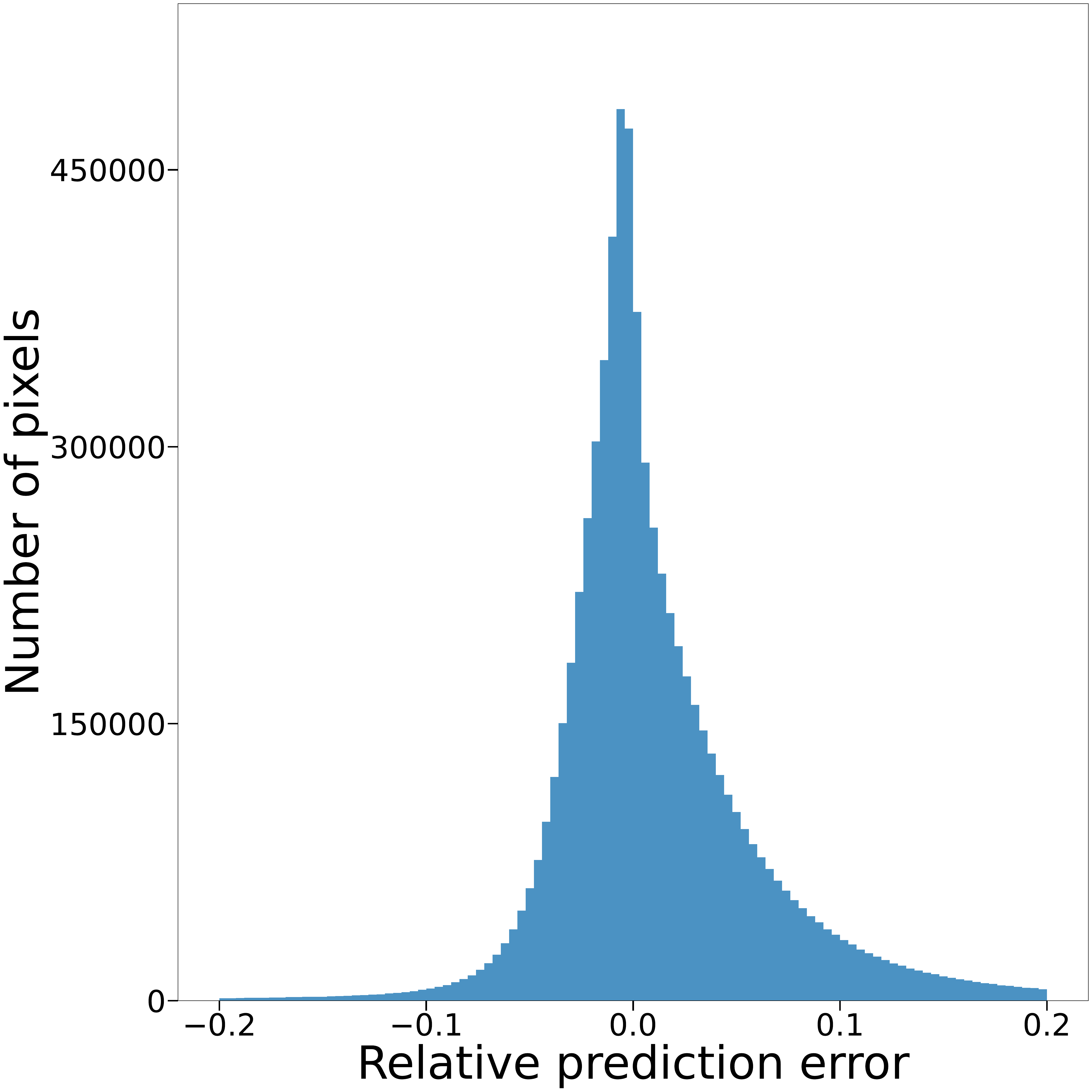}
    \end{subfigure}
    \begin{subfigure}[b]{0.19\textwidth}
        \captionsetup{labelformat=empty}
        \includegraphics[width=\textwidth]{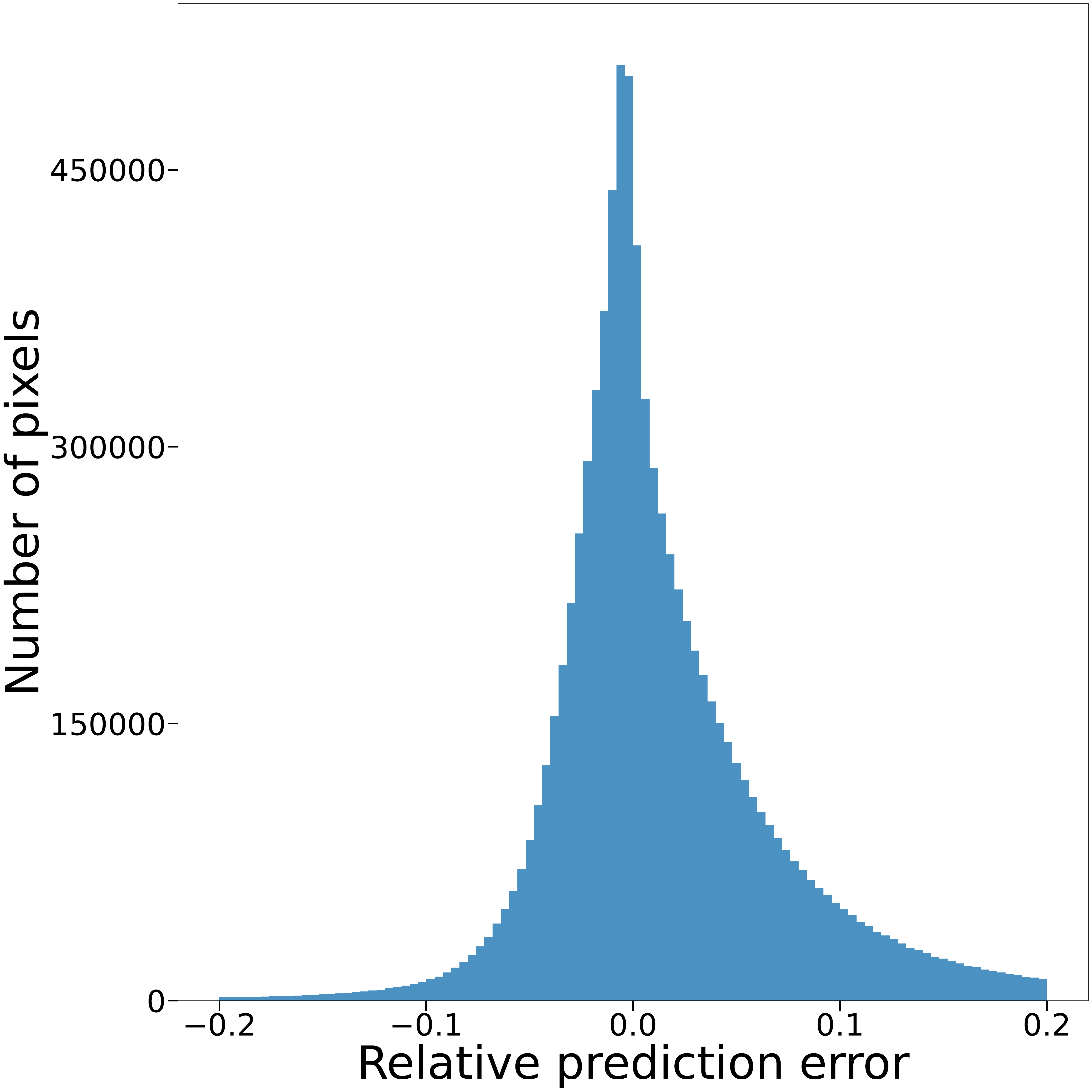}
    \end{subfigure}
    \begin{subfigure}[b]{0.19\textwidth}
        \captionsetup{labelformat=empty}
        \includegraphics[width=\textwidth]{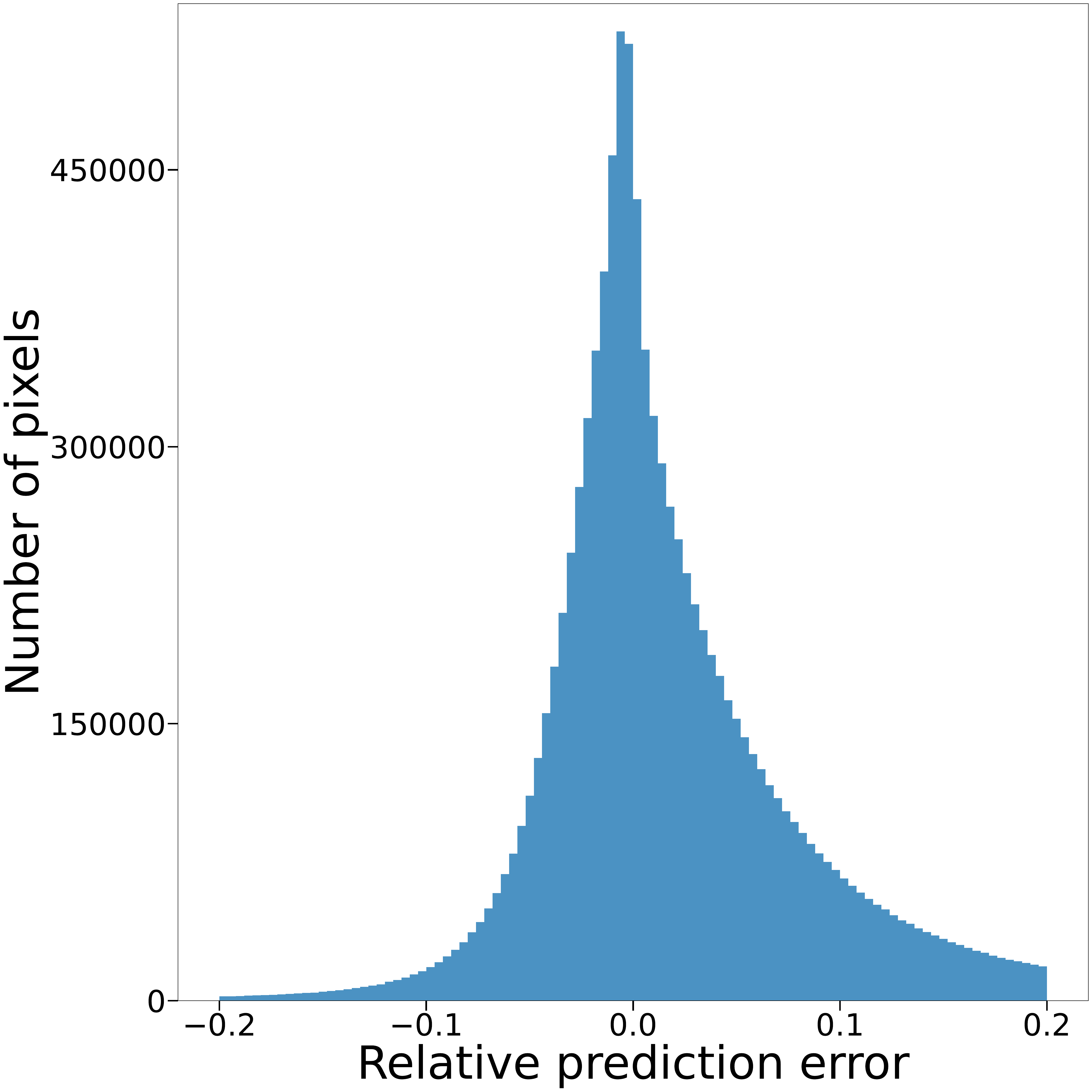}
    \end{subfigure}
    }
    \begin{tikzpicture}
        \def\arrowLength{10}
        \def\arrowHeight{0.15}
        \def\arrowTip{0.2} 
    
        \draw[fill=black!30, draw=black] 
            (0, -\arrowHeight/2) -- 
            ({\arrowLength - \arrowTip}, -\arrowHeight/2) -- 
            ({\arrowLength - \arrowTip}, -\arrowHeight) -- 
            (\arrowLength, 0) -- 
            ({\arrowLength - \arrowTip}, \arrowHeight) -- 
            ({\arrowLength - \arrowTip}, \arrowHeight/2) -- 
            (0, \arrowHeight/2) -- 
            cycle;
        \node at (\arrowLength/2, 0.25) {\normalsize \textbf{Temporal evolution}};
    \end{tikzpicture}

        \caption{Temporal evolution of pixelwise residuals (upper block) and relative errors (lower block) computed over all test samples: Timesteps proceed from top left to bottom right. 
        }
    \label{fig:histos_3d_time}
\end{figure}

Figure~\ref{fig:histos_3d_time} shows the pixelwise relative deviations at each time step of the recurrent radiative intensity predictions, computed over all test samples.
The relative deviations remain generally low across all timesteps. Even in the final time step, the majority of pixels exhibit relative deviations below 10\%.
However, a gradual decrease in accuracy over time is noticeable, caused by the accumulation of errors inherent to the recurrent prediction approach.
As stated in the main part, the average per-pixel relative error for next-state predictions is 2.9\%. 
During inference, however, the model uses its own previous outputs as inputs to predict the full temporal evolution.
Consequently, predictions are based on states that already deviate from the numerical reference.
The bias towards more positive residuals suggests that this causes a slightly faster evolution of structures in the predictions compared to the reference.
Nevertheless, the overall prediction accuracy remains high.

Note that in all histograms presenting relative deviations, we only include pixels for which the preprocessed numerically computed radiative intensity is unequal to zero. Hence, in Figure \ref{fig:histos_3d_time}, the histograms for later time steps contain a larger number of entries. Likewise, when computing the relative errors reported in the main part, pixels for which the preprocessed numerically computed radiative intensity is zero are excluded.





\section{Extended Cross-Sectional Analysis of Model Predictions}
\label{extended_crosssections}
\begin{figure}[p]
    \centering
    \begin{subfigure}[b]{1.0\textwidth}
        \captionsetup{labelformat=empty}
        \includegraphics[width=\textwidth]{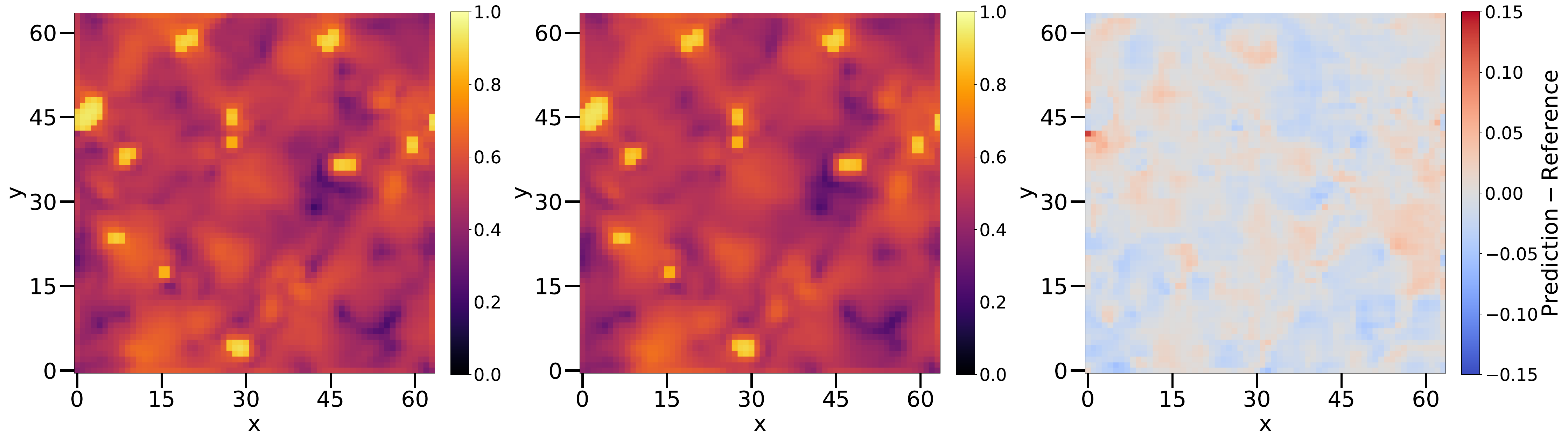}
    \end{subfigure}

    \begin{subfigure}[b]{1.0\textwidth}
        \captionsetup{labelformat=empty}
        \includegraphics[width=\textwidth]{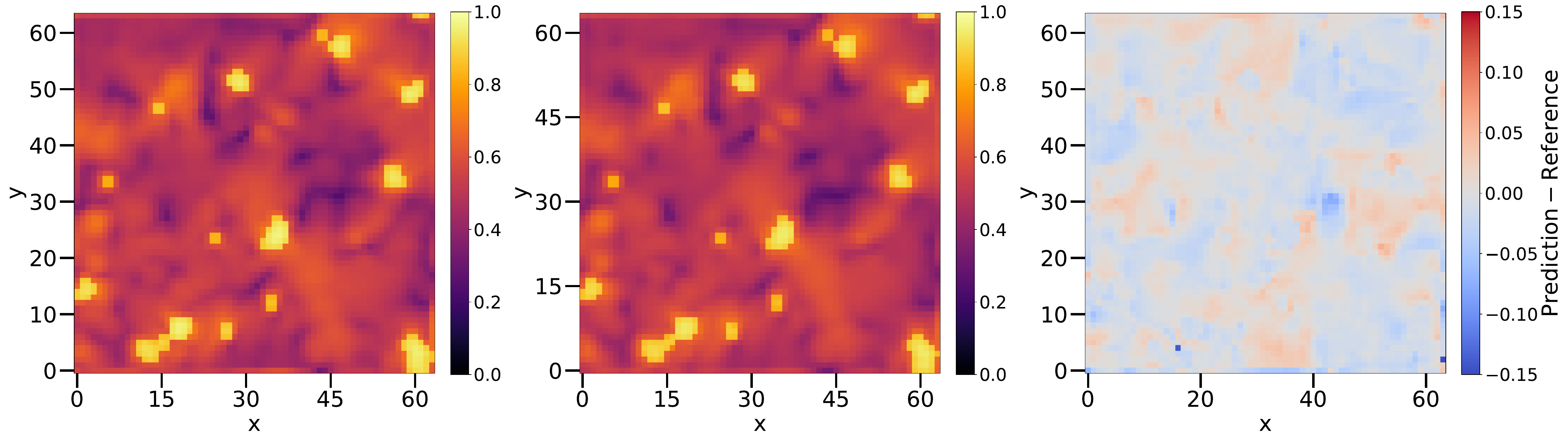}
    \end{subfigure}

    \begin{subfigure}[b]{1.0\textwidth}
        \captionsetup{labelformat=empty}
        \includegraphics[width=\textwidth]{graphics/3d_XY_plane.pdf}
    \end{subfigure}
    
    \caption{Comparison of the steady-state radiative intensity at cross-section x=32 (first row), y=32 (second row), and z=32 (third row): Respectively, in each row the left image shows the preprocessed numerical reference, the middle image the model prediction, and the right image the corresponding residual.}
    \label{fig:complete_3d}
\end{figure}

In addition to the z=32 cross-section displayed in Figure~\ref{fig:3d}, Figure~\ref{fig:complete_3d} for the same test sample presents three cross-sections, at x=32, y=32, and z=32, of the preprocessed (log-transformed and min-max normalized) numerical reference and the corresponding predicted radiative intensity.
These are shown together with the corresponding residuals to illustrate the model’s performance across all spatial directions and indicate that predictions are consistently accurate. Residuals in the x=32 and y=32 cross-sections are of similar magnitude to those in the z=32 cross-section that was already presented. This confirms the steady-state model’s ability to determine the radiative intensity distribution across all spatial dimensions.

\begin{figure}[p]
    \centering

\vspace{1cm}

\makebox[\textwidth][c]{
    \begin{subfigure}[b]{0.0882\textwidth}
        \captionsetup{labelformat=empty}
        \includegraphics[width=\textwidth]{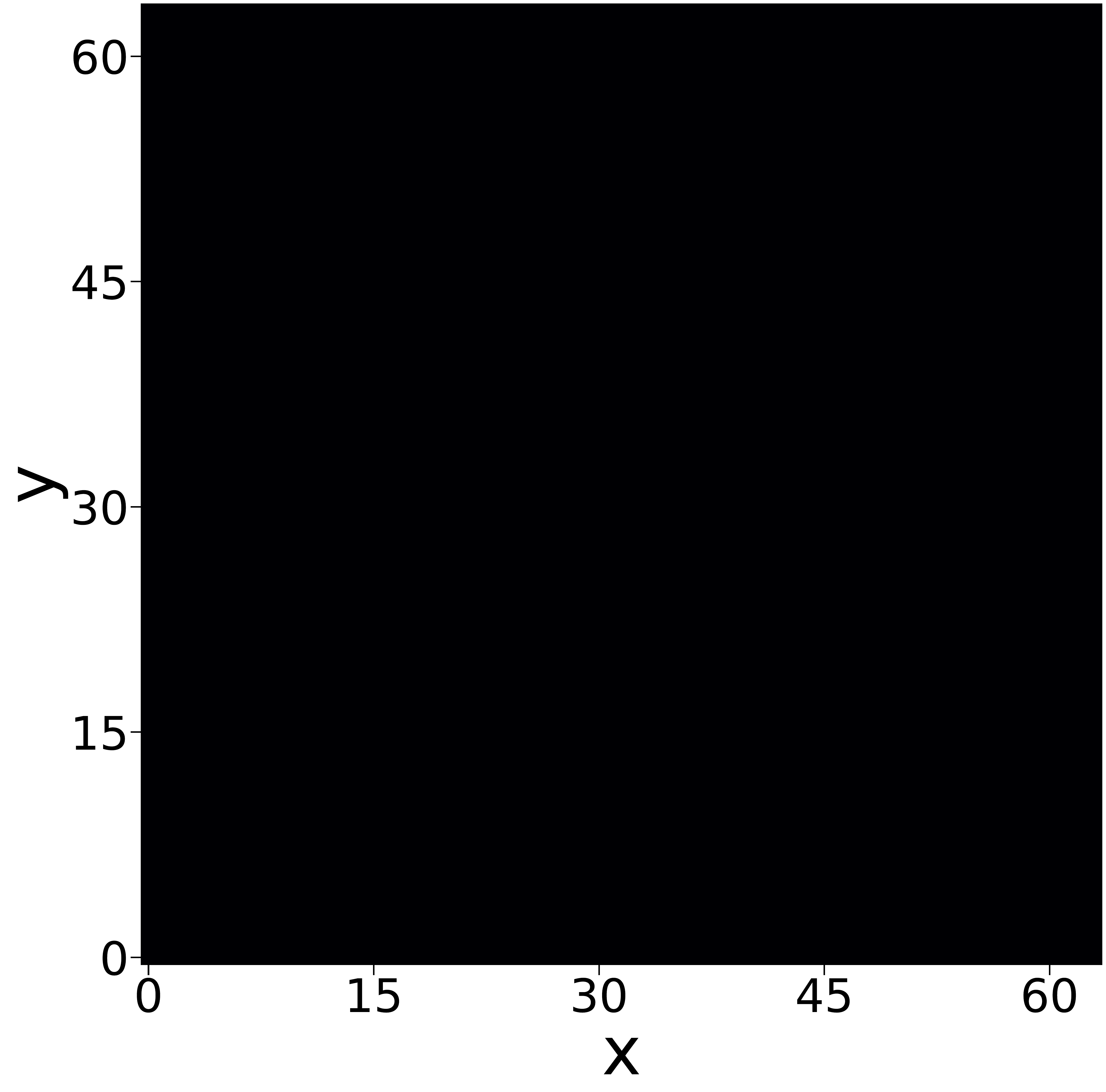}
    \end{subfigure}
    \begin{subfigure}[b]{0.0882\textwidth}
        \captionsetup{labelformat=empty}
        \includegraphics[width=\textwidth]{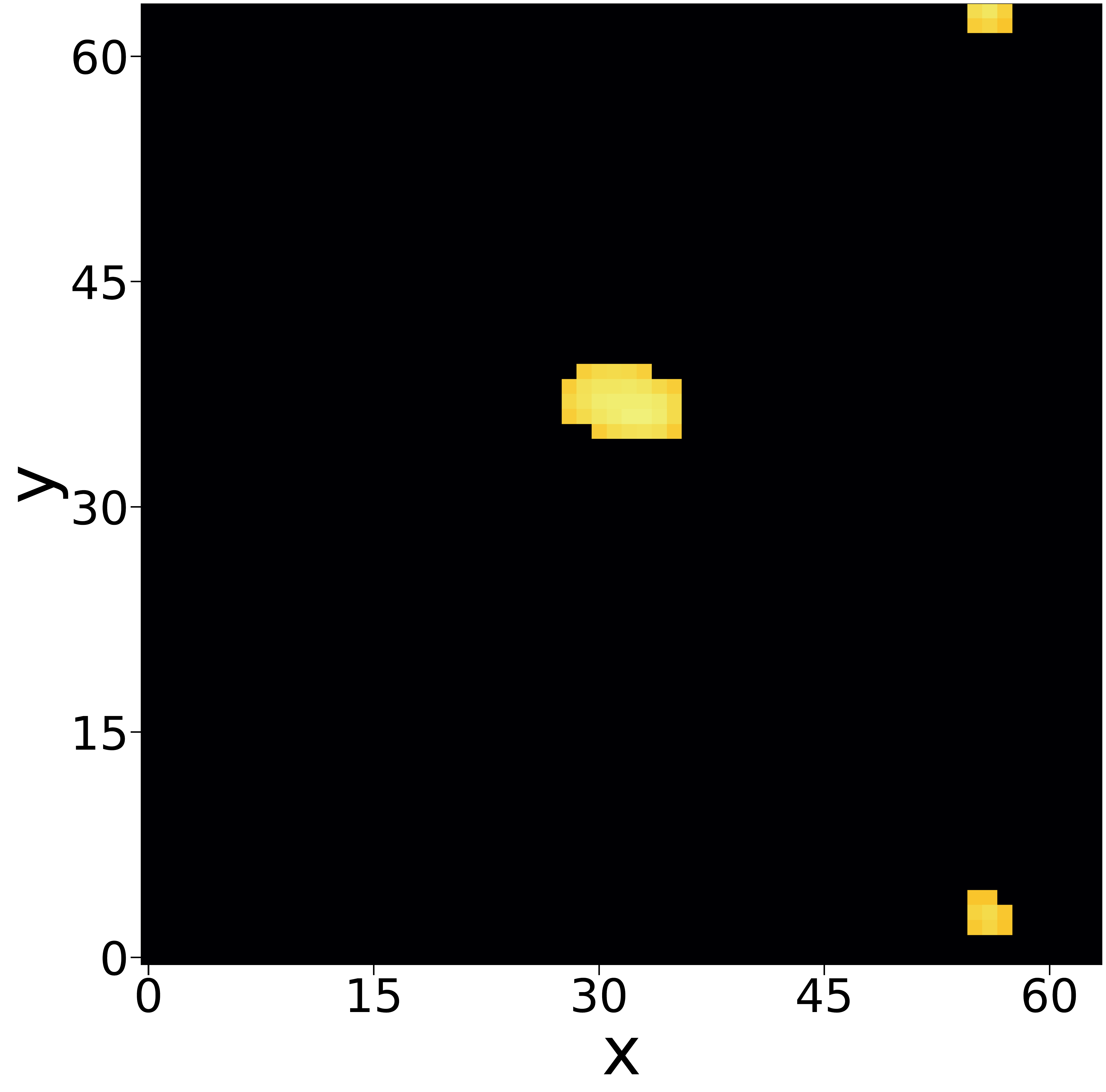}
    \end{subfigure}
    \begin{subfigure}[b]{0.0882\textwidth}
        \captionsetup{labelformat=empty}
        \includegraphics[width=\textwidth]{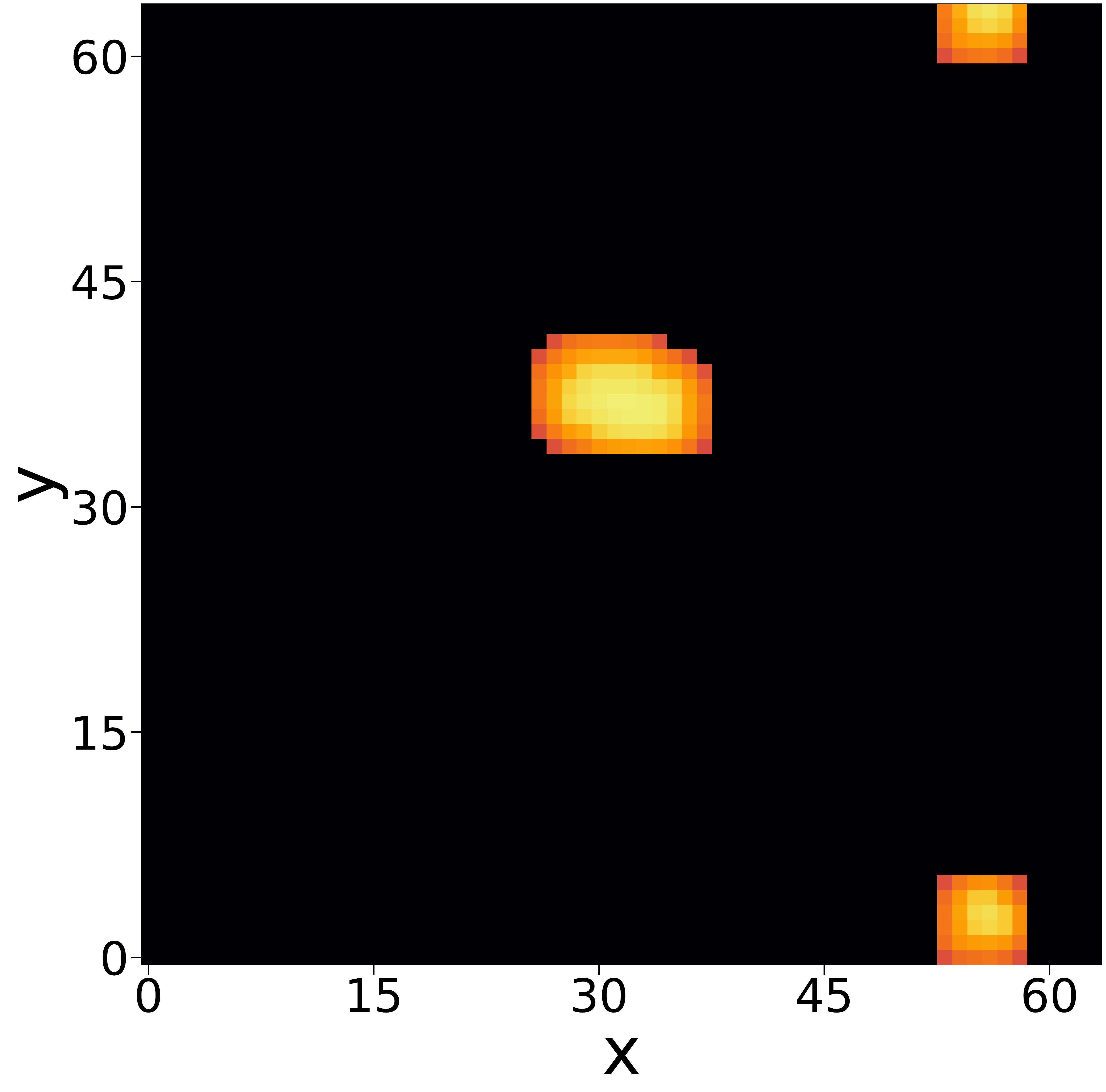}
    \end{subfigure}
    \begin{subfigure}[b]{0.0882\textwidth}
        \captionsetup{labelformat=empty}
        \includegraphics[width=\textwidth]{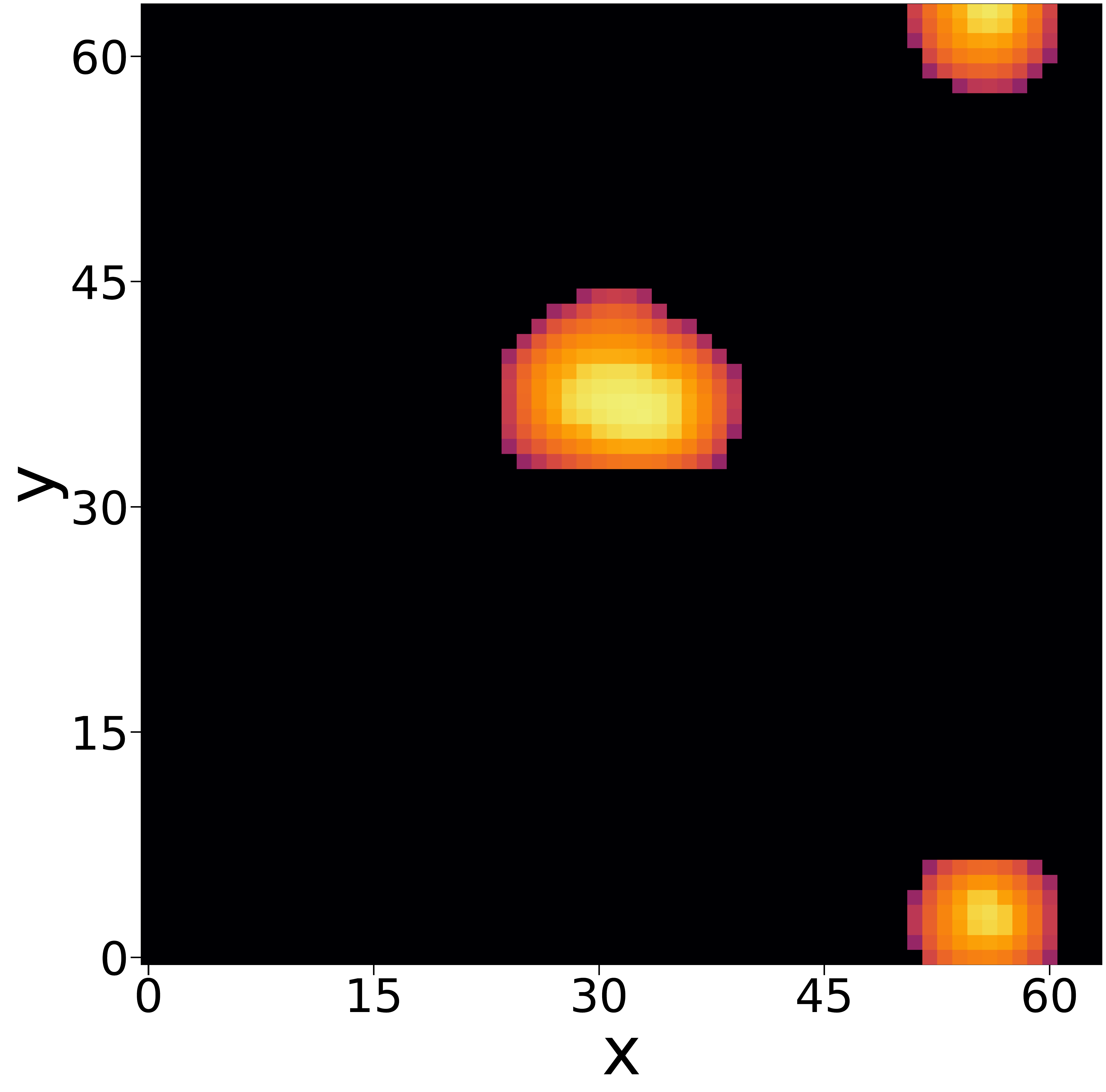}
    \end{subfigure}
    
    \begin{subfigure}[b]{0.0882\textwidth}
        \captionsetup{labelformat=empty}
        \includegraphics[width=\textwidth]{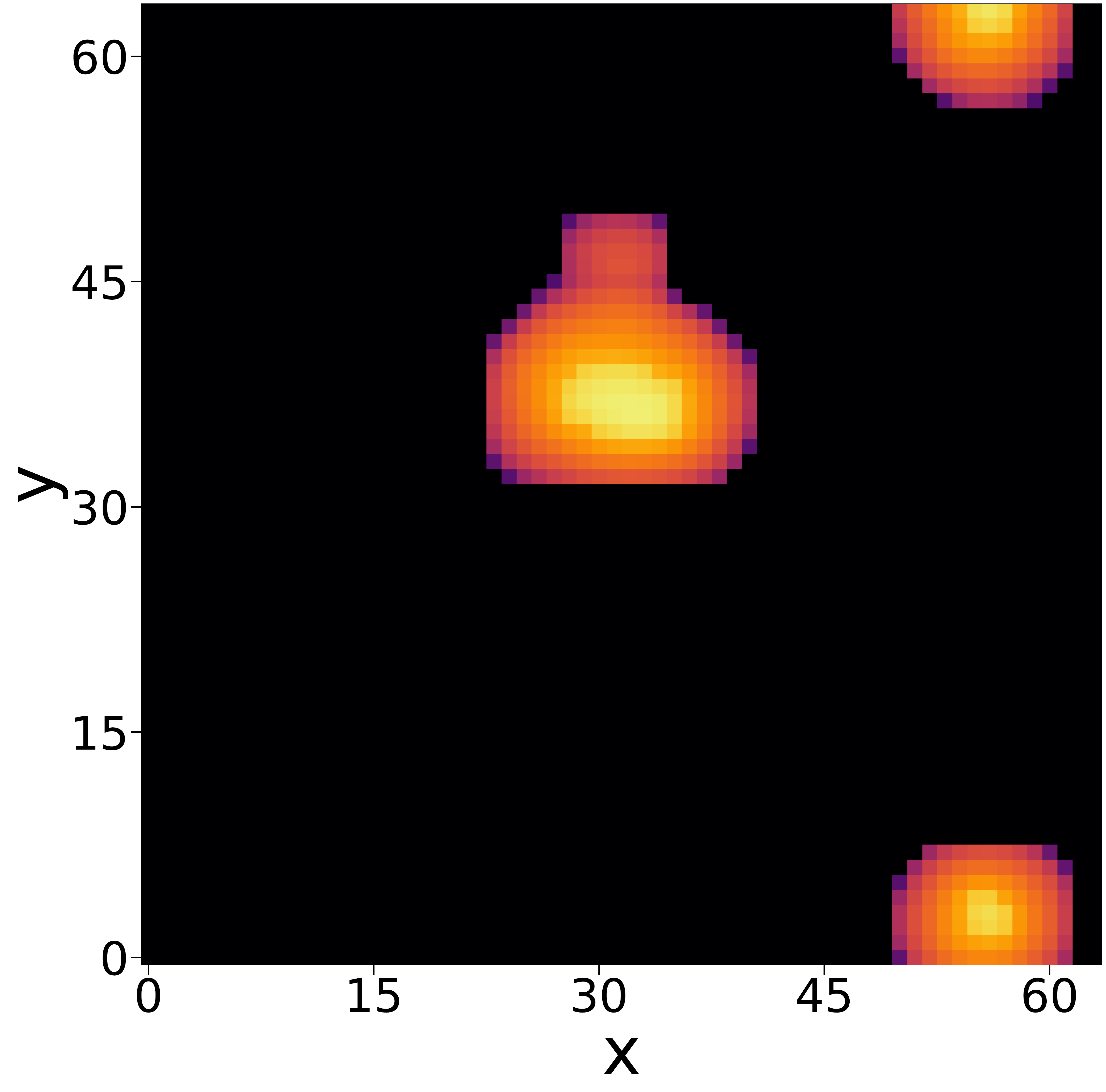}
    \end{subfigure}

    \begin{subfigure}[b]{0.0882\textwidth}
        \captionsetup{labelformat=empty}
        \includegraphics[width=\textwidth]{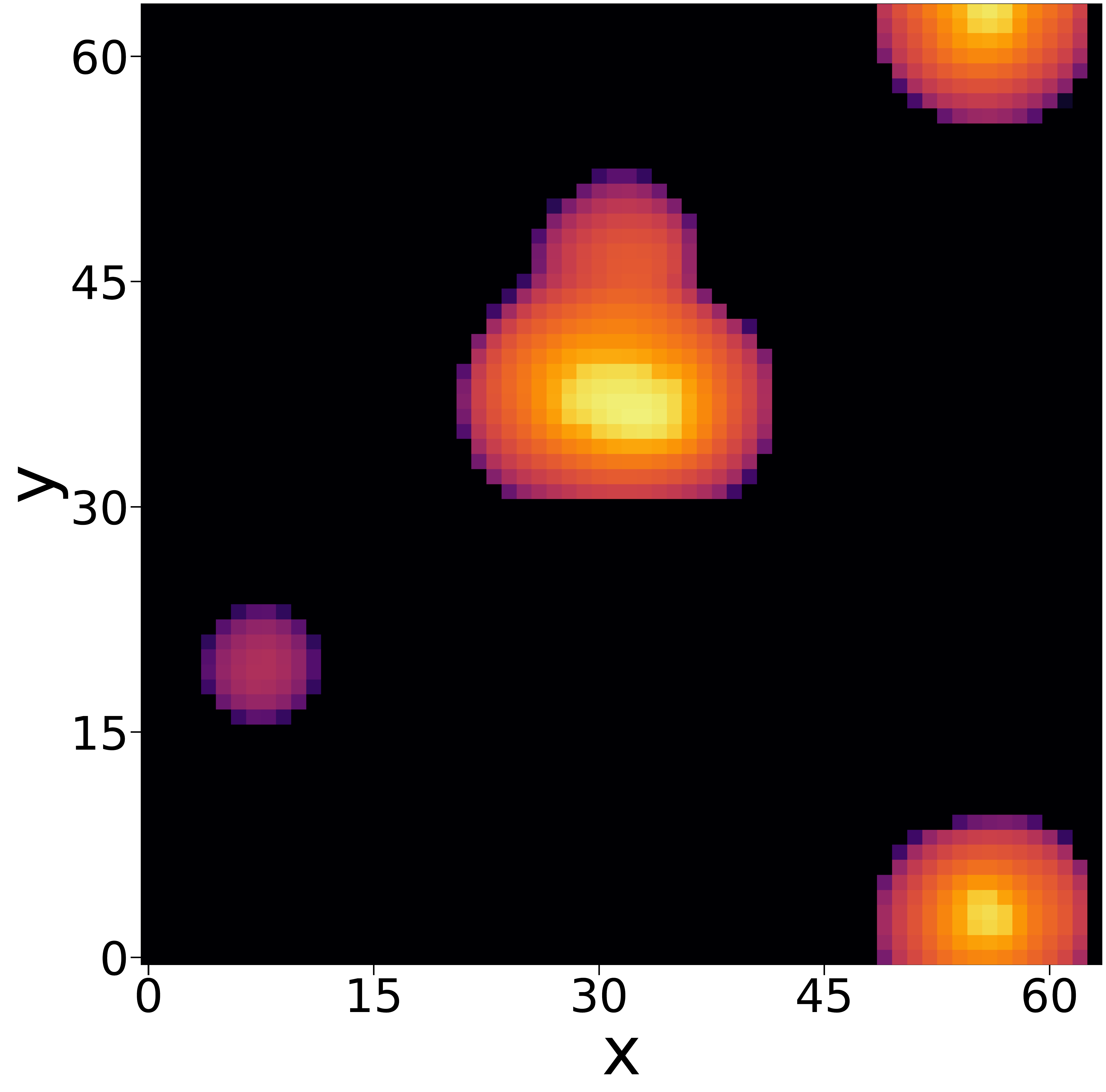}
    \end{subfigure}
    \begin{subfigure}[b]{0.0882\textwidth}
        \captionsetup{labelformat=empty}
        \includegraphics[width=\textwidth]{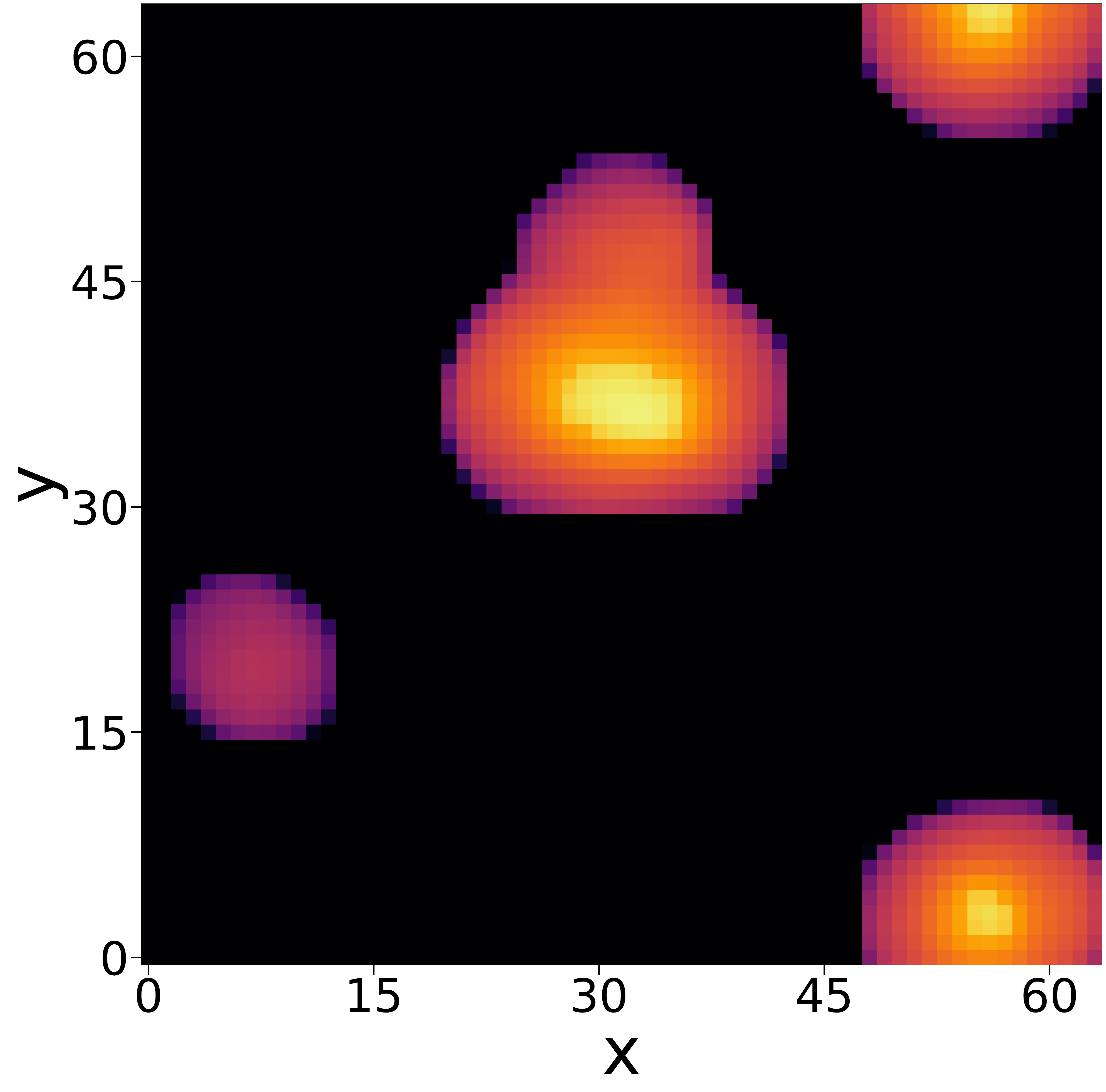}
    \end{subfigure}
    \begin{subfigure}[b]{0.0882\textwidth}
        \captionsetup{labelformat=empty}
        \includegraphics[width=\textwidth]{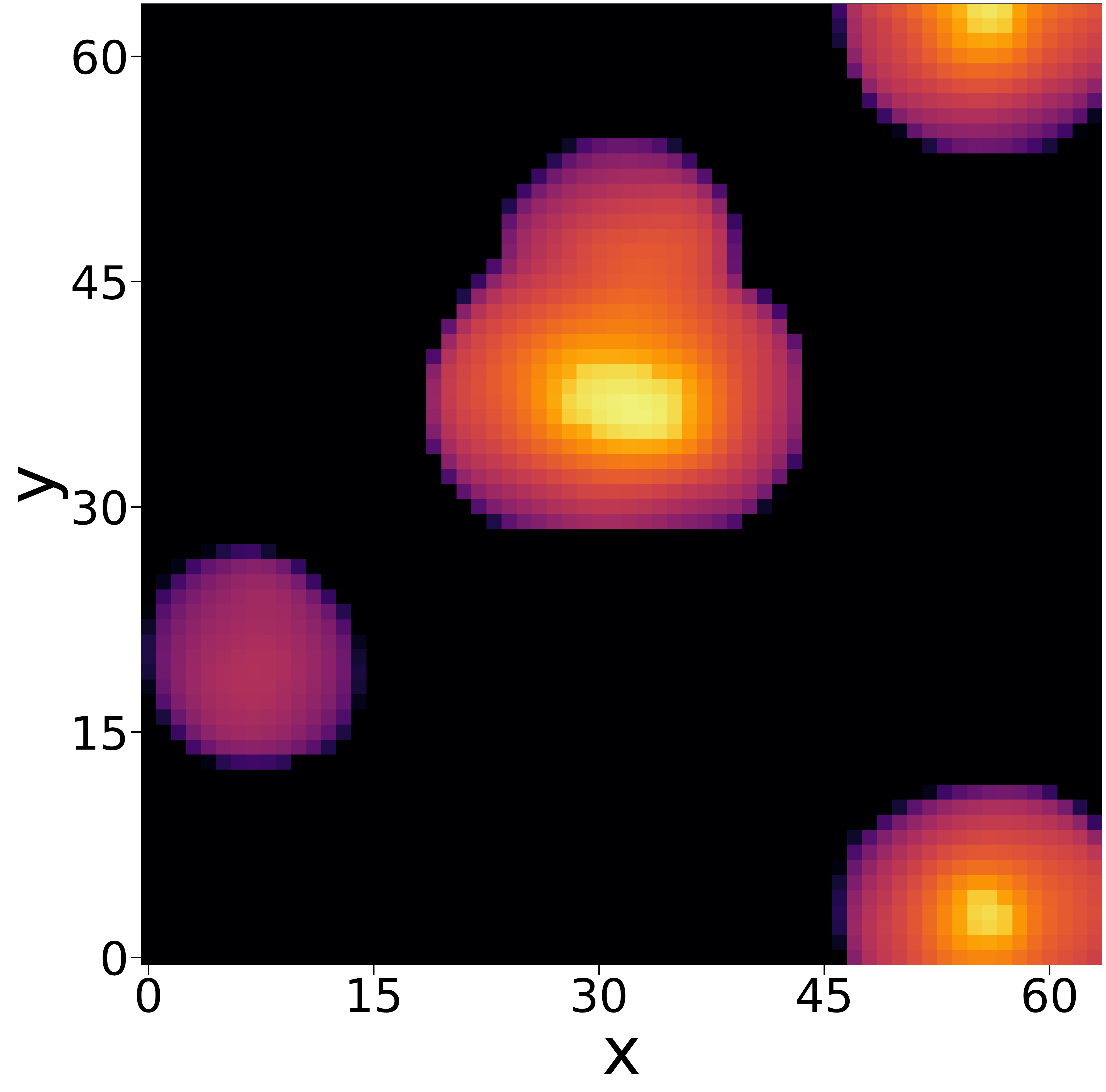}
    \end{subfigure}
    \begin{subfigure}[b]{0.0882\textwidth}
        \captionsetup{labelformat=empty}
        \includegraphics[width=\textwidth]{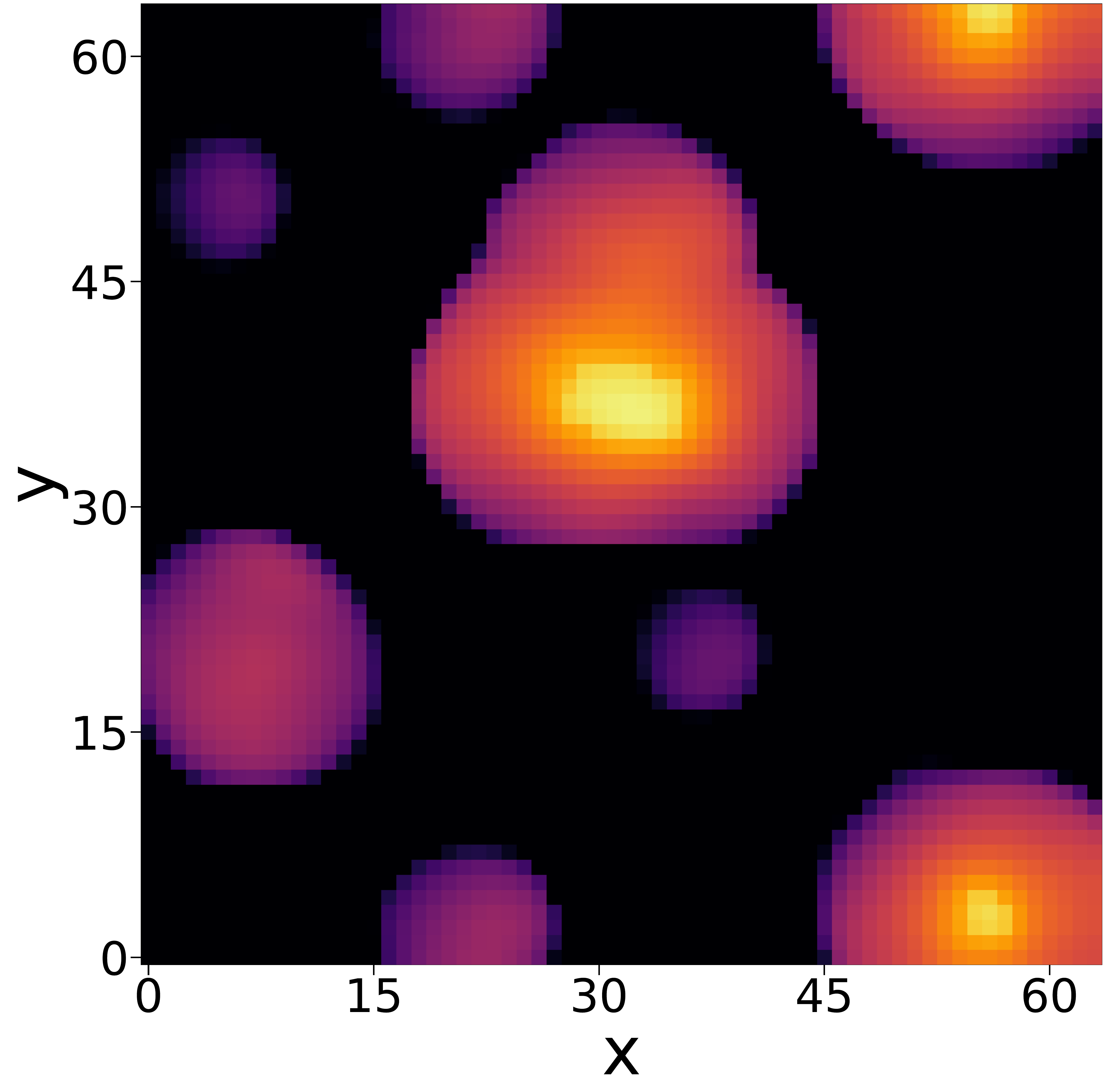}
    \end{subfigure}
    \begin{subfigure}[b]{0.0882\textwidth}
        \captionsetup{labelformat=empty}
        \includegraphics[width=\textwidth]{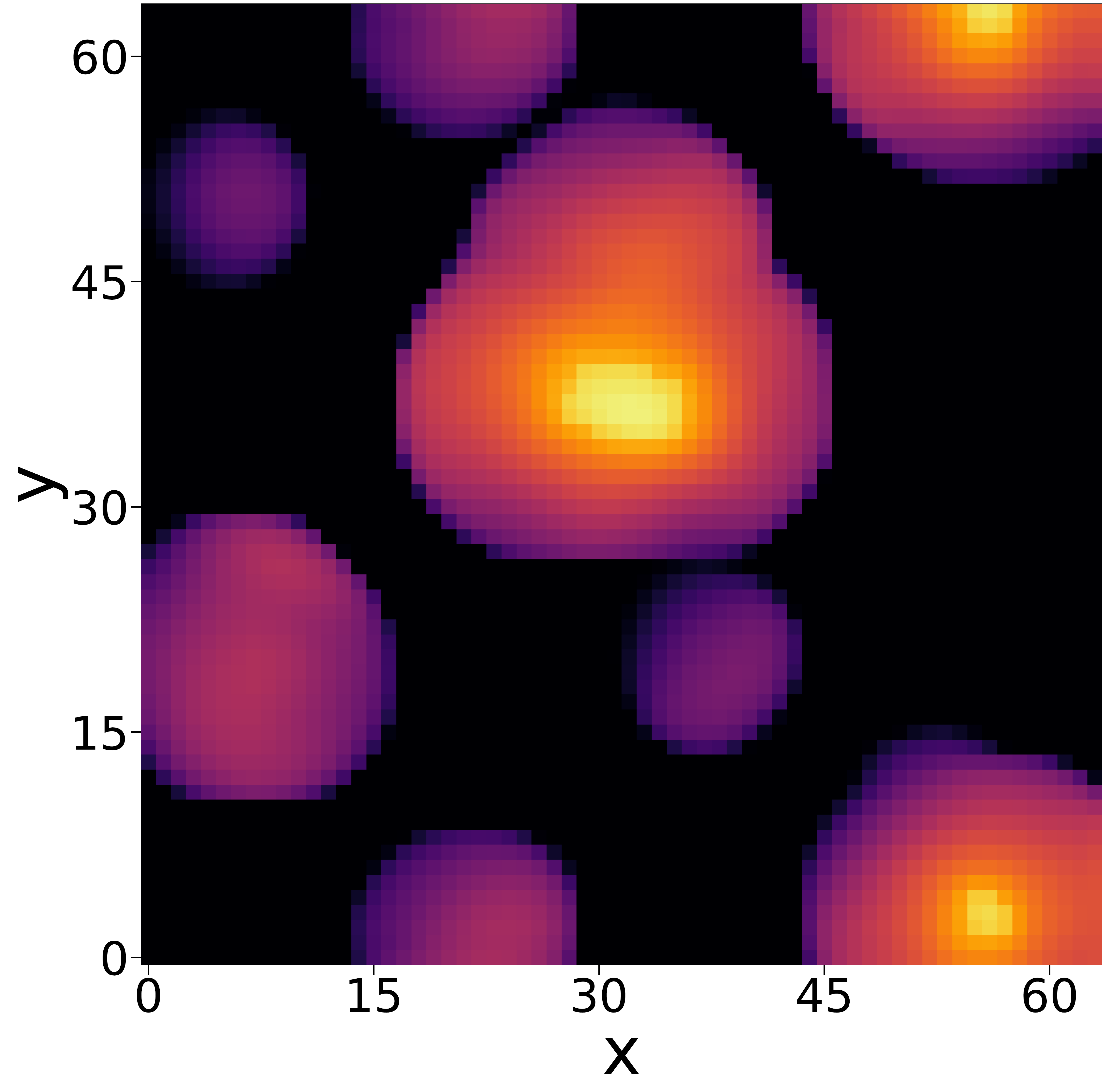}
    \end{subfigure}

    \begin{subfigure}[b]{0.02346\textwidth}
        \captionsetup{labelformat=empty}
        \includegraphics[width=\textwidth]{graphics/colorbar_pred.pdf}
    \end{subfigure}

    }
    \makebox[\textwidth][c]{
    \begin{subfigure}[b]{0.0882\textwidth}
        \captionsetup{labelformat=empty}
        \includegraphics[width=\textwidth]{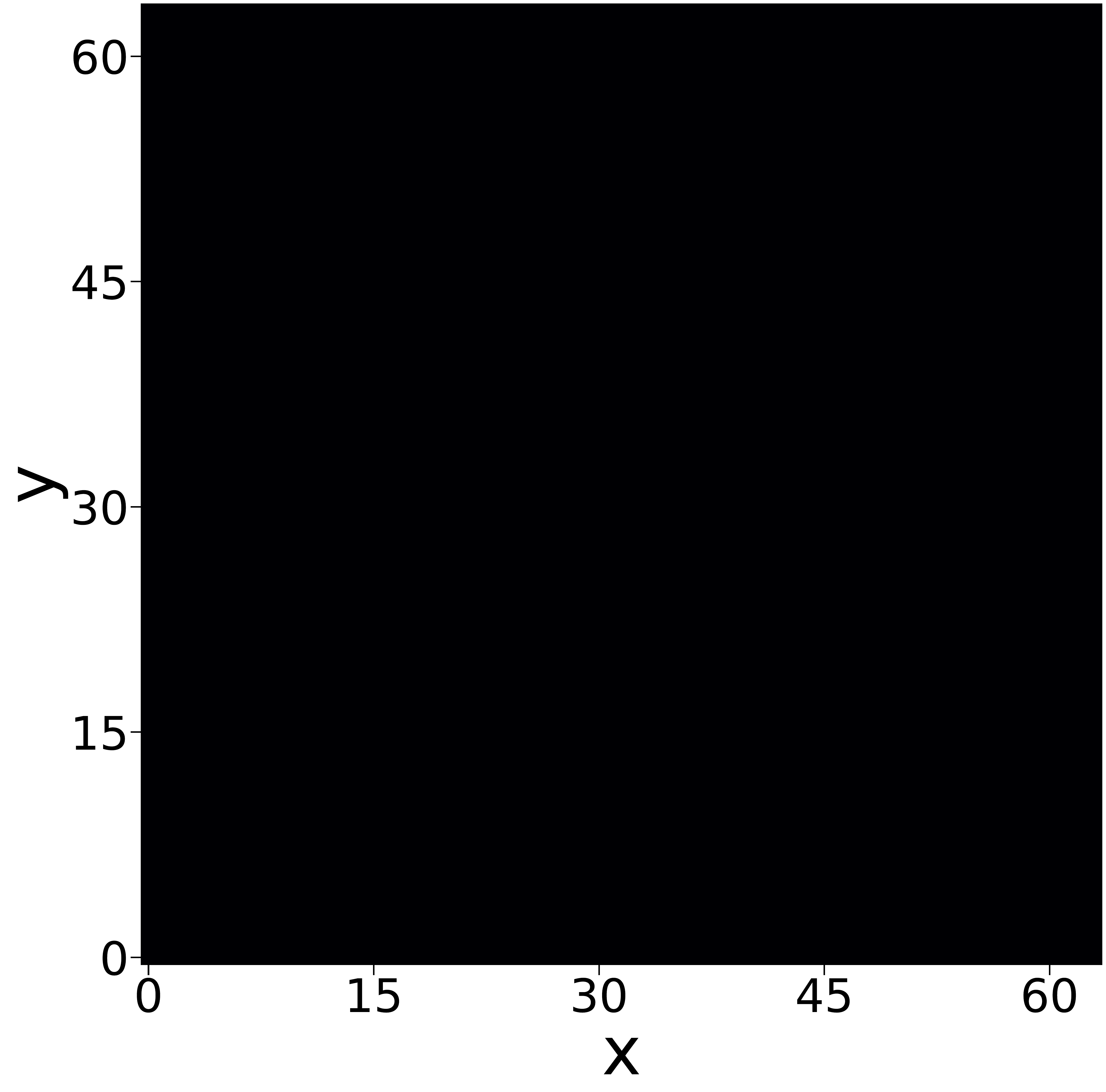}
    \end{subfigure}
    \begin{subfigure}[b]{0.0882\textwidth}
        \captionsetup{labelformat=empty}
        \includegraphics[width=\textwidth]{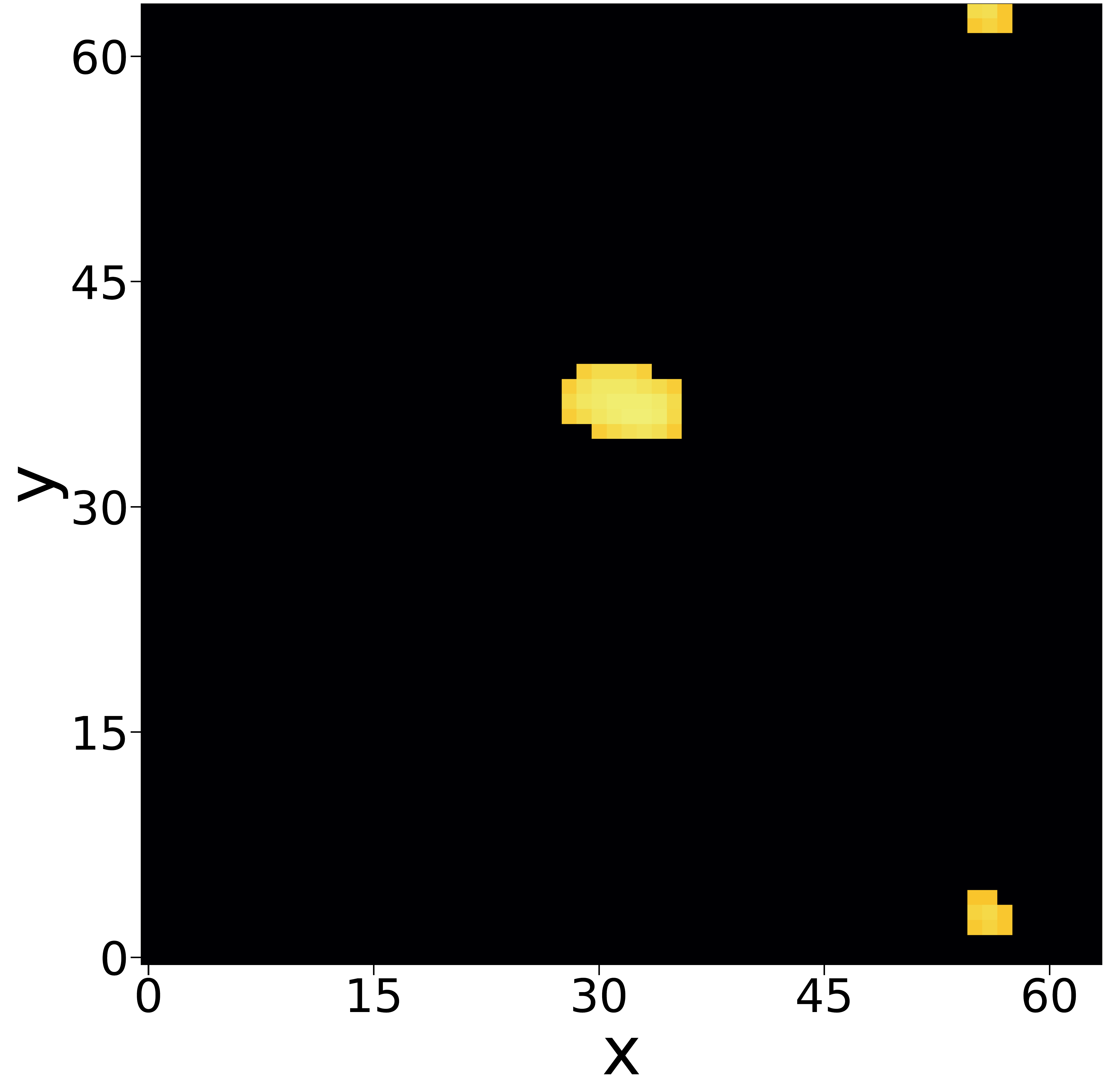}
    \end{subfigure}
    \begin{subfigure}[b]{0.0882\textwidth}
        \captionsetup{labelformat=empty}
        \includegraphics[width=\textwidth]{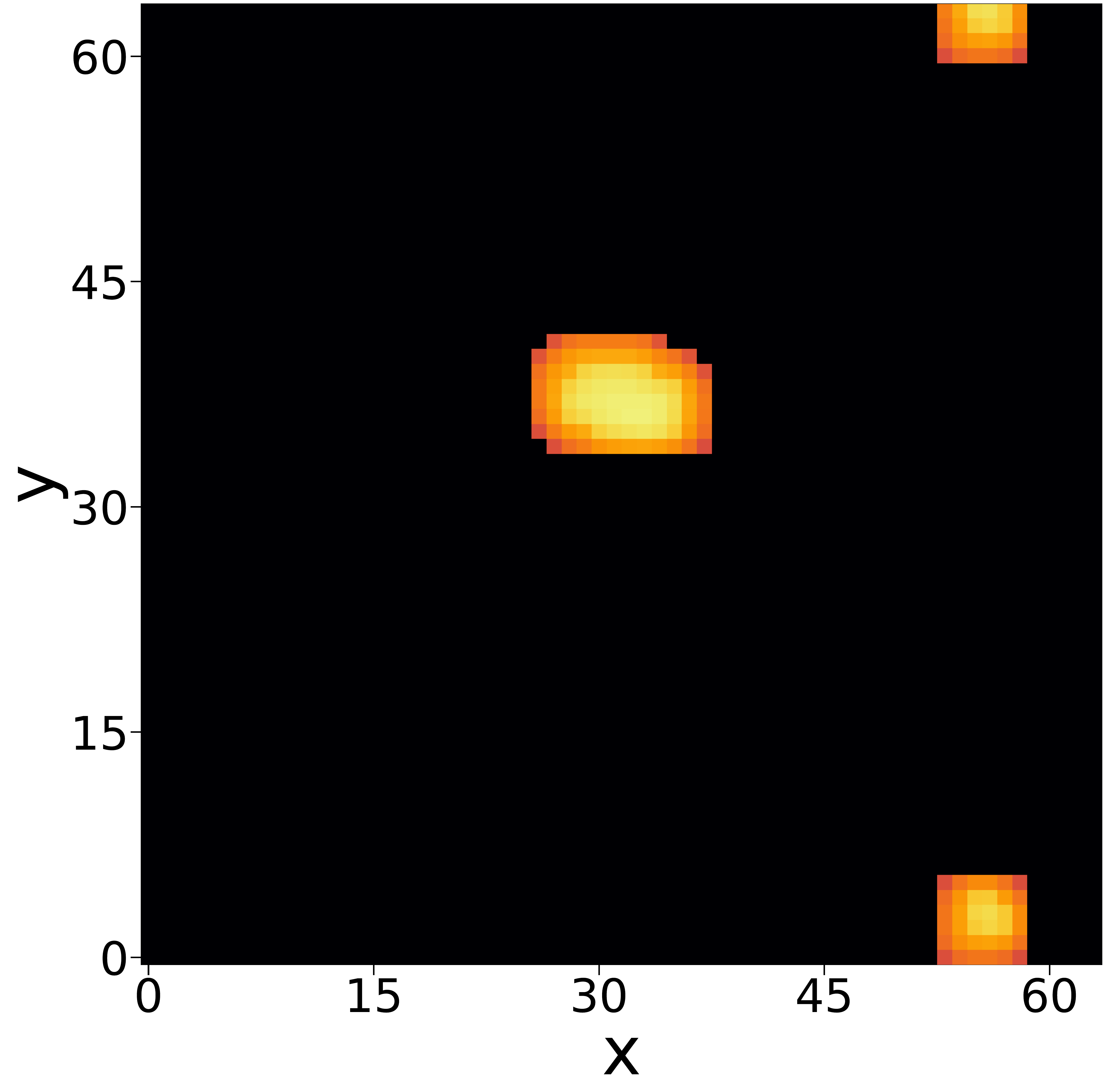}
    \end{subfigure}
    \begin{subfigure}[b]{0.0882\textwidth}
        \captionsetup{labelformat=empty}
        \includegraphics[width=\textwidth]{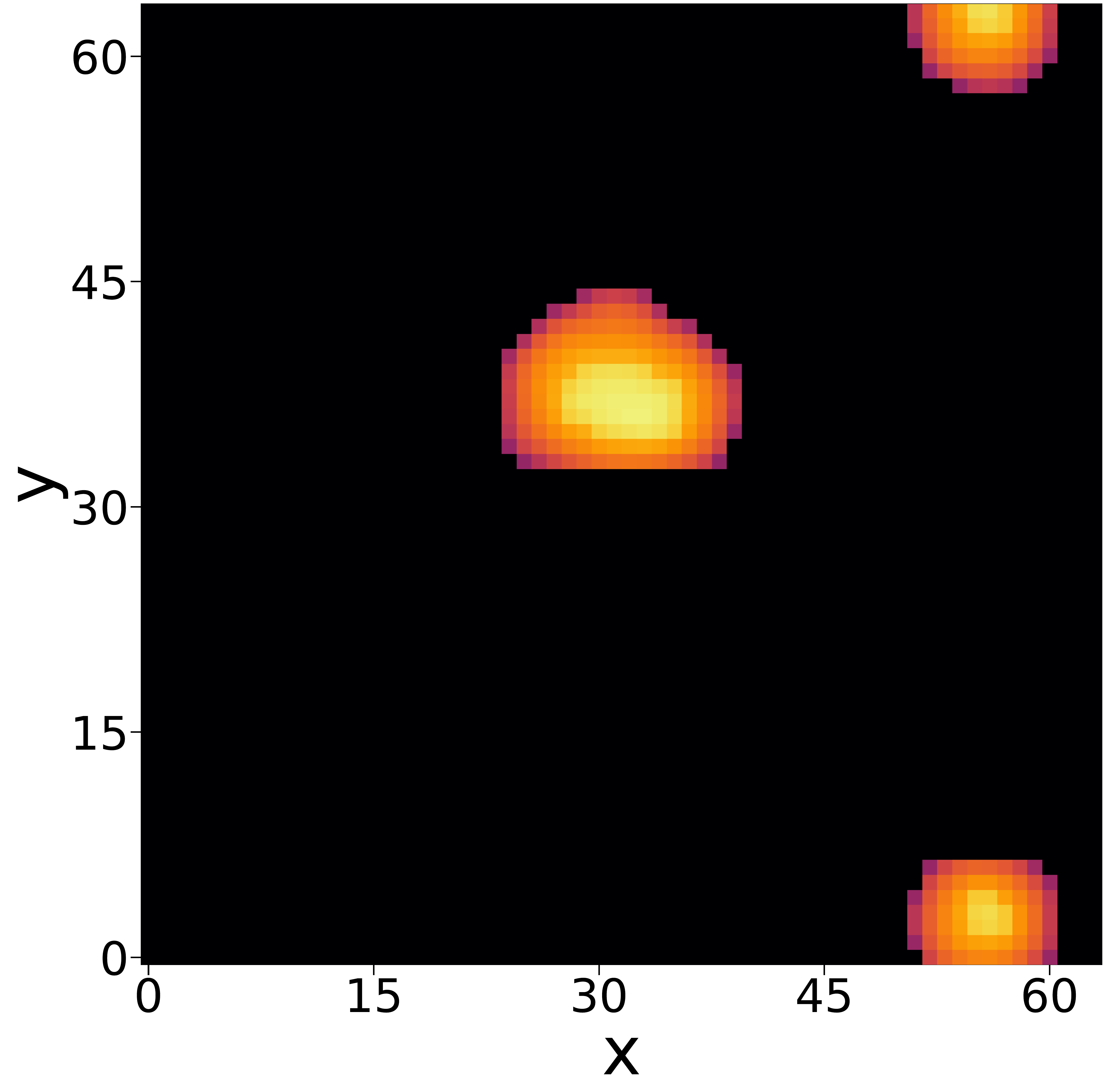}
    \end{subfigure}
    
    \begin{subfigure}[b]{0.0882\textwidth}
        \captionsetup{labelformat=empty}
        \includegraphics[width=\textwidth]{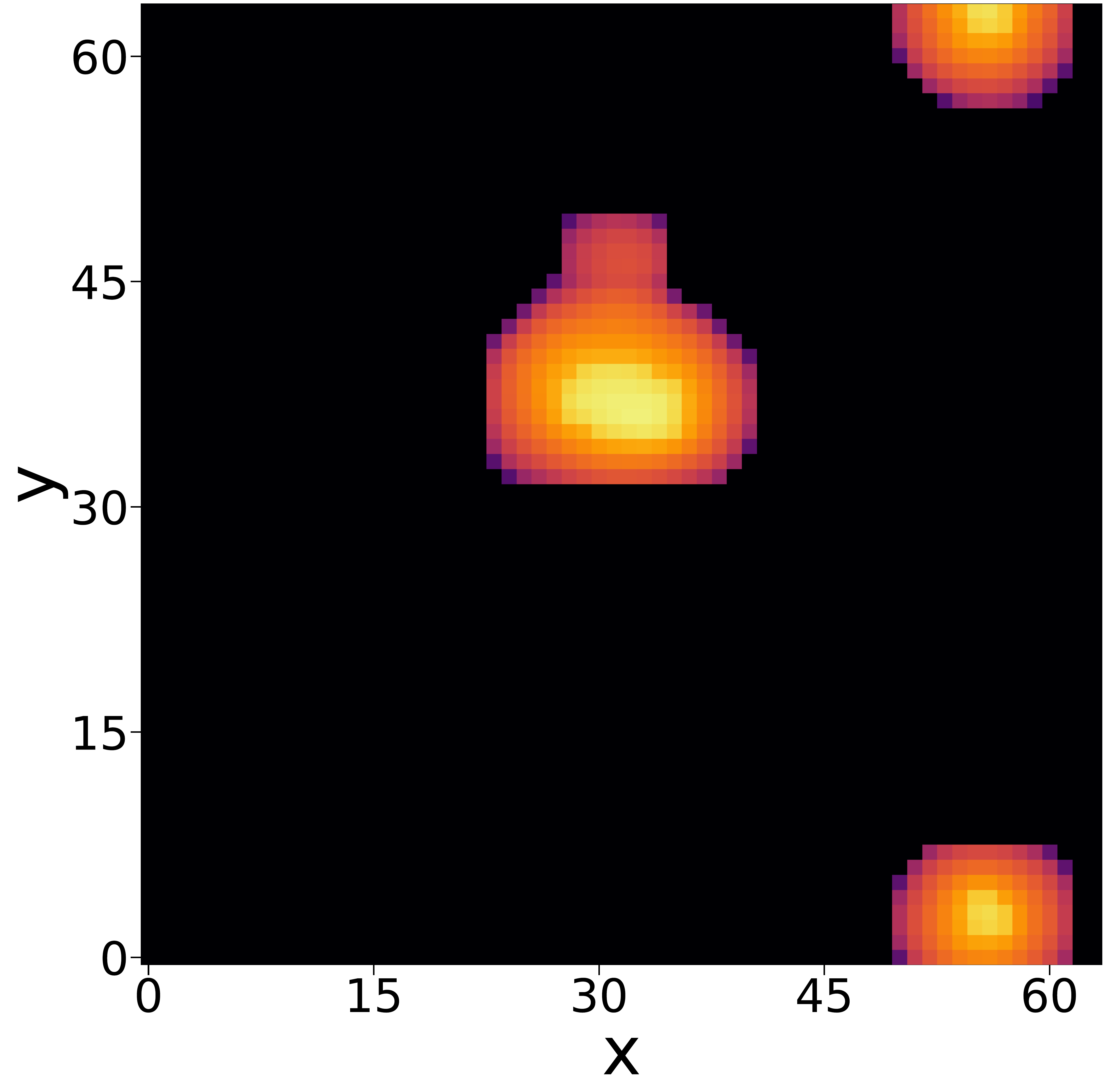}
    \end{subfigure}

    \begin{subfigure}[b]{0.0882\textwidth}
        \captionsetup{labelformat=empty}
        \includegraphics[width=\textwidth]{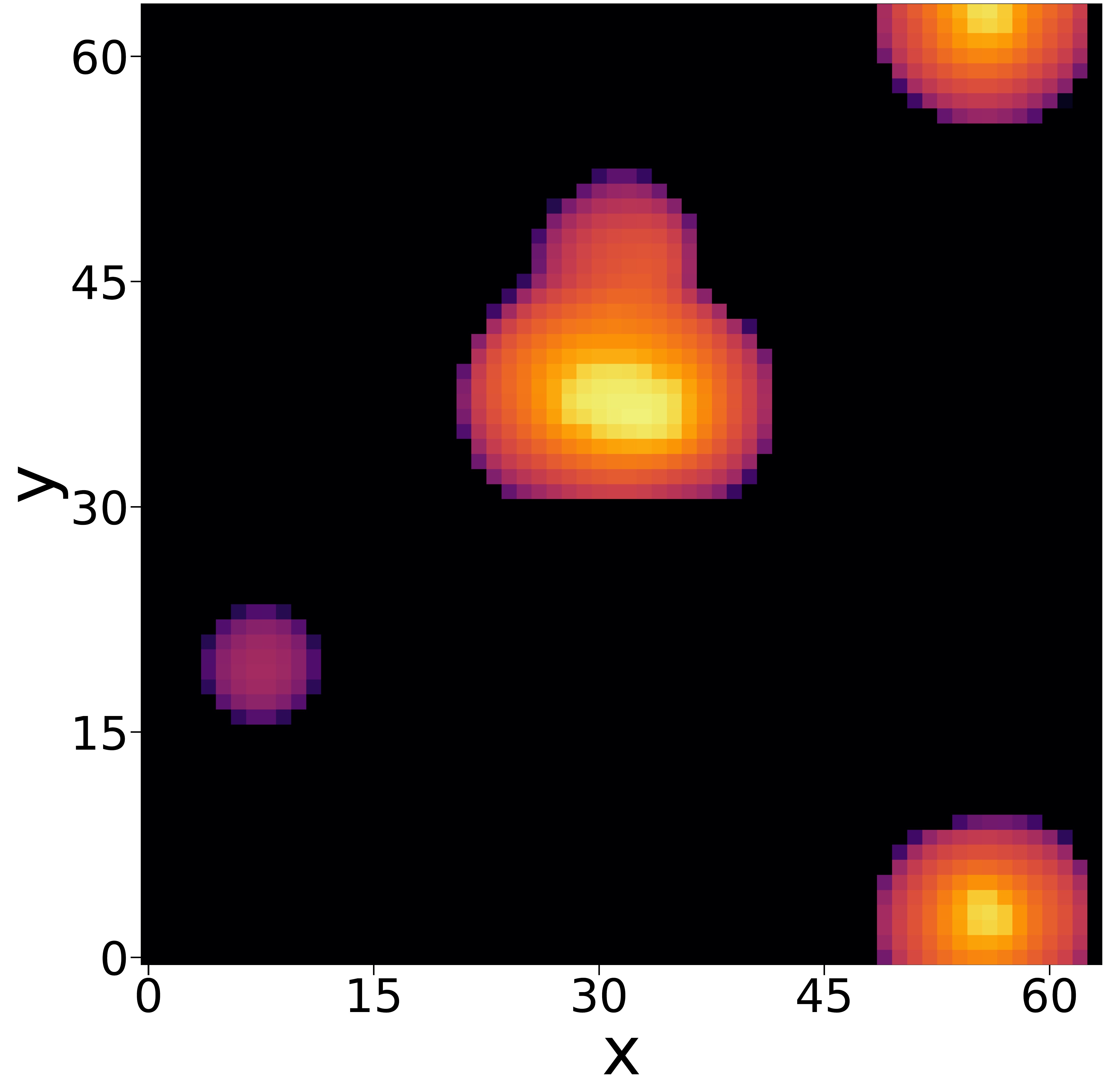}
    \end{subfigure}
    \begin{subfigure}[b]{0.0882\textwidth}
        \captionsetup{labelformat=empty}
        \includegraphics[width=\textwidth]{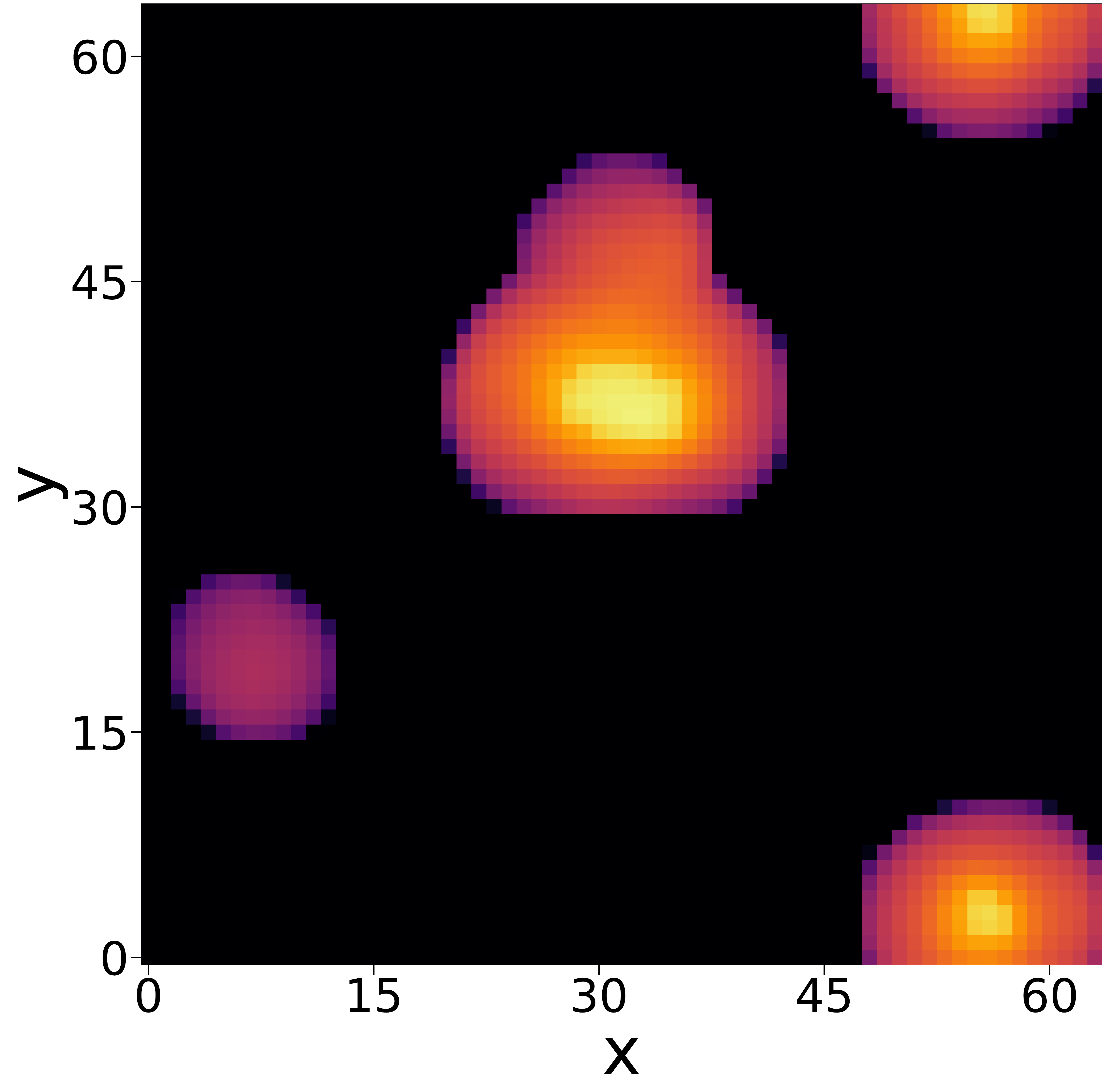}
    \end{subfigure}
    \begin{subfigure}[b]{0.0882\textwidth}
        \captionsetup{labelformat=empty}
        \includegraphics[width=\textwidth]{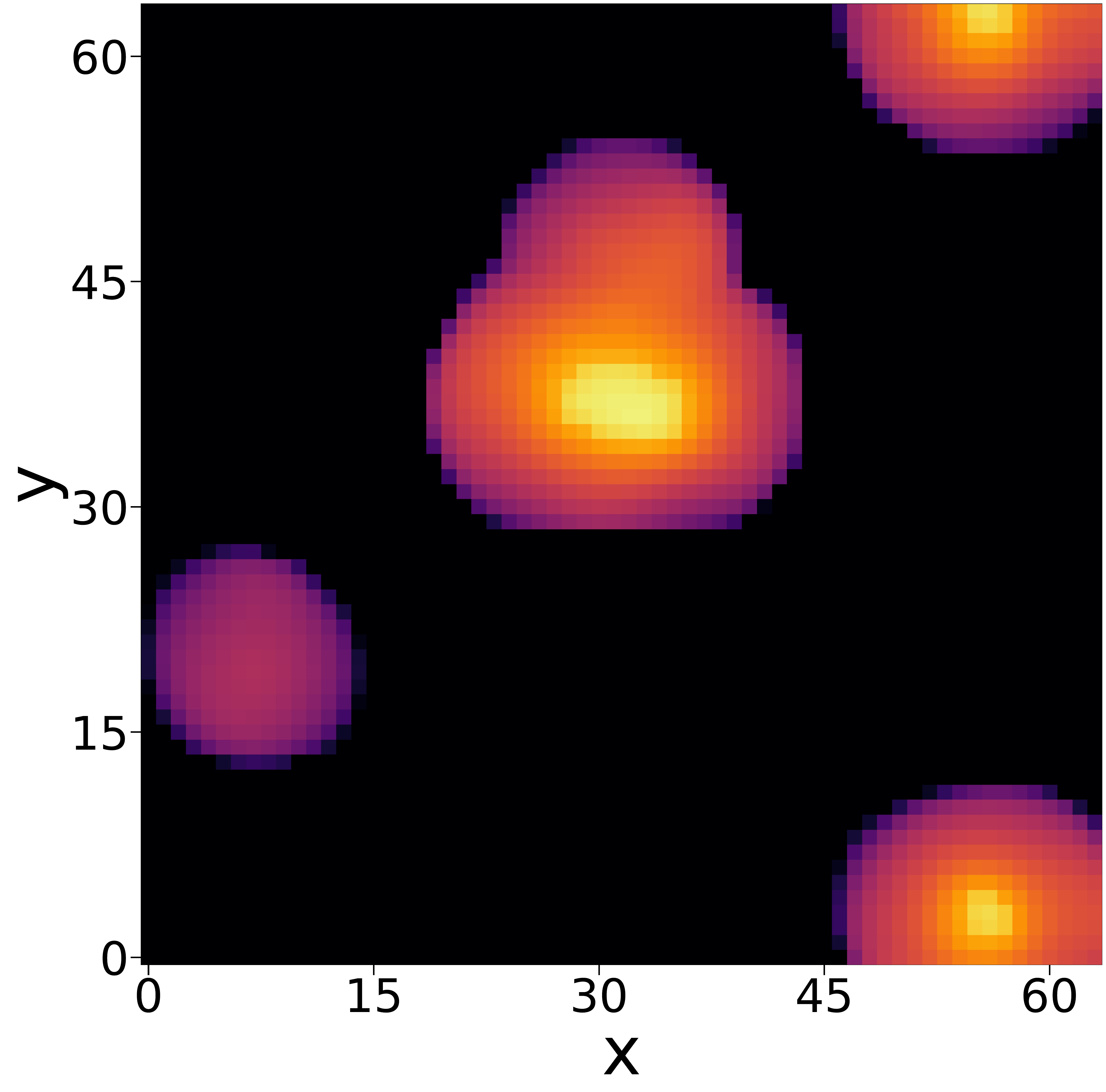}
    \end{subfigure}
    \begin{subfigure}[b]{0.0882\textwidth}
        \captionsetup{labelformat=empty}
        \includegraphics[width=\textwidth]{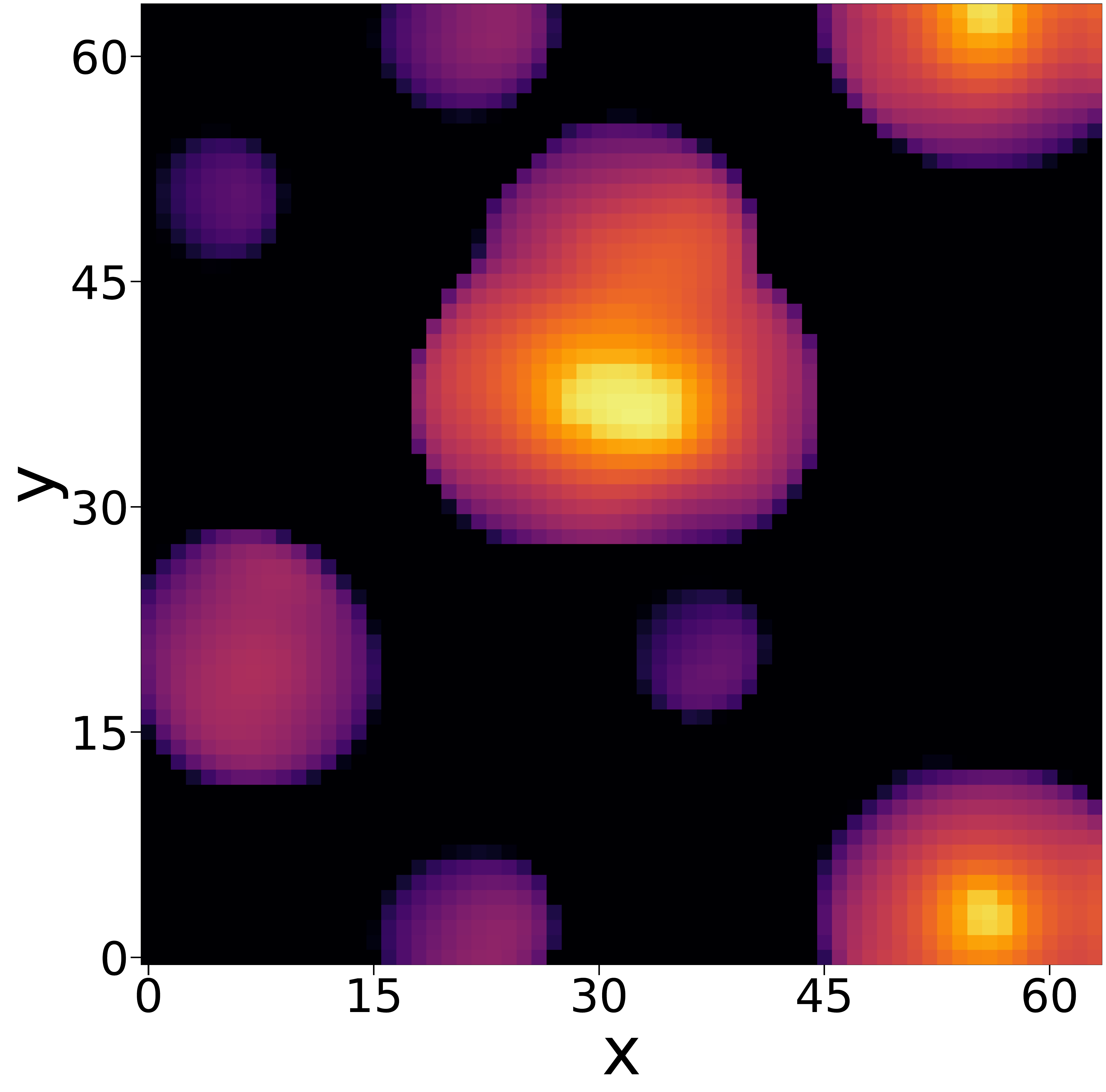}
    \end{subfigure}
    \begin{subfigure}[b]{0.0882\textwidth}
        \captionsetup{labelformat=empty}
        \includegraphics[width=\textwidth]{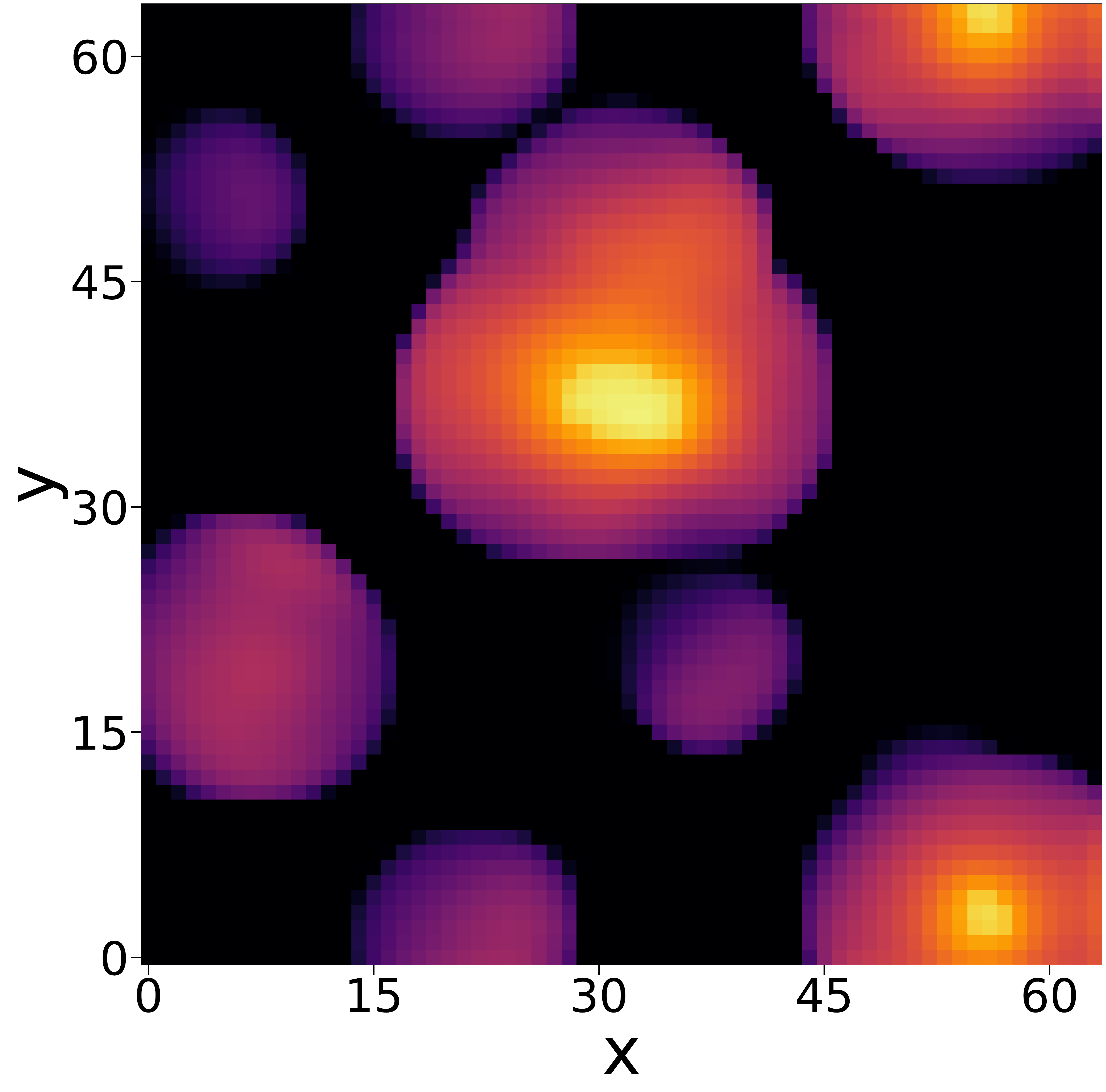}
    \end{subfigure}

    \begin{subfigure}[b]{0.02346\textwidth}
        \captionsetup{labelformat=empty}
        \includegraphics[width=\textwidth]{graphics/colorbar_num.pdf}
    \end{subfigure}

    }
    \makebox[\textwidth][c]{
    \begin{subfigure}[b]{0.0882\textwidth}
        \captionsetup{labelformat=empty}
        \includegraphics[width=\textwidth]{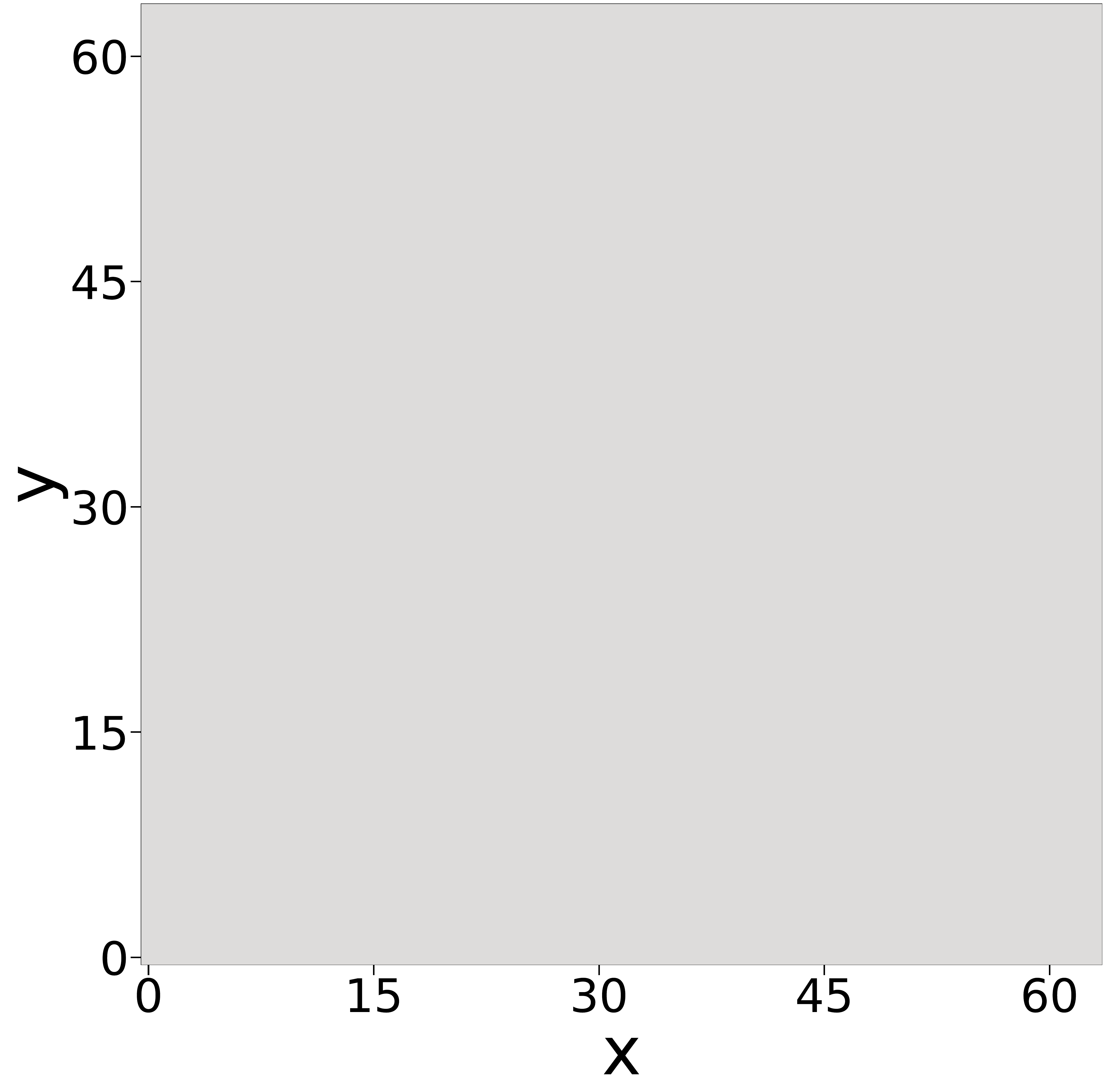}
    \end{subfigure}
    \begin{subfigure}[b]{0.0882\textwidth}
        \captionsetup{labelformat=empty}
        \includegraphics[width=\textwidth]{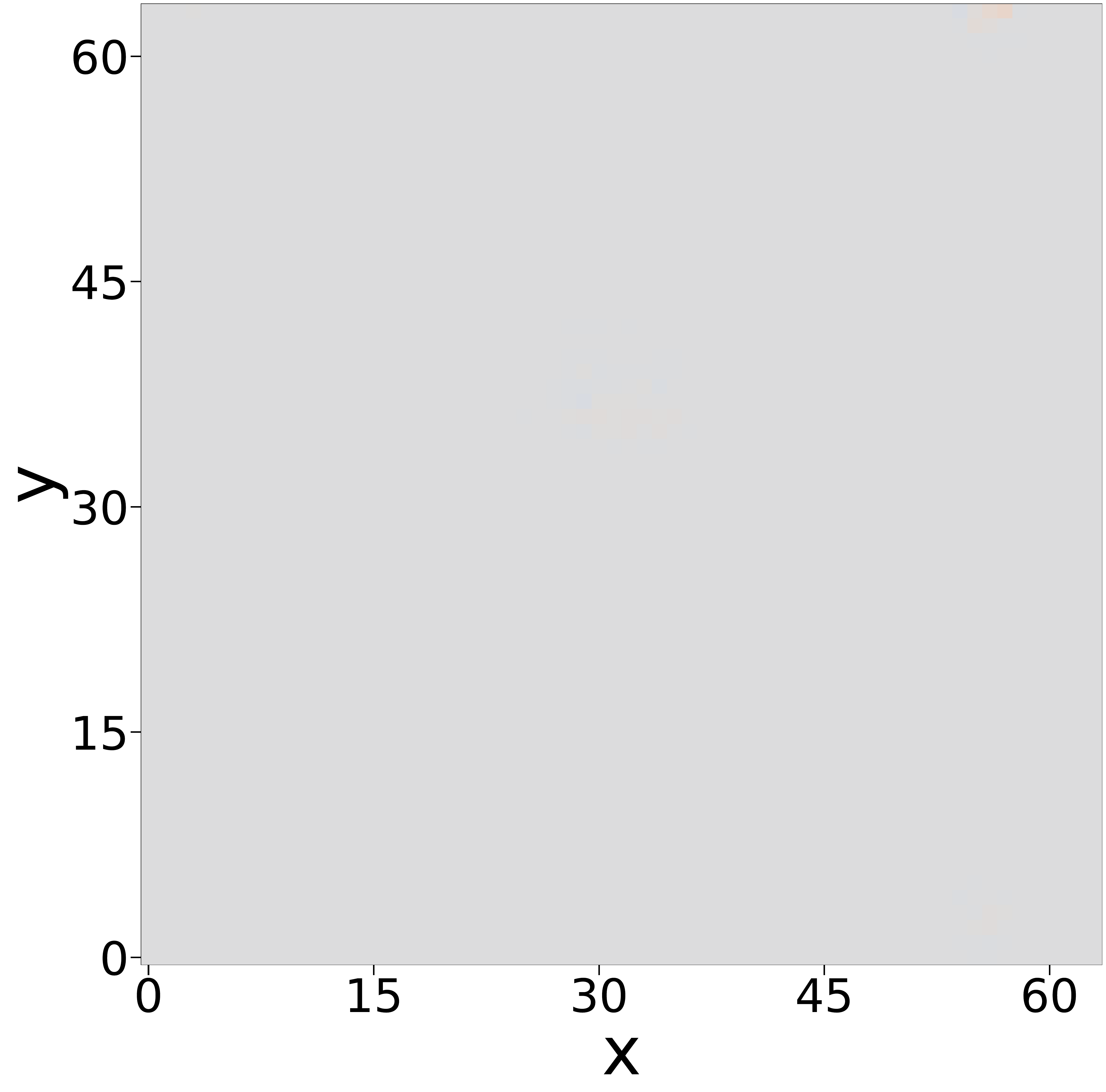}
    \end{subfigure}
    \begin{subfigure}[b]{0.0882\textwidth}
        \captionsetup{labelformat=empty}
        \includegraphics[width=\textwidth]{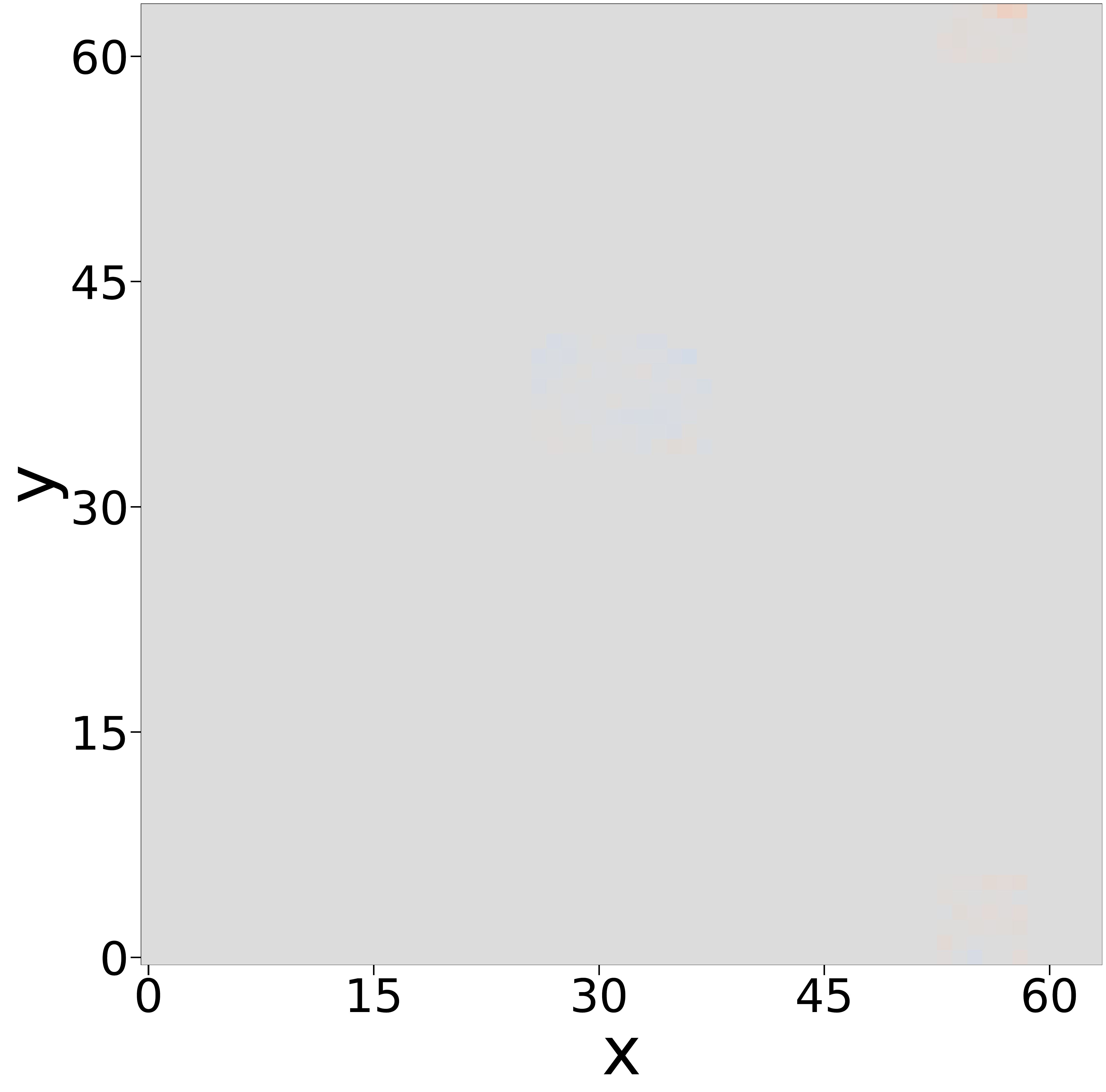}
    \end{subfigure}
    \begin{subfigure}[b]{0.0882\textwidth}
        \captionsetup{labelformat=empty}
        \includegraphics[width=\textwidth]{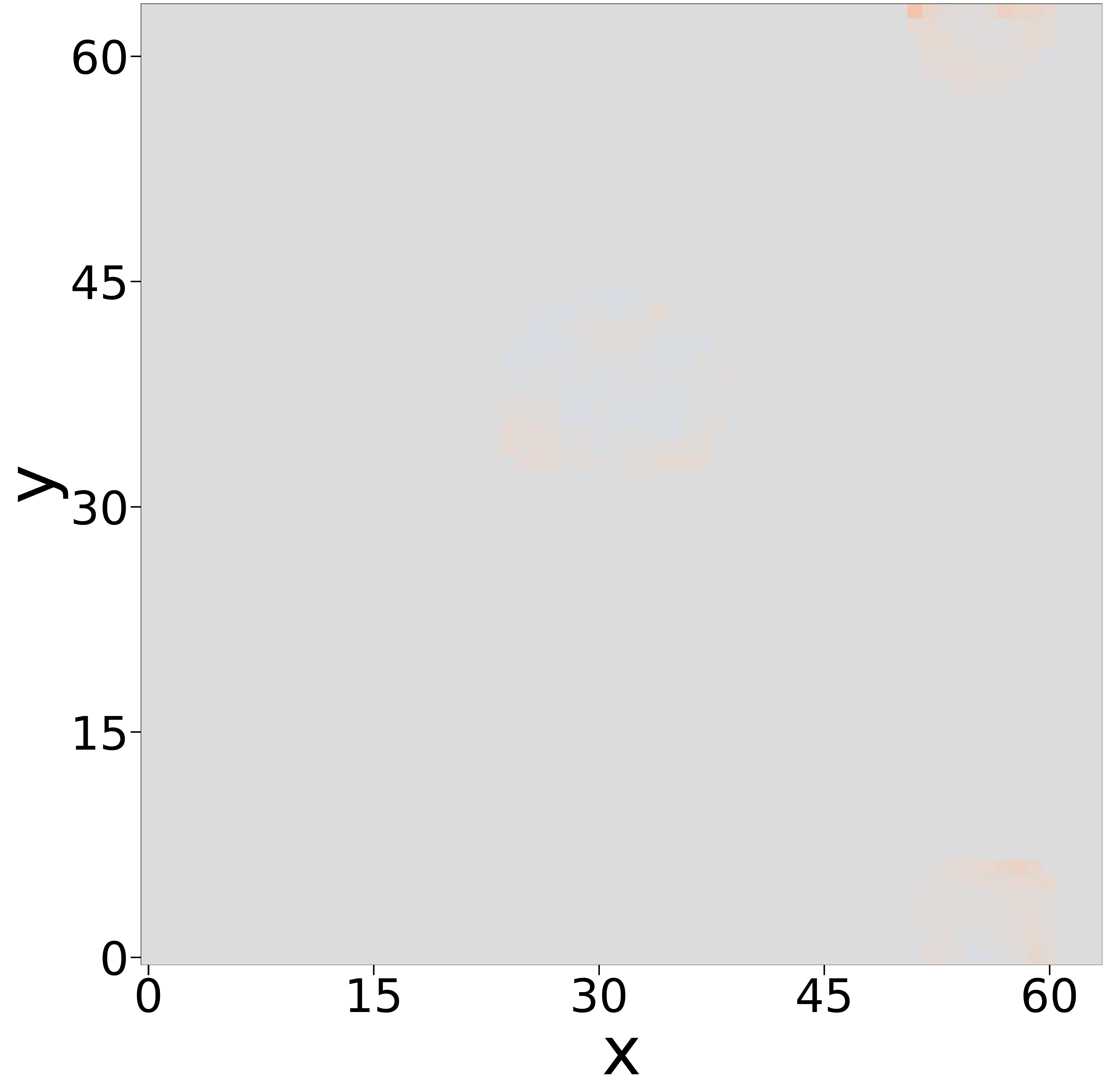}
    \end{subfigure}
    
    \begin{subfigure}[b]{0.0882\textwidth}
        \captionsetup{labelformat=empty}
        \includegraphics[width=\textwidth]{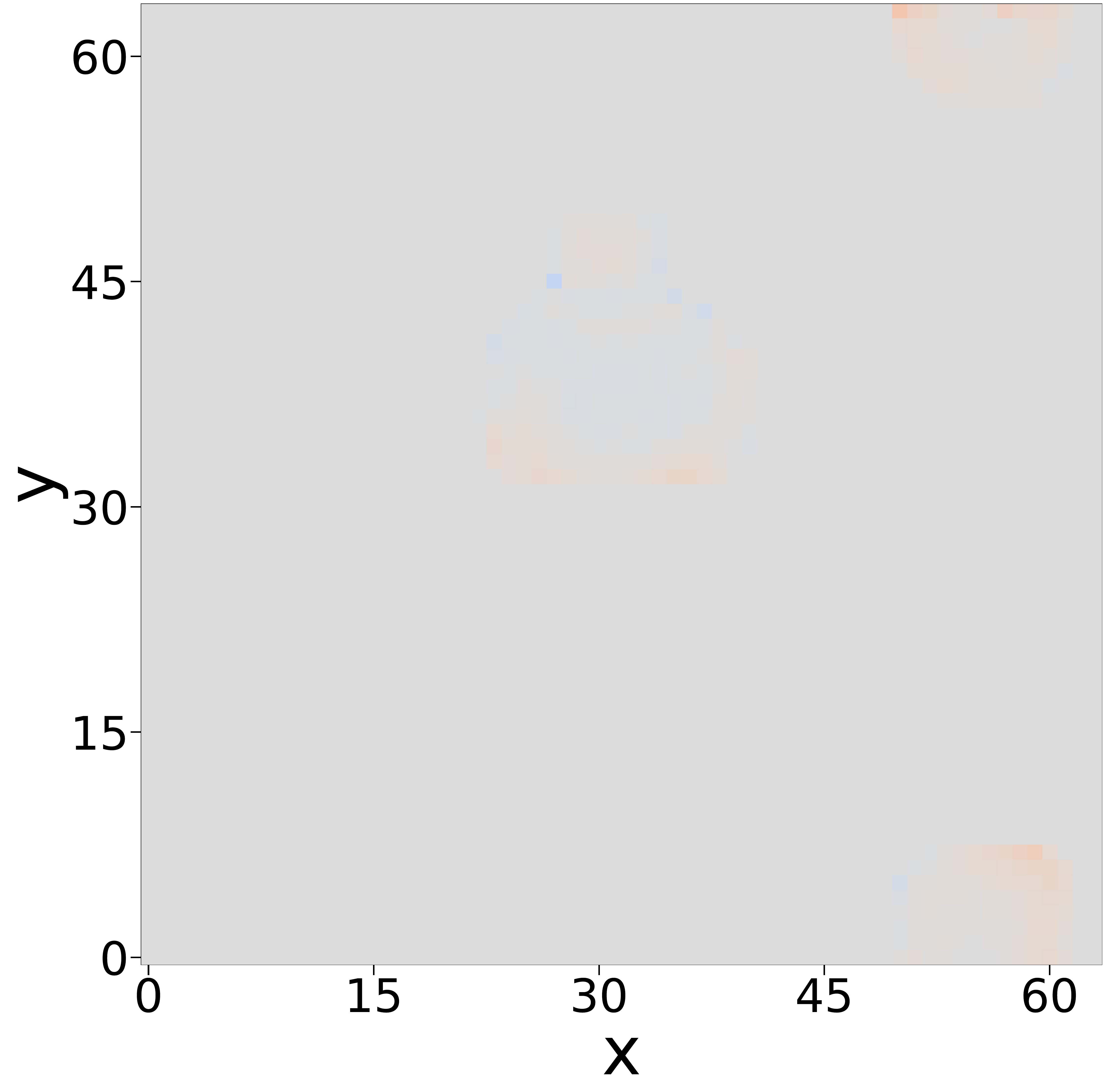}
    \end{subfigure}

    \begin{subfigure}[b]{0.0882\textwidth}
        \captionsetup{labelformat=empty}
        \includegraphics[width=\textwidth]{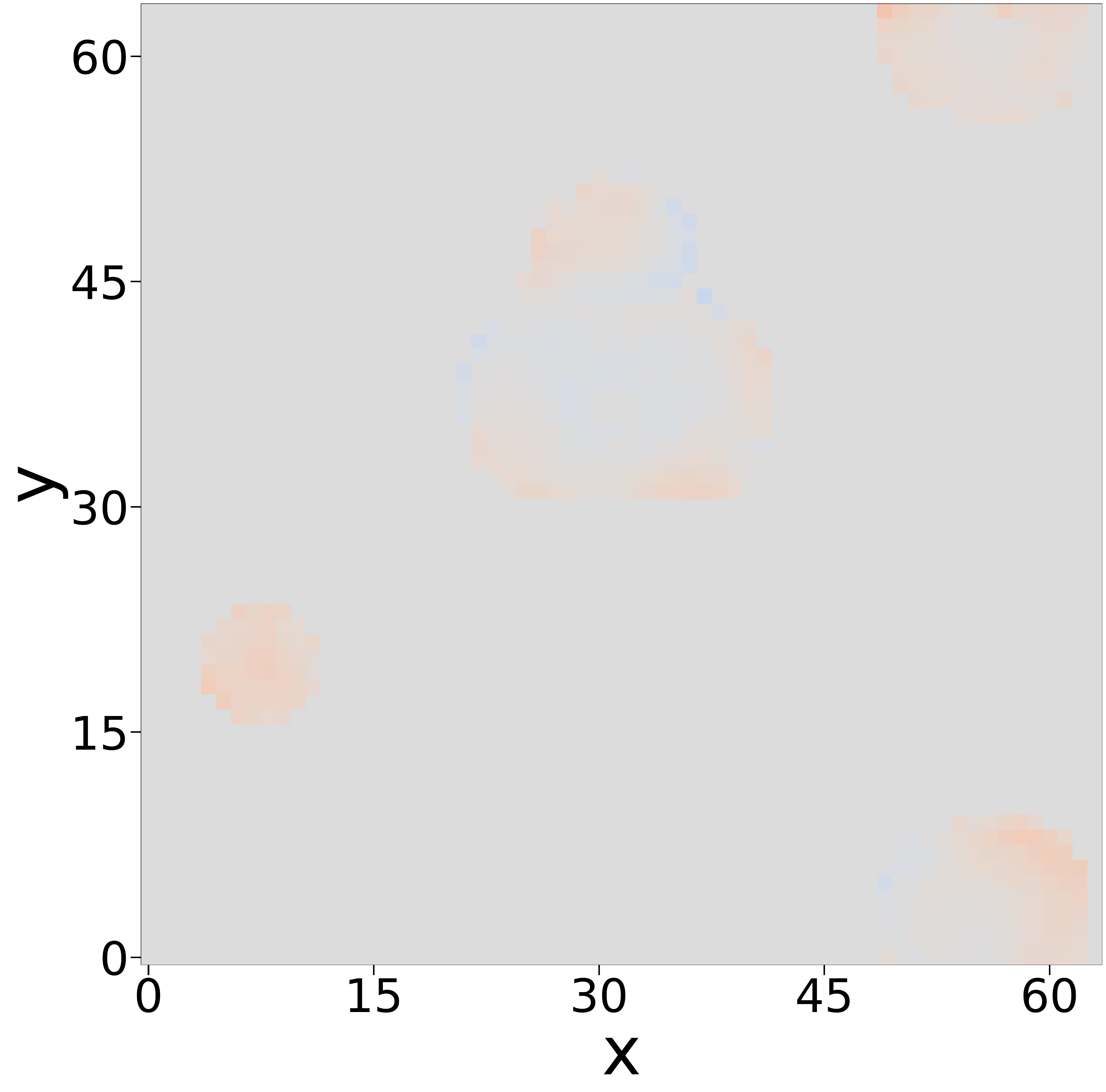}
    \end{subfigure}
    \begin{subfigure}[b]{0.0882\textwidth}
        \captionsetup{labelformat=empty}
        \includegraphics[width=\textwidth]{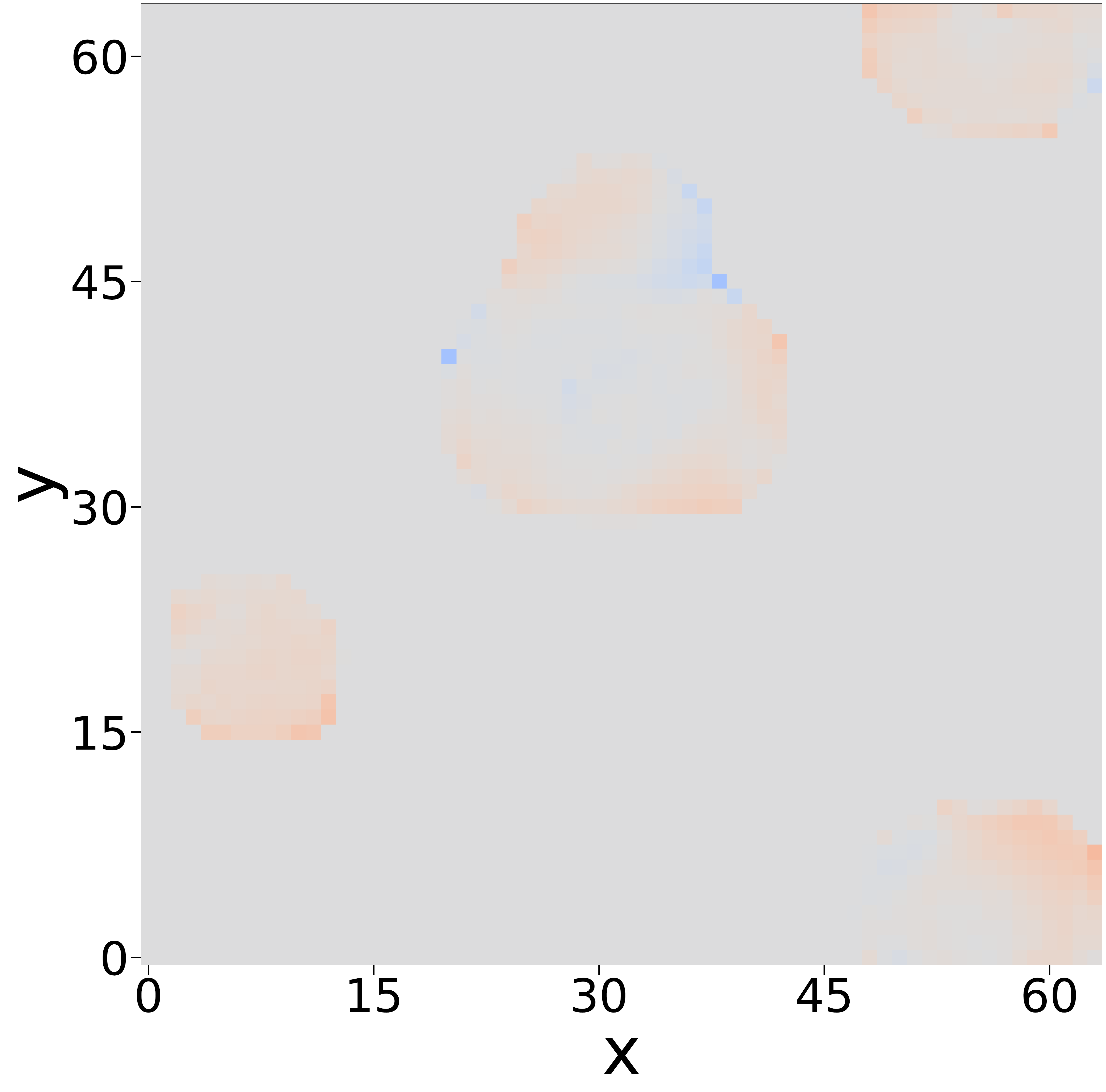}
    \end{subfigure}
    \begin{subfigure}[b]{0.0882\textwidth}
        \captionsetup{labelformat=empty}
        \includegraphics[width=\textwidth]{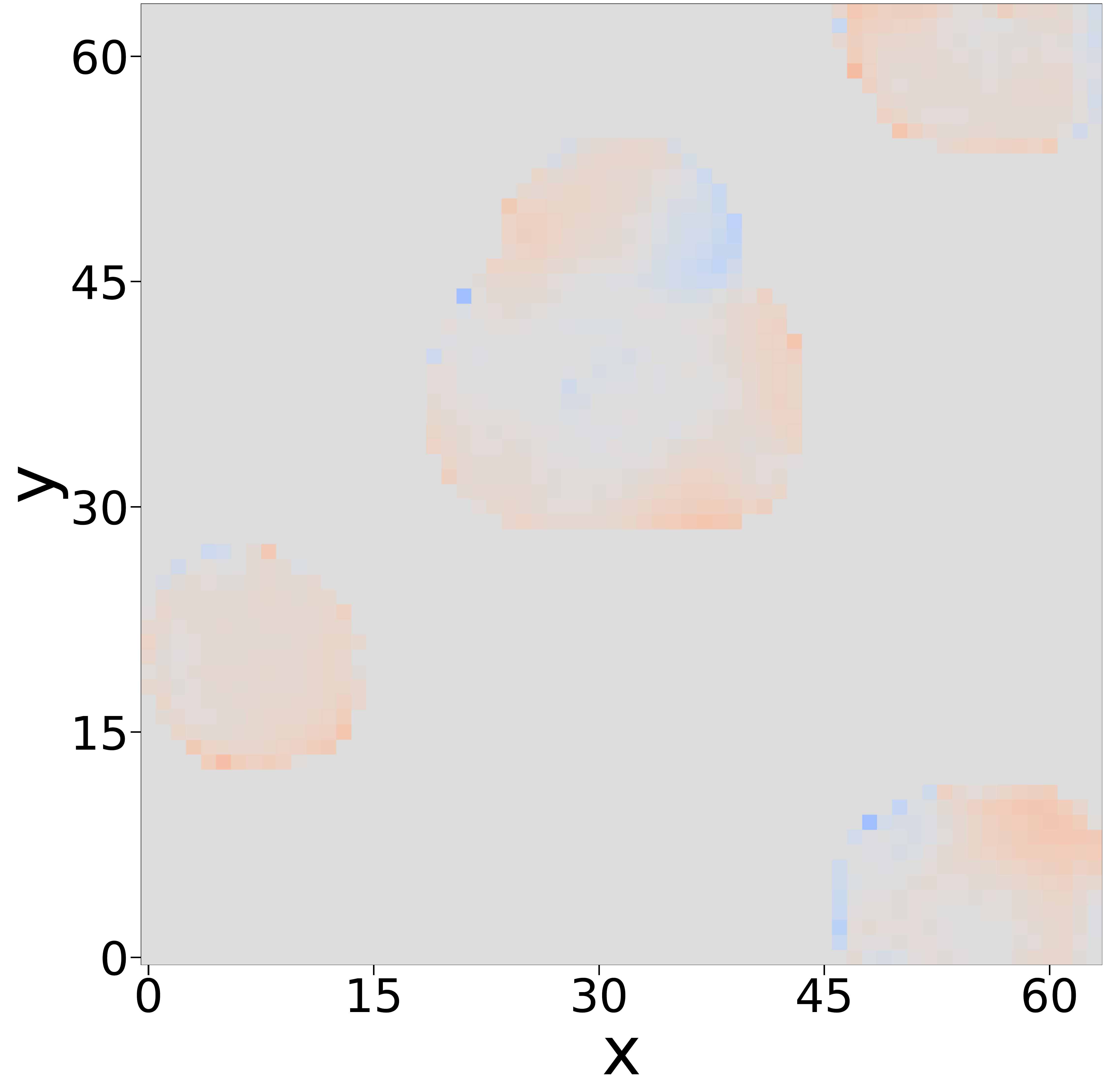}
    \end{subfigure}
    \begin{subfigure}[b]{0.0882\textwidth}
        \captionsetup{labelformat=empty}
        \includegraphics[width=\textwidth]{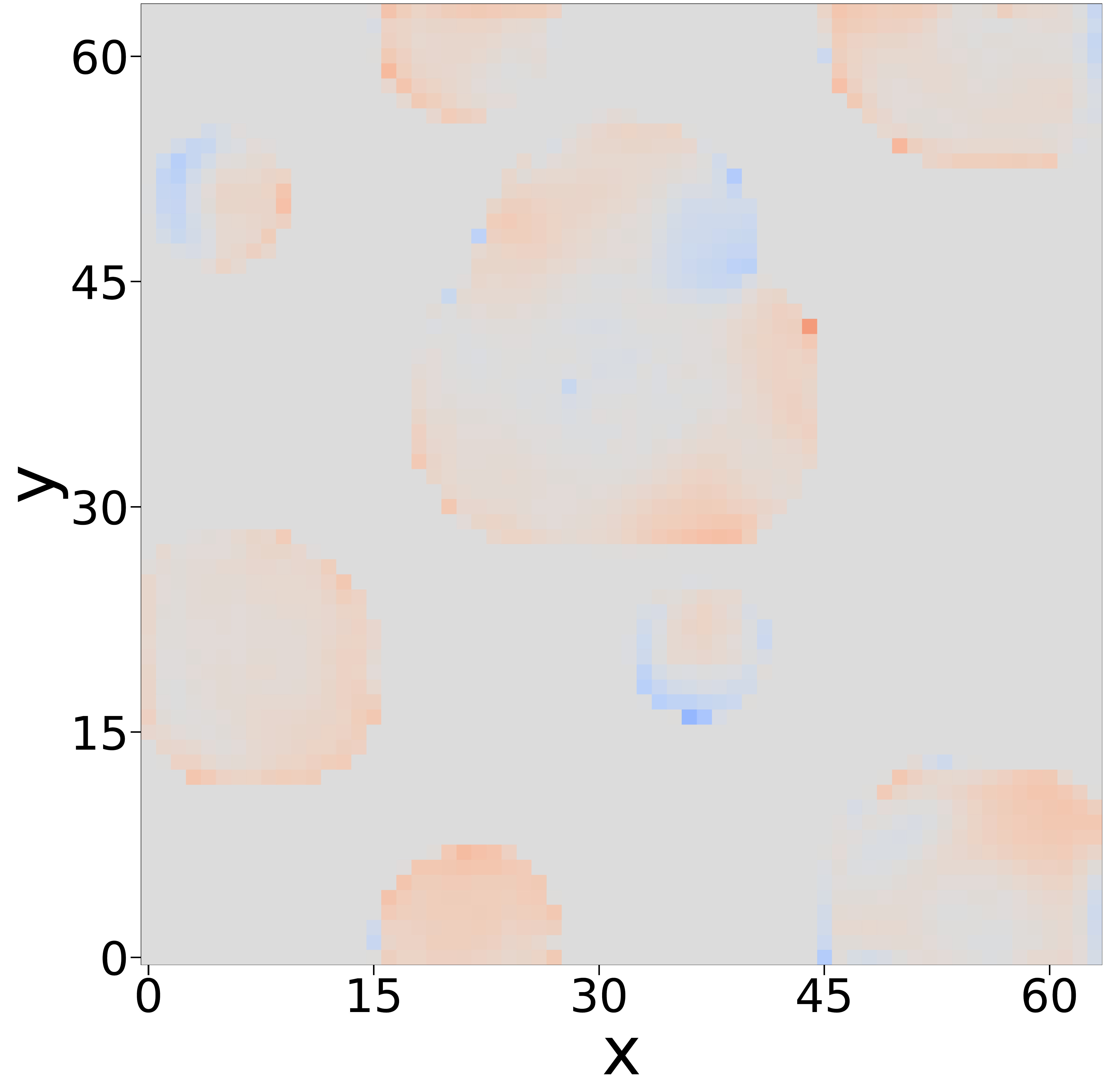}
    \end{subfigure}
    \begin{subfigure}[b]{0.0882\textwidth}
        \captionsetup{labelformat=empty}
        \includegraphics[width=\textwidth]{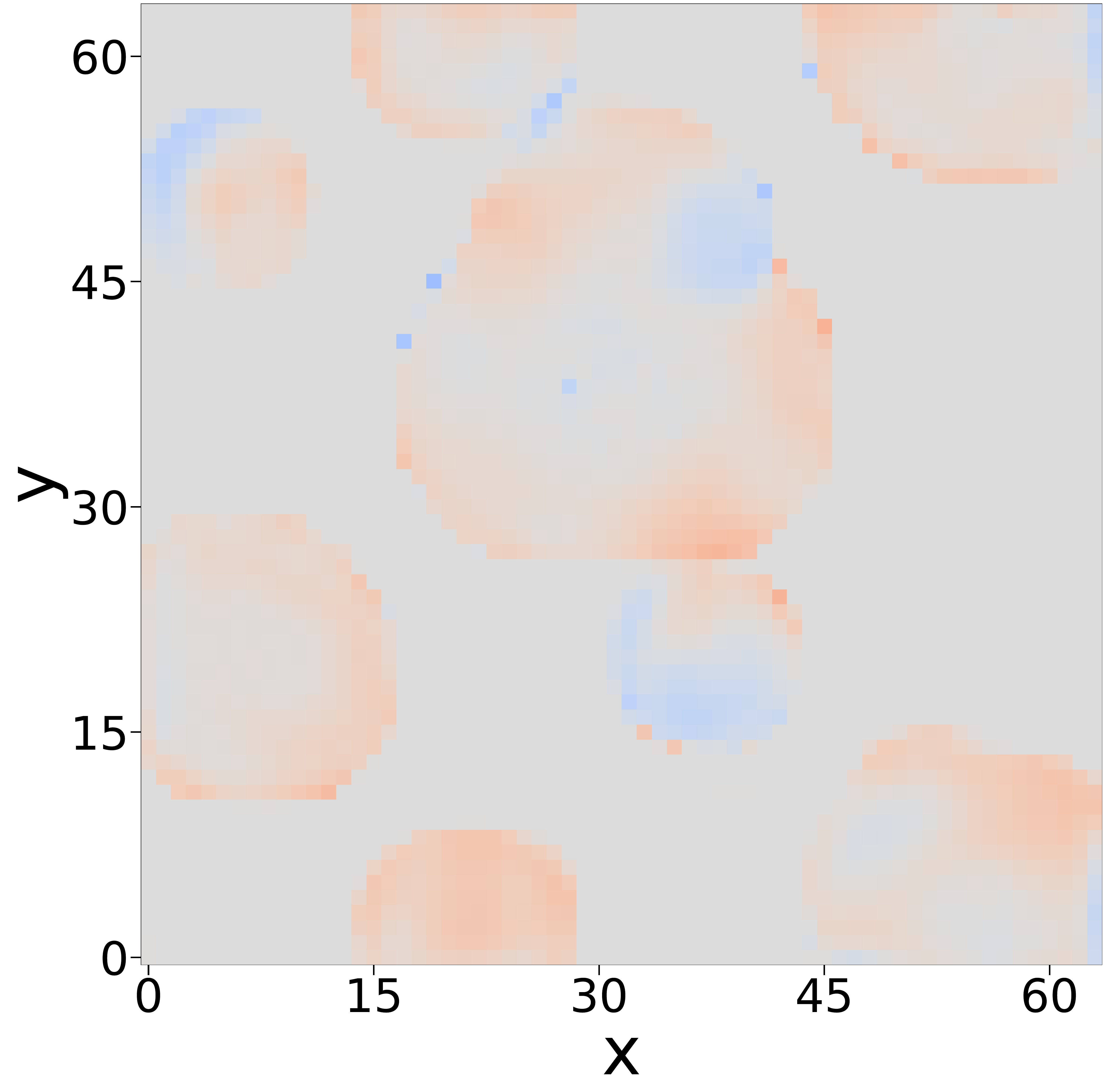}
    \end{subfigure}

    \begin{subfigure}[b]{0.02346\textwidth}
        \captionsetup{labelformat=empty}
        \includegraphics[width=\textwidth]{graphics/colorbar_res.pdf}
    \end{subfigure}

    }

    \vspace{1cm}

\makebox[\textwidth][c]{
    \begin{subfigure}[b]{0.0882\textwidth}
        \captionsetup{labelformat=empty}
        \includegraphics[width=\textwidth]{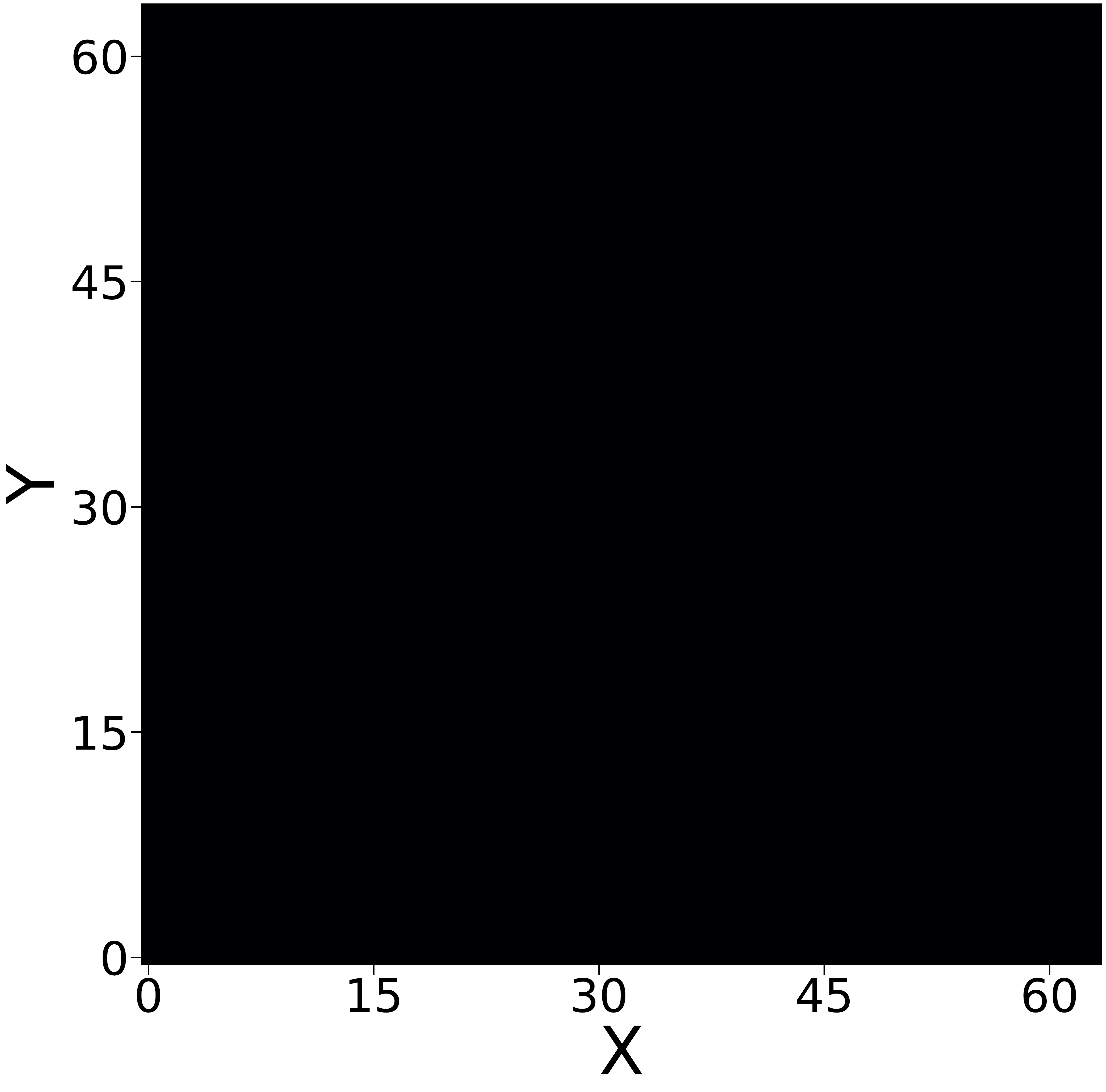}
    \end{subfigure}
    \begin{subfigure}[b]{0.0882\textwidth}
        \captionsetup{labelformat=empty}
        \includegraphics[width=\textwidth]{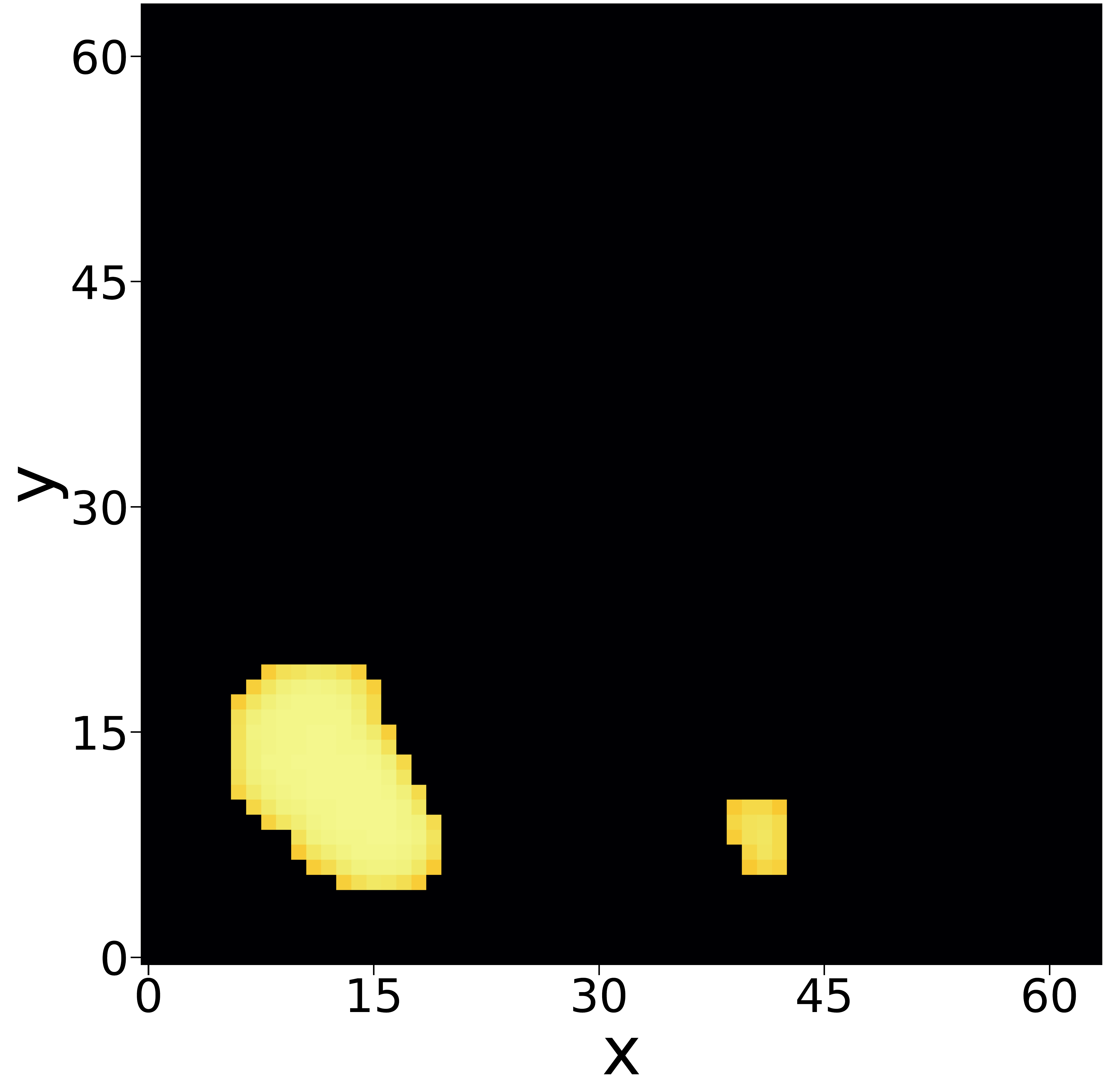}
    \end{subfigure}
    \begin{subfigure}[b]{0.0882\textwidth}
        \captionsetup{labelformat=empty}
        \includegraphics[width=\textwidth]{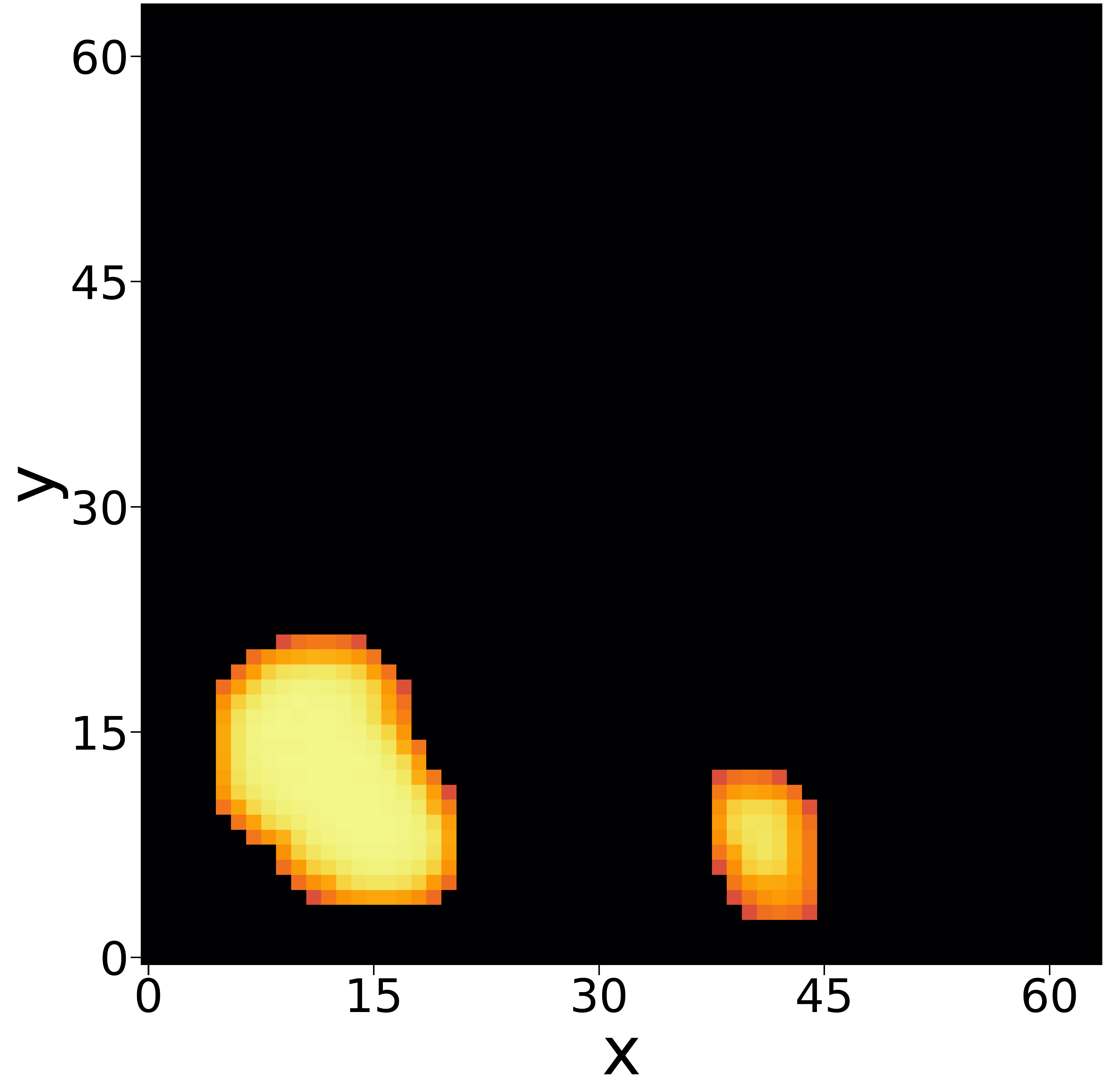}
    \end{subfigure}
    \begin{subfigure}[b]{0.0882\textwidth}
        \captionsetup{labelformat=empty}
        \includegraphics[width=\textwidth]{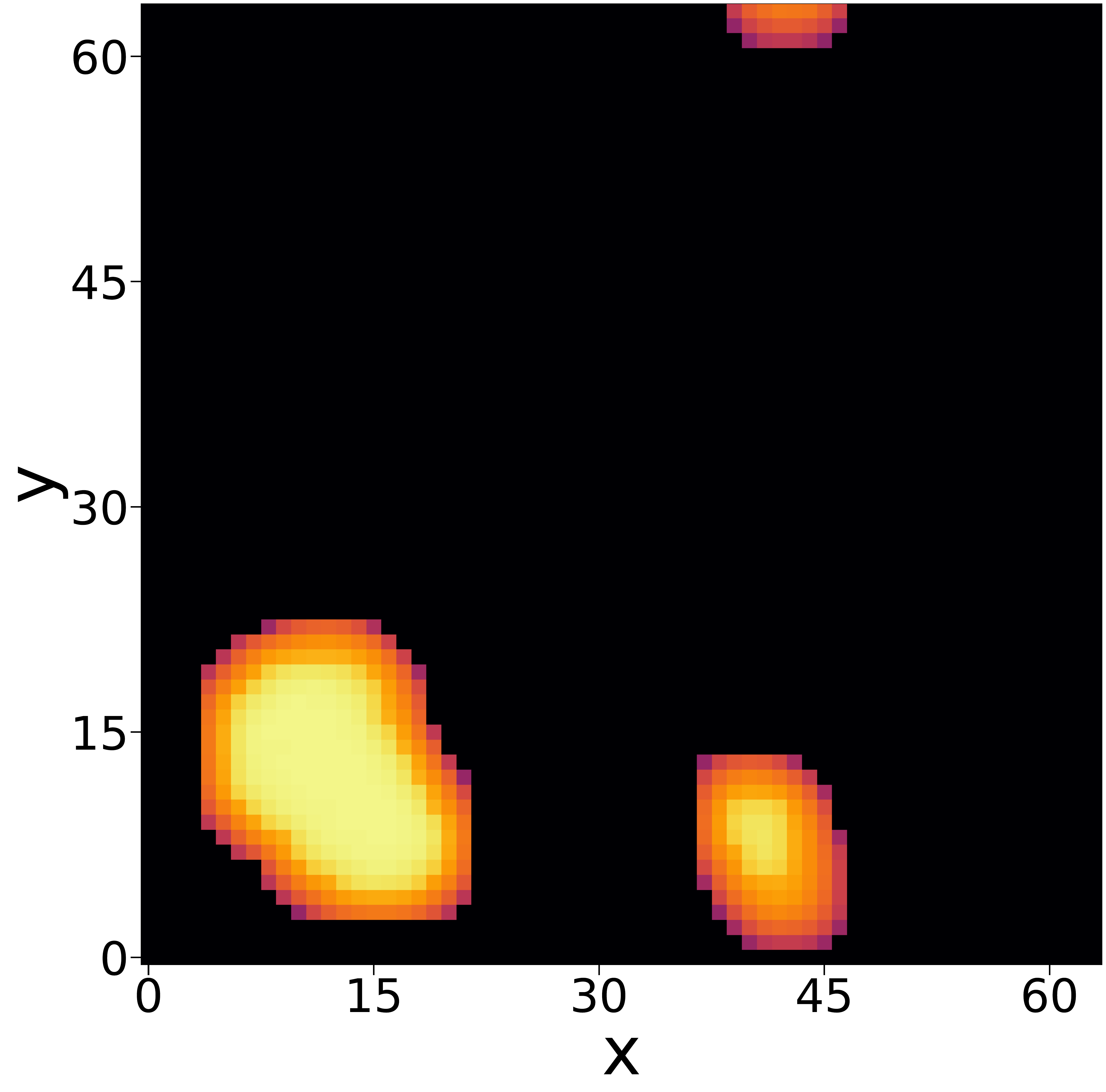}
    \end{subfigure}
    
    \begin{subfigure}[b]{0.0882\textwidth}
        \captionsetup{labelformat=empty}
        \includegraphics[width=\textwidth]{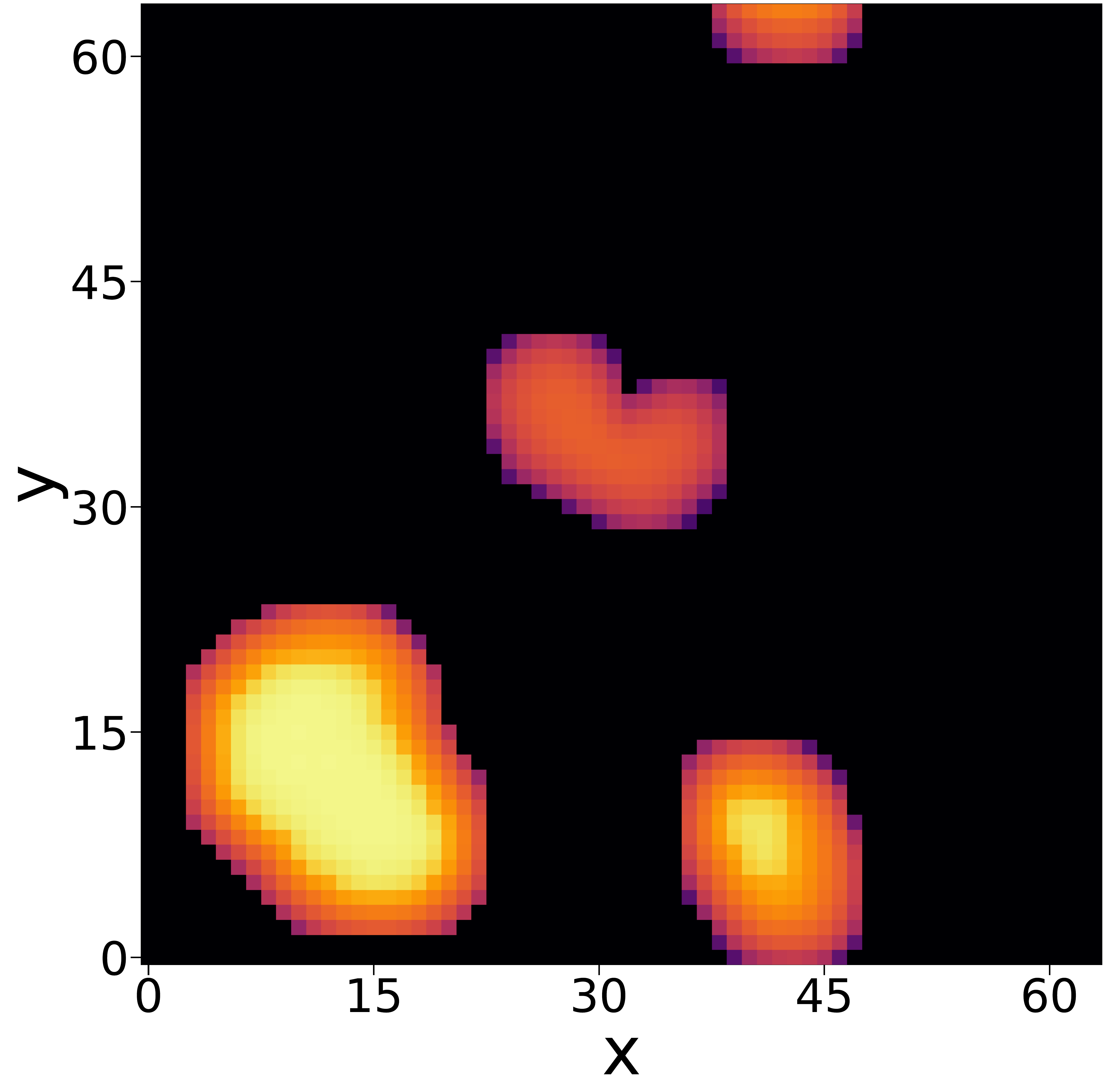}
    \end{subfigure}

    \begin{subfigure}[b]{0.0882\textwidth}
        \captionsetup{labelformat=empty}
        \includegraphics[width=\textwidth]{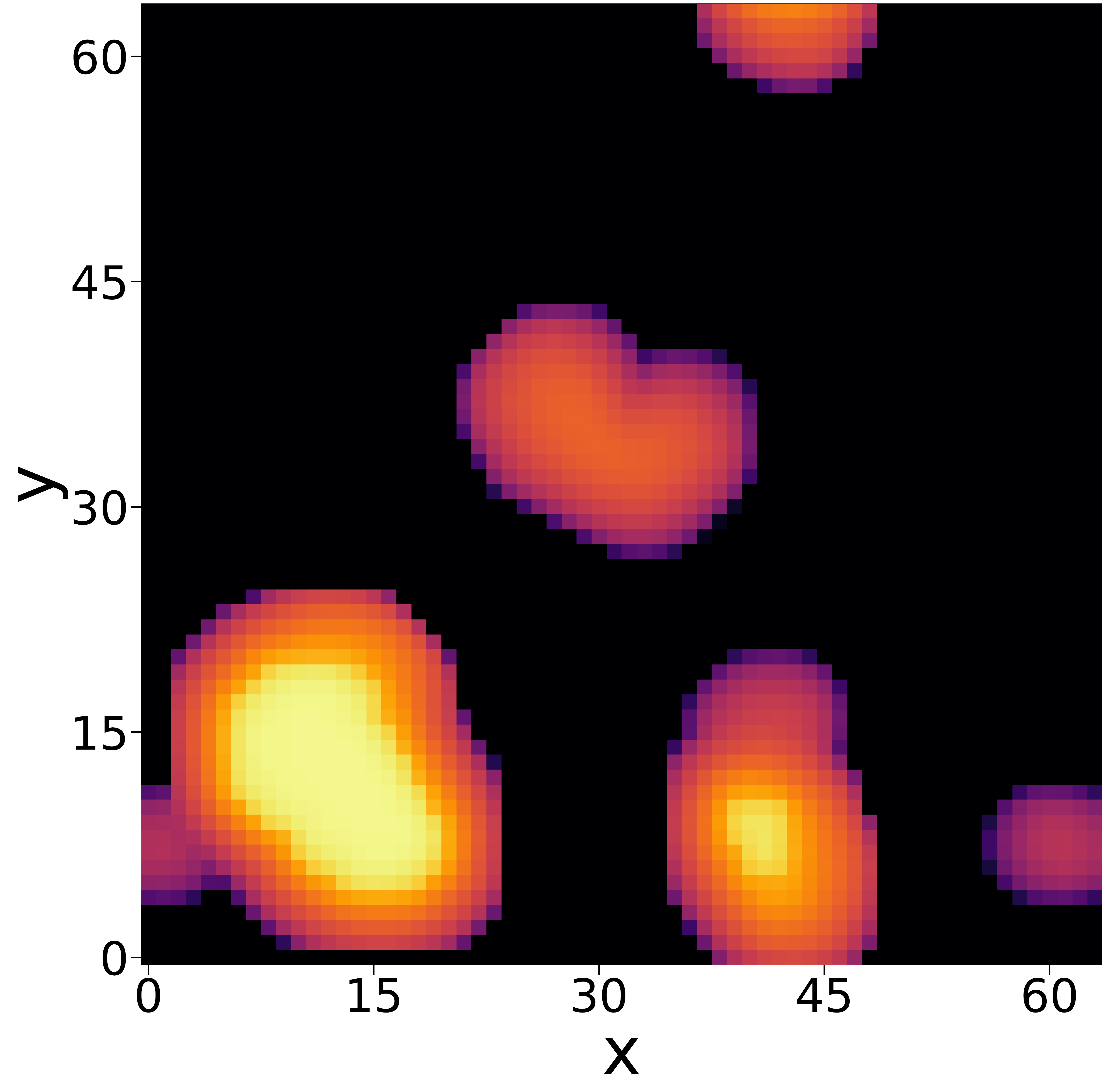}
    \end{subfigure}
    \begin{subfigure}[b]{0.0882\textwidth}
        \captionsetup{labelformat=empty}
        \includegraphics[width=\textwidth]{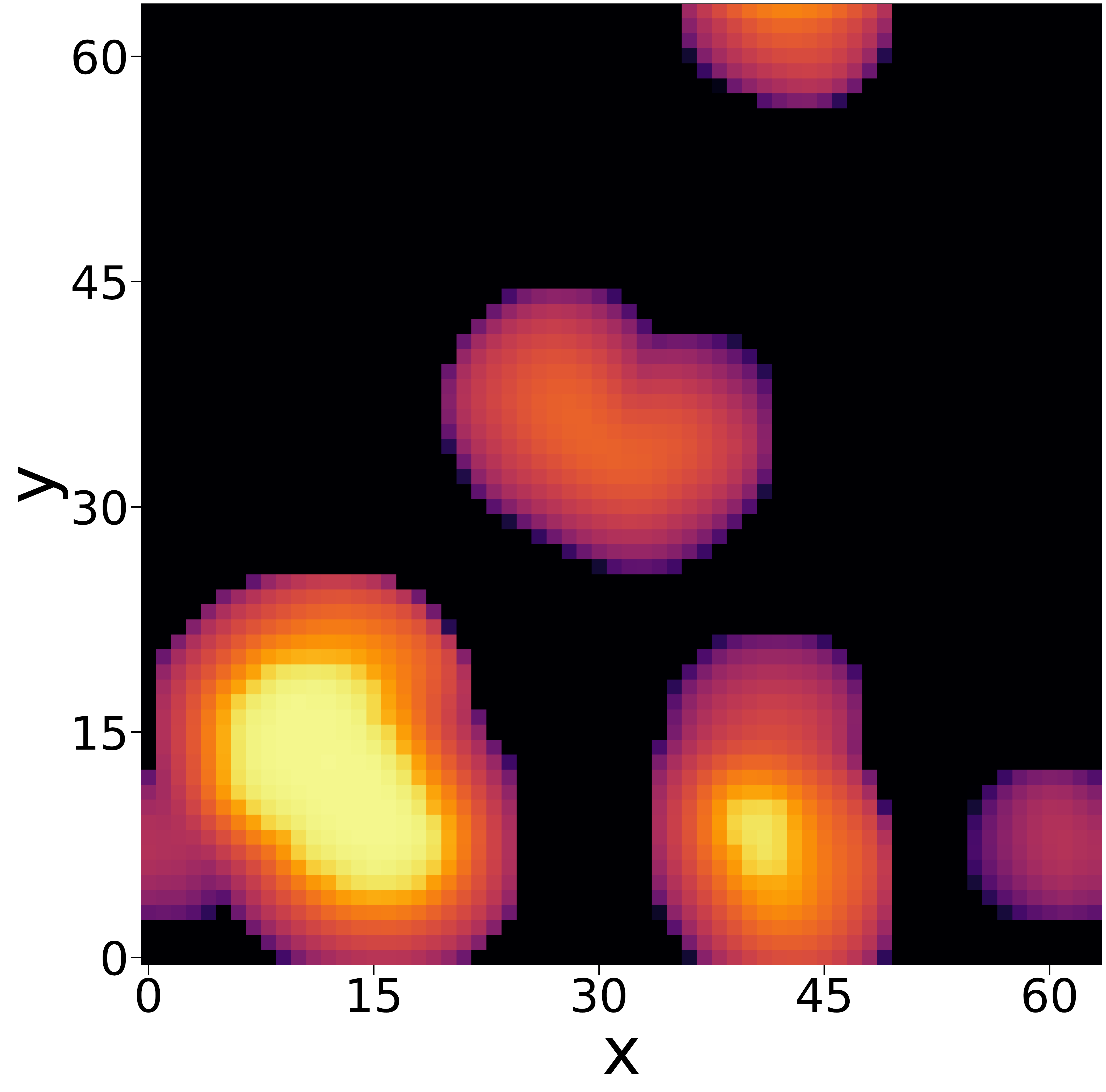}
    \end{subfigure}
    \begin{subfigure}[b]{0.0882\textwidth}
        \captionsetup{labelformat=empty}
        \includegraphics[width=\textwidth]{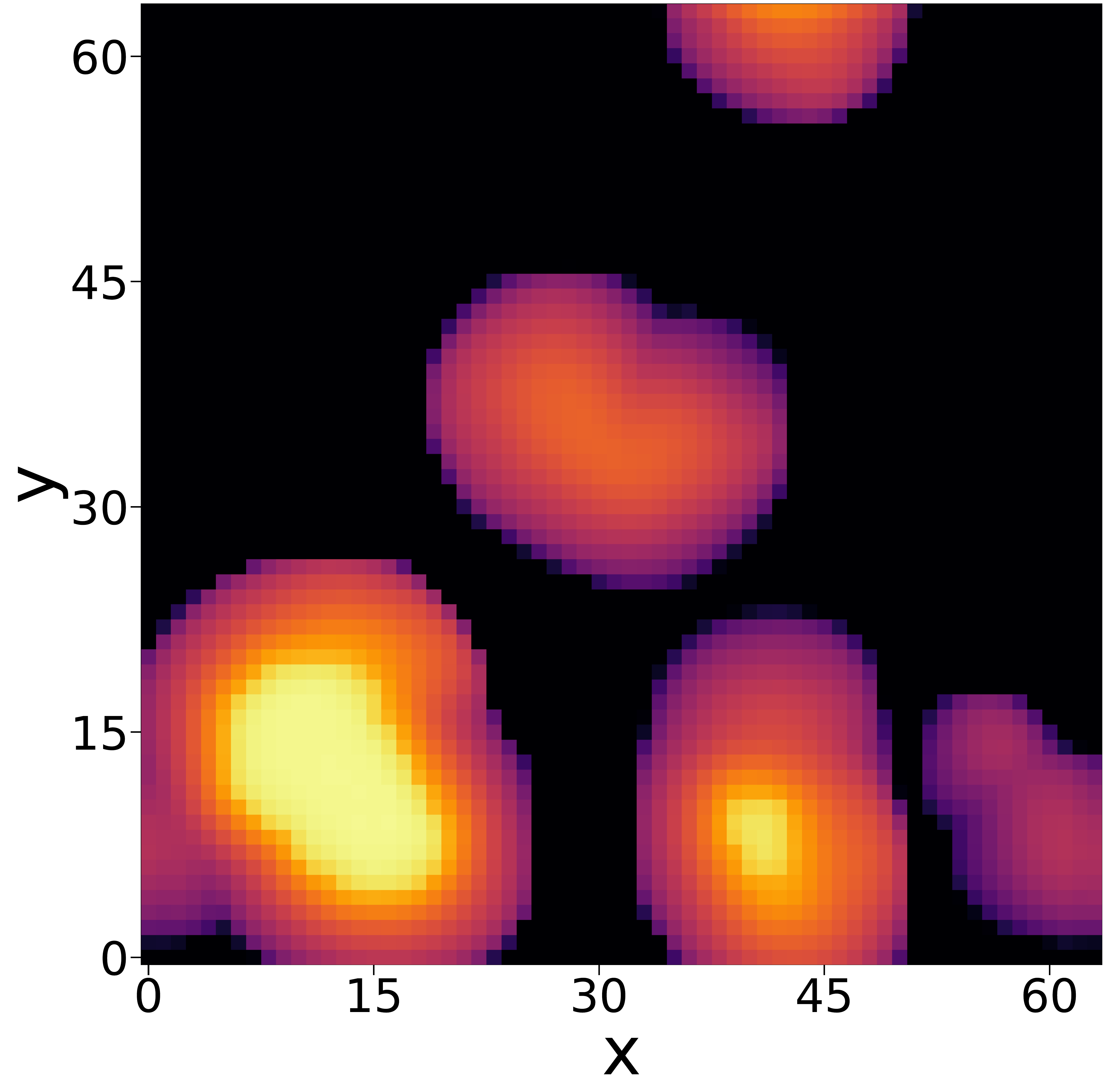}
    \end{subfigure}
    \begin{subfigure}[b]{0.0882\textwidth}
        \captionsetup{labelformat=empty}
        \includegraphics[width=\textwidth]{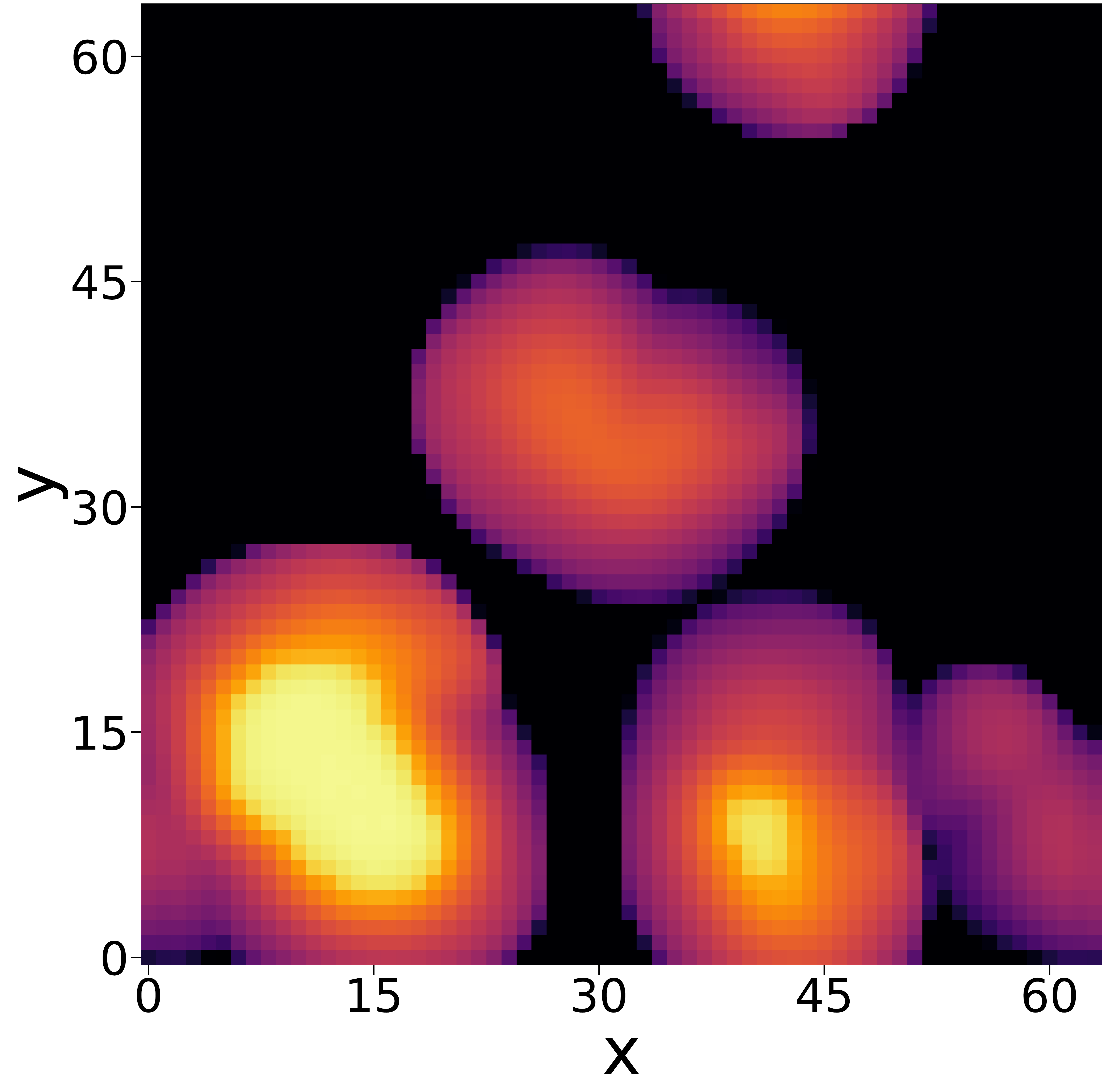}
    \end{subfigure}
    \begin{subfigure}[b]{0.0882\textwidth}
        \captionsetup{labelformat=empty}
        \includegraphics[width=\textwidth]{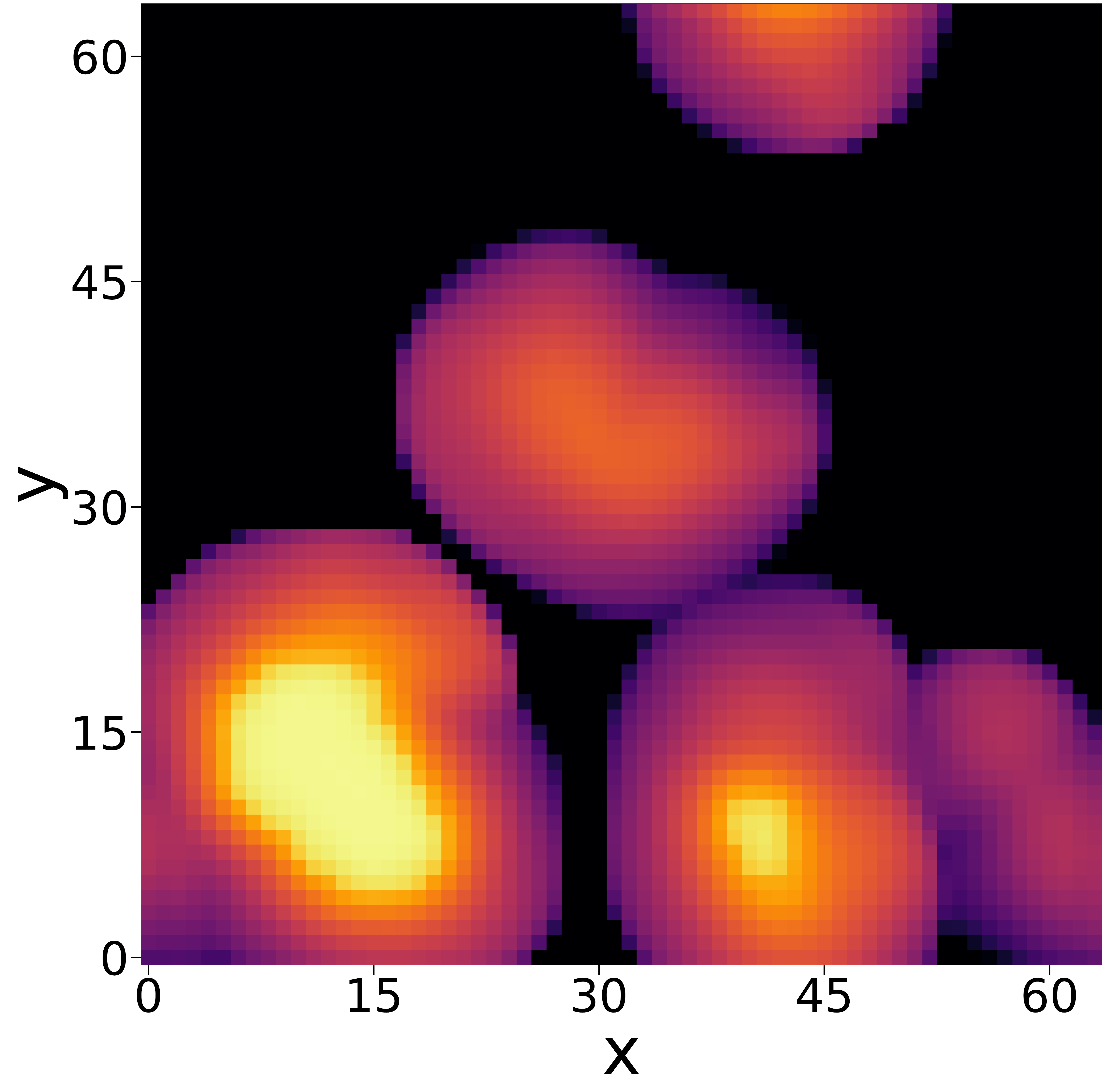}
    \end{subfigure}

    \begin{subfigure}[b]{0.02346\textwidth}
        \captionsetup{labelformat=empty}
        \includegraphics[width=\textwidth]{graphics/colorbar_pred.pdf}
    \end{subfigure}

    }
    \makebox[\textwidth][c]{
    \begin{subfigure}[b]{0.0882\textwidth}
        \captionsetup{labelformat=empty}
        \includegraphics[width=\textwidth]{graphics/true_XZ_init.pdf}
    \end{subfigure}
    \begin{subfigure}[b]{0.0882\textwidth}
        \captionsetup{labelformat=empty}
        \includegraphics[width=\textwidth]{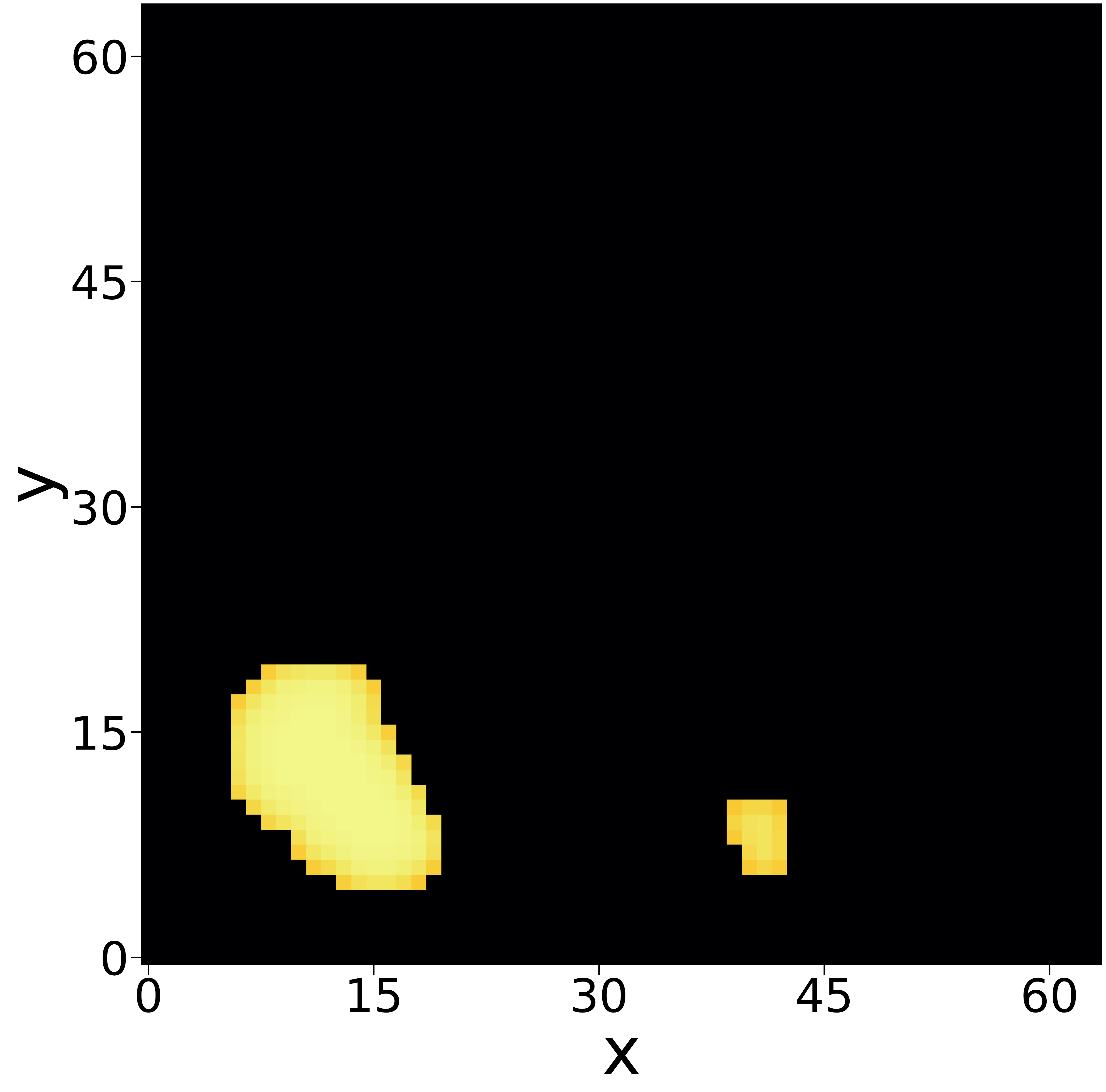}
    \end{subfigure}
    \begin{subfigure}[b]{0.0882\textwidth}
        \captionsetup{labelformat=empty}
        \includegraphics[width=\textwidth]{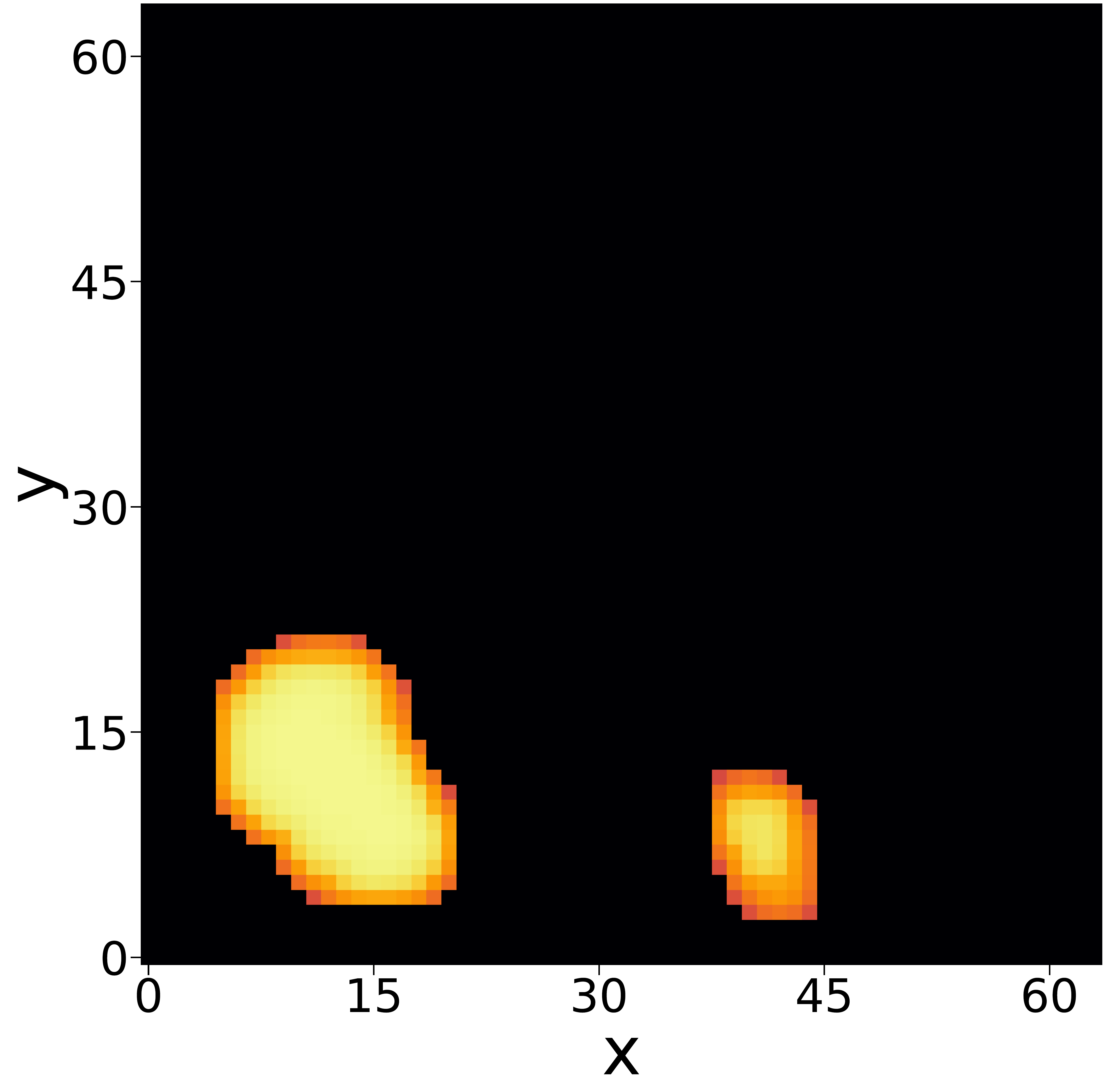}
    \end{subfigure}
    \begin{subfigure}[b]{0.0882\textwidth}
        \captionsetup{labelformat=empty}
        \includegraphics[width=\textwidth]{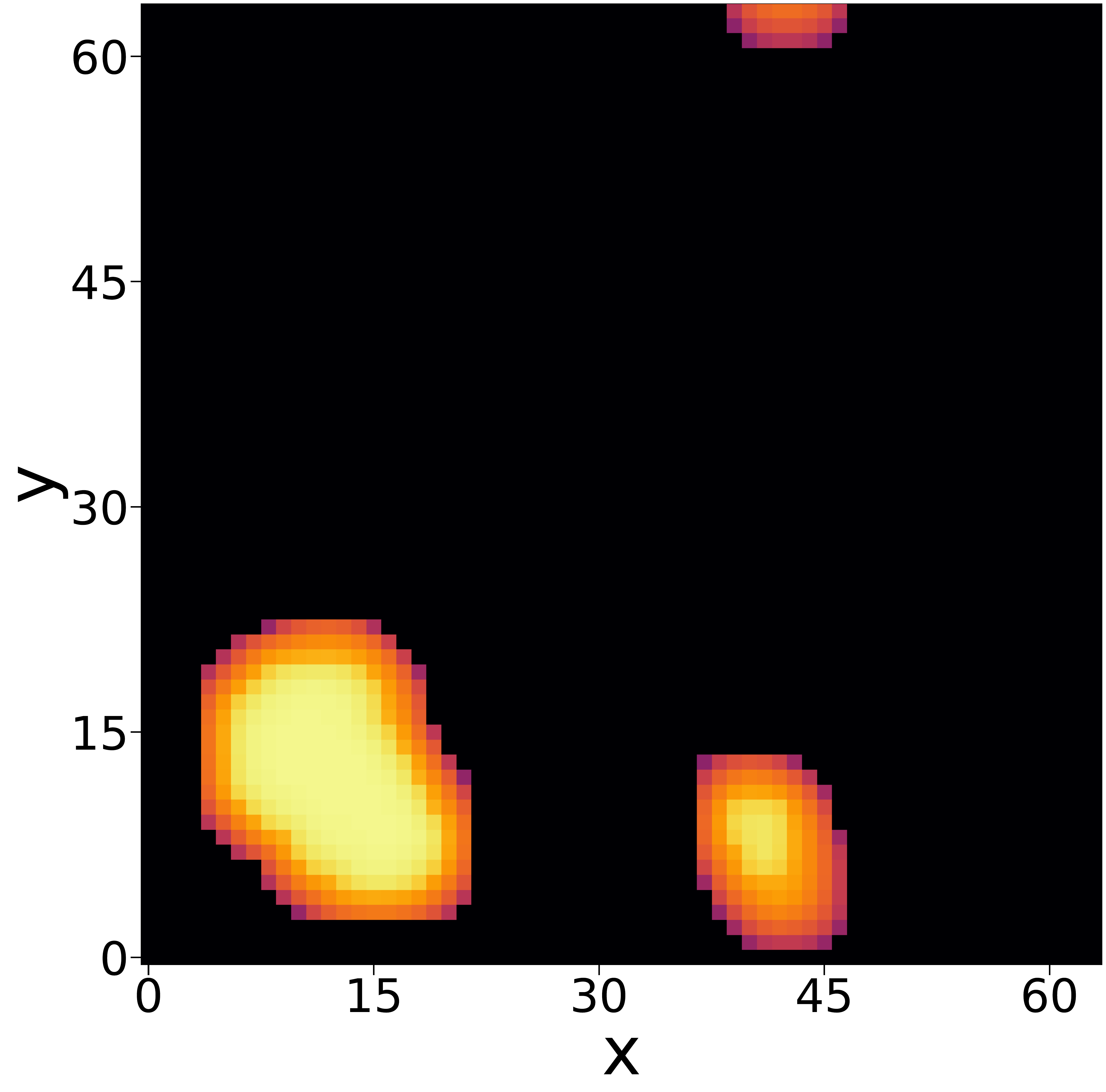}
    \end{subfigure}
    
    \begin{subfigure}[b]{0.0882\textwidth}
        \captionsetup{labelformat=empty}
        \includegraphics[width=\textwidth]{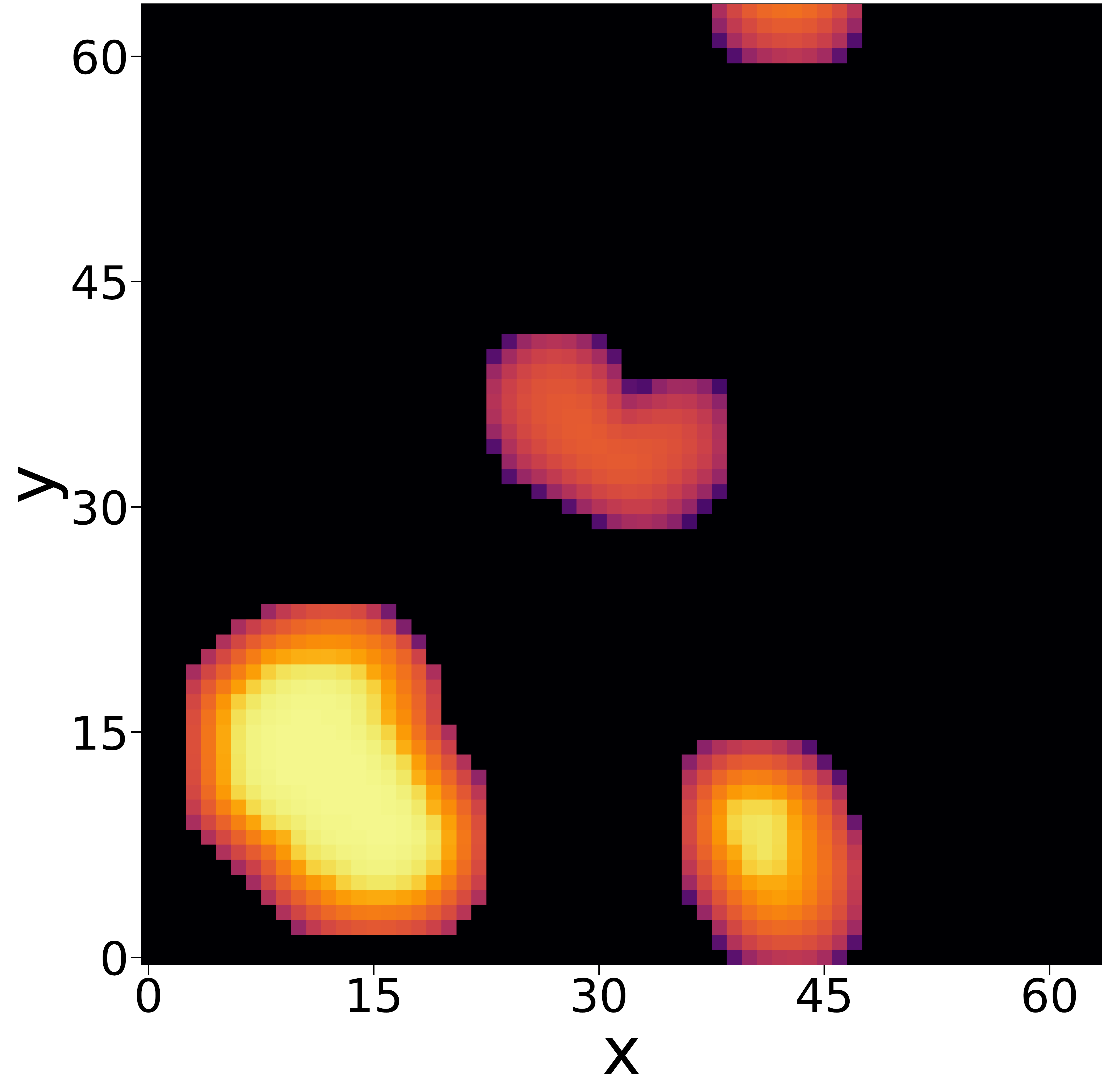}
    \end{subfigure}

    \begin{subfigure}[b]{0.0882\textwidth}
        \captionsetup{labelformat=empty}
        \includegraphics[width=\textwidth]{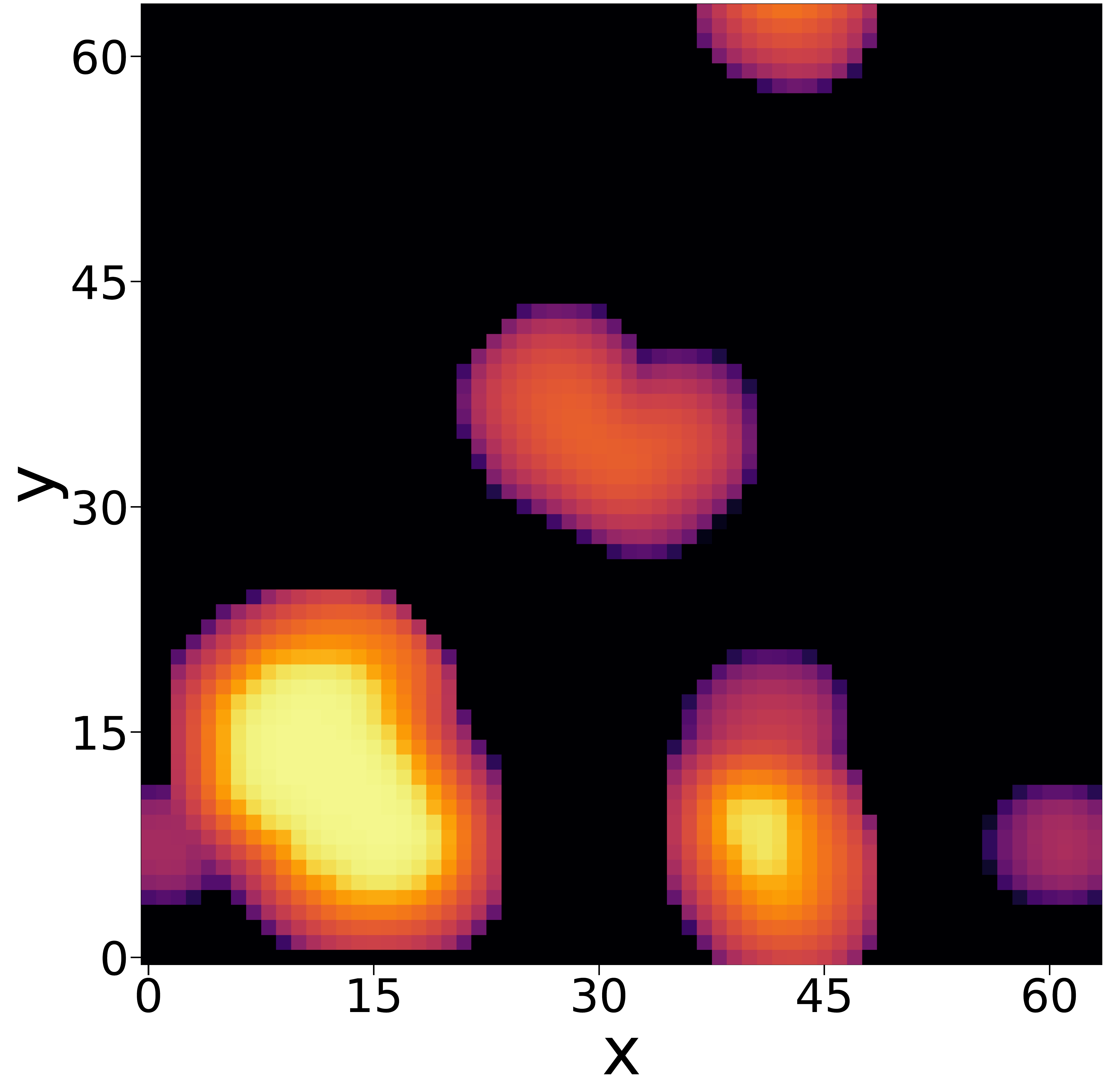}
    \end{subfigure}
    \begin{subfigure}[b]{0.0882\textwidth}
        \captionsetup{labelformat=empty}
        \includegraphics[width=\textwidth]{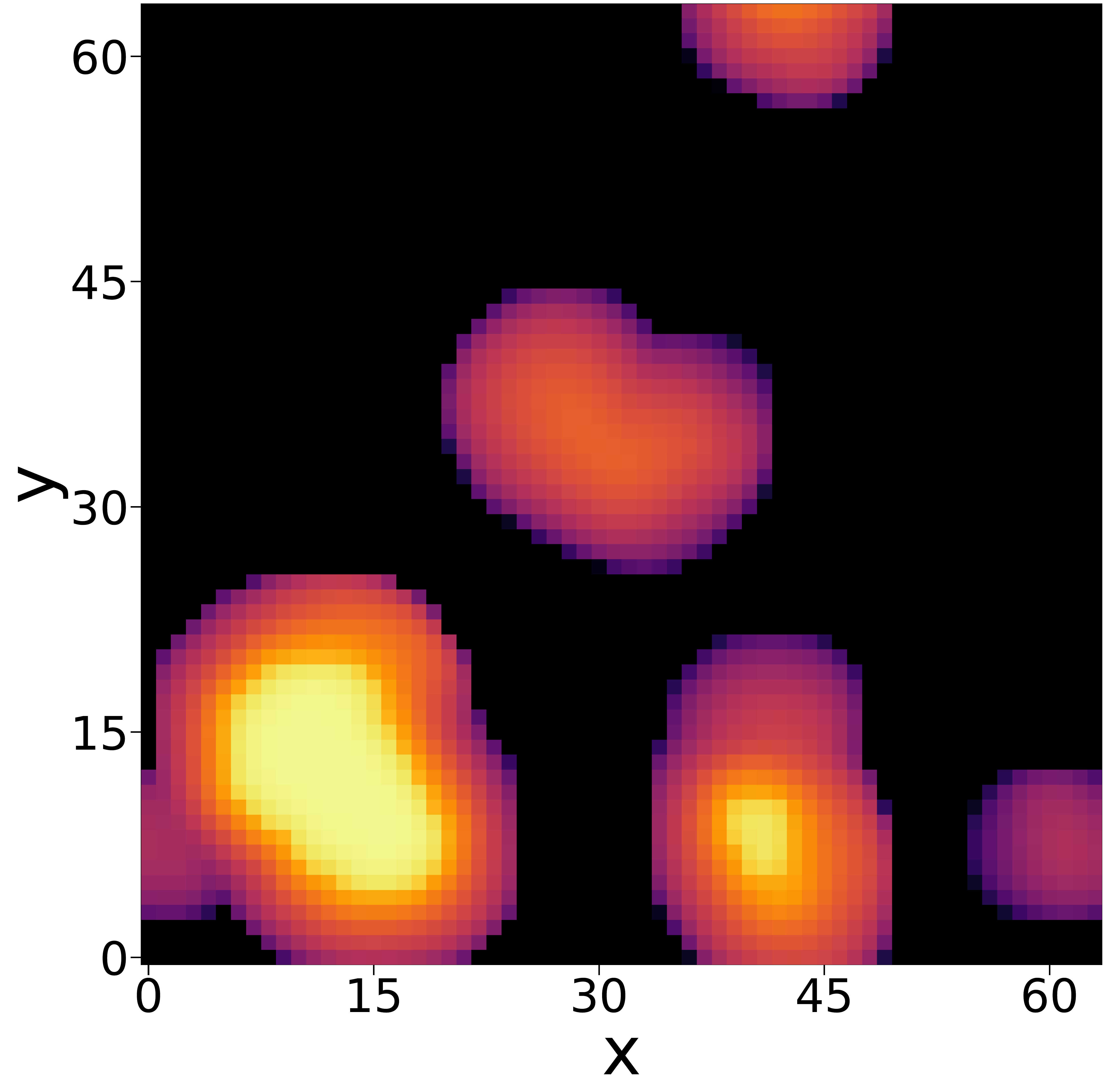}
    \end{subfigure}
    \begin{subfigure}[b]{0.0882\textwidth}
        \captionsetup{labelformat=empty}
        \includegraphics[width=\textwidth]{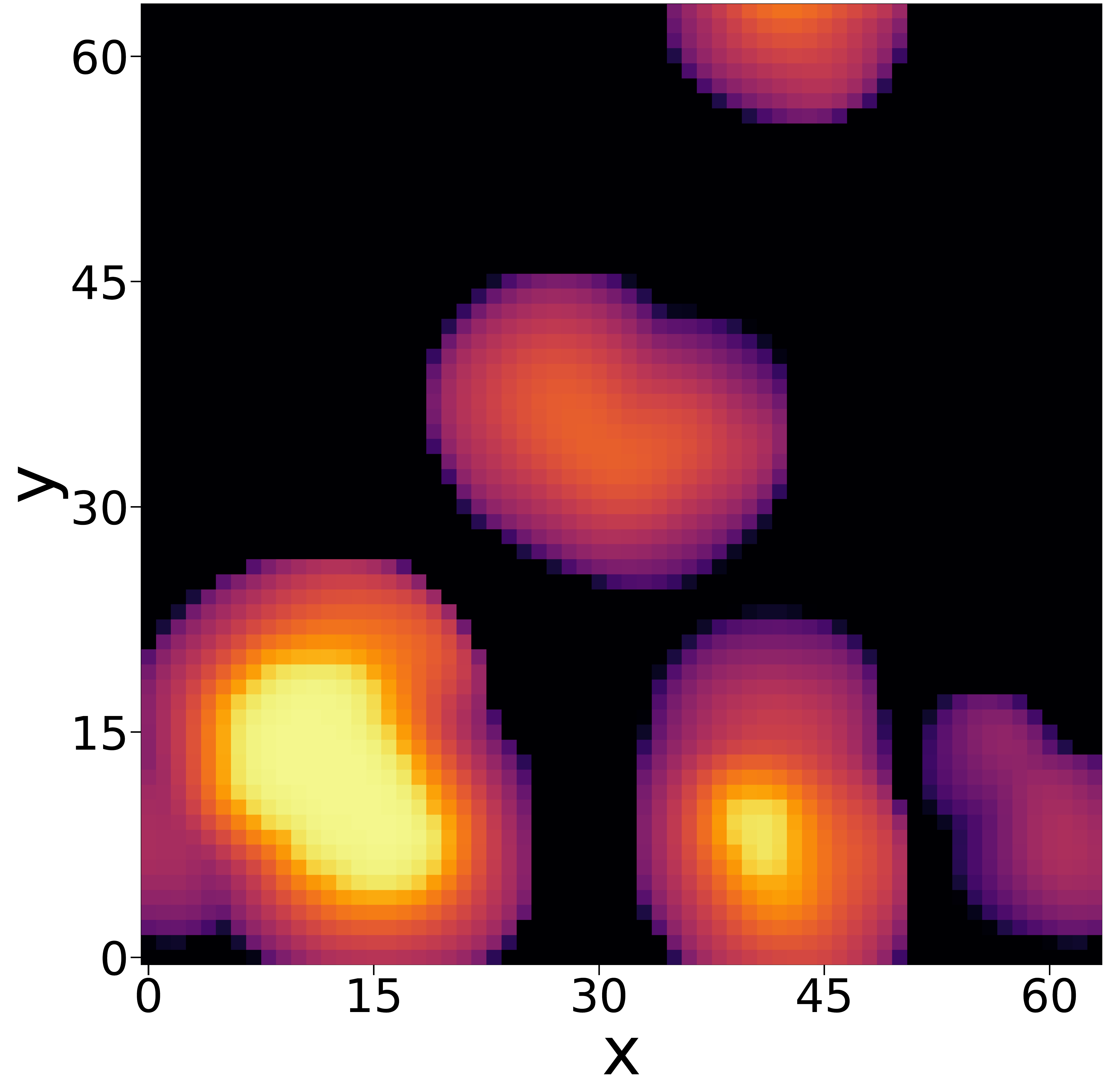}
    \end{subfigure}
    \begin{subfigure}[b]{0.0882\textwidth}
        \captionsetup{labelformat=empty}
        \includegraphics[width=\textwidth]{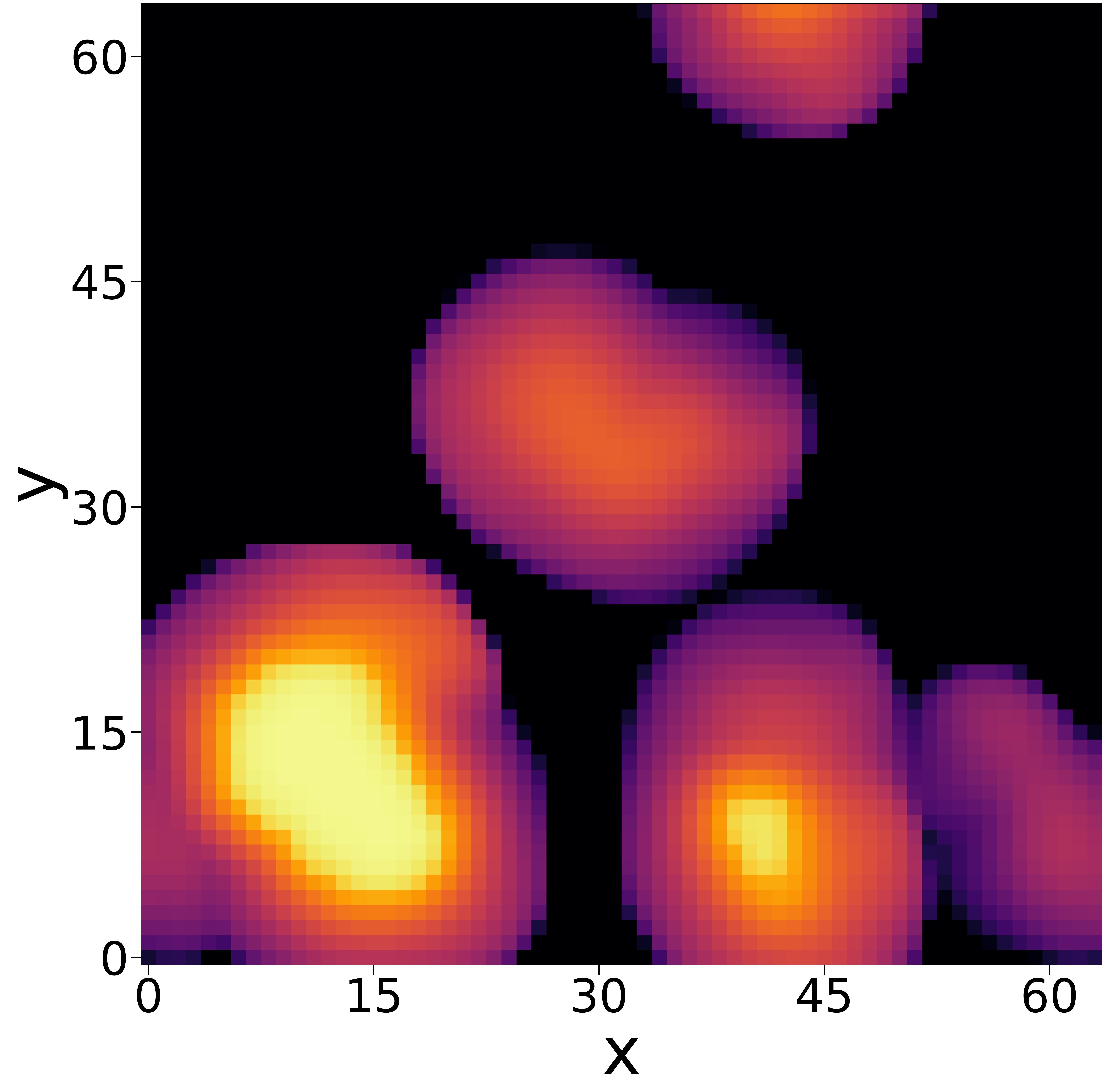}
    \end{subfigure}
    \begin{subfigure}[b]{0.0882\textwidth}
        \captionsetup{labelformat=empty}
        \includegraphics[width=\textwidth]{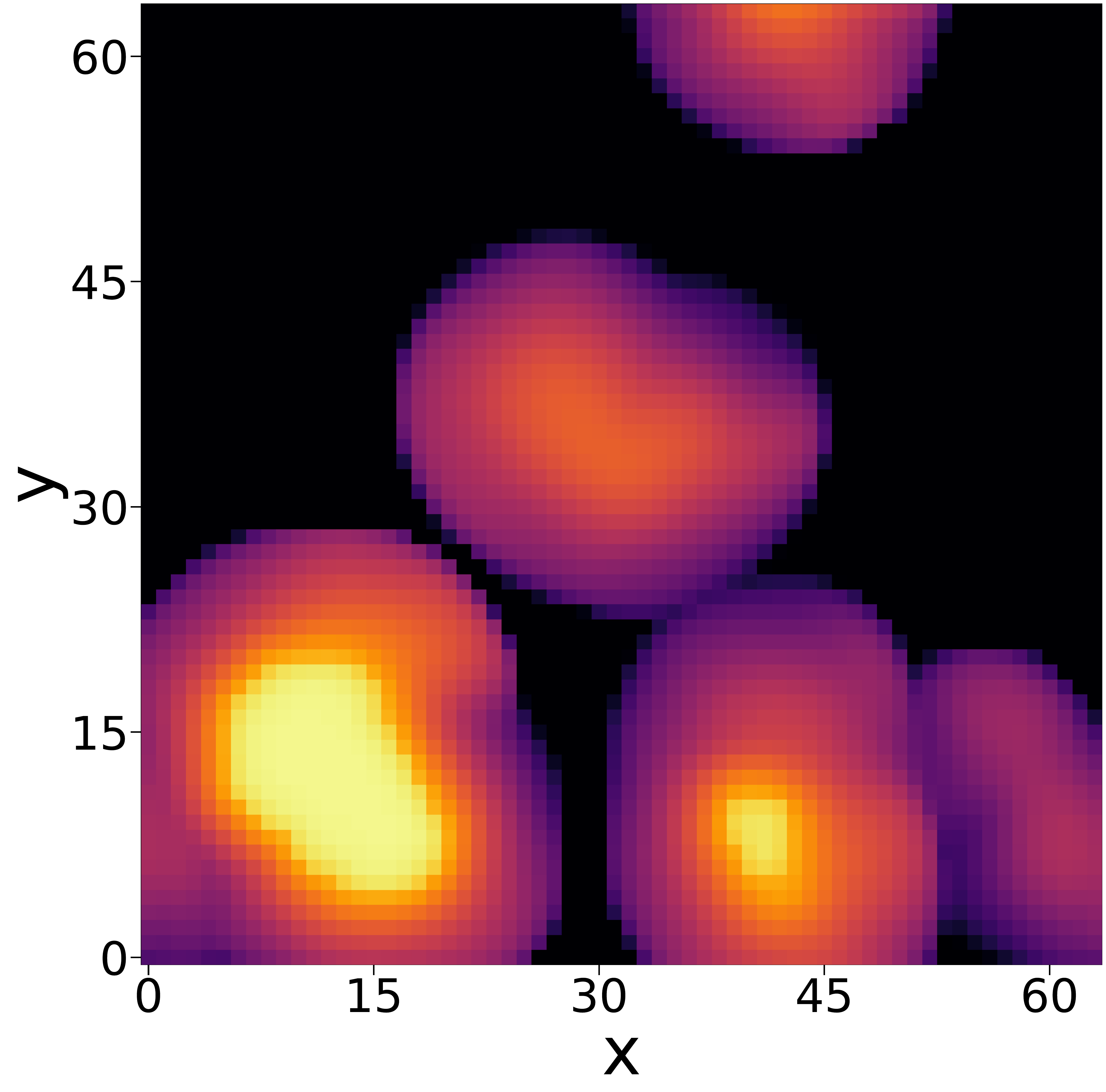}
    \end{subfigure}

    \begin{subfigure}[b]{0.02346\textwidth}
        \captionsetup{labelformat=empty}
        \includegraphics[width=\textwidth]{graphics/colorbar_num.pdf}
    \end{subfigure}

    }
    \makebox[\textwidth][c]{
    \begin{subfigure}[b]{0.0882\textwidth}
        \captionsetup{labelformat=empty}
        \includegraphics[width=\textwidth]{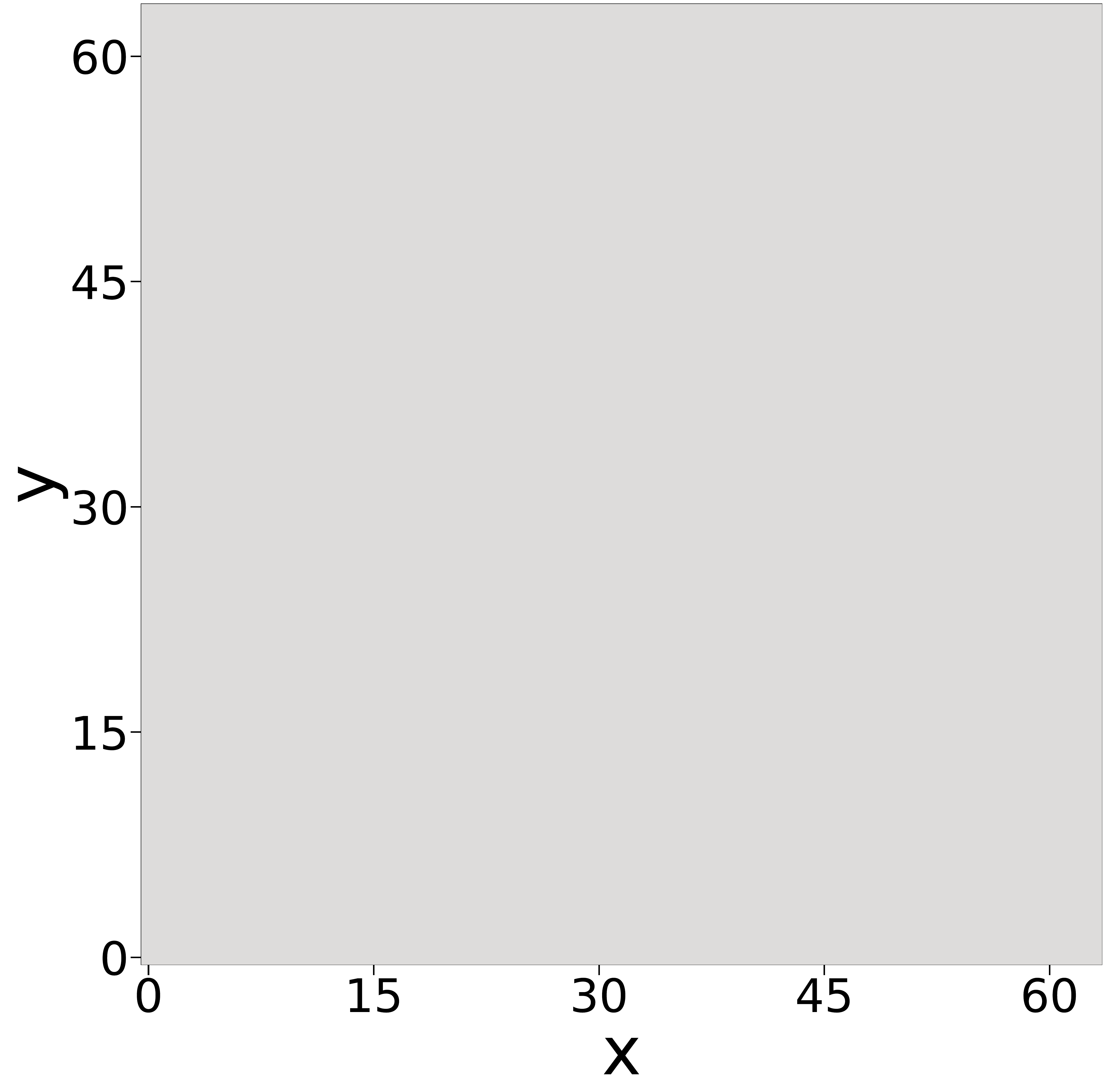}
    \end{subfigure}
    \begin{subfigure}[b]{0.0882\textwidth}
        \captionsetup{labelformat=empty}
        \includegraphics[width=\textwidth]{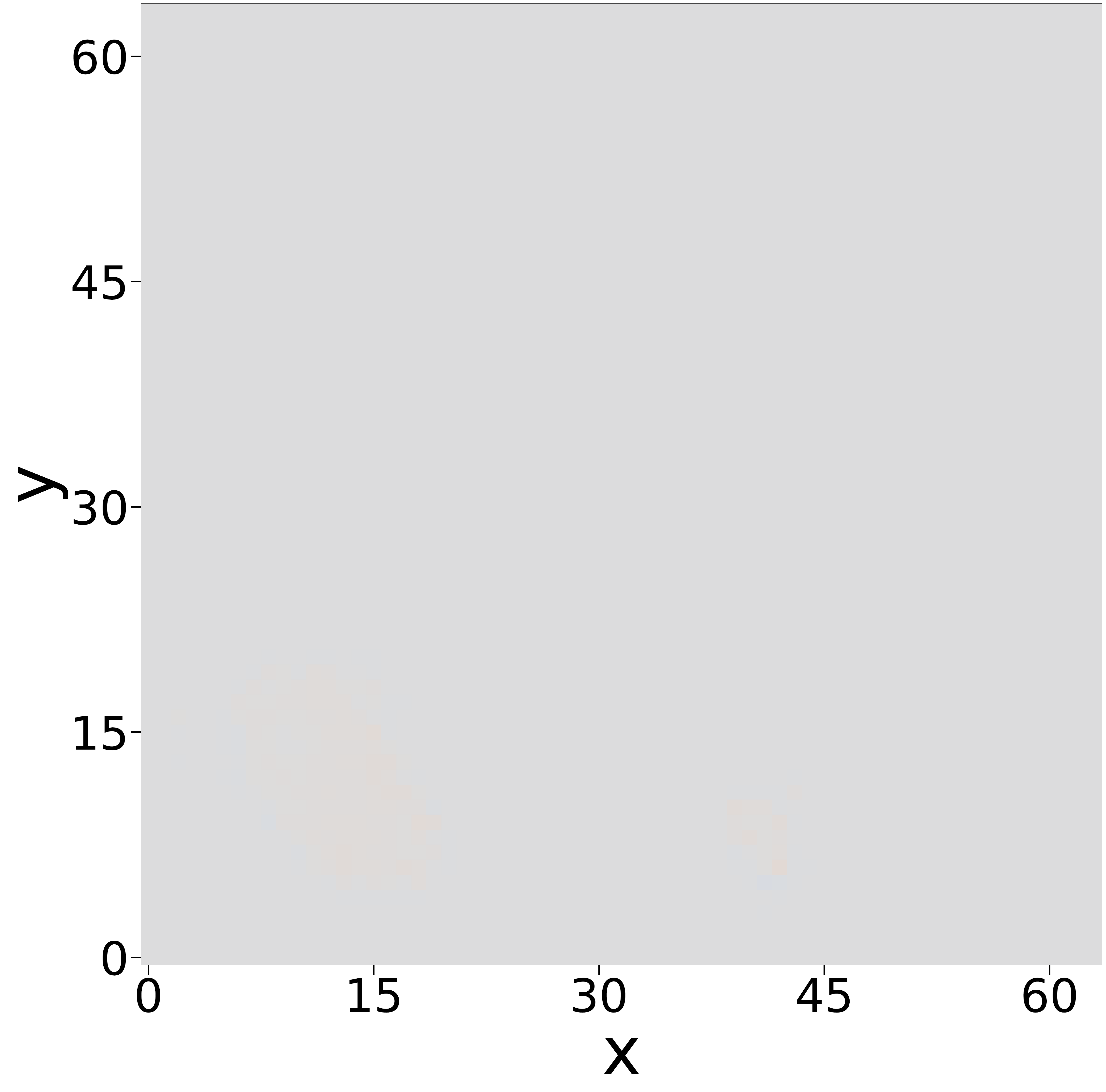}
    \end{subfigure}
    \begin{subfigure}[b]{0.0882\textwidth}
        \captionsetup{labelformat=empty}
        \includegraphics[width=\textwidth]{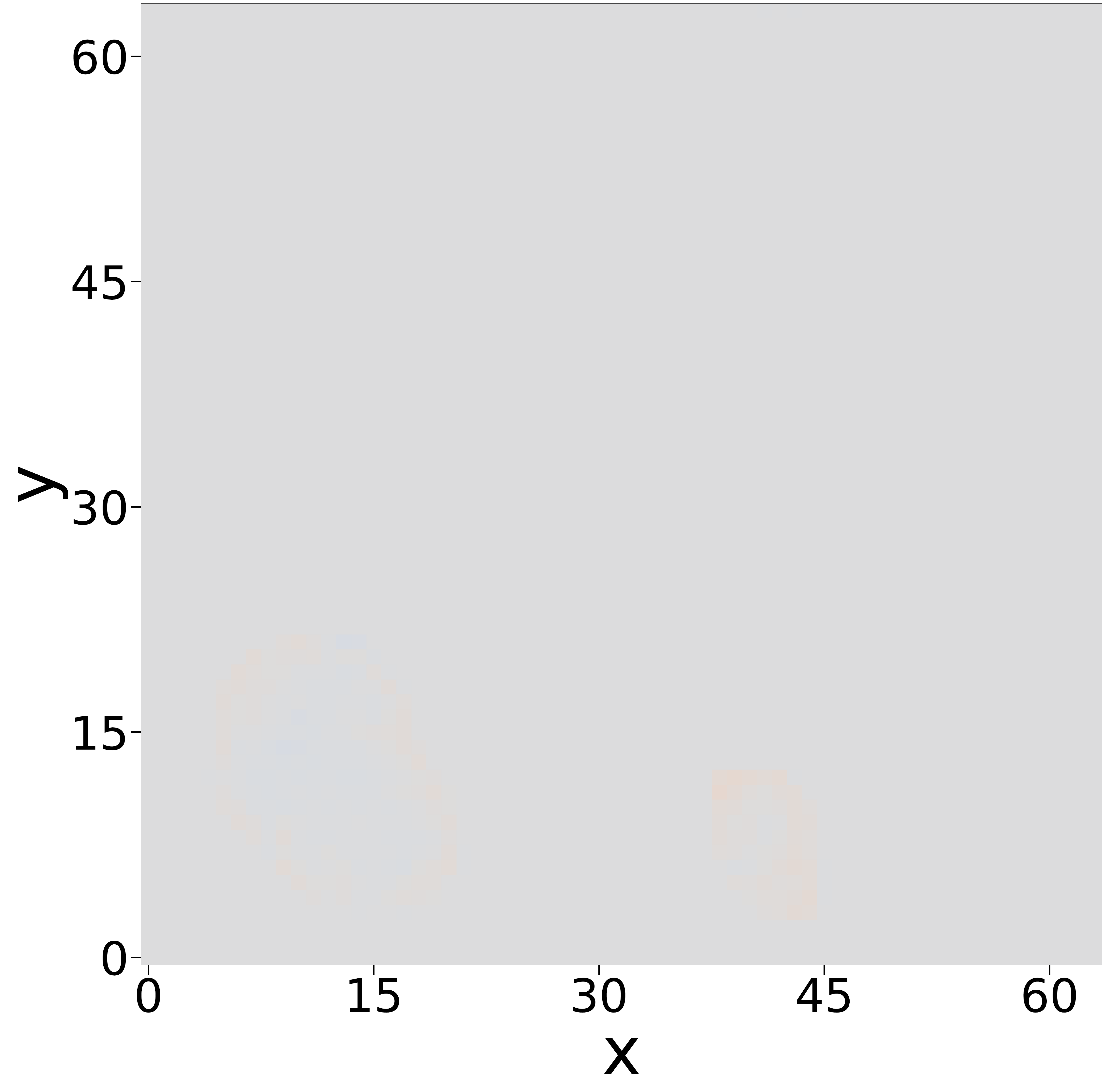}
    \end{subfigure}
    \begin{subfigure}[b]{0.0882\textwidth}
        \captionsetup{labelformat=empty}
        \includegraphics[width=\textwidth]{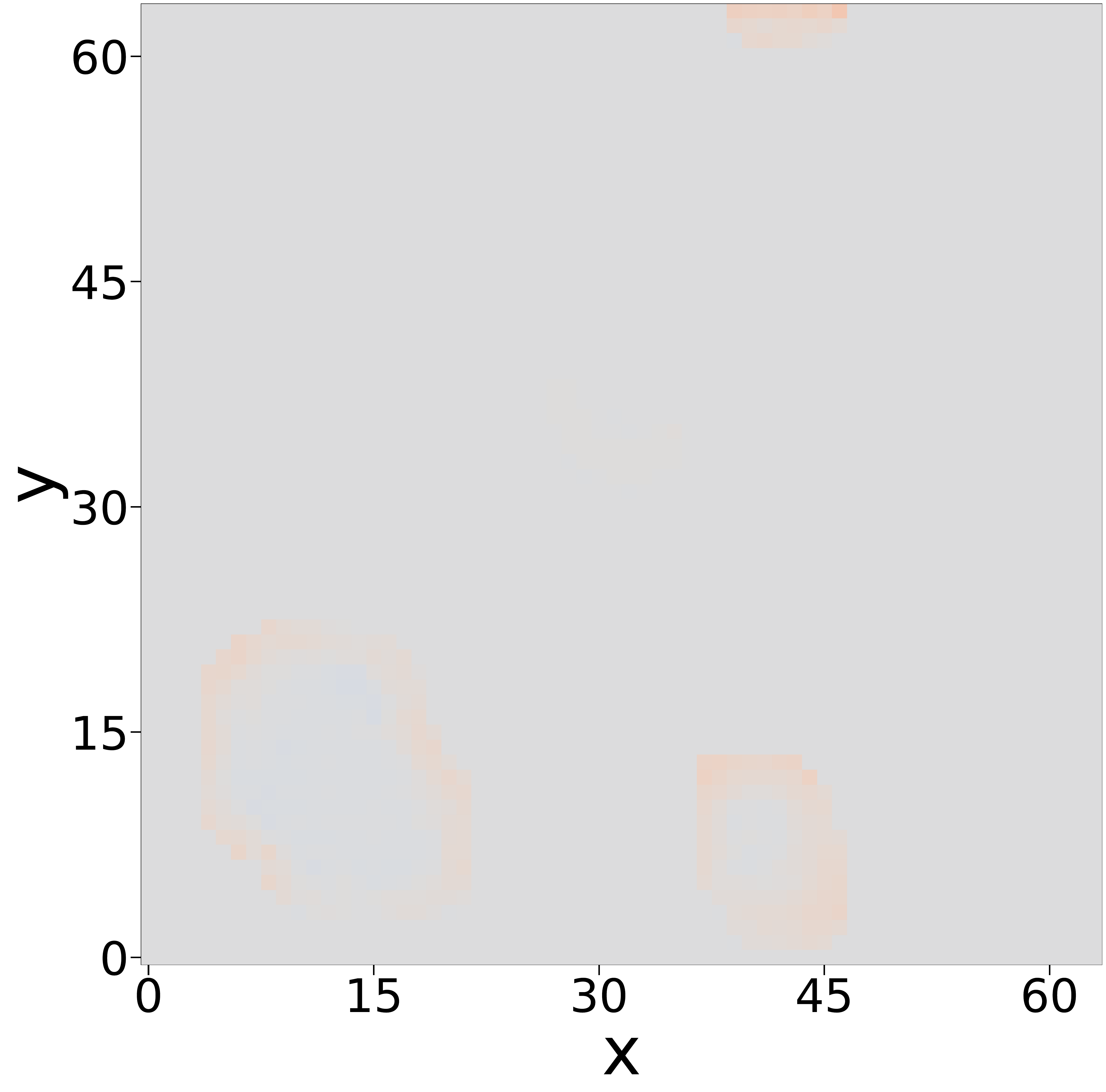}
    \end{subfigure}
    
    \begin{subfigure}[b]{0.0882\textwidth}
        \captionsetup{labelformat=empty}
        \includegraphics[width=\textwidth]{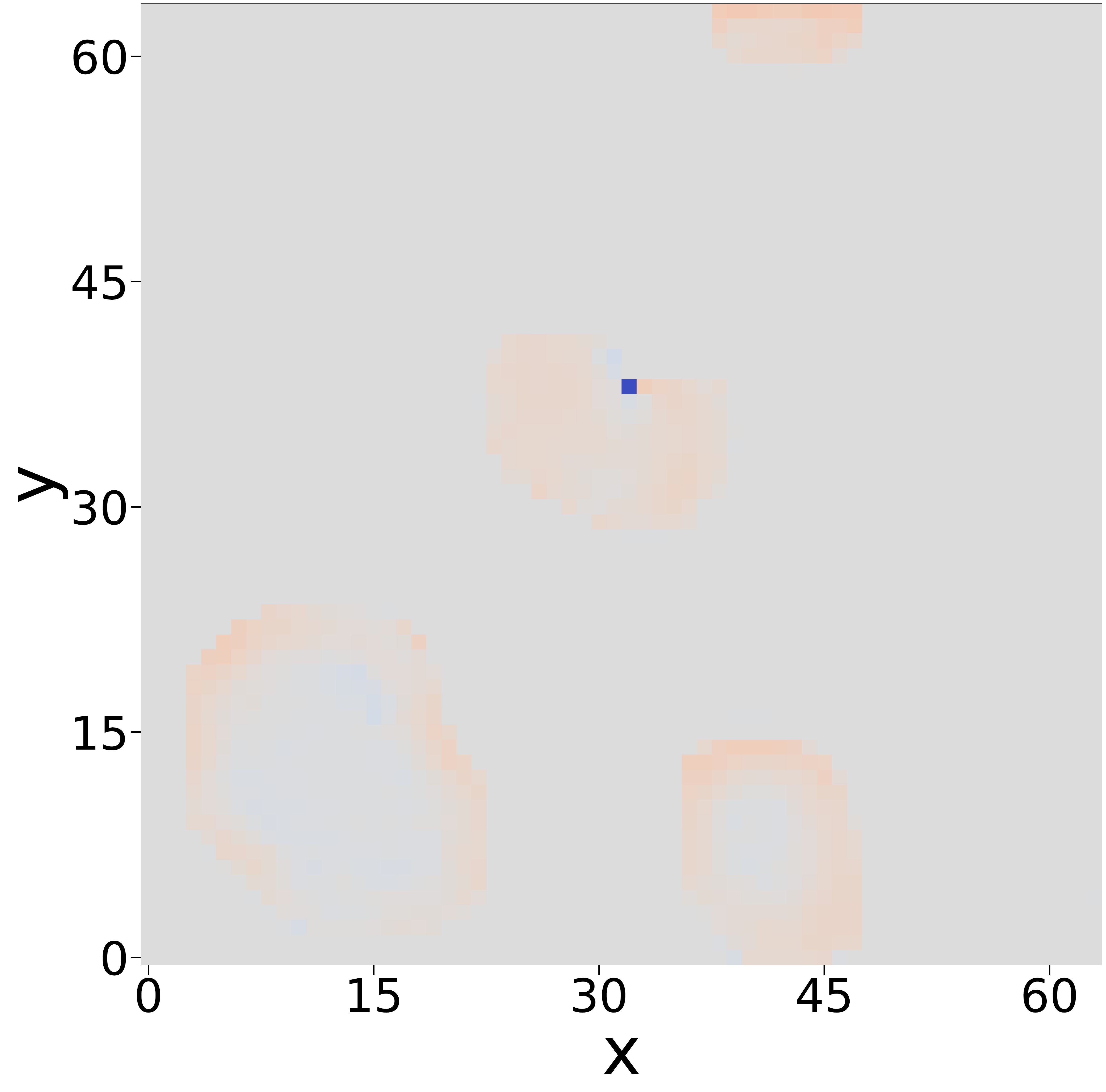}
    \end{subfigure}

    \begin{subfigure}[b]{0.0882\textwidth}
        \captionsetup{labelformat=empty}
        \includegraphics[width=\textwidth]{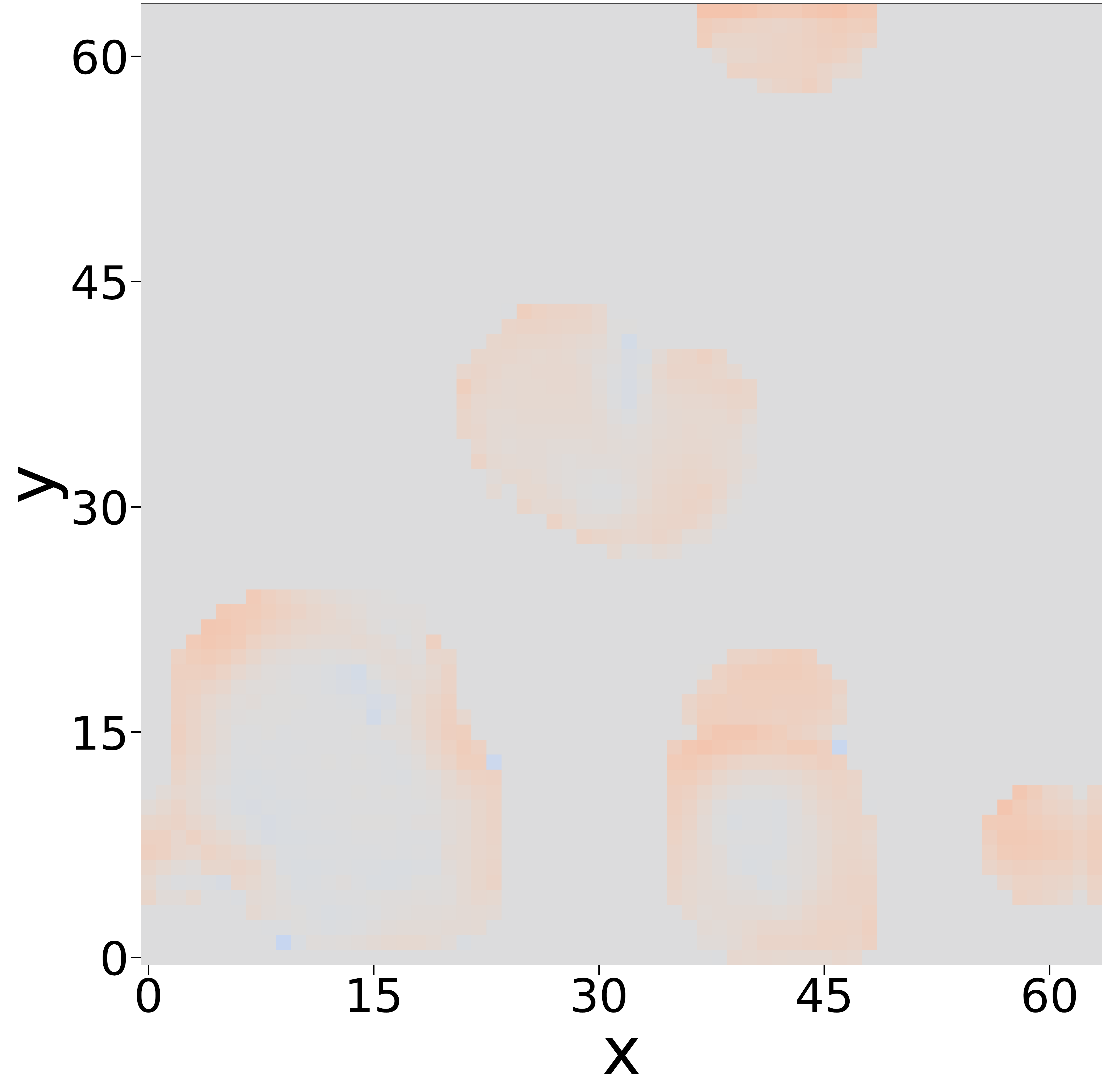}
    \end{subfigure}
    \begin{subfigure}[b]{0.0882\textwidth}
        \captionsetup{labelformat=empty}
        \includegraphics[width=\textwidth]{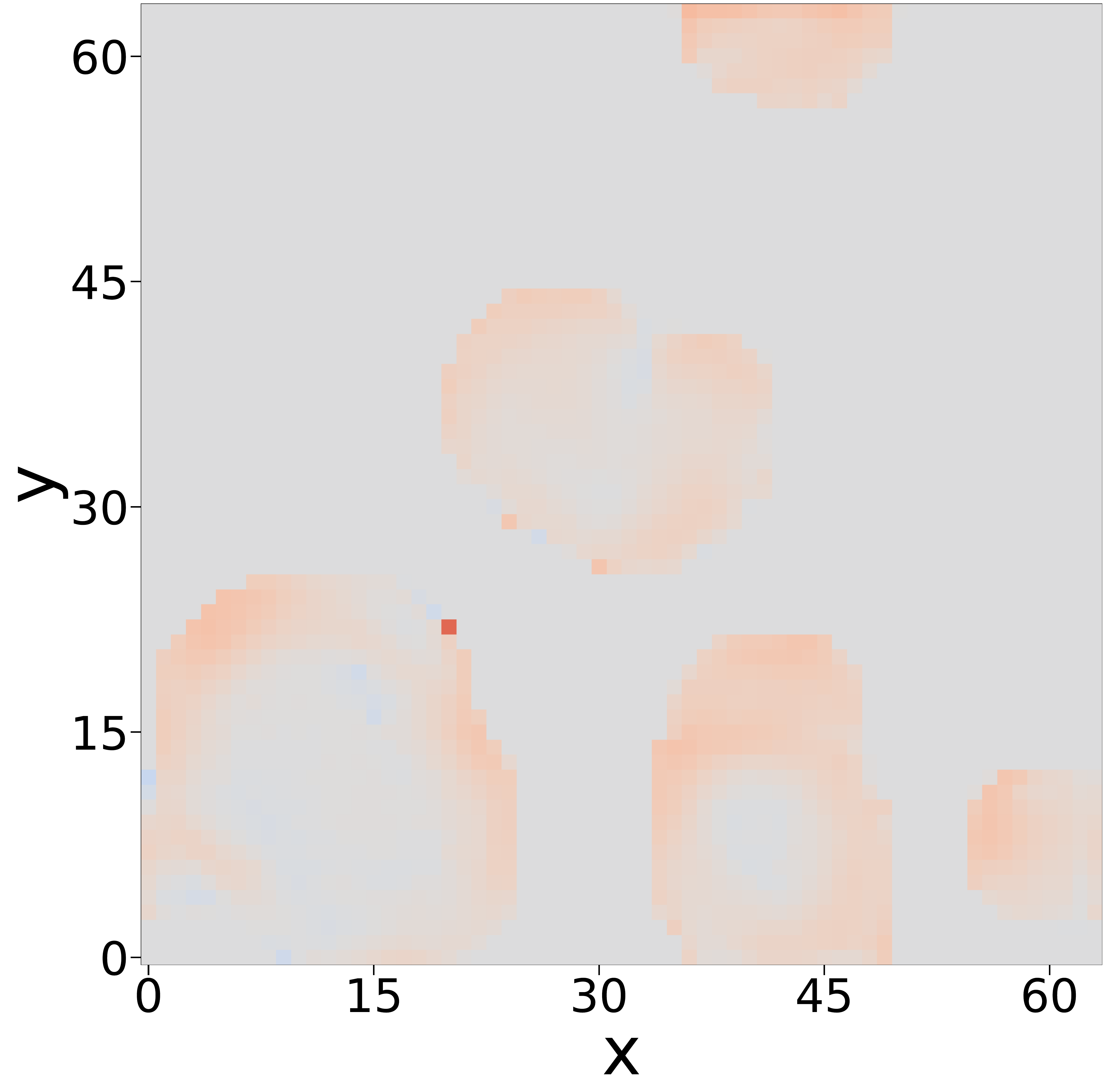}
    \end{subfigure}
    \begin{subfigure}[b]{0.0882\textwidth}
        \captionsetup{labelformat=empty}
        \includegraphics[width=\textwidth]{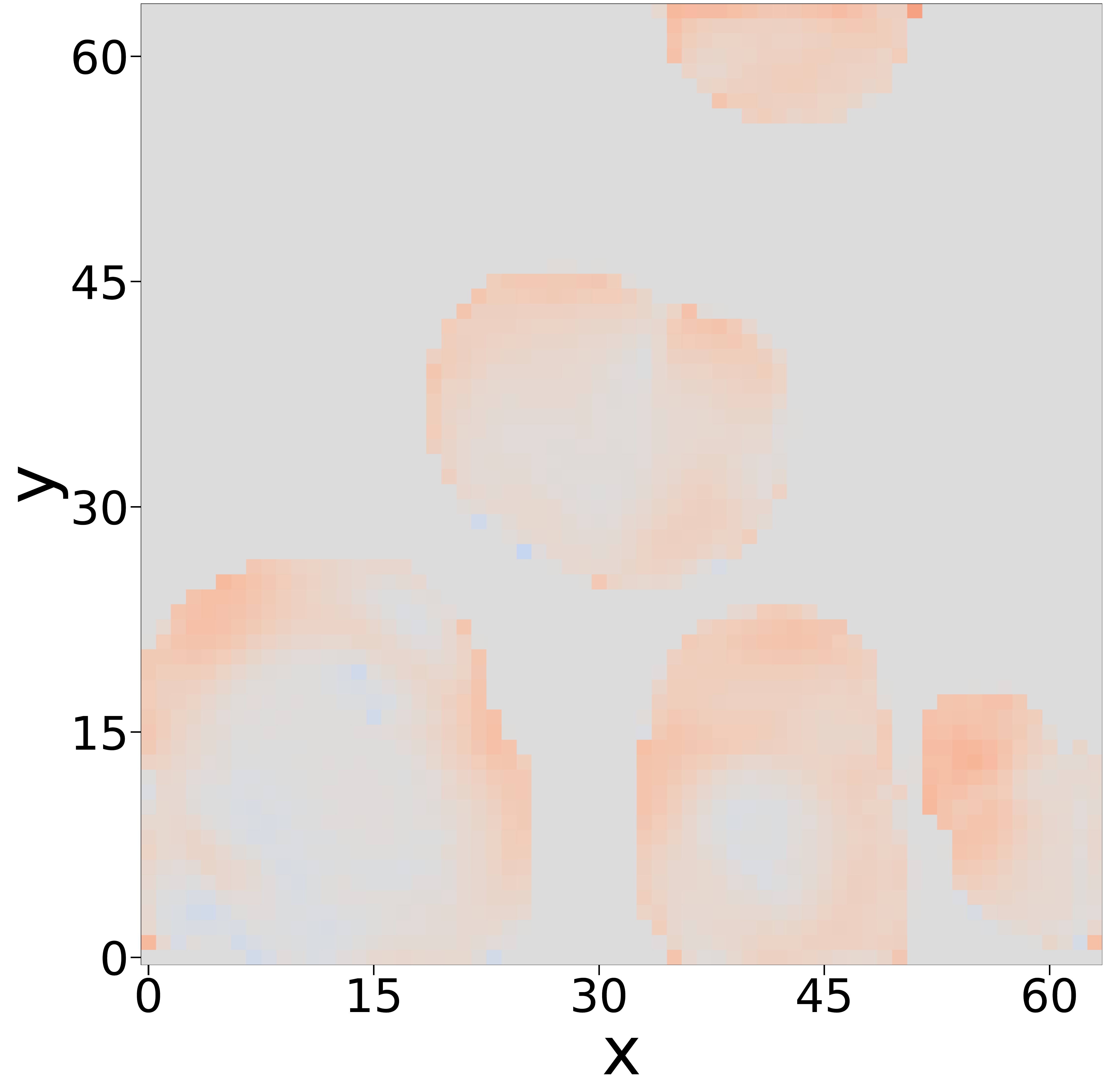}
    \end{subfigure}
    \begin{subfigure}[b]{0.0882\textwidth}
        \captionsetup{labelformat=empty}
        \includegraphics[width=\textwidth]{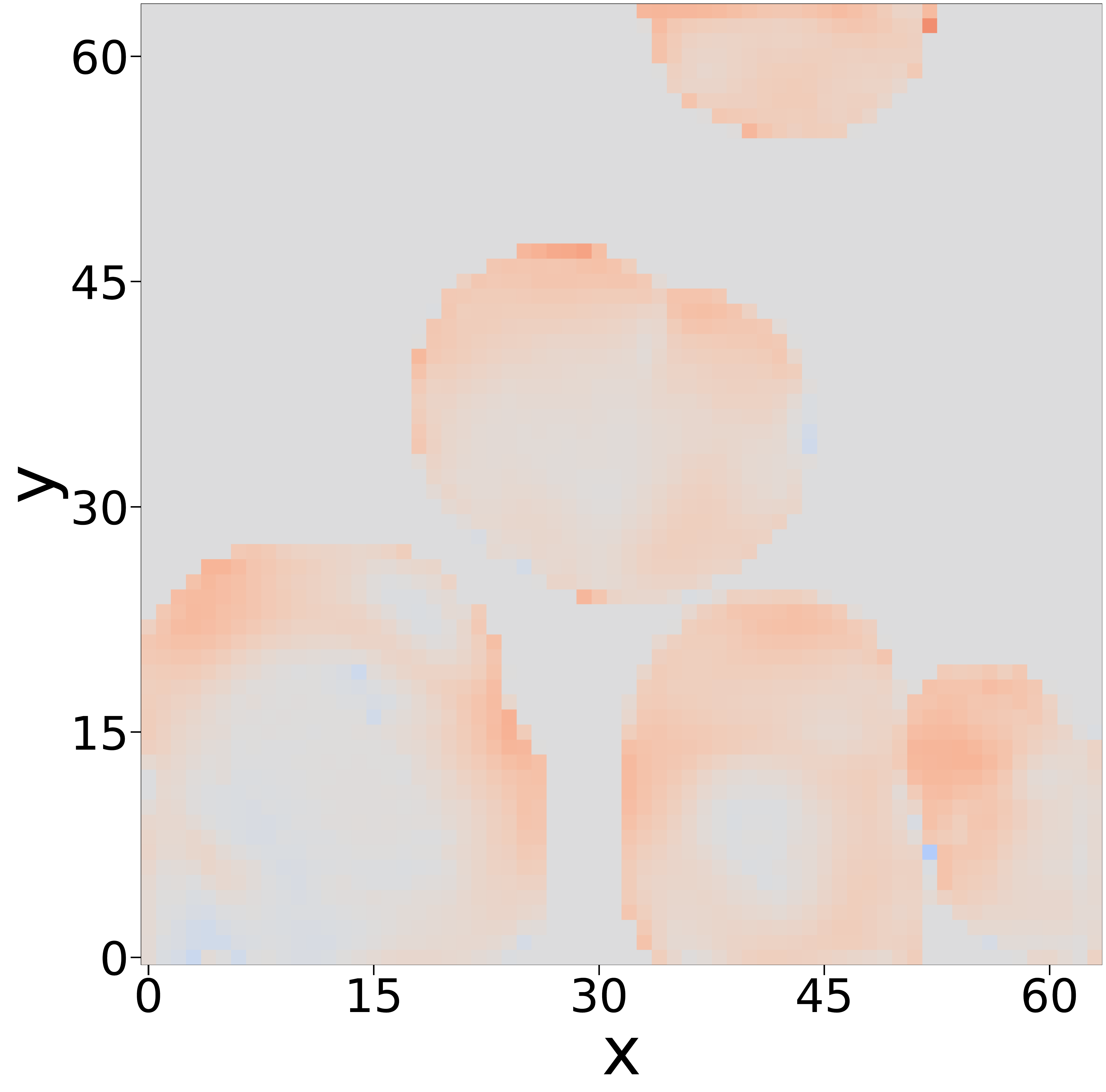}
    \end{subfigure}
    \begin{subfigure}[b]{0.0882\textwidth}
        \captionsetup{labelformat=empty}
        \includegraphics[width=\textwidth]{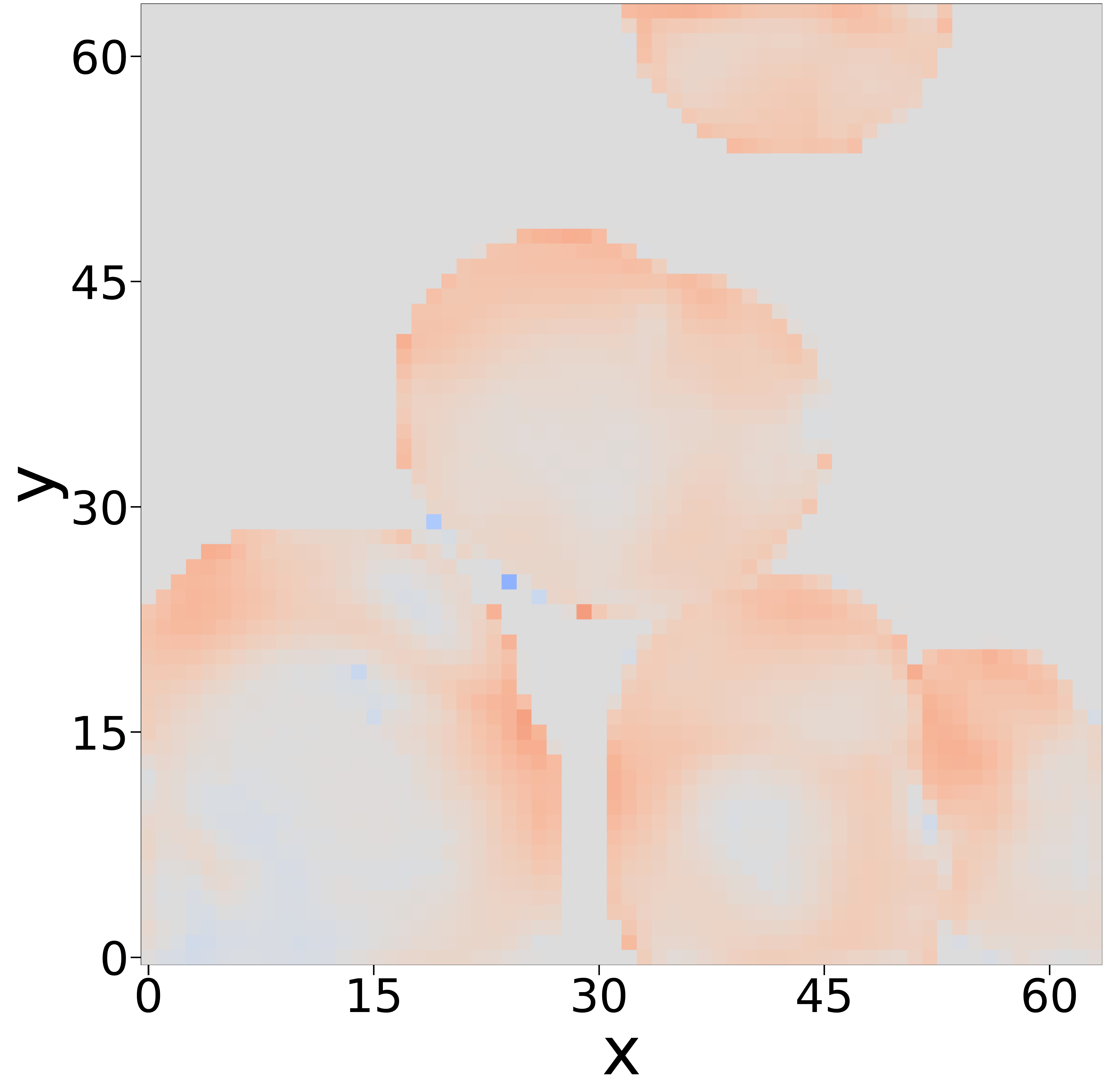}
    \end{subfigure}

    \begin{subfigure}[b]{0.02346\textwidth}
        \captionsetup{labelformat=empty}
        \includegraphics[width=\textwidth]{graphics/colorbar_res.pdf}
    \end{subfigure}

    }

    \vspace{1cm}    

    \makebox[\textwidth][c]{
    \begin{subfigure}[b]{0.0882\textwidth}
        \captionsetup{labelformat=empty}
        \includegraphics[width=\textwidth]{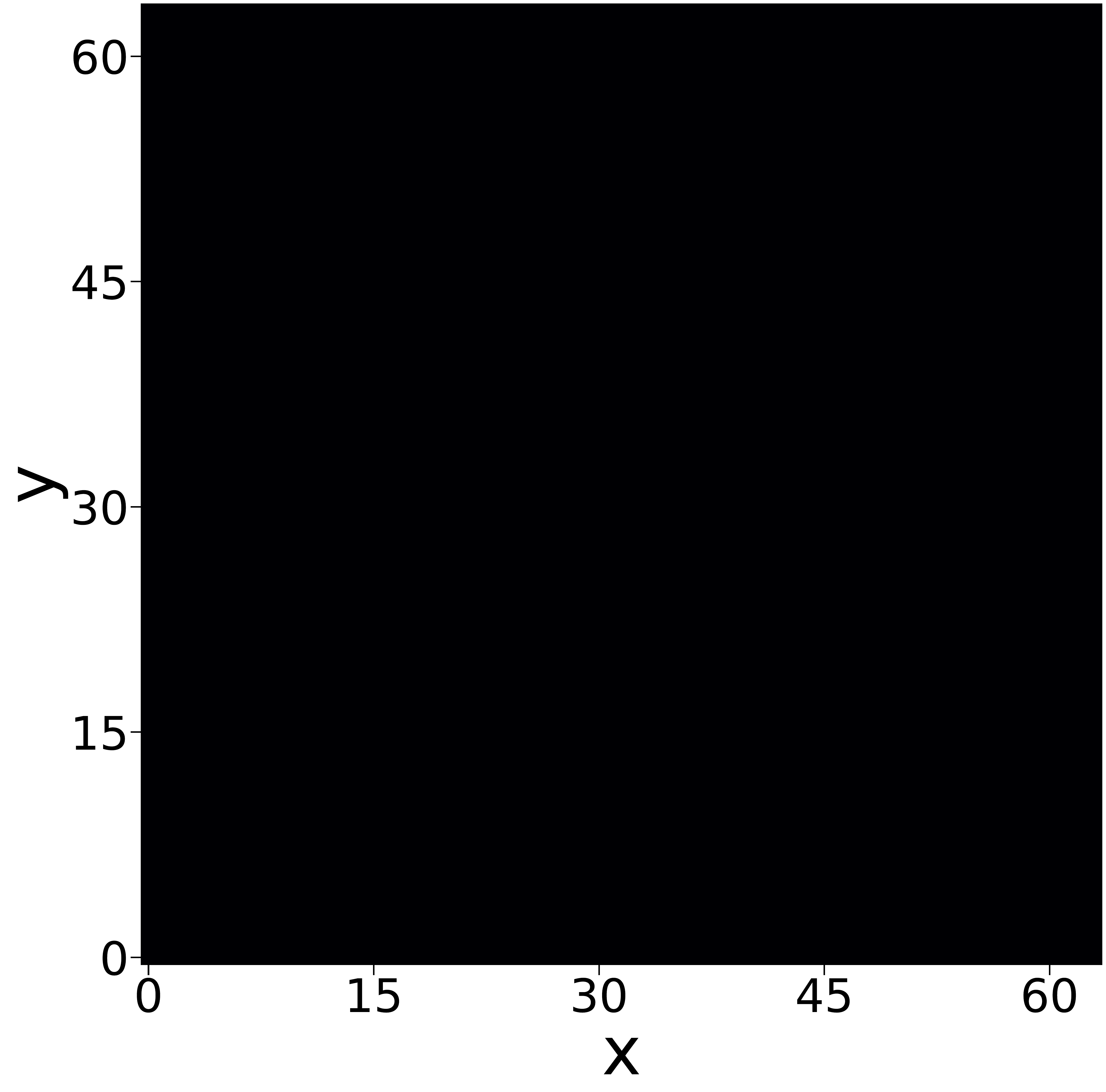}
    \end{subfigure}
    \begin{subfigure}[b]{0.0882\textwidth}
        \captionsetup{labelformat=empty}
        \includegraphics[width=\textwidth]{graphics/pred_XY_000.pdf}
    \end{subfigure}
    \begin{subfigure}[b]{0.0882\textwidth}
        \captionsetup{labelformat=empty}
        \includegraphics[width=\textwidth]{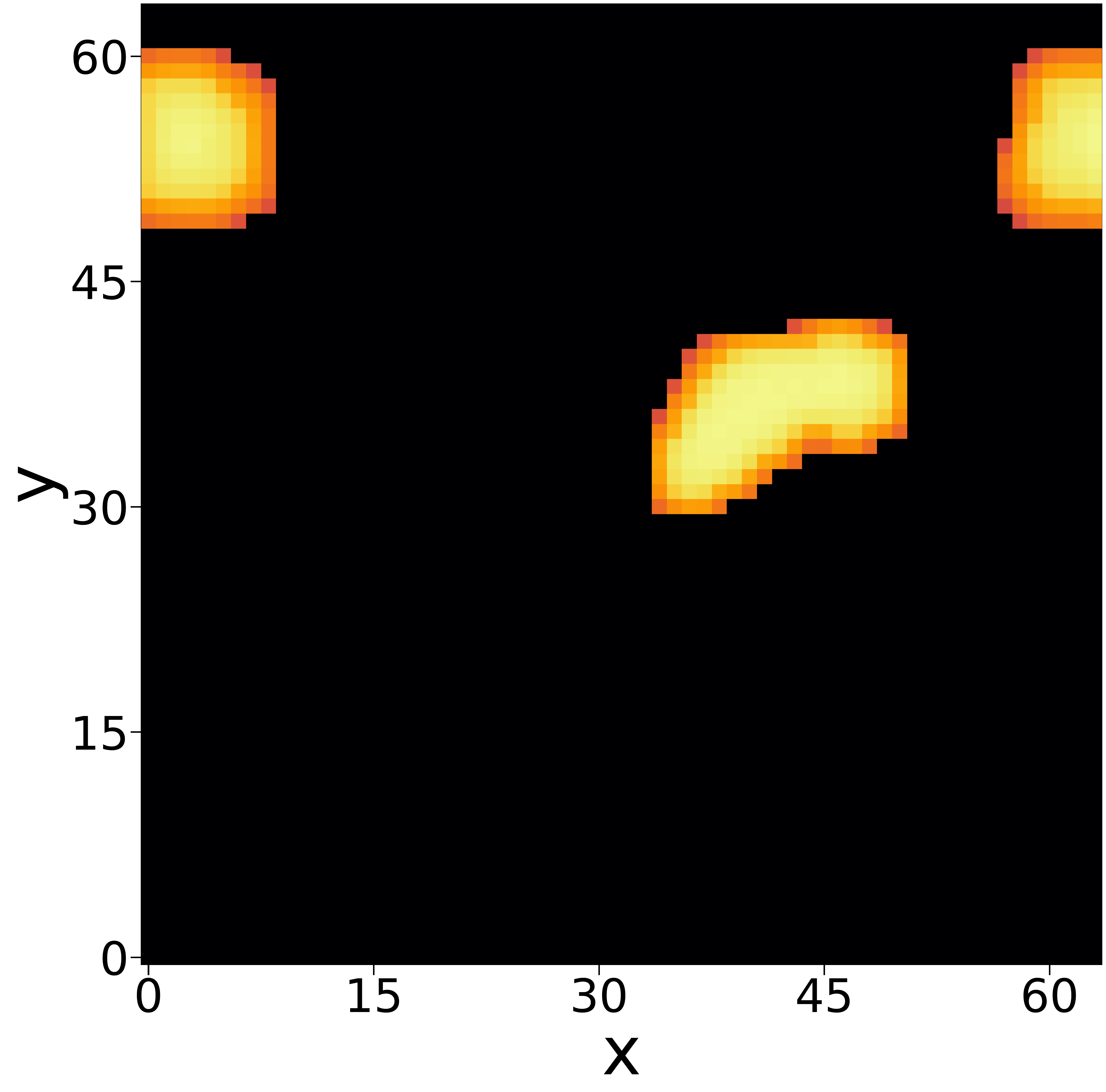}
    \end{subfigure}
    \begin{subfigure}[b]{0.0882\textwidth}
        \captionsetup{labelformat=empty}
        \includegraphics[width=\textwidth]{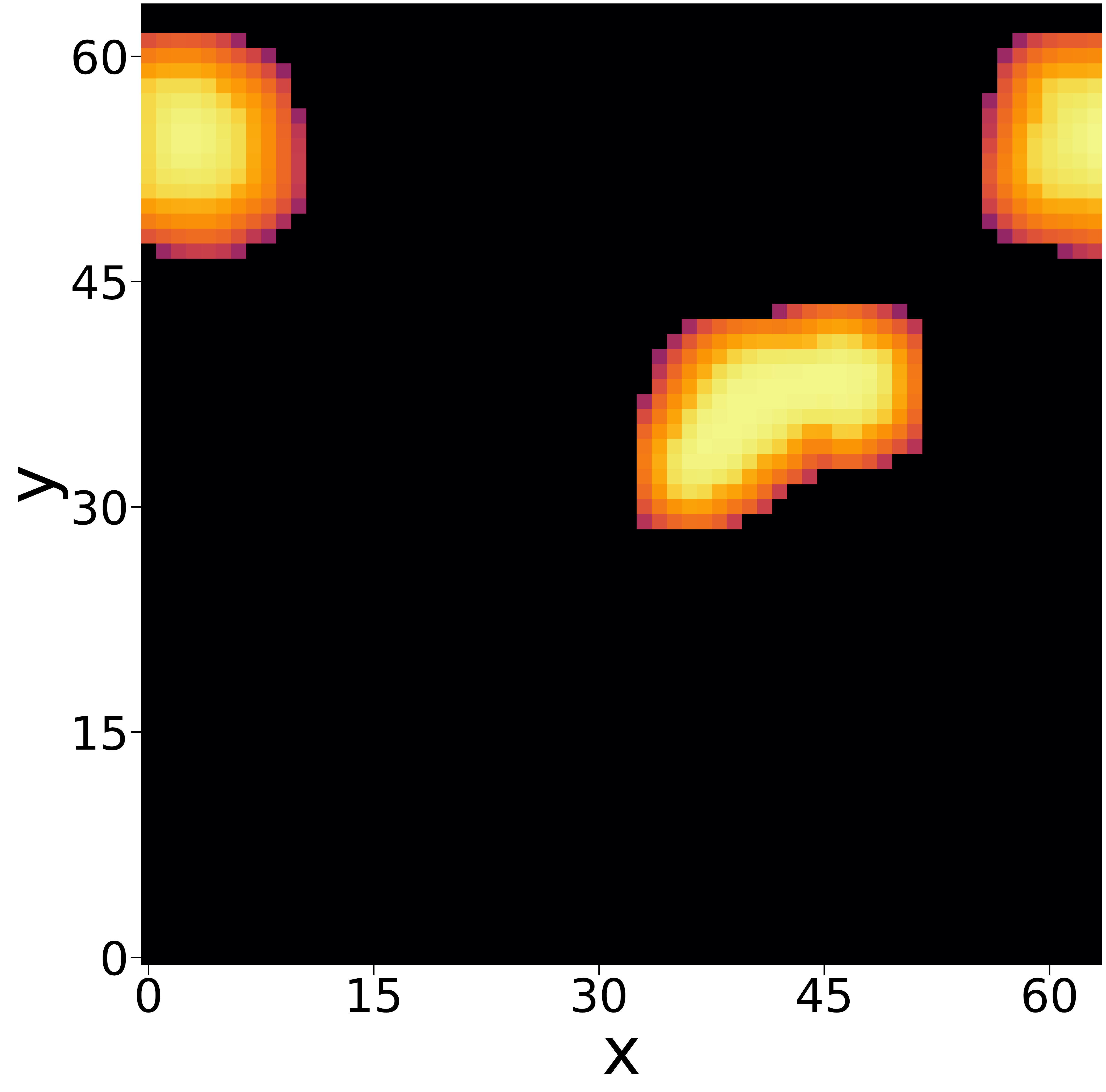}
    \end{subfigure}
    
    \begin{subfigure}[b]{0.0882\textwidth}
        \captionsetup{labelformat=empty}
        \includegraphics[width=\textwidth]{graphics/pred_XY_003.pdf}
    \end{subfigure}

    \begin{subfigure}[b]{0.0882\textwidth}
        \captionsetup{labelformat=empty}
        \includegraphics[width=\textwidth]{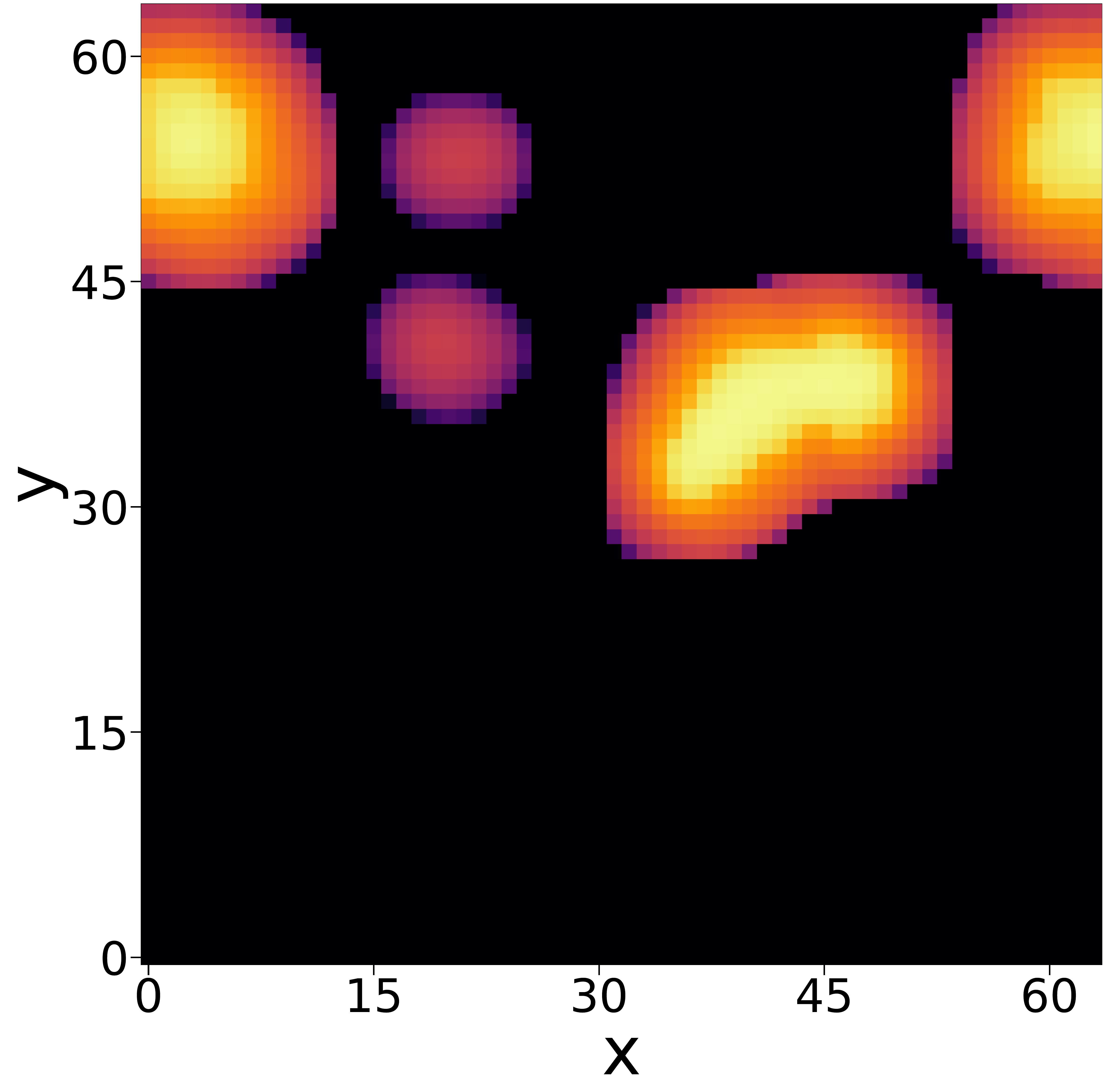}
    \end{subfigure}
    \begin{subfigure}[b]{0.0882\textwidth}
        \captionsetup{labelformat=empty}
        \includegraphics[width=\textwidth]{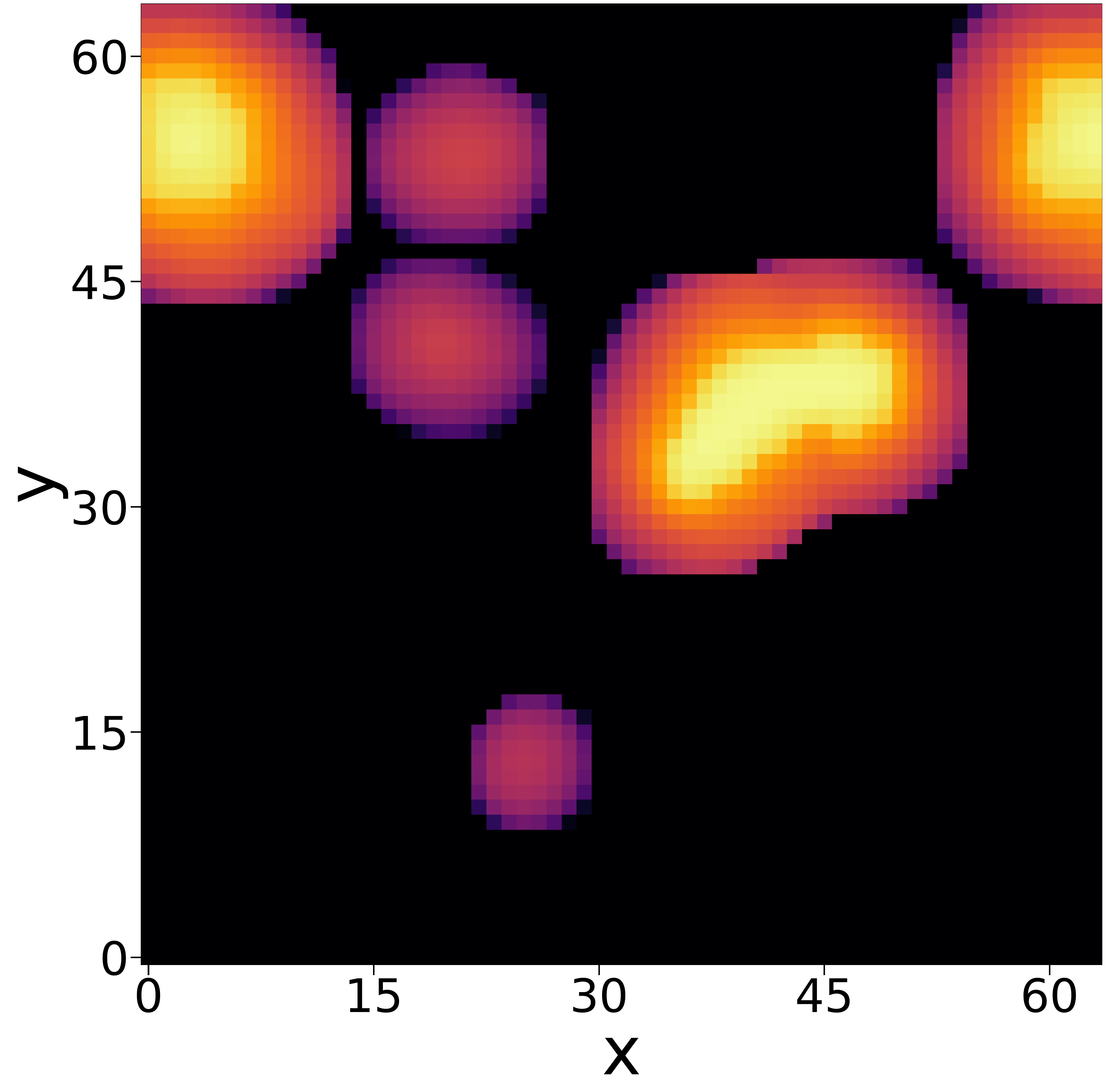}
    \end{subfigure}
    \begin{subfigure}[b]{0.0882\textwidth}
        \captionsetup{labelformat=empty}
        \includegraphics[width=\textwidth]{graphics/pred_XY_006.pdf}
    \end{subfigure}
    \begin{subfigure}[b]{0.0882\textwidth}
        \captionsetup{labelformat=empty}
        \includegraphics[width=\textwidth]{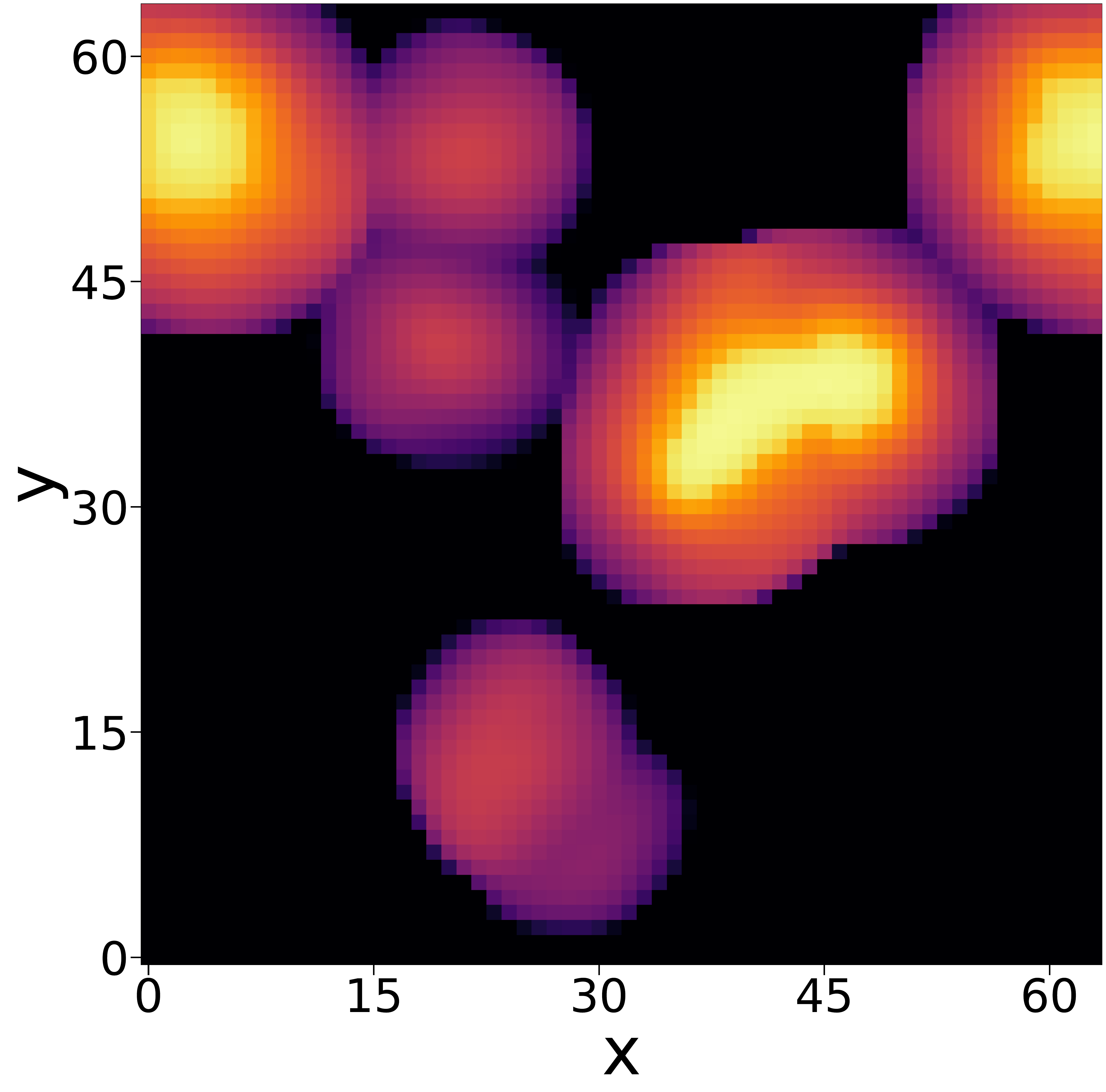}
    \end{subfigure}
    \begin{subfigure}[b]{0.0882\textwidth}
        \captionsetup{labelformat=empty}
        \includegraphics[width=\textwidth]{graphics/pred_XY_008.pdf}
    \end{subfigure}

    \begin{subfigure}[b]{0.02346\textwidth}
        \captionsetup{labelformat=empty}
        \includegraphics[width=\textwidth]{graphics/colorbar_pred.pdf}
    \end{subfigure}

    }
    \makebox[\textwidth][c]{
    \begin{subfigure}[b]{0.0882\textwidth}
        \captionsetup{labelformat=empty}
        \includegraphics[width=\textwidth]{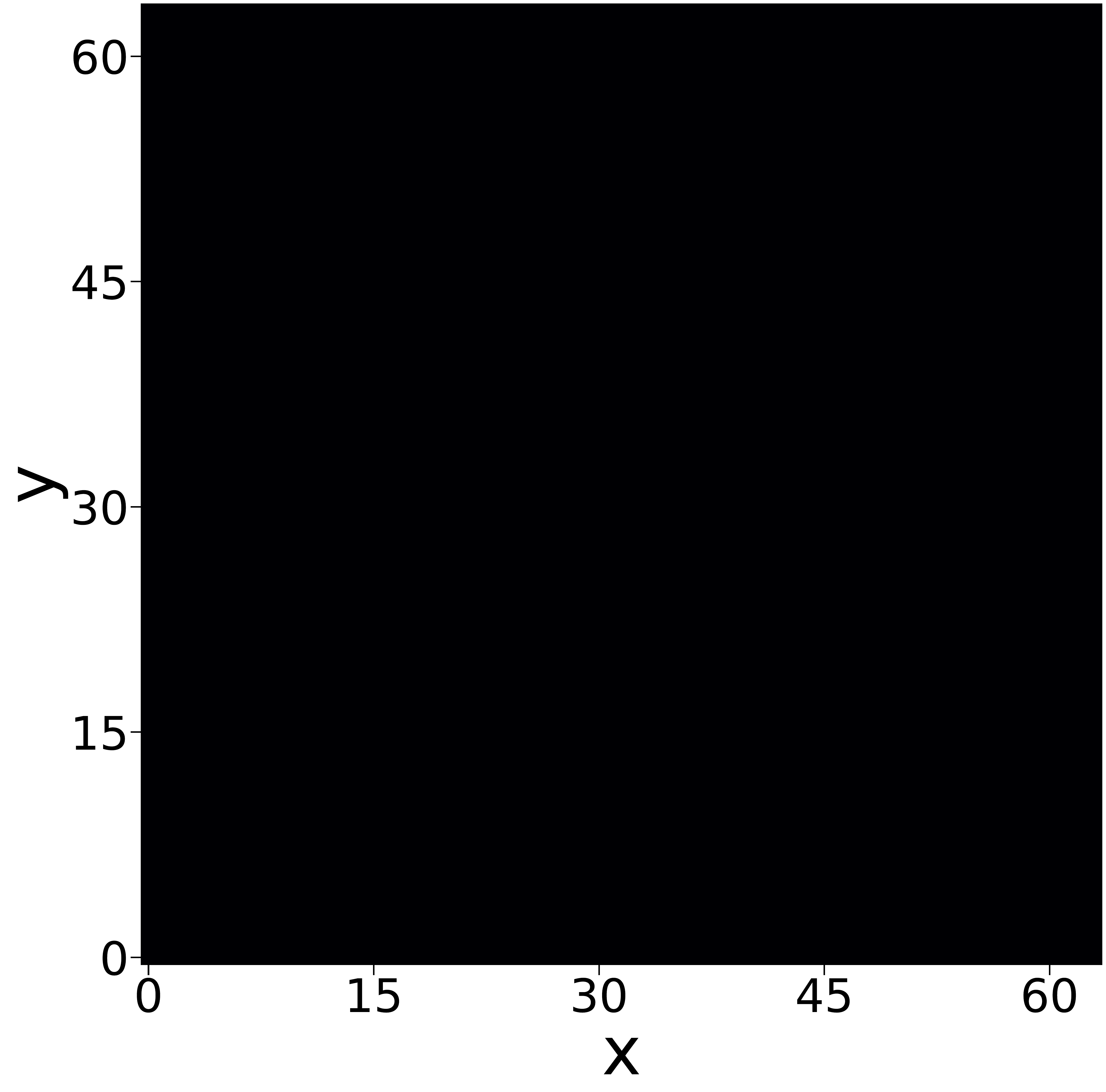}
    \end{subfigure}
    \begin{subfigure}[b]{0.0882\textwidth}
        \captionsetup{labelformat=empty}
        \includegraphics[width=\textwidth]{graphics/true_XY_000.pdf}
    \end{subfigure}
    \begin{subfigure}[b]{0.0882\textwidth}
        \captionsetup{labelformat=empty}
        \includegraphics[width=\textwidth]{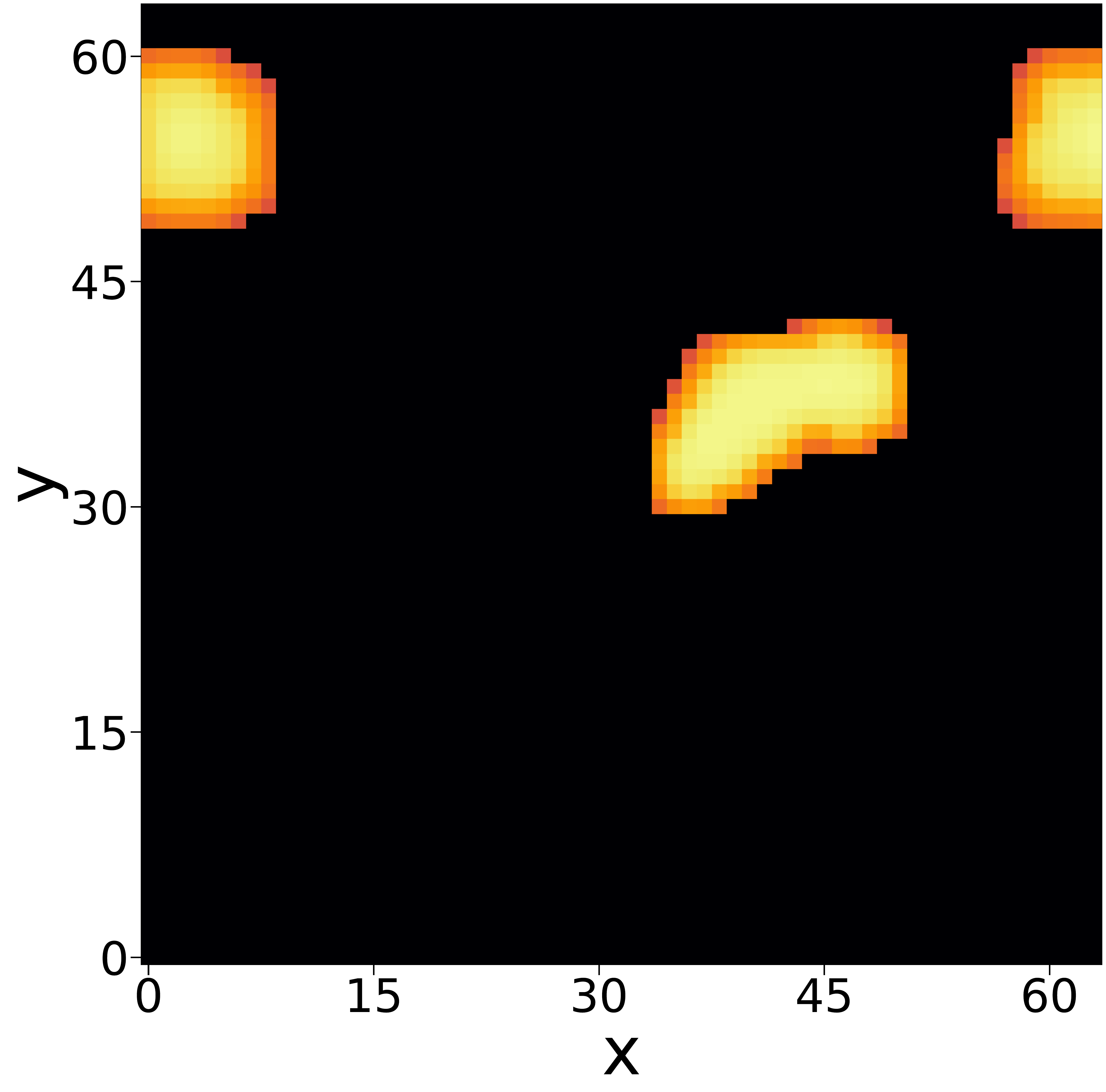}
    \end{subfigure}
    \begin{subfigure}[b]{0.0882\textwidth}
        \captionsetup{labelformat=empty}
        \includegraphics[width=\textwidth]{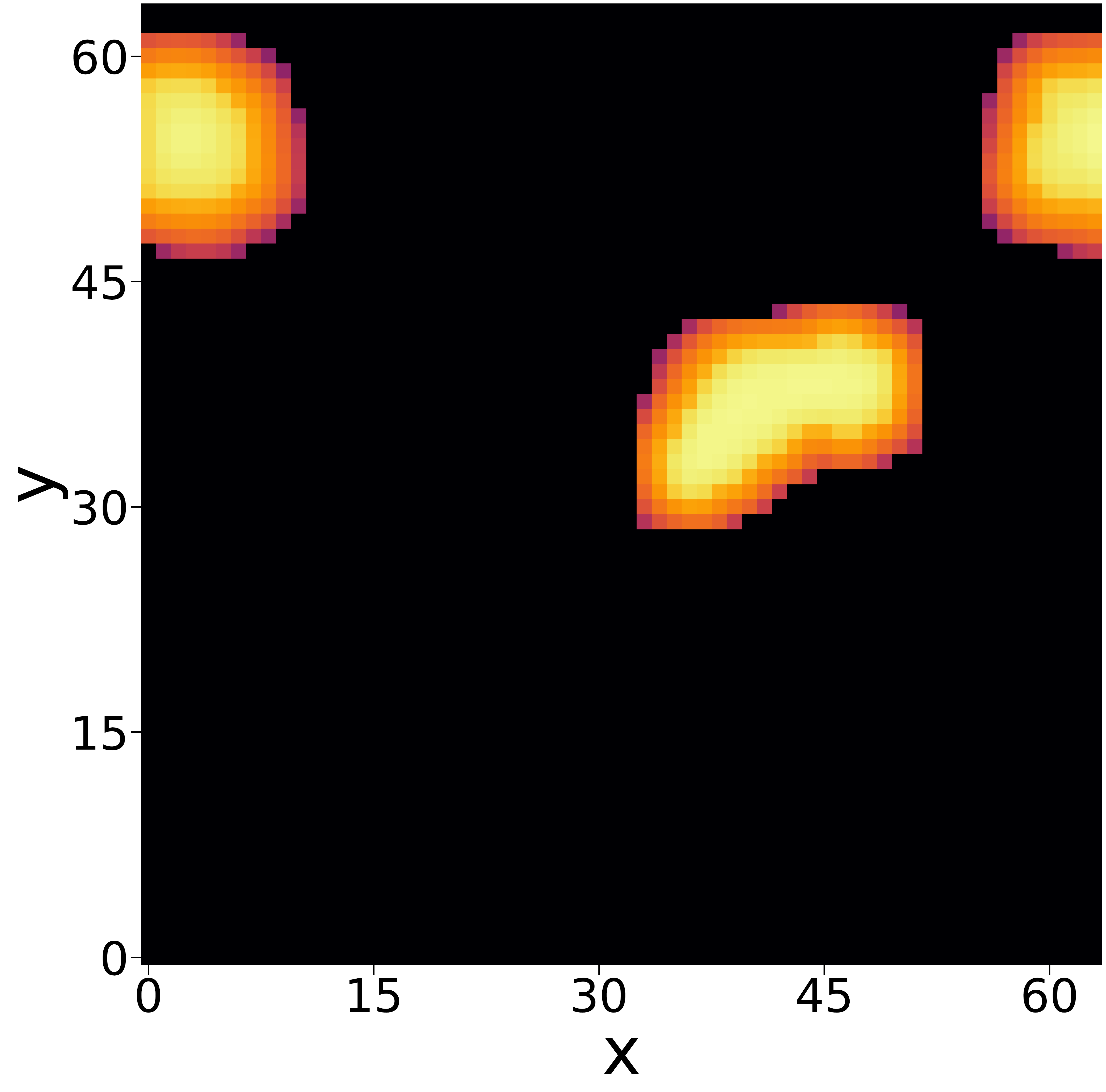}
    \end{subfigure}
    
    \begin{subfigure}[b]{0.0882\textwidth}
        \captionsetup{labelformat=empty}
        \includegraphics[width=\textwidth]{graphics/true_XY_003.pdf}
    \end{subfigure}

    \begin{subfigure}[b]{0.0882\textwidth}
        \captionsetup{labelformat=empty}
        \includegraphics[width=\textwidth]{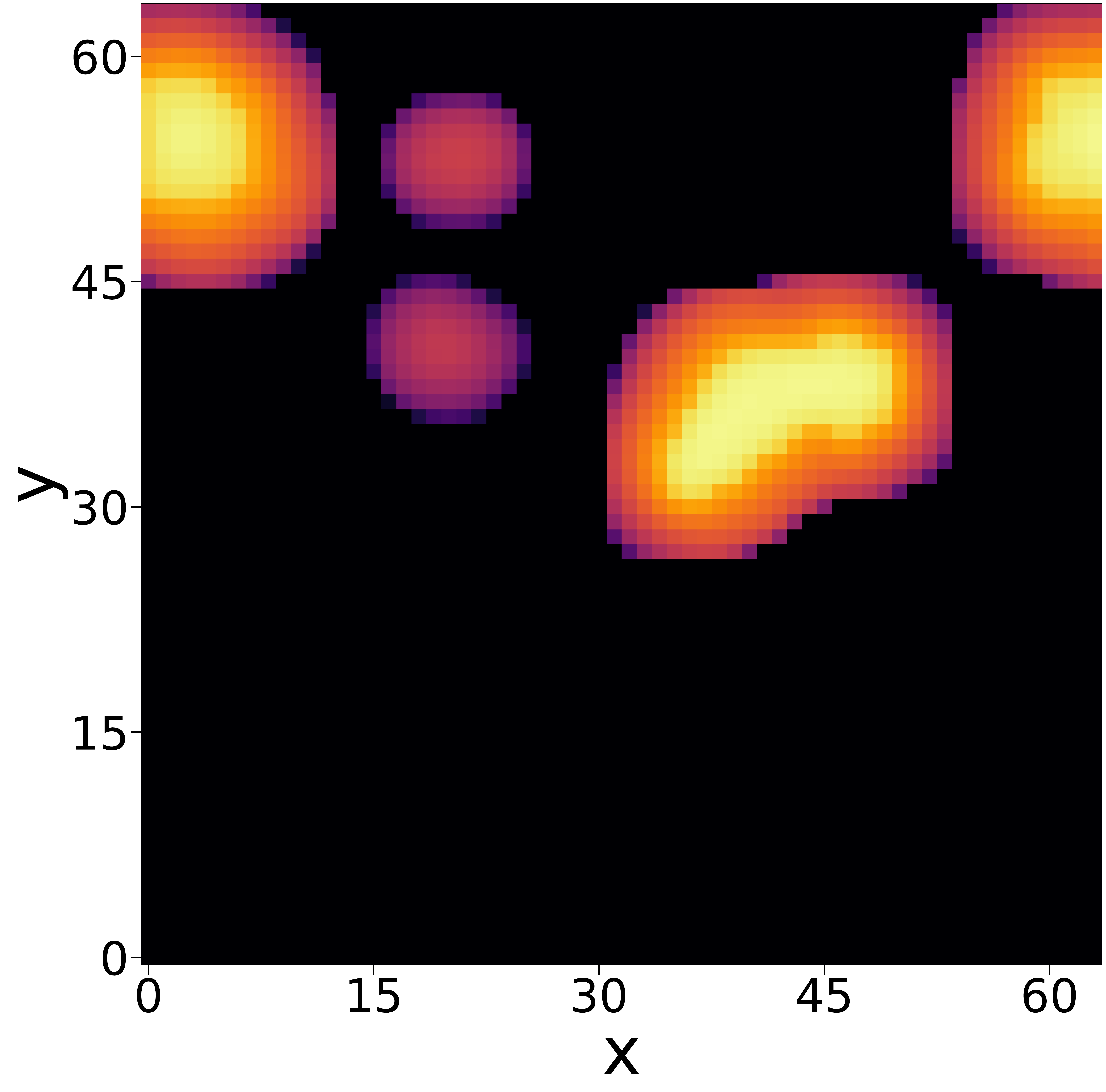}
    \end{subfigure}
    \begin{subfigure}[b]{0.0882\textwidth}
        \captionsetup{labelformat=empty}
        \includegraphics[width=\textwidth]{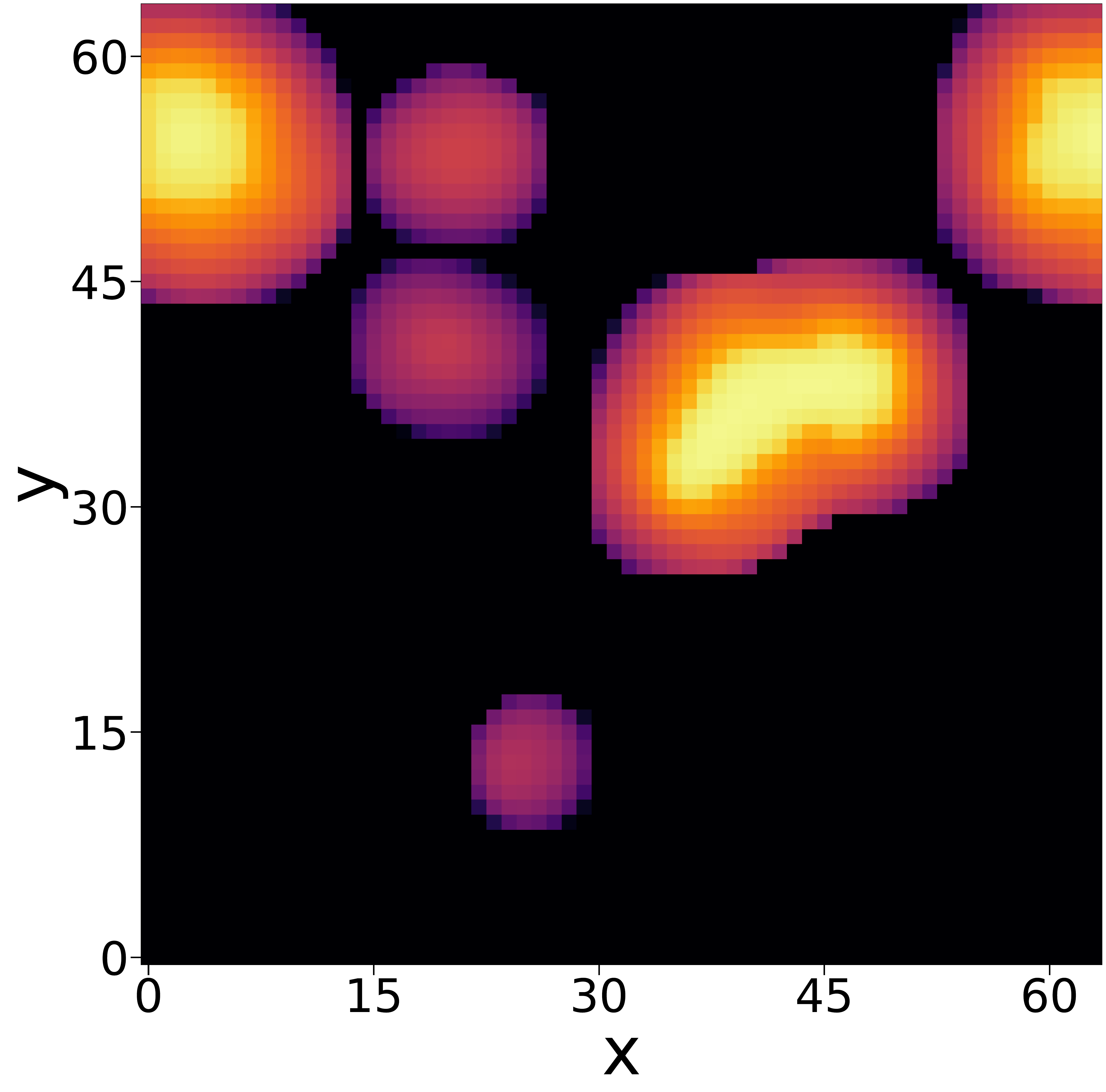}
    \end{subfigure}
    \begin{subfigure}[b]{0.0882\textwidth}
        \captionsetup{labelformat=empty}
        \includegraphics[width=\textwidth]{graphics/true_XY_006.pdf}
    \end{subfigure}
    \begin{subfigure}[b]{0.0882\textwidth}
        \captionsetup{labelformat=empty}
        \includegraphics[width=\textwidth]{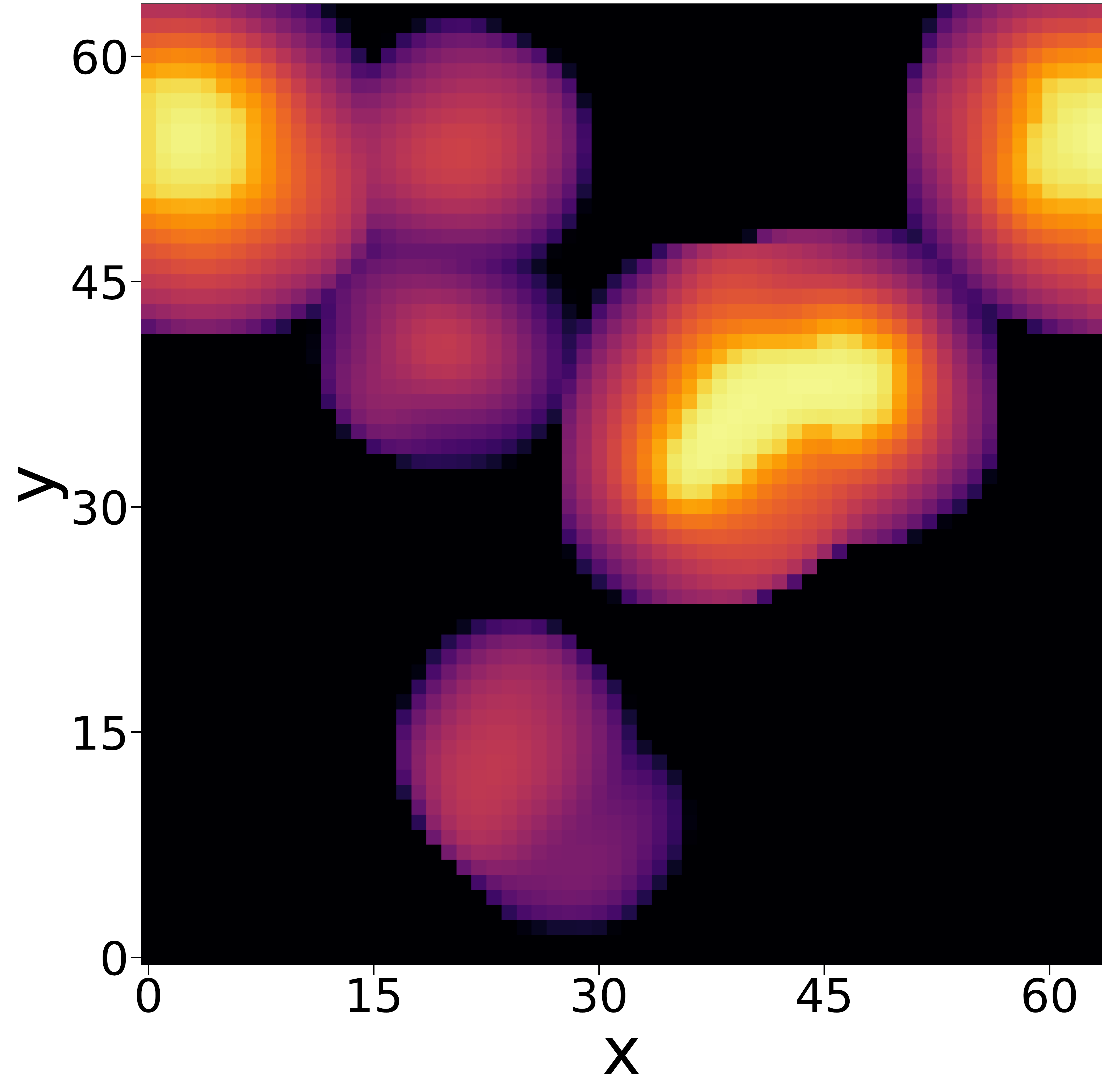}
    \end{subfigure}
    \begin{subfigure}[b]{0.0882\textwidth}
        \captionsetup{labelformat=empty}
        \includegraphics[width=\textwidth]{graphics/true_XY_008.pdf}
    \end{subfigure}

    \begin{subfigure}[b]{0.02346\textwidth}
        \captionsetup{labelformat=empty}
        \includegraphics[width=\textwidth]{graphics/colorbar_num.pdf}
    \end{subfigure}

    }
    \makebox[\textwidth][c]{
    \begin{subfigure}[b]{0.0882\textwidth}
        \captionsetup{labelformat=empty}
        \includegraphics[width=\textwidth]{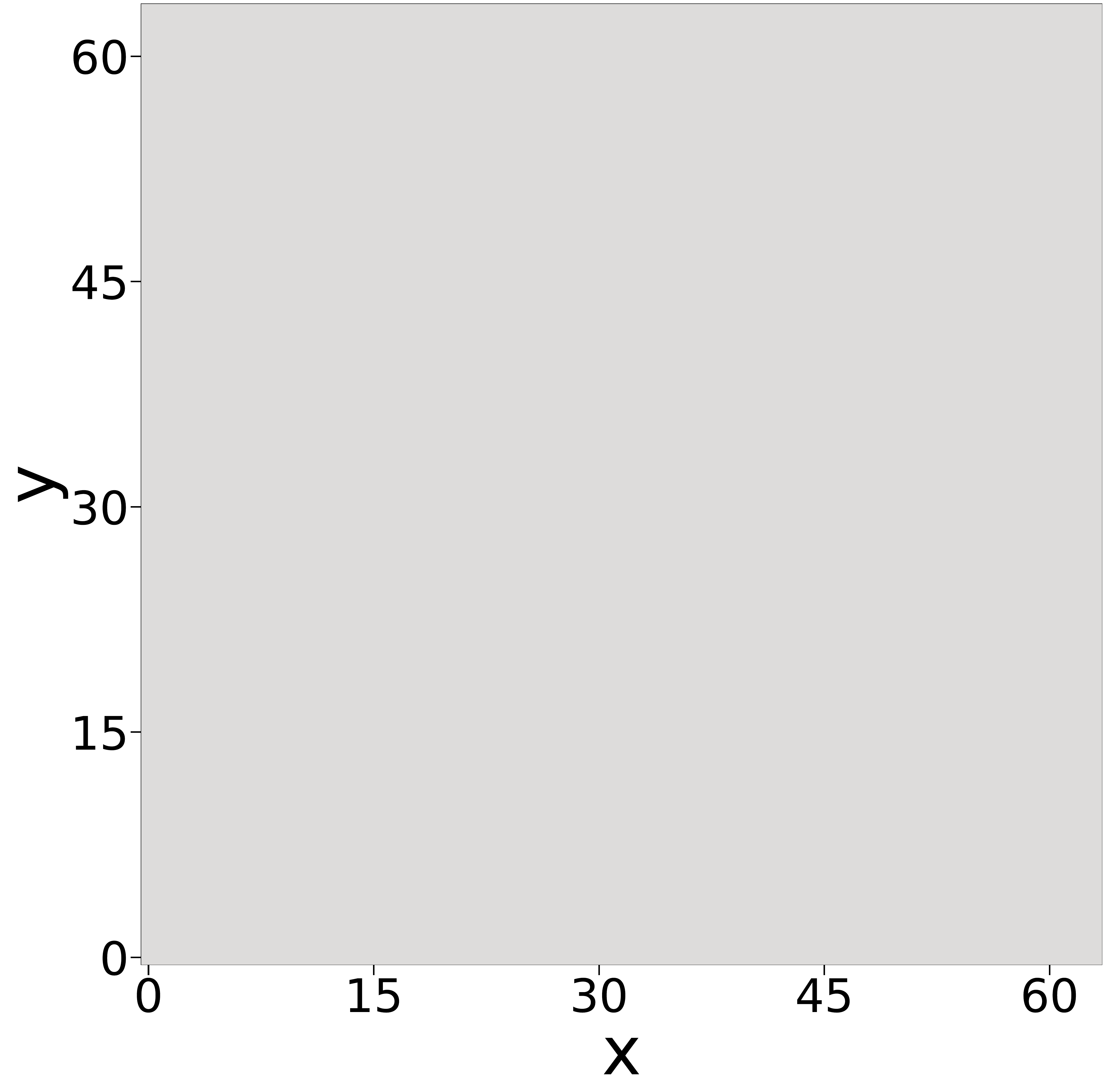}
    \end{subfigure}
    \begin{subfigure}[b]{0.0882\textwidth}
        \captionsetup{labelformat=empty}
        \includegraphics[width=\textwidth]{graphics/res_XY_000.pdf}
    \end{subfigure}
    \begin{subfigure}[b]{0.0882\textwidth}
        \captionsetup{labelformat=empty}
        \includegraphics[width=\textwidth]{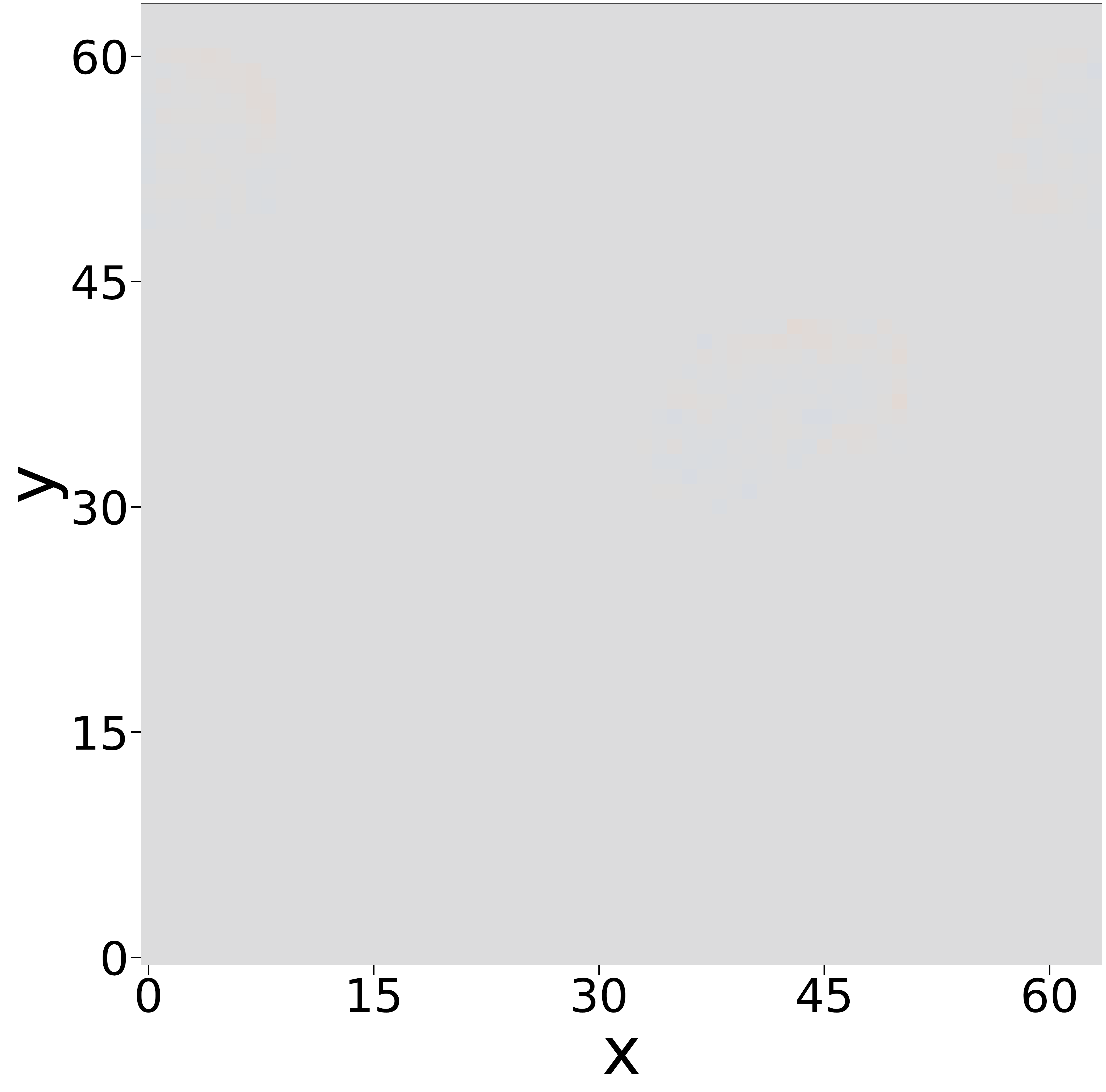}
    \end{subfigure}
    \begin{subfigure}[b]{0.0882\textwidth}
        \captionsetup{labelformat=empty}
        \includegraphics[width=\textwidth]{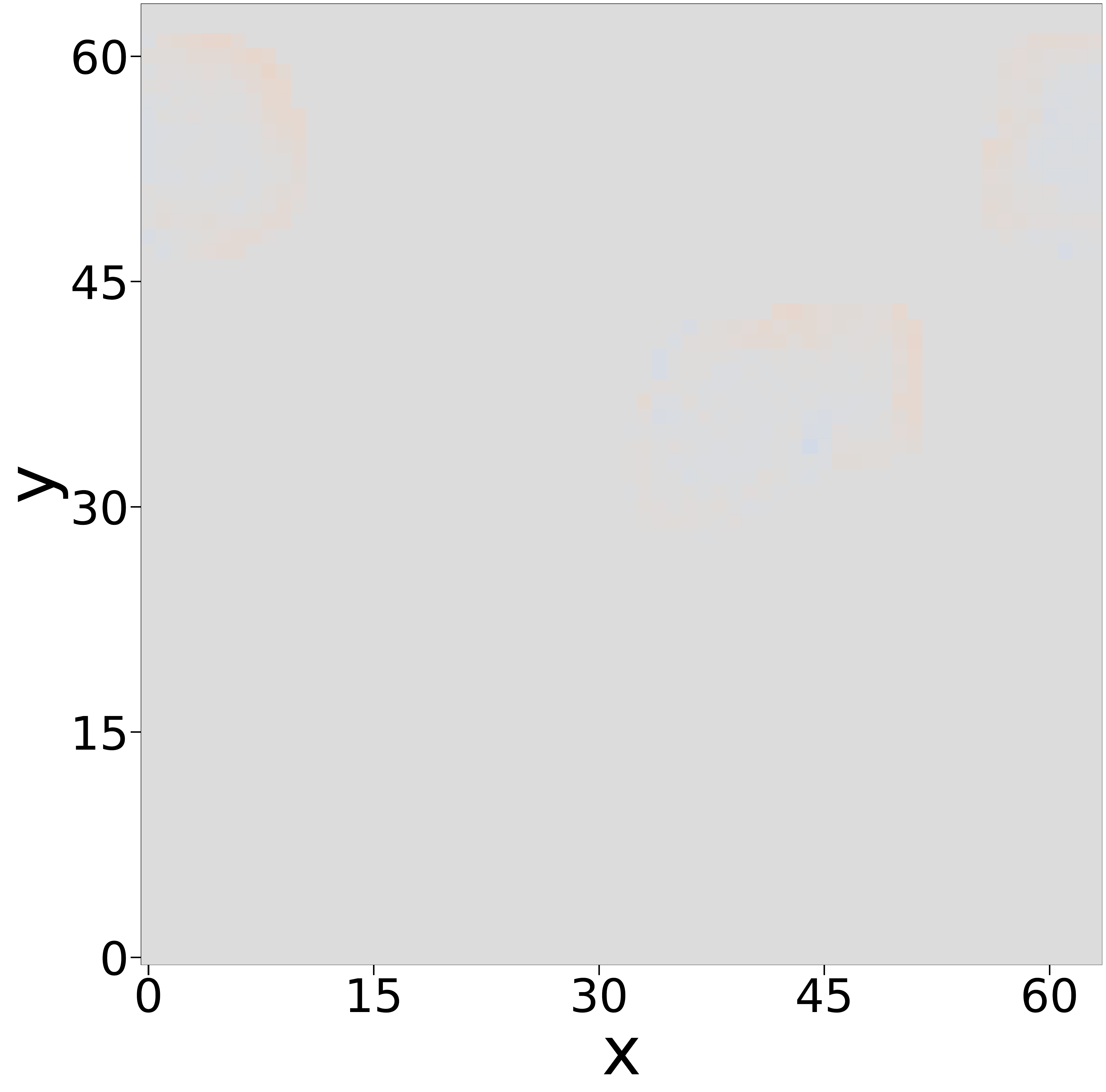}
    \end{subfigure}
    
    \begin{subfigure}[b]{0.0882\textwidth}
        \captionsetup{labelformat=empty}
        \includegraphics[width=\textwidth]{graphics/res_XY_003.pdf}
    \end{subfigure}

    \begin{subfigure}[b]{0.0882\textwidth}
        \captionsetup{labelformat=empty}
        \includegraphics[width=\textwidth]{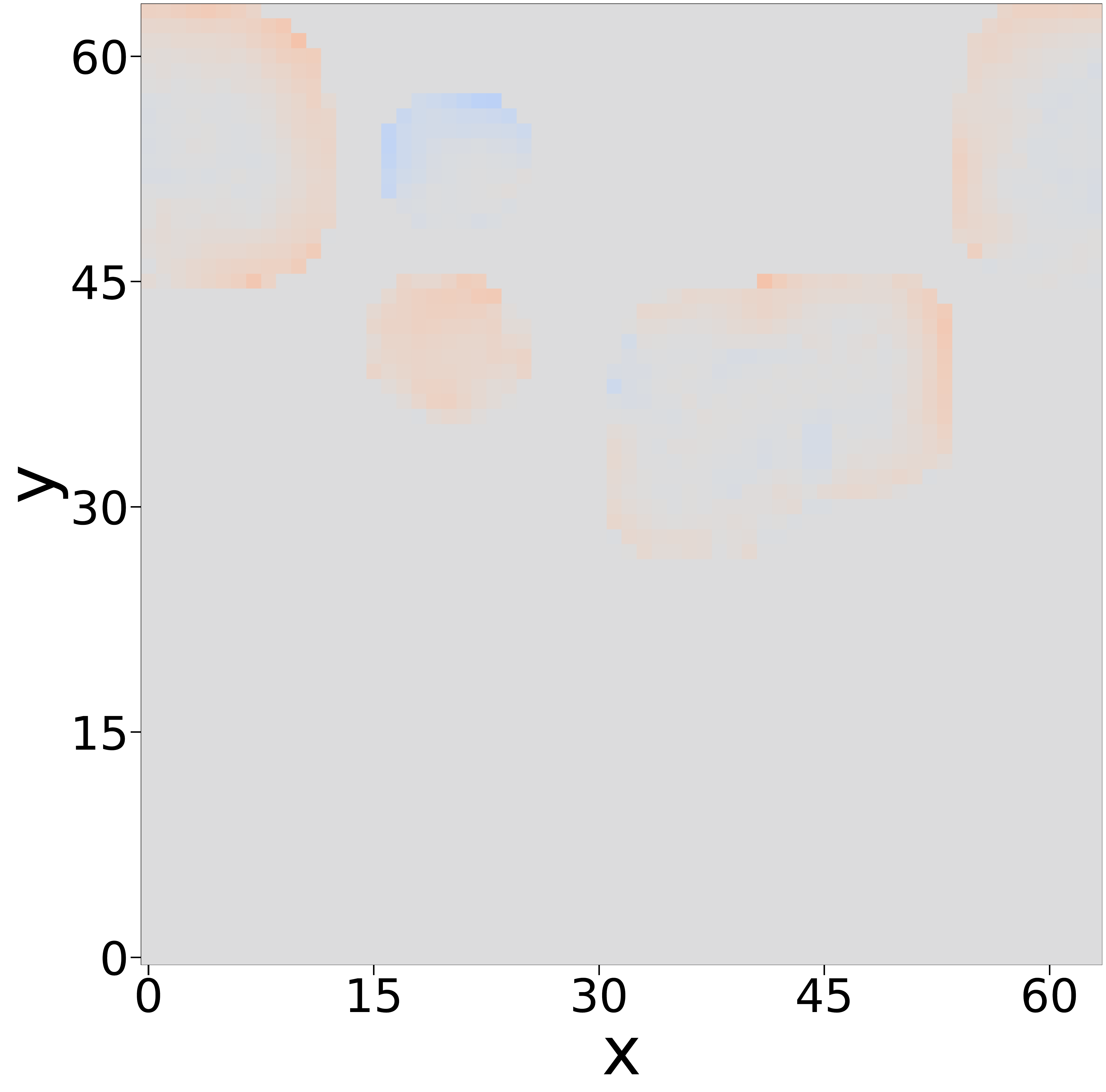}
    \end{subfigure}
    \begin{subfigure}[b]{0.0882\textwidth}
        \captionsetup{labelformat=empty}
        \includegraphics[width=\textwidth]{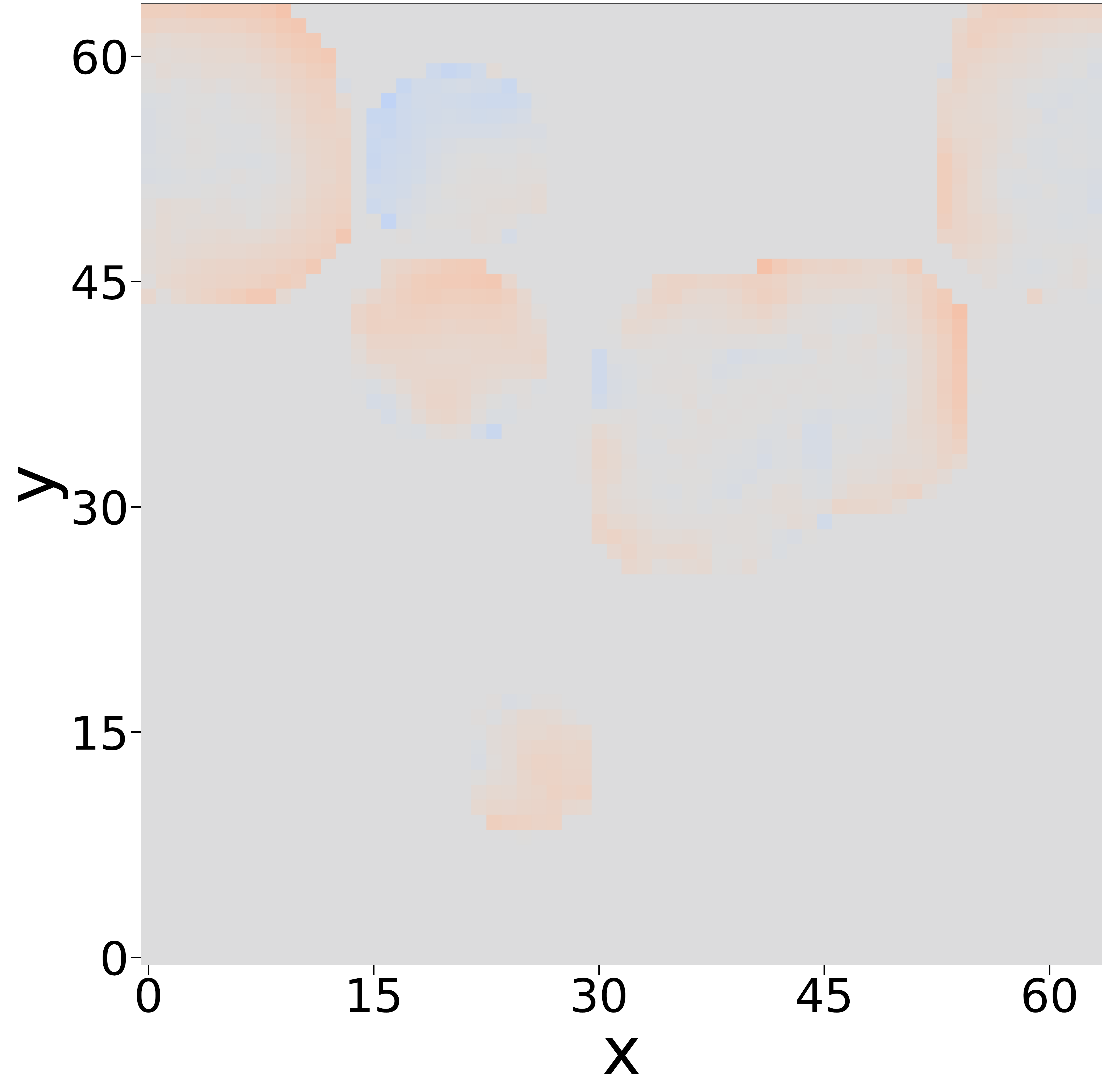}
    \end{subfigure}
    \begin{subfigure}[b]{0.0882\textwidth}
        \captionsetup{labelformat=empty}
        \includegraphics[width=\textwidth]{graphics/res_XY_006.pdf}
    \end{subfigure}
    \begin{subfigure}[b]{0.0882\textwidth}
        \captionsetup{labelformat=empty}
        \includegraphics[width=\textwidth]{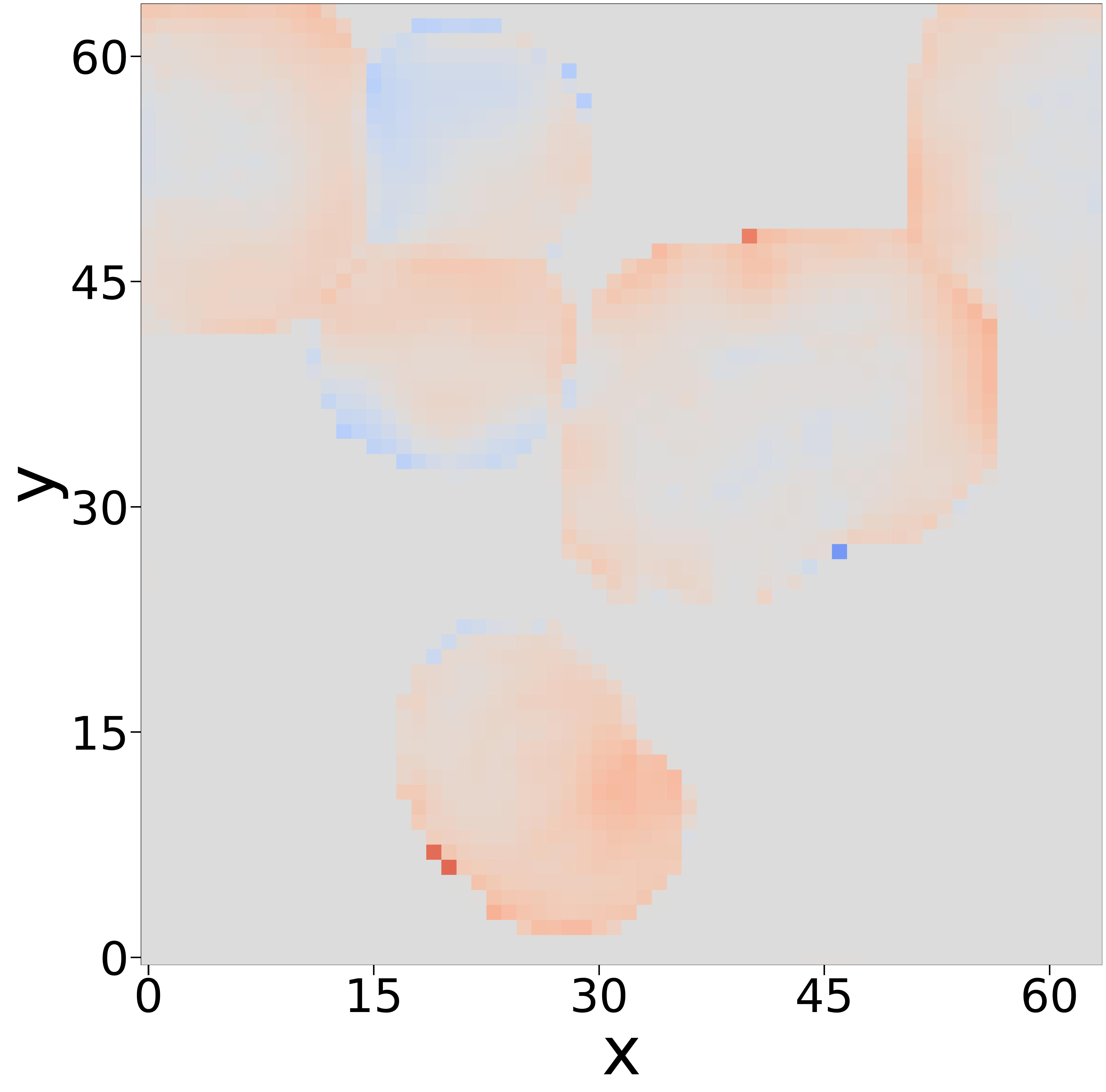}
    \end{subfigure}
    \begin{subfigure}[b]{0.0882\textwidth}
        \captionsetup{labelformat=empty}
        \includegraphics[width=\textwidth]{graphics/res_XY_008.pdf}
    \end{subfigure}

    \begin{subfigure}[b]{0.02346\textwidth}
        \captionsetup{labelformat=empty}
        \includegraphics[width=\textwidth]{graphics/colorbar_res.pdf}
    \end{subfigure}

    }

    \vspace{1cm}

    \begin{tikzpicture}
        \def\arrowLength{8}
        \def\arrowHeight{0.2}
        \def\arrowTip{0.4} 
    
        \draw[fill=black!30, draw=black] 
            (0, -\arrowHeight/2) -- 
            ({\arrowLength - \arrowTip}, -\arrowHeight/2) -- 
            ({\arrowLength - \arrowTip}, -\arrowHeight) -- 
            (\arrowLength, 0) -- 
            ({\arrowLength - \arrowTip}, \arrowHeight) -- 
            ({\arrowLength - \arrowTip}, \arrowHeight/2) -- 
            (0, \arrowHeight/2) -- 
            cycle;
        \node at (\arrowLength/2, 0.45) {\large \textbf{Temporal evolution}};
    \end{tikzpicture}

    \caption{All timesteps of the temporal evolution of radiative intensity at cross-section x=32 (first block), y=32 (second block), and z=32 (third block): Respectively, in each block the top row shows the preprocessed numerical reference, the middle row the model prediction, and the bottom row the residual.
    }
    \label{fig:3d_time_complete}
\end{figure}

Finally, in addition to  Figure~\ref{fig:3d_time} shown in the main part, Figure~\ref{fig:3d_time_complete} for the same test sample shows the complete temporal evolution of radiative intensity along all time steps for cross-sections at x=32, y=32, and z=32. This involves the preprocessed (log-transformed and min-max normalized) numerical reference, the predicted radiative intensity, as well as the corresponding residuals.
Similar to the previous figure, this demonstrates the recurrent model's ability to precisely predict the temporal evolution of radiative intensity across all spatial dimensions.


\clearpage

\clearpage

\newpage
\newpage

\end{document}